\documentclass[]{fairmeta}

\title{
    \resizebox{\linewidth}{!}{\ours: Open and Efficient Video Watermarking}
}

\author[*]{Pierre Fernandez}
\author[*]{Hady Elsahar}

\author{I. Zeki Yalniz}
\author{Alexandre Mourachko}

\affiliation[]{Meta FAIR}

\contribution[*]{Equal contribution}

\abstract{
The proliferation of AI-generated content and sophisticated video editing tools has made it both important and challenging to moderate digital platforms.
Video watermarking addresses these challenges by embedding imperceptible signals into videos, allowing for identification.
However, the rare open tools and methods often fall short on efficiency, robustness, and flexibility.
To reduce these gaps, this paper introduces \ours, a comprehensive framework for neural video watermarking
and a competitive open-sourced model.
Our approach jointly trains an embedder and an extractor, while ensuring the watermark robustness by applying transformations in-between, \eg, video codecs.
This training is multistage and includes image pre-training, hybrid post-training and extractor fine-tuning.
We also introduce temporal watermark propagation, a technique to convert any image watermarking model to an efficient video watermarking model without the need to watermark every high-resolution frame.
We present experimental results demonstrating the effectiveness of the approach in terms of speed, imperceptibility, and robustness. 
\ours\ achieves higher robustness compared to strong baselines especially under challenging distortions combining geometric transformations and video compression.
Additionally, we provide new insights such as the impact of video compression during training, and how to compare methods operating on different payloads.
Contributions in this work -- including the codebase, models, and a public demo -- are open-sourced under permissive licenses to foster further research and development in the field.
}

\correspondence{\email{pfz,hadyelsahar@meta.com}}

\metadata[Code]{\url{https://github.com/facebookresearch/videoseal}}
\metadata[Demo]{\url{https://aidemos.meta.com/videoseal}}

\usepackage{amsmath, amsthm, amsfonts, amssymb, bbm}

\def\eqref#1{equation~\ref{#1}}

\def\1{\bm{1}}

\DeclareMathAlphabet{\mathsfit}{\encodingdefault}{\sfdefault}{m}{sl}
\SetMathAlphabet{\mathsfit}{bold}{\encodingdefault}{\sfdefault}{bx}{n}

\newcommand{\R}{\mathbb{R}}

\usepackage{ifthen}  %
\newboolean{set_me_to_remove_todos}
\setboolean{set_me_to_remove_todos}{false} %

\ifthenelse{\boolean{set_me_to_remove_todos}}{
    \newcommand{\pierre}[1]{}
    \newcommand{\hady}[1]{}
    \newcommand{\todo}[1]{}
    \newcommand{\alex}[1]{}
}{
    \newcommand{\pierre}[1]{{\color{blue} [\textbf{Pierre}: #1]}}
    \newcommand{\hady}[1]{{\color{purple} [\textbf{Hady}: #1]}}
    \newcommand{\alex}[1]{{\color{olive} [\textbf{Alex}: #1]}}
    \newcommand{\todo}[1]{{\color{red} [\textbf{TODO}: #1]}}
}

\newcommand{\abs}[1]{\lvert{#1}\rvert}

\def\1{\mathbbm{1}}

\def\ind#1{\1_{#1}}

\newcommand{\eg}{e.g.}
\newcommand{\ie}{i.e.}

\newcommand{\NA}{-}

\newcommand{\nbits}{ n_{\text{bits}} }
\newcommand{\msgdim}{ d_{\text{msg}} }
\newcommand{\latentdim}{ d_{\text{z}} }

\newcommand{\msgdec}{\hat{m}}

\newcommand{\bitacc}{\text{bit accuracy}}
\newcommand{\img}{\mathbf{x}}
\newcommand{\msg}{m}
\newcommand{\w}{w}

\newcommand{\logpval}{\log_{10}(p)}
\newcommand{\pval}{p\textrm{-value}}

\definecolor{Gray}{gray}{0.95}

\newlength\savewidth
\definecolor{metablue}{HTML}{0064E0}

\newcommand{\ours}{Video\,Seal}
\usepackage{float} %

\begin{document}

\maketitle

\begin{figure}[h!]
    \centering
\includegraphics[width=0.98\textwidth, clip, trim={0 6.1in 0in 0}]{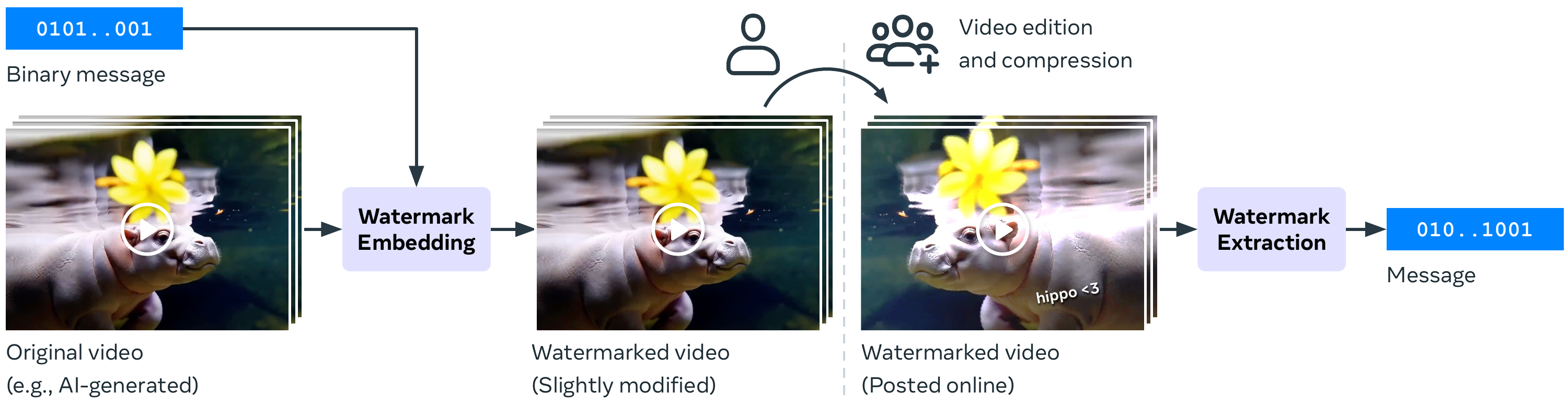}
    \caption{
        Overview of digital video watermarking. 
        A binary message is embedded into an original video (\eg, an AI-generated video), producing an imperceptible change in the pixels.
        This watermarked video may be compressed or edited when saved or shared online.
        Despite these transformations, the watermark extraction process should retrieve the embedded message. 
        The two primary challenges in this process are 
        (1) the speed of embedding and extraction, which must be computationally efficient to handle the large number of frames in a video, and 
        (2) robustness to common video codecs that often degrade the watermark to the point of being undetectable.
    }\label{fig:fig1}
\end{figure}

\section{Introduction}

Within digital media, video watermarking has always been a very active field of research. 
The film industry, including Hollywood studios and streaming websites, has been particularly invested in developing robust video watermarking techniques to fight against piracy. 
However, with the rapid advancement of technology, new challenges and applications have emerged. 
For instance, the development of generative AI models for images, like DALL·E~\citep{ramesh2022dalle2} or Stable Diffusion~\citep{rombach2022high}, and videos like Sora~\citep{brooks2024video} or MovieGen~\citep{polyak2024movie}, raises concerns about the spread of misinformation and general misuse of such technology.
Regulators~\citep{ChineseAIGovernance, EuropeanAIAct, USAIAnnouncement} are now pushing generative model providers to embed watermarks into the generated content to ease detection and attribution of said content.
Additionally, they also encourage hardware providers to watermark real data at the physical device level~\citep{ca_ab3211_2024}, which requires fast embedding and detection. 
All this requires the development of robust and efficient video watermarking techniques that can keep pace with the rapidly evolving landscape of digital media and AI-generated content.

It may seem logical to simply decompose videos into their constituent frames and leverage well-established image watermarking techniques to embed watermarks into each frame separately.
This approach, however, is hindered by two significant limitations. Firstly, the computational load of watermarking every frame is prohibitively high, particularly for high-resolution videos with high frame rates. 
Processing videos as clips (chunks of frames) for embedding or extraction can help with parallelization, but large clips exceed memory limits, while smaller clips introduce synchronization issues, complicating watermark extraction.
Secondly, the widespread use of video compression codecs such as AV1~\citep{AV12018} and H.264~\citep{richardson2010} along with the ease of access to free video editing software 
and social media filters poses a significant challenge to video watermarking. Whenever a video is downloaded, or shared on social media platforms, these codecs are often automatically applied, storing videos as keyframes, intraframes, and optical flows that enable frame decoding through interpolation. This process substantially reduces redundancy in videos, resulting in a strong decrease in the watermark signal. Consequently, even when computational efficiency is no longer a concern, image watermarking models may still struggle to remain effective in the face of these codecs and video editing tools, underscoring the need for video-specific watermarking solutions.

There have been some works on neural video watermarking addressing the aforementioned challenges. For instance, DVMark~\citep{luo2023dvmark} employ a compression network to simulate video compression, while VHNet~\citep{shen2023vhnet} leverages a similar trick for steganography applications. 
ItoV~\citep{ye2023itov} adapts architectures from image models to video watermarking by merging the temporal dimension of the videos with the channel dimension, enabling deep neural networks to treat videos as images.
It also employs a straight-through estimator to allow for gradient flow on compression augmentation\footnote{see Sec.~\ref{sec:related} for a comprehensive literature review.}. 
However, despite these efforts, several limitations persist. 
Notably, most existing models are restricted to low-resolution videos (\eg, 128$\times$128) or short clips (\eg, 64 frames), rendering them impractical for real-world applications. 

Most importantly, there is a lack of reproducibility in existing research on video watermarking. 
To our knowledge, none of the existing video watermarking models have been publicly released, hindering fair comparisons and reproducibility. 
This omission not only undermines the validity of the reported results but also stifles progress in the field.

In this paper, we introduce \ours, a state-of-the-art video watermarking model that sets a new standard for efficiency and robustness. 
By leveraging temporal watermark propagation, a novel technique that converts any image watermarking model into an efficient video watermarking model, \ours\ eliminates the need to watermark every frame in a video.
\ours\ also employs a multistage training that includes image pre-training, hybrid post-training, and extractor fine-tuning. 
This training regimen is supplemented with a range of differentiable augmentations, including the popular H.264 codec, allowing \ours\ to withstand common video transformations and high compression rates.

Due to the scarcity of reproducible baselines for video watermarking, we adapt state-of-the-art image watermarking models to create strong baselines using the temporal watermark propagation technique. 
This adaptation is a significant contribution of this paper, as it provides a much-needed foundation for evaluating and comparing video watermarking techniques. 
\ours\ outperforms strong image baselines, including MBRS~\citep{jia2021mbrs}, TrustMark~\citep{bui2023trustmark} and WAM~\citep{sander2024watermark}, in terms of robustness under basic geometric transformations such as cropping, small rotations, and perspective changes.
Although MBRS and TrustMark offer higher message capacities (256 and 100 bits, respectively), their design and training limitations make them vulnerable to degradation under these common transformations, which limits their real-world applicability. 

We also conduct ablation studies to investigate the impact of each component of the video inference and of our model training, including multistage training, differentiable compressions, and extractor fine-tuning. 
Our results show that extractor fine-tuning allows for extra gains in bit accuracy and increased robustness without compromising the quality measure through PSNR. Furthermore, we find that the most effective multistage training involves pre-training on images, followed by video training with the differentiable compression augmentation, which yields significant improvements in bit accuracy, particularly at higher compression rates.

To facilitate future research and development in video watermarking, we release several artifacts under a permissive license: model checkpoints, training and evaluation code, as well as a demo endpoint to test the models in action. 
We hope that the released models, along with the experiments, insights, and baselines, will serve the community and boost research in video watermarking.
As a summary, our contributions are:
\begin{itemize}[itemsep=1pt, topsep=0pt]
\item We introduce \ours, an open-source video watermarking model that sets a new standard for efficiency and robustness.
Using a novel temporal watermark propagation technique, \ours\ enables fast inference times by eliminating the need to individually watermark each frame in a video.
\item We release a comprehensive and easy-to-use codebase for training and evaluation, as well as a demo that enables effortless testing of our models.
\item We propose a multistage training that includes image pre-training, hybrid post-training, and extractor fine-tuning, supplemented with a range of differentiable augmentations, including multiple video codecs, allowing \ours\ to withstand common video transformations and high compression rates.
\item Through extensive experimentation, we gain valuable insights into the impact of video compression during training, the role of image and video data in training video watermarking models, and other key factors influencing model performance. These findings contribute to a deeper understanding of video watermarking and inform the development of more effective models.
\end{itemize}

\section{Method}\label{sec:method}

We adopt the embedder/extractor framework originally developed for image watermarking by~\citet{zhu2018hidden} and extend it to videos in a similar way as~\cite{ye2023itov}.
We focus on speed and practicality.
Our approach operates in 2D to ensure streamability, simplify extraction, and maintain flexibility. 
This design also enables a unified embedder-extractor mechanism for both images and videos. 
Our models are based on state-of-the-art architectures trained on longer schedules with a comprehensive set of augmentations that include video codecs.
They are effective at any resolution and for videos of any length.

\subsection{Embedder \& extractor architectures}

Our architectures are kept voluntarily small and efficient to facilitate inference and to possibly run on mobile devices.
The embedder is based on an efficient U-Net architecture 
with $16$M parameters in total, 
while the extractor is based on a vision transformer 
with $24$M parameters.
The number of bits $\nbits$ is set to $96$.

\subsubsection{Embedder}

The embedder takes as input a frame $x \in \R^{3 \times 256 \times 256}$ and a binary message $\msg \in \{0, 1\}^{\nbits}$, and outputs a watermarked frame $x_w \in \R^{3 \times 256 \times 256}$ that slightly differs from the original.
Its architecture is detailed in Tab.~\ref{tab:app-unet}.
It is based on a shrunk U-Net architecture~\citep{ronneberger2015u, bui2023trustmark}, with modifications taken from the ``Efficient U-Net'' of Imagen~\citep{saharia2022photorealistic}.
The message embedding happens in the bottleneck which operates at a lower resolution.
It is done through a binary message lookup table $\mathcal{T}$ structured to facilitate the embedding of binary messages into the latent representation of the frame, as previously presented by~\citet{san2024proactive, sander2024watermark}.

The U-Net consists of an encoder-decoder structure with skip connections, allowing to preserve the image information throughout the network, while doing most of the operations at a lower resolution. 
The encoder path begins with an initial residual block ``ResNetBlock'' that processes the input image of shape $3 \times 256 \times 256$ into a feature map of shape $\latentdim/8 \times 256 \times 256$. 
This is followed by a series of downsampling blocks ``DBlocks'', which progressively reduce the spatial dimensions and increase the feature depth, resulting in feature maps of shapes $\latentdim/4 \times 128 \times 128$, $\latentdim/2 \times 64 \times 64$, and $\latentdim \times 32 \times 32$. 
Each DBlock is made of a bilinear downsampling of factor $2$ followed by a ResNet block.
The message processor, described in the following paragraph, then integrates the message into the deepest feature map, producing a tensor of shape $(\latentdim+\msgdim) \times 32 \times 32$. 
The bottleneck consists of multiple residual blocks which merge the message and the image representations. 
The decoder path mirrors the encoder, using ``UBlocks'' to upsample the feature maps back to the original spatial dimensions, with shapes $\latentdim/2 \times 64 \times 64$, $\latentdim/4 \times 128 \times 128$, and $\latentdim/8 \times 256 \times 256$. 
In particular, we choose not to use deconvolution layers (ConvTranspose2D) because of the checkerboard patterns they introduce~\citep{odena2016deconvolution}, and use bilinear interpolation instead.
Each UBlock incorporates skip connections from the corresponding encoder layers, preserving information from the original image. 
The final output is produced by a Conv2D layer, resulting in an image of shape $C \times 256 \times 256$. 
Each ResNetBlock is composed of two convolutional layers with RMSNorm~\citep{zhang2019root} and SiLU~\citep{elfwing2018sigmoid} activation, and includes a linear residual connection implemented as a Conv2D layer with a kernel size of $1$.

The binary message lookup table $\mathcal{T}$ has a shape of $(\nbits, 2, \msgdim)$. 2 accounts for the binary values (0 or 1) each bit can take, and $\msgdim$ is the dimensionality of the embedding space.
For each bit $\msg_k$ in the message, indexed by $k \in \{1, \ldots, \nbits\}$, the table maps the bit to an embedding vector $\mathcal{T}(k, m_k, \cdot) \in \mathbb{R}^{\msgdim}$.  
These embeddings are averaged to produce a single vector of size $\msgdim$, capturing the overall message. 
This averaged vector is then repeated to match the spatial dimensions of the latent space $(\msgdim, 32, 32)$. 
The resulting message tensor is concatenated with the latent representation of the frame, yielding an activation tensor of shape $(\latentdim + \msgdim) \times 32 \times 32$. 
For our embedder, we use $\latentdim=128$ and $\msgdim=192$.

\begin{table}[t!]
    \centering
    \caption{
        High-level architecture of the encoder and decoder of the watermark embedder.
    }\label{tab:app-unet}
    {\small
    \begin{tabular}{c|c}
      Encoder path & Bottleneck and decoder path \\
      \toprule
      $x \in \mathbb{R}^{3 \times H \times W}$ & $(z, z_{\text{msg}}) \in \mathbb{R}^{(\latentdim+\msgdim) \times 32 \times 32}$ \\
      Interpolate, ResNetBlock $\to \mathbb{R}^{\latentdim/8 \times 256 \times 256}$ & Bottleneck Residual Blocks $\to \mathbb{R}^{\latentdim \times 32 \times 32}$ \\
    DBlock $\to \mathbb{R}^{\latentdim/4 \times 128 \times 128}$ & UBlock $\to \mathbb{R}^{\latentdim/2 \times 64 \times 64}$ \\
    DBlock $\to \mathbb{R}^{\latentdim/2 \times 64 \times 64}$ & UBlock $\to \mathbb{R}^{\latentdim/4 \times 128 \times 128}$ \\
    DBlock $\to \mathbb{R}^{\latentdim \times 32 \times 32}$ & UBlock $\to \mathbb{R}^{\latentdim/8 \times 256 \times 256}$ \\
    Message embedding, Repeat $\to \mathbb{R}^{\msgdim \times 32 \times 32}$ & Final Conv2D $\to \mathbb{R}^{3 \times 256 \times 256}$ \\
      \bottomrule
    \end{tabular}
    }
  \end{table}

\begin{table}[t!]
    \centering
    \caption{
        High-level architecture of the watermark extractor.
    }\label{tab:app-extractor}
    {\small
    \begin{tabular}{c|c}
      Image encoder (ViT) & Patch decoder (CNN) \\
      \toprule
      $x \in \mathbb{R}^{3 \times H \times W}$     & $z \in \mathbb{R}^{ d' \times 16 \times 16}$ \\
      Interpolation $\to \mathbb{R}^{3 \times 256 \times 256 } $ &  Residual Block $\to \mathbb{R}^{d' \times 16 \times 16} $  \\
      Patch Embed (Conv2D), Pos. Embed $\to \mathbb{R}^{d \times 16 \times 16}$ & Average pooling $\to \mathbb{R}^{d'} $  \\
      $L \times $ $\{$ Transformer Block $\}$ $\to \mathbb{R}^{d \times 16 \times 16} $  & Linear $\to \mathbb{R}^{\nbits} $ \\
      LayerNorm, GELU, Conv2D $\to \mathbb{R}^{d' \times 16 \times 16} $  &  Sigmoid (optional) $\to \mathbb{R}^{\nbits}$  \\
      \bottomrule
    \end{tabular}
    }
\end{table}

\subsubsection{Extractor}

The extractor takes as input a frame $x \in \R^{3 \times 256 \times 256}$ and outputs a ``soft'' message $\tilde{\msg} \in \R^{\nbits}$ which can be thresholded to recover a ``hard'' binary message $\msgdec \in \{0, 1\}^{\nbits}$ (soft because continuous, hard because binary).
Its architecture is detailed in Tab.~\ref{tab:app-extractor}.
It is based on a vision transformer (ViT)~\citep{dosovitskiy2020image} followed by a patch decoder and an average pooling layer that maps to a $\nbits$ dimensional vector.

The ViT consists of a series of attention blocks to process image patches into a high-dimensional feature space.
We use the 
ViT-Small architecture~\citep{touvron2021training} ($22$M parameters), with patch size $16$, with $d=d'=384$.
The patch embeddings are processed by a residual block, which is made of a Conv2D with kernel size of $3$ and stride of $1$, a LayerNorm, and a GELU activation, and with the number of channels equals to the one of input channels.
We obtain a latent map of shape $(d', 256, 256)$, which is average-pooled and mapped to $\nbits$-dimensional pixel features by a linear layer.
Finally, a Sigmoid layer scales the outputs to $[0,1]$ (this is in fact only done at inference time, since the training objective implicitly applies it in PyTorch).

\subsection{Video inference}\label{sec:video-inference}

\begin{figure}[b]
    \centering
    \includegraphics[width=0.99\linewidth, trim=0 7in 3in 0, clip]{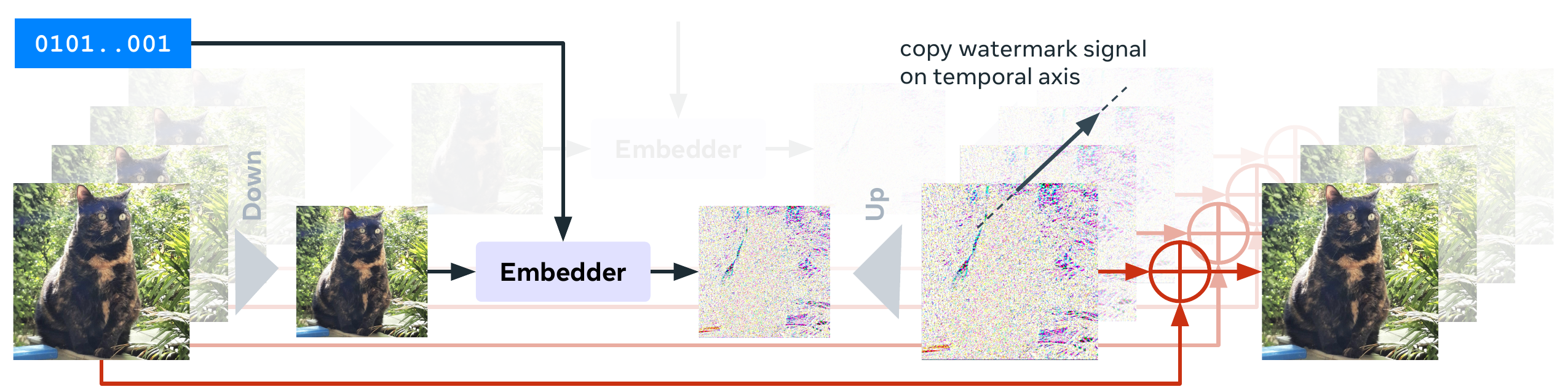}
    \caption{
        Illustration of the embedding process for video watermarking including temporal watermark propagation. 
        To minimize computational overhead, the embedder processes every $k$ frames of the video independently, producing a watermark signal that is copied along the temporal axis to the $k$ neighboring frames. 
        Additionally, the embedding is performed on a downscaled version of the video and the watermark is later upscaled to match the original resolution. This approach helps balance efficiency and robustness.
    }
    \label{fig:video_inference}
\end{figure}

Our embedder and extractor are designed to work on individual frames of fixed resolution ($256 \times 256$).
To operate in an efficient manner on videos, we use a few tricks to speed up the embedding and extraction processes.
Namely, we downscale frames to the fixed resolution, embed the watermark every $k$ frames, upscale the watermark to the original resolution, and propagate the watermark signal to the $k-1$ neighboring frames.
This is illustrated in Fig.~\ref{fig:video_inference} and detailed in the following paragraphs.

\subsubsection{High-resolution and scaling factor}\label{subsec:highres}
Our embedder and extractor are trained at a fixed resolution of $256 \times 256$.
To extend it to higher resolution, we use the same trick as presented by \citet{bui2023trustmark, sander2024watermark}.

Given a frame $x$ of size $H \times W$, we first downscale it to $256 \times 256$ using bilinear interpolation.
The embedder takes the downsampled frame and the message as input and produces the watermark distortion $\w$.
We then upscale $\w$ to the original resolution -- again using bilinear interpolation -- and add it to the original frame to obtain the watermarked frame:
\begin{equation}\label{eq:highres}
    x_w = x + \alpha_w \cdot \text{resize}(\w), \quad \w = \text{Embedder}(\text{resize}(x), m).
\end{equation}
$\alpha_w$ is called the scaling factor and controls the strength of the watermark.
It may be adjusted at inference time to trade quality for robustness.
In the following sections of the paper, we say that $\alpha_w$ is ``nominal'' at inference when it is set to the same value as during training. 

We proceed similarly for extraction and we resize all frames to $256 \times 256$ before giving them to the extractor.

\subsubsection{Temporal watermark propagation}\label{subsec:video-inference} 

Watermarking each frame of a video can be computationally costly. 
To mitigate this, a trick suggested in the codebase by~\citet{xian2024raw} is to watermark every $k$ frames instead. 
However, this approach complicates the extraction process.
Indeed, leaving some frames unwatermarked can compromise the robustness of the watermark under temporal editing and video compression algorithms. 
Even without any video edition, the extractor signal will be mixed with a lot of signal coming from unwatermarked frames, which will reduce the accuracy of the extraction.

In our approach, called temporal watermark propagation, the video is divided into segments of $k$ frames, the first frame of each segment is passed through the embedder to generate a watermark distortion which is then copied to the $k-1$ subsequent frames within the segment.
More rigorously, let $\img_i \in \mathbb{R}^{3 \times 256 \times 256}$ denote the $i^{th}$ frame of the video, and $\mathbf\w_{i} \in \mathbb{R}^{3 \times 256 \times 256}$ denote the watermark distortion of $\img_i$. 
Let $\msg \in \{0, 1\}^{\nbits}$ denote the binary message to be embedded. 
Temporal watermark propagation can be formulated as follows:
\begin{equation}\label{eq:video-embed}
    \mathbf\w_{i} = 
    \begin{cases} 
    \text{Embedder}(\img_i, \msg), & \text{if } i \bmod k = 0, \\
    \mathbf\w_{i-1}, & \text{otherwise.}
    \end{cases}
\end{equation}
In practice, if $k$ is set to $1$, the watermark is applied to every frame of the video, and temporal watermark propagation is equivalent to watermarking each frame independently.
When $k$ increases the efficiency of the embedding increases. 
At the same time, it introduces some noise in the extraction process because we approximate the watermark signal in the unwatermarked frames.
It may also introduce ``shadow'' artifacts if the video contains a lot of motion as the distortion often follows the image content.
In practice $k$ is set small enough for these two reasons, $k=4$ in this work.
Note that this operation is fully differentiable, allowing for the optimization of both imperceptibility and robustness during training.

\subsubsection{Extraction}\label{subsec:extraction}

The watermark extraction processes each frame $\img_i$ independently before aggregating the soft messages $\tilde{\mathbf{m}_i}$ over the entire video.
For aggregation, we simply average the soft messages across all frames, and threshold the average to obtain the hard message contained in the video $\msgdec$:
\begin{equation}\label{eq:video-extraction}
    \msgdec_k =
    \left\{
    \begin{array}{ll}
    1 & \text{if } \left( \frac{1}{T} \sum_{i=1}^T \tilde{\mathbf{m}_i}_{,k} \right) > 0 \\
    0 & \text{otherwise}
    \end{array}
    \right.
    \text{, with } \msgdec_k \text{ the bit at position } k.
\end{equation}
where $T$ is the number of frames on which the extraction is done.
In particular, one may choose to extract the watermark on certain frames -- for instance the first ones only or the whole video -- to increase robustness or to speed up the extraction process.
This aggregation is chosen for simplicity and speed, but more advanced aggregation methods could be used, as studied in Sec.~\ref{sec:ablation-video-inference}.

\begin{figure}[b!]
    \includegraphics[width=1.0\linewidth, trim=0 0.7in 0 0, clip]{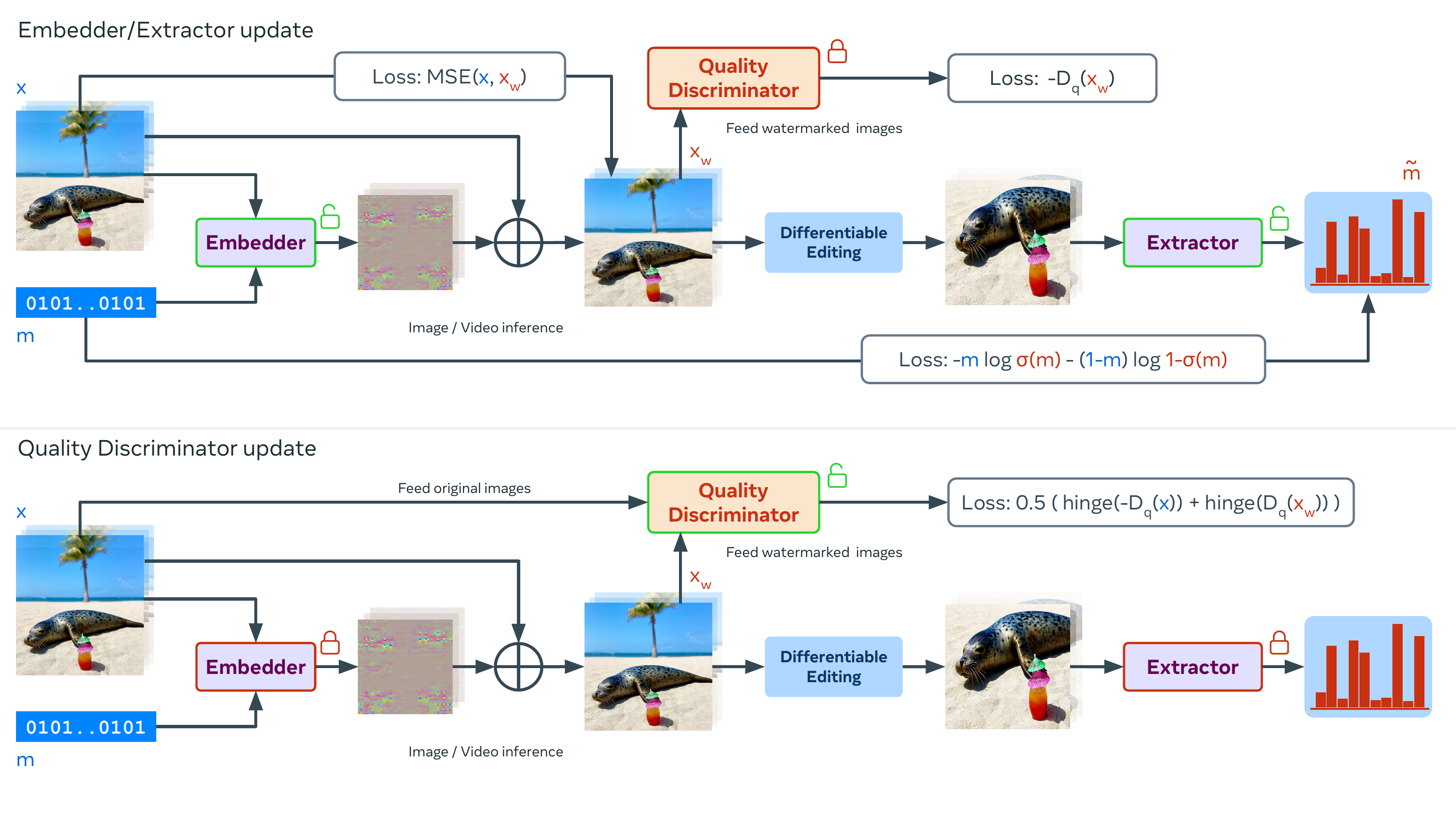}
    \caption{ 
    Detailed optimization pipeline of \ours.
    The embedder takes a batch of input images or a sequence of video frames $x$ and random binary messages $m$, and outputs a batch of watermarked images or frames $x_w$. 
    Differentiable transformations are randomly applied to $x_w$ to simulate real-world transmissions, such as crops, brightness changes, or video compression. 
    The extractor then processes these transformed images to estimate the embedded messages $\tilde{m}$.
    The watermark embedder and extractor are trained jointly to minimize two objectives: the message reconstruction loss and the mean squared error (MSE) between the original images $x$ and the watermarked images $x_w$. 
    Additionally, they are trained to maximize the adversarial loss against a quality discriminator.
    In a separate optimization step, the quality discriminator $D_q$ itself is trained to distinguish between the watermarked and original images, while keeping the embedder and extractor parameters fixed.
    }\label{fig:app-training}
\end{figure}

\subsection{Training pipeline}
In this section, we describe our method in detail, including image pre-training, mixed training with videos and images, and embedder freezing. 
Our training pipeline follows the traditional embedder/extractor approach~\citep{zhu2018hidden}, illustrated in Fig.~\ref{fig:app-training}. 
The embedder takes as input a batch of images or video frames and a binary message and produces watermarked images or frames. 
The extractor then attempts to recover the original message from them. 
We adopt a multistage training strategy that combines the benefits of image and video training. 
The following paragraphs detail these stages.

\subsubsection{Training objectives}

The training process involves minimizing a combination of perceptual losses and an extraction loss. 
The perceptual losses ensure that the watermark is imperceptible, while the extraction losses ensure that the extractor's output is close to the original message.
The optimizer minimizes the following objective function:
\begin{equation}\label{eq:objective}
    \mathcal{L} = \lambda_{\text{disc}} \mathcal{L}_{\text{disc}} + \lambda_{\text{i}} \mathcal{L}_{\text{i}} + \lambda_{\text{w}} \mathcal{L}_{\text{w}},
\end{equation}
where $\lambda_{\text{disc}}$, $\lambda_{\text{i}}$, and $\lambda_{\text{w}}$ are the weights of the discriminative loss, the image perceptual loss, and the watermark extraction loss, defined in the following paragraphs.

\paragraph{Extraction loss.}
The watermark extraction loss ensures that the extracted message $\tilde m$ is as close as possible to the original message $m$. 
We use the average binary cross-entropy (BCE) loss:
\begin{equation}
\mathcal{L}_{\text{w}} 
= -\frac{1}{\nbits} \sum_{k=1}^{\nbits} \text{BCE}(m_k, \tilde m_k), 
\textrm{ with } \text{BCE}(m_k, \tilde m_k) = m_k \log(\tilde m_k) + (1 - m_k) \log(1 - \tilde m_k).
\end{equation}

\paragraph{Perceptual losses.}
Additionally we compute the Mean Squared Error (MSE) between the original image $x$ and the watermarked image $x_w$, given by:
\begin{equation}
    \mathcal{L}_{\text{i}} = \frac{1}{N} \sum_{i=1}^{N} (x_i - x_{w,i})^2,
\end{equation}
where $N$ is the number of pixels in the image.
Although we experimented with more advanced perceptual models such as LPIPS~\citep{zhang2018perceptual} and Watson perceptual models~\citep{czolbe2020loss}, gains were not significant enough to justify their complexity.

\paragraph{Quality discriminator loss.}
We use an adversarial training with a patch-based discriminator $D$~\citep{isola2017image,rombach2022high}, and the update rules presented by~\citet{lim2017geometric}.

During the embedder-extractor update, we optimize the adversarial loss to ensure that the watermarked image \(x_w\) is indistinguishable from real images. This loss is given by:
\begin{equation*}
\mathcal{L}_{\text{disc}} = -D_q(x_w),
\end{equation*}
where \(D_q(\cdot)\) represents the quality discriminator's output in raw logits.

In a separate optimization step, the quality discriminator itself is being optimized through minimizing the Dual-Hinge Discriminator Loss, \(\mathcal{L}_{\text{disc'}}\), which enforces the quality discriminator to correctly classify both original images \(x\) and watermarked images \(x_w\) and therefore present a strong challenge to the embedder. This loss is defined as:
\begin{equation*}
\mathcal{L}_{\text{disc'}} = \frac{1}{2} \left( \max(0, 1 - D_q(x)) + \max(0, 1 + D_q(x_w)) \right),
\end{equation*}
where the hinge function \(\max(0, 1 - z)\) penalizes incorrect classifications.

\paragraph{Balancer.}

To balance the different loss components and to stabilize training, we compute adaptive weights as done in previous works~\citep{defossez2022high, rombach2022high}.
Our balancer is based on the norm of the gradients of each loss with respect to the last layer of the embedder (in the case of the U-Net this corresponds to the weights of the final convolution that maps to $\R^{3 \times 256 \times 256}$).
Each loss $\mathcal{L}_k$, where $k \in \{ \text{disc}, \text{i}, \text{w} \}$, is rescaled by the norm of its gradient:
\begin{equation}
    \tilde{\lambda}_k = \frac{\lambda_k}{\sum_{k'} \lambda_{k'}} \cdot \frac{
    R
    }{
    \| \nabla_{\theta}(\mathcal{L}_k) \| + \epsilon
    },
\end{equation}
where 
$R$ is a constant representing the total gradient norm -- set to $1$ as in EnCodec~\citep{defossez2022high} --,
$\theta$ represents the parameters of the last layer and $\epsilon$ is a small constant to avoid division by zero.
Eventually, we backpropagate $
    \tilde{\mathcal{L}} = \tilde \lambda_{\text{disc}} \mathcal{L}_{\text{disc}} + \tilde \lambda_{\text{i}} \mathcal{L}_{\text{i}} + \tilde \lambda_{\text{w}} \mathcal{L}_{\text{w}}
$ instead of $\mathcal{L}$ in Eq.~\ref{eq:objective}.

\subsubsection{Multistage training}\label{sec:multistage}

\paragraph{Image pre-training and hybrid post-training.}

Our approach employs a multistage training strategy, where we first pre-train our model on images and then continue training on a mix of images and videos using a scheduled approach. %
This approach has few benefits, first it allows us to leverage the faster training times of image-based models while still adapting to video-specific distortions. 
Second, as we show in Sec.~\ref{ablation:video_training}, this approach provides more stable training and yields significant improvements in terms of bit accuracy, and robustness to higher compression rates. 
During the pre-training phase, we train our model solely on images for a specified number of epochs. 
We then transition to a hybrid training phase, where we alternate between training on images and videos according to a predefined schedule, with a proportion of epochs for each modality fixed in advance.

\paragraph{Embedder freeze and extractor fine-tuning.} 
To further improve the robustness of our model, we employ a two-stage training process where we first train the entire model to convergence and then fine-tune the extractor while freezing the generator. This approach allows us to break free from the trade-off between imperceptibility and robustness, as we can focus solely on improving the extractor's performance without affecting the generated watermark. As we show in Sec.~\ref{ablation:extractor_finetuning} this allows us to gain extra points for robustness without compromising the watermark imperceptibility.

\subsubsection{Transformations}

\begin{table}[t!]
    \centering
    \caption{
        List of transformations used during training.
        A wide range of operations is covered, from valuemetric changes like brightness, contrast and video compressions, to more complex geometric transformations like perspective distortion.
    }\label{tab:transformations}
{ \small
  \begin{tabular}{ l *{3}{l}}
      \toprule
      Transformation  & Type & Parameter                      & Choice at training                      \\
      \midrule
      Identity & \NA   & \NA    & \NA                                                         \\
      Brightness      & Valuemetric & from torchvision        & Random between 0.5 and 2.0    \\
      Contrast        & Valuemetric & from torchvision        & Random between 0.5 and 2.0    \\
      Hue             & Valuemetric & from torchvision        & Random between -0.5 and 0.5   \\
      Saturation      & Valuemetric & from torchvision        & Random between 0.5 and 2.0    \\
      Gaussian blur   & Valuemetric & kernel size $k$         & Random odd between 3 and 17   \\
      Median filter   & Valuemetric & kernel size $k$         & Random odd between 3 and 7    \\
      JPEG            & Valuemetric & quality $Q$             & Random between 40 and 80      \\
      H.264            & Valuemetric & constant rate factor    & Random between 9 and 27  \\
      Horizontal flip & Geometric   & \NA                     & \NA                           \\
      Crop            & Geometric   & edge size ratio $r$     & Random between 0.7 and 1.0    \\
      Resize          & Geometric   & edge size ratio $r$     & Random between 0.7 and 1.5    \\
      Rotation        & Geometric   & angle $\theta$          & Random between -10 and 10     \\
      Perspective     & Geometric   & distortion scale $d$    & Random between 0.1 and 0.5    \\
      \bottomrule
  \end{tabular}
}
\end{table}

We use a comprehensive set of transformations during training, which are detailed in Tab.~\ref{tab:transformations}.
Most of them are applied at the frame level.
We categorize them into two main groups: 
\emph{valuemetric}, which change the pixel values;
\emph{geometric}, which modify the image's geometry -- and are unfortunately absent from many recent works on both image and video watermarking~\citep{jia2021mbrs, ma2022towards, ye2023itov}.

\paragraph{Frame transformations.}
We include crop, resize, rotation, perspective, brightness, contrast, hue, saturation, Gaussian blur, median filter, JPEG compression.
The strengths of these transformations are randomly sampled from a predefined range during training, and applied the same way to all images of the mini-batch.
For crop and resize, each new edge size is selected independently, which means that the aspect ratio can change (because the extractor resizes the image).
Moreover, an edge size ratio of $0.33$ means that the new area of the image is $0.33^2 \approx 10\%$ times the original area.
For brightness, contrast, saturation, and sharpness, the parameter is the default factor used in the PIL and Torchvision~\citep{marcel2010torchvision} libraries.
For JPEG, we use the \href{https://pillow.readthedocs.io/en/stable/releasenotes/8.0.0.html#jpeg-quality}{Pillow} library.

\paragraph{Video transformations.}
When applied on videos, frame transformations are applied to their whole content.
Additionally, we train and evaluate on common video codecs (\eg, H.264, H.265), as implemented in the \href{https://github.com/PyAV-Org/PyAV}{PyAV} wrapper around \href{https://ffmpeg.org/}{FFmpeg}.

\paragraph{About non-differentiable transformations.}
Non-differentiability or lack of backpropagatable implementations in PyTorch prevents us from backpropagating through video codecs.
This poses a challenge since the gradients of the objective function cannot be backpropagated through the compression back to the embedder.
One common solution is to use a differentiable approximation of the augmentation instead of the real one.
For instance, \citet{zhu2018hidden, zhang2023novel} use a differentiable JPEG compression and \citet{luo2023dvmark,shen2023vhnet} use a neural network trained to mimick video codec artifacts.
We choose a second option for its ease of implementation and its popularity~\citep{zhang2021asl, ye2023itov, sander2024watermark}.
It involves using a straight-through estimator that approximates the gradient of the non-differentiable operation with the identity function~\citep{bengio2013estimating}:
\begin{equation}
    x_{\textrm{aug}} = x_w + \mathrm{nograd} \left( T(x_w) - x_w \right),
\end{equation}
where $\mathrm{no grad}$ does not propagate gradients and $T$ is the non-differentiable transformation.

\section{Experimental Setup and Implementation Details}\label{sec:experiments}

\subsection{Metrics}\label{sec:metrics}

Watermarking is subject to a trade-off between imperceptibility, \ie, how much the watermarking degrades the video, and robustness, \ie, how much image or video transformations affect the recovery of the input binary message.
We therefore use two main categories of evaluation metrics.

\paragraph{Metrics for image and video quality.}
We evaluate the quality of the watermarked videos using per-pixel and perceptual metrics.
The PSNR (peak-signal-to-noise ratio) measures the difference between the original and watermarked videos in terms of mean squared error (MSE), and is defined as $\mathrm{PSNR} = 10 \log_{10} \left( 255^2 / \mathrm{MSE}\right)$.
SSIM~\citep{wang2004image} (structural similarity index measure) measures the similarity between the original and watermarked videos in terms of luminance, contrast, and structure.
LPIPS~\citep{zhang2018perceptual} is better at evaluating how humans perceive similarity. 
It is calculated by comparing the features extracted from the two frames using a pre-trained neural network.
On videos, SSIM and LPIPS metrics are computed frame-wise and averaged over the entire video. 

The above metrics do not take into account the temporal consistency of the video.
VMAF~\citep{vmaf} (video multi-method assessment fusion) is, on the contrary, designed specifically for video quality assessment. 
It uses a neural network to predict the subjective quality of a video based on various objective metrics such as PSNR, SSIM, and motion vectors.

\paragraph{Metrics for robustness of extraction.}
The main metric to evaluate the robustness of the watermarking in a multi-bit setting is the bit accuracy. 
Given an input message $m \in \{0, 1\}^{\nbits}$ and an output message $\msgdec$, the bit accuracy is defined as the percentage of bits that are correctly decoded, \ie, 
\begin{equation}
    \bitacc(m, \msgdec) = \frac{1}{\nbits} \sum_{k=1}^{\nbits} \ind{(m_k = \msgdec_k)}.
\end{equation}

The biggest issue with bit accuracy is that it is agnostic to the number of bits being hidden, and does not account for the total capacity of the watermarking method.
For instance, a method with average bit accuracy $p=0.9$ and $\nbits=128$ has a total capacity bigger than a method with bit accuracy $p=0.99$ and $\nbits=64$, in an information-theoretic sense~\citep{cover1999elements}\footnote{
    Assuming that bit errors are independent and distributed as Bernoulli variables with probability of failure $p$, the capacity is defined as $c(p) = 1 - \left( -p \log_2 p - (1-p) \log_2 p \right)$ and the total capacity as $C(p, \nbits) = \nbits * c(p)$.
    For $\nbits = 64, p = 0.99$, $C(p, \nbits) = 58.8$, and for $\nbits = 128, p = 0.9$, $C(p, \nbits) = 68.0$
    (see App.~\ref{app:payload} for more details).
}.

To account for this and to be able to properly compare methods, we thus introduce the $p$-value associated to a given bit accuracy.
Given the two messages and the observed $\bitacc(m, \msgdec)$, it is defined as the probability of observing, by chance, a bit accuracy greater than the one obtained.
Assuming that the $\nbits$ bits are independent and distributed as Bernoulli variables with probability of failure $0.5$, it is given by:
\begin{align}
    \pval(m, \msgdec) &= \mathbb{P}
    \big[
        \bitacc(m, m') \geq \bitacc(m, \msgdec) \mid m' \sim \mathcal{B}(0.5)^{\nbits}
    \big] \nonumber \\
    &= \sum_{k \geq \nbits \times \bitacc(m , \msgdec) }^{\nbits} \binom{\nbits}{k} 1/2^{\nbits}.
\end{align}
We report the log $p$-value, denoted as $\logpval$, which is more interpretable.
Given an observed bit accuracy $\bitacc(m, \msgdec)$, the $p$-value represents the confidence that the observed bit accuracy is due to chance\footnote{If the p-value is $10^{-6}$, it also means that we would need to set the threshold in such a way to have a false positive rate of $10^{-6}$ to flag the image or video as containing a watermark.}.
Another way to interpret the $p$-value is to link it to the false positive rate (FPR) when using the watermarking for a detection test. 
The FPR is the probability of falsely detecting a watermark when there is none.
In practice, if we want to have FPR$ < 10^{-6}$, we would need to set the threshold at $\logpval < -6$ to flag the image or video as containing a watermark.
We refer the interested reader to App.~\ref{app:payload} for more details.

\subsection{Datasets}\label{sec:datasets}

We use two main datasets for training and evaluation across video and image domains. 
For image training, we use the SA-1B dataset~\citep{kirillov2023segment}, from which we randomly select 500k images resized to $256 \times 256$.
For evaluation we use 1k random images at their original image resolution (with an average resolution of $1500 \times 2250$).
To keep a fair comparison with existing image watermarking models, we also evaluate on 1k images from the COCO validation dataset~\citep{lin2014microsoft}, which are of slightly lower resolution.

For video training we use the SA-V dataset~\citep{ravi2024sam2} which comprises 51k diverse videos captured across multiple countries, with resolutions ranging from 240p to 4K and an average duration of 14 seconds at 24 fps. 
We randomly select 1.3-second clips (32 frames) from each video resized to a resolution of $256 \times 256$, while evaluation uses the first 5 seconds at the original resolution, unless stated otherwise. 

\subsection{Training}

We first train the model on 16 GPUs, using the AdamW optimizer~\citep{loshchilov2017decoupled}. 
For the first $800$ epochs, we only use images from the SA-1b dataset (see Sec.~\ref{sec:datasets} for details on datasets), with a batch size of $16$ per GPU, with $1500$ steps per epoch. 
The learning rate is linearly increased from $10^{-6}$ to $10^{-5}$ over the first $50$ epochs, and then follows a cosine schedule~\citep{loshchilov2016sgdr} down to $10^{-7}$ until epoch 800.
For the last $300$ epochs, we also use the SA-V dataset, with $200$ steps per epoch.
We only forward one $32$-frame clip per GPU, randomly chosen at every step.
The learning rate is linearly increased from $10^{-7}$ to $10^{-6}$ over the first $10$ epochs, and then follows a cosine schedule down to $10^{-8}$ until the last epoch.
At epoch $250$, we freeze the embedder and only optimize the extractor (see Sec.~\ref{sec:multistage}).
The objectives are weighted with $\lambda_{\text{w}} = 1.0$, $\lambda_{\text{i}} = 0.5$, $\lambda_{\text{disc}} = 0.1$.

\subsection{Baselines}

In the absence of an established open-source video watermarking baselines, we leverage state-of-the-art image watermarking models as foundational baselines for video watermarking.
HiDDeN~\citep{zhu2018hidden} is one of the earliest deep-learning watermarking methods. 
We trained it on 48 bits with the same augmentations for fairer comparison. 
MBRS~\citep{jia2021mbrs} is based on the same architecture, but embeds 256-bit watermarks, with a training using mini-batches of real and simulated JPEG compression.
CIN~\citep{ma2022towards} combines invertible and non-invertible mechanisms to embed 30-bit watermarks. 
TrustMark~\citep{bui2023trustmark} also uses a U-Net architecture trained similarly to HiDDeN, embedding 100 bits.
Finally, WAM~\citep{sander2024watermark} embeds $32$ bits (in addition to one bit of detection which we do not use in this study), and offers robustness to splicing and inpainting. 
We use the original open weights for all baselines, except for HiDDeN, for which the authors do not provide weights.
\ours\ operates with $\nbits=96$, with $\alpha_w = 2.0$, unless stated otherwise.

\paragraph{Inference.}
All methods operate at resolution $256 \times 256$, except CIN, which is at $128\times 128$. 
We extend them to arbitrary resolutions as presented in Sec.~\ref{subsec:highres} (when the networks directly predict an image $x_w$ and not a watermark distortion $\w$, we retrieve it by doing $\w = x_w - x$).
By default, we use the original watermark strength $\alpha_w$ of Eq.~\ref{eq:highres} ($1.0$ for most methods), except in Sec.~\ref{sec:adaptable-inference} where we study the imperceptibility/robustness trade-off.
When evaluating the baselines on videos, we apply the image watermarking model with the same inference strategy as our models, \ie, get the watermark distortion every $k=4$ frames, and propagate the watermark to the other $3$ frames as described in Sec.~\ref{subsec:video-inference}.
For watermark extraction, we aggregate the soft bit predictions across the frames, and average the outputs to retrieve the global message (see Sec.~\ref{subsec:extraction}).

\subsection{Evaluated transformations}

\begin{figure}[b]
    \centering
    \centering
    \scriptsize
    \newcommand{\imwidth}{0.19\textwidth}
    \setlength{\tabcolsep}{0pt}
    \begin{tabular}{c@{\hskip 2pt}c@{\hskip 2pt}c@{\hskip 2pt}c@{\hskip 2pt}c}
    \toprule
    Original & H.264 & Crop & Brightness & Combined \\
    \midrule
    \includegraphics[width=\imwidth]{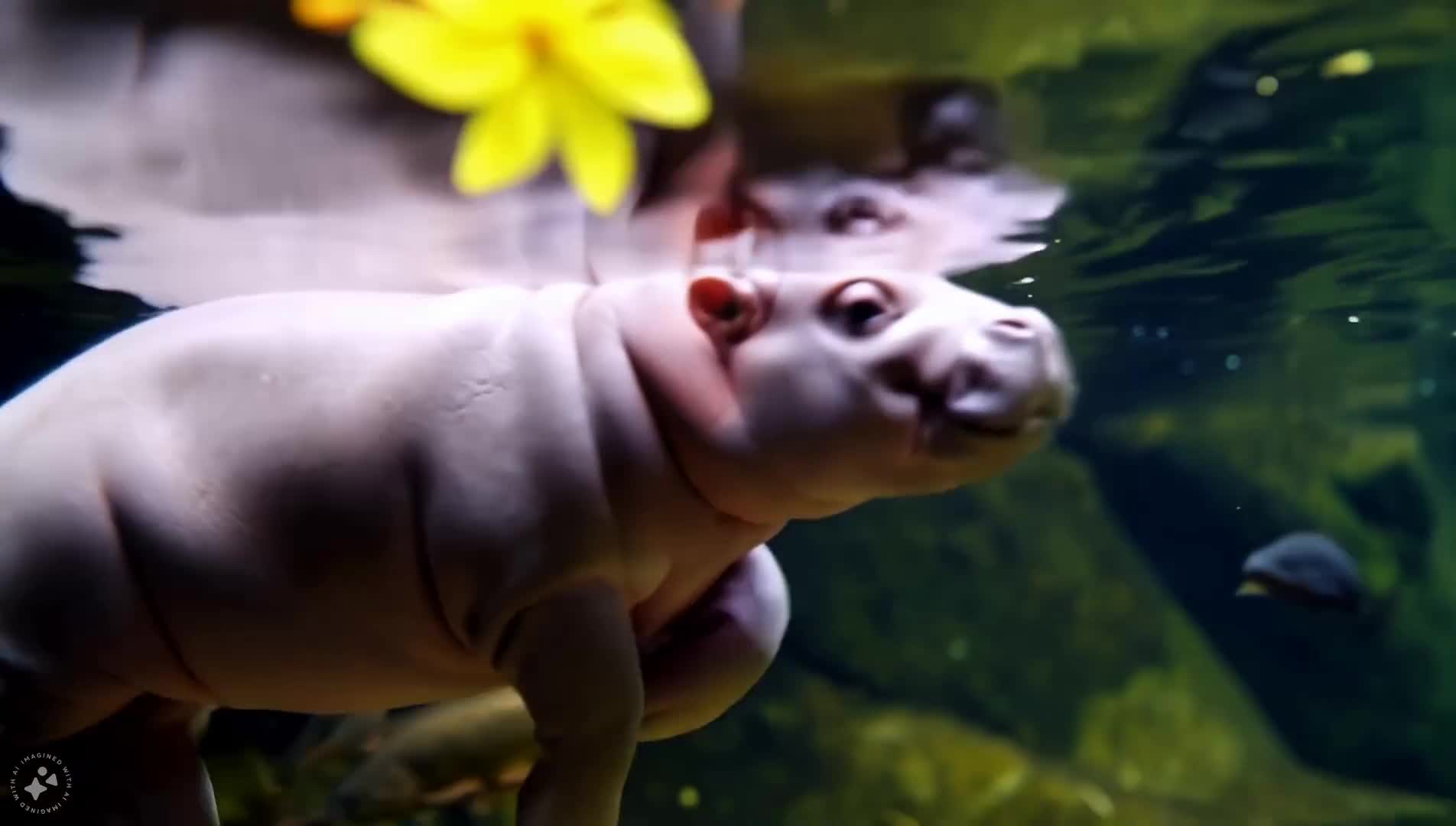} & 
    \includegraphics[width=\imwidth]{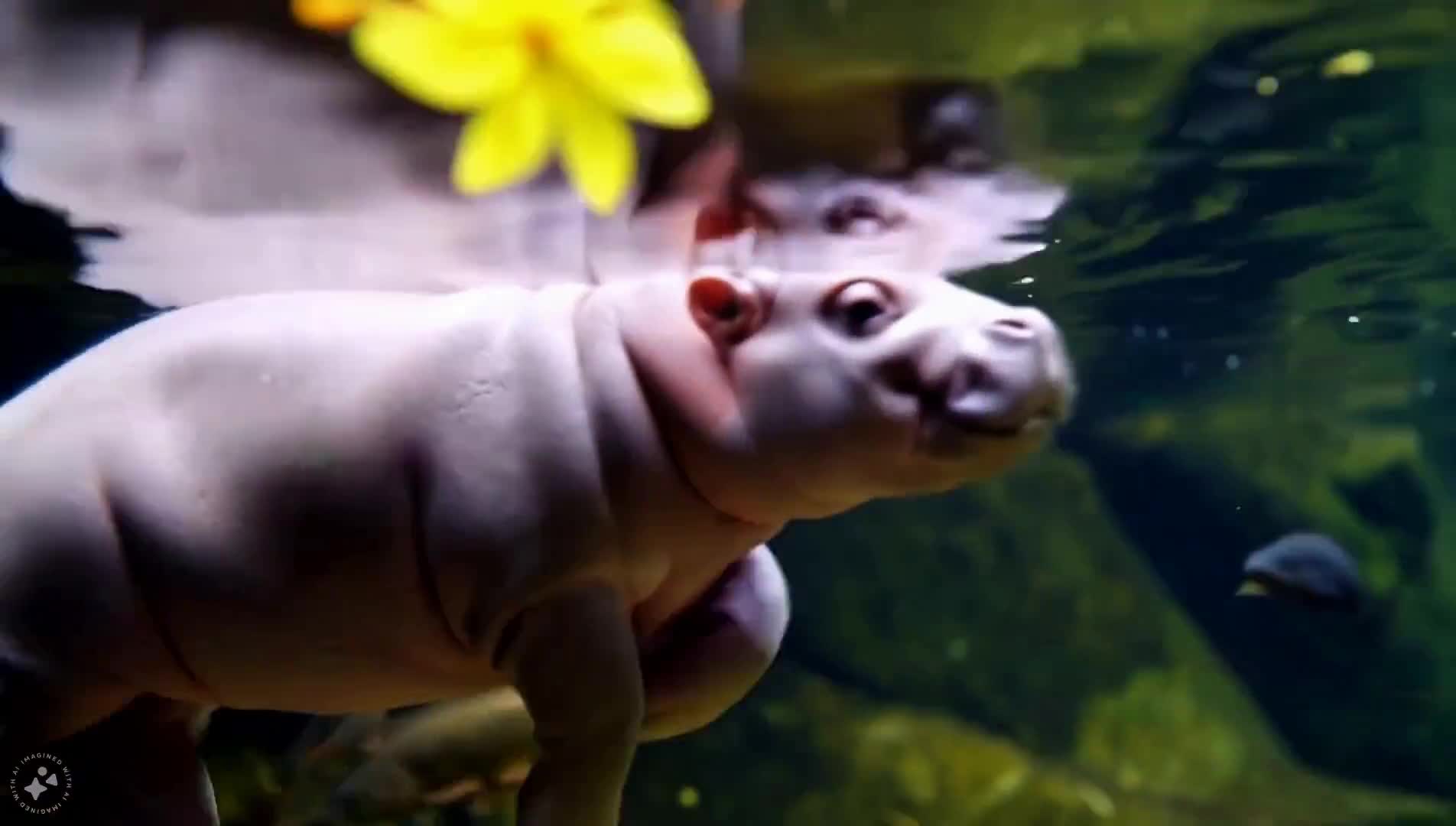} & 
    \includegraphics[width=\imwidth]{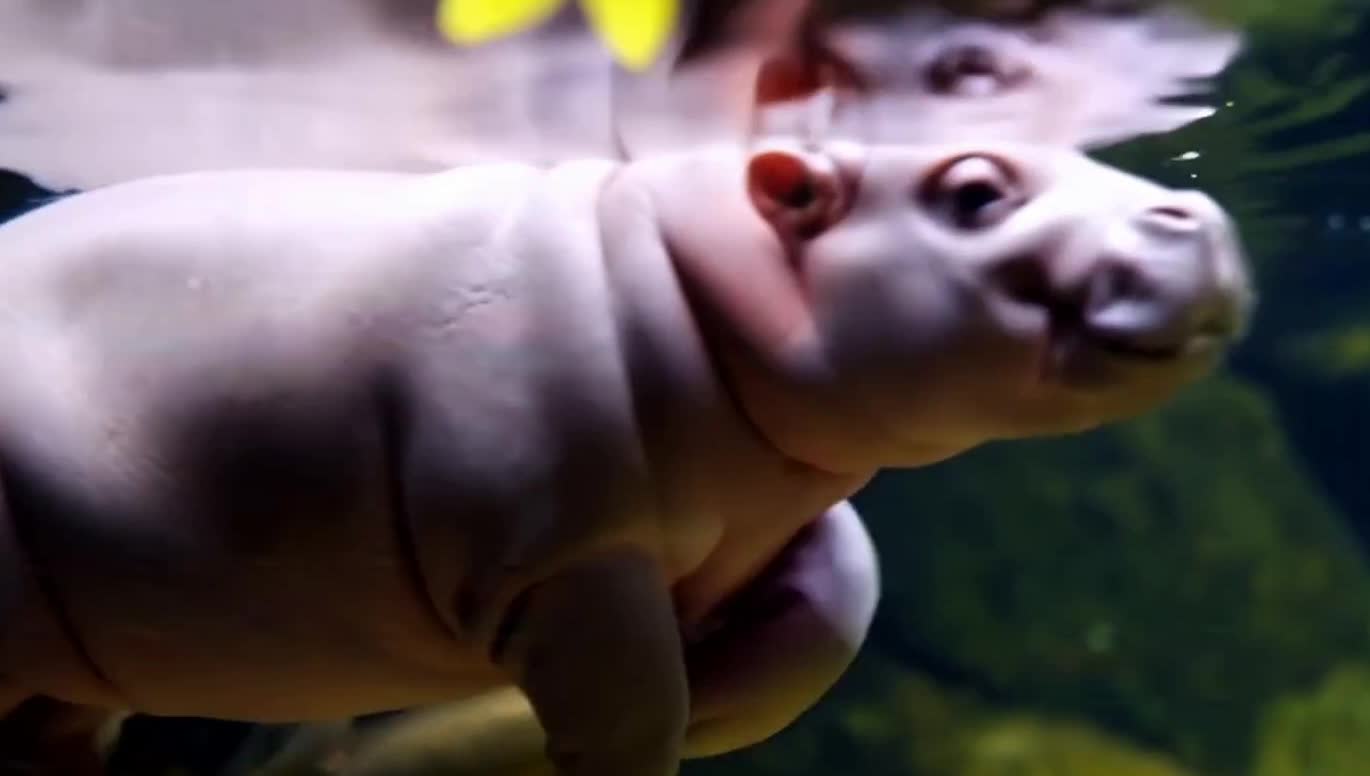} & 
    \includegraphics[width=\imwidth]{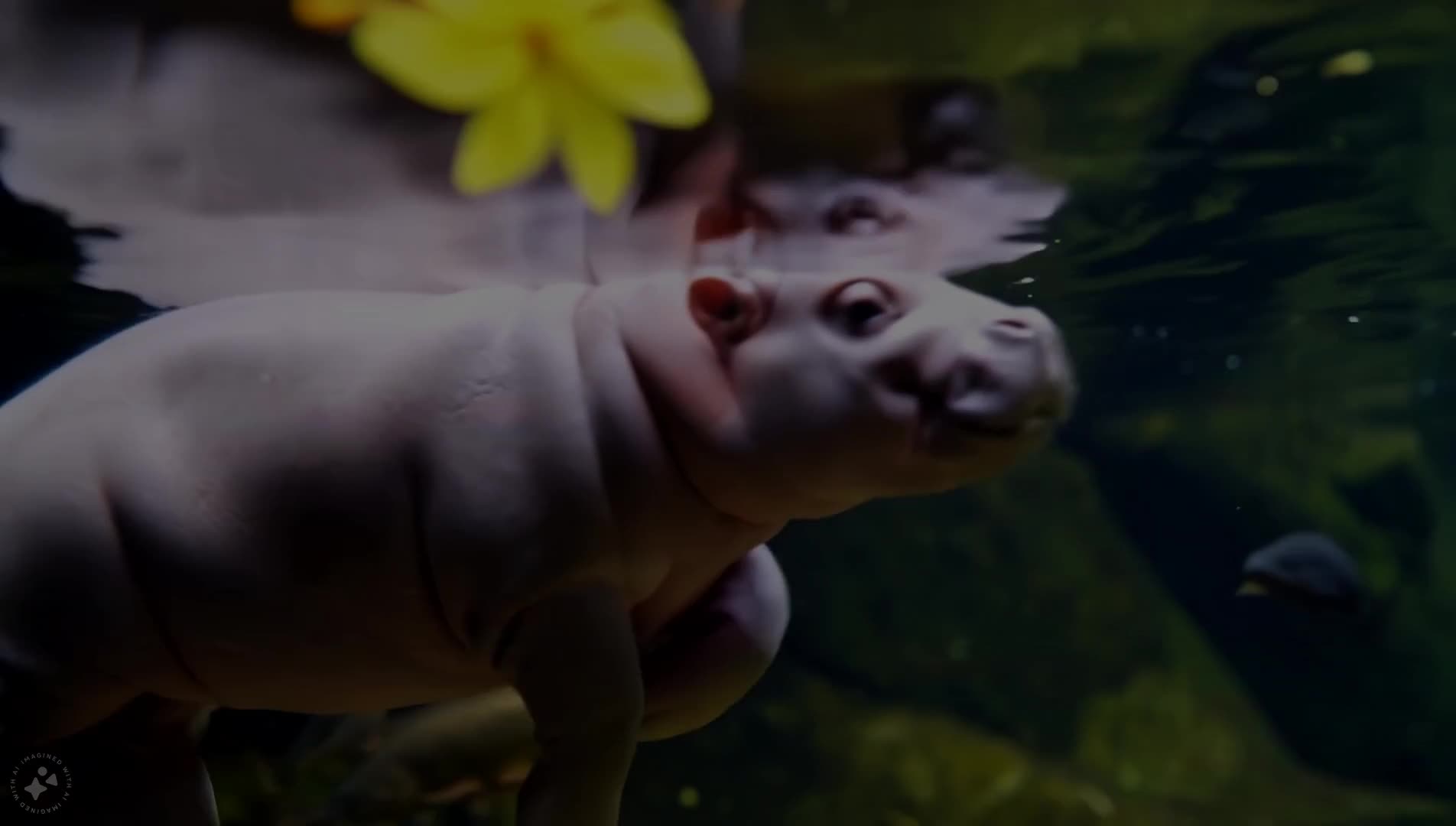} & 
    \includegraphics[width=\imwidth]{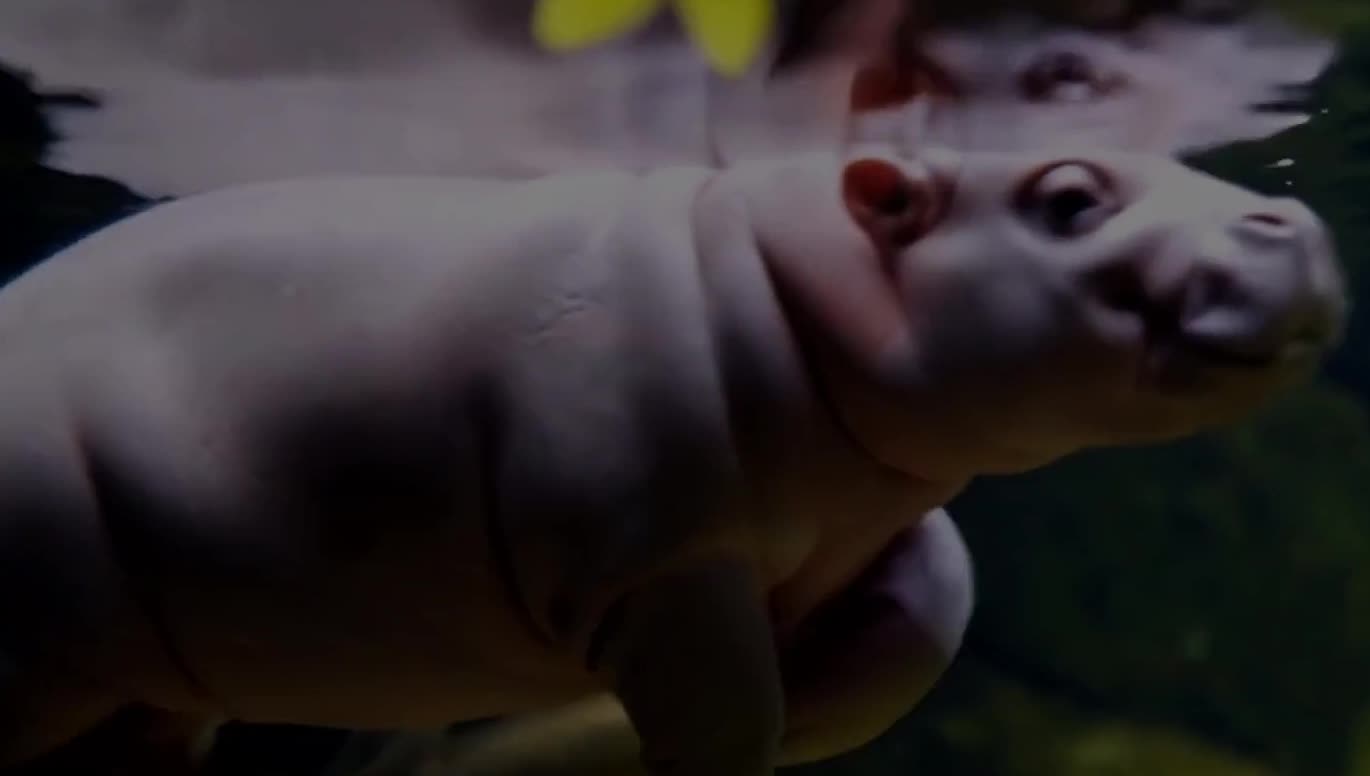} \\
    \end{tabular}
    \caption{
      Examples of transformations used for robustness evaluation, \eg, in Fig.~\ref{fig:trade-off} (we show the 20$\textsuperscript{th}$ frame of a 10-second video).
      We choose H.264 (CRF=30), crop (50\% area-wise), brightness with factor 0.5, as representative of video compression codecs, geometric transformations and valuemetric transformations, respectively.
    }\label{fig:transformations}
\end{figure}

We evaluate the robustness of our method to many transformations for different strengths.
For simplicity, we aggregate the transformations into five categories: no transformation, geometric, valuemetric, compression, and combined transformations.
For instance, geometric transformations include rotations, crops and perspective, while valuemetric transformations include brightness, contrast, and saturation changes, all with different ranges.
The combined augmentations are realistic augmentations applied sequentially, \eg, an H.264 compression followed by a crop and a brightness change.
We show some examples of these transformations in Fig.~\ref{fig:transformations}.
Full results and details on which transformations constitute each group are given in App.~\ref{app:transformations}.

\section{Results}

\subsection{Robustness}

We report in Tab.~\ref{tab:robustness-sa-1b-sa-v} the robustness of watermark extraction across many transformations and for various models, on the SA-1b and the SA-V datasets.
Full results, detailed by transformation type and strength, are available in App.~\ref{app:robustness}.
We also report results on the COCO dataset, to test the generalization of the models to unseen distributions.

We first observe that many of the image models are already strong baselines for video watermarking (as suggested by~\citet{ye2023itov}, although this seems to be even more the case when working on high resolution videos).
Most of them achieve high bit accuracy both for image and video transformations, even against video codecs.
It must be noted that MBRS and CIN were trained with augmentations that do not change the geometry of the image\footnote{In particular the crop considered by MBRS and CIN is simply a black mask applied on the image, keeping the original pixels at their exact location.}.
Therefore, their robustness against valuemetric transformations and video codecs is particularly strong, but at the same time their robustness on geometric transformations is particularly weak.

We also observe that \ours\ is overall the most robust model when considering transformations, especially against combinations of geometric transformations and video codecs.
For instance, under a combined transformation of H.264 compression (CRF=30), brightness adjustment (strength 0.5), and cropping (50\% area-wise), \ours\ achieves $\logpval  = -6.1$ on average.
This means that if one were to use \ours\ in a detection scenario, most of the transformed watermarked video would be detected as watermarked, even at low false positive rates ($<10^{-6}$).

\begin{table}[t!]
    \caption{
      Evaluation of the watermark robustness for various models.
      Models hide different number of bits, therefore, in addition to the bit accuracy  we also report $\logpval$, which takes into accounts $\nbits$ (and reflects that a bit accuracy of $1.0$ for WAM which hides $32$ bits is different than \ours\ which hides $96$ bits).
      Embedding is done either on the SA-1b (image) or the SA-V (video) dataset at their original resolution with the downscaling/upscaling inference trick presented in Sec.~\ref{subsec:highres}. 
      For video, the embedding is done with $k=4$ (see Eq.~\ref{eq:video-embed}) and extraction is performed on the first $3$s (see Eq.~\ref{eq:video-extraction}).
      The results are averaged under transformations of different types (more details in App.~\ref{app:robustness}).
    }
    \label{tab:robustness-sa-1b-sa-v}
    \resizebox{\linewidth}{!}{
        
\begin{tabular}{rr *{2}{>{\centering\arraybackslash}p{1.3cm}} *{2}{>{\centering\arraybackslash}p{1.3cm}} *{2}{>{\centering\arraybackslash}p{1.3cm}} *{2}{>{\centering\arraybackslash}p{1.3cm}} *{2}{>{\centering\arraybackslash}p{1.3cm}} *{2}{>{\centering\arraybackslash}p{1.3cm}}}
\toprule
& & \multicolumn{2}{c}{\shortstack{HiDDeN}} & \multicolumn{2}{c}{\shortstack{MBRS}} & \multicolumn{2}{c}{\shortstack{CIN}} & \multicolumn{2}{c}{\shortstack{TrustMark}} & \multicolumn{2}{c}{\shortstack{WAM}} & \multicolumn{2}{c}{\shortstack{Video Seal (ours)}} \\
\cmidrule(lr){3-4} \cmidrule(lr){5-6} \cmidrule(lr){7-8} \cmidrule(lr){9-10} \cmidrule(lr){11-12} \cmidrule(lr){13-14}
& & \multicolumn{2}{p{1cm}}{Bit~acc.~\footnotesize{($\uparrow$)}/~$\logpval$\footnotesize{($\downarrow$)}} & \multicolumn{2}{p{1cm}}{Bit~acc.~\footnotesize{($\uparrow$)}/~$\logpval$\footnotesize{($\downarrow$)}} & \multicolumn{2}{p{1cm}}{Bit~acc.~\footnotesize{($\uparrow$)}/~$\logpval$\footnotesize{($\downarrow$)}} & \multicolumn{2}{p{1cm}}{Bit~acc.~\footnotesize{($\uparrow$)}/~$\logpval$\footnotesize{($\downarrow$)}} & \multicolumn{2}{p{1cm}}{Bit~acc.~\footnotesize{($\uparrow$)}/~$\logpval$\footnotesize{($\downarrow$)}} & \multicolumn{2}{p{1cm}}{Bit~acc.~\footnotesize{($\uparrow$)}/~$\logpval$\footnotesize{($\downarrow$)}} \\
\midrule
\multirow{5}{*}{\rotatebox[origin=c]{90}{SA-1b}} & Identity & 1.00 & -14.2 & 0.99 & -70.6 & 1.00 & -9.0 & 1.00 & -29.9 & 1.00 & -9.6 & 0.99 & -27.3 \\ 
& Valuemetric & 0.88 & -10.8 & 0.95 & -59.8 & 0.91 & -8.1 & 0.98 & -27.4 & 0.95 & -8.7 & 0.93 & -23.4 \\ 
& Geometric & 0.76 & -5.5 & 0.52 & -3.3 & 0.52 & -0.7 & 0.65 & -8.5 & 0.81 & -5.5 & 0.83 & -16.4 \\ 
& Compression & 1.00 & -14.2 & 0.99 & -69.9 & 1.00 & -9.0 & 1.00 & -29.7 & 1.00 & -9.6 & 0.99 & -27.1 \\ 
& Combined & 0.70 & -2.6 & 0.50 & -0.4 & 0.50 & -0.4 & 0.53 & -0.8 & 0.86 & -5.9 & 0.91 & -18.4 \\ 
 \midrule
\multirow{5}{*}{\rotatebox[origin=c]{90}{SA-V}} & Identity & 0.99 & -14.0 & 1.00 & -77.1 & 1.00 & -9.0 & 1.00 & -30.1 & 1.00 & -9.6 & 0.99 & -26.8 \\ 
& Valuemetric & 0.88 & -9.1 & 0.89 & -54.3 & 0.93 & -7.5 & 0.93 & -24.7 & 0.92 & -7.7 & 0.90 & -19.9 \\ 
& Geometric & 0.68 & -2.9 & 0.50 & -0.4 & 0.50 & -0.4 & 0.60 & -5.5 & 0.81 & -5.5 & 0.85 & -17.0 \\ 
& Compression & 0.83 & -7.2 & 0.79 & -34.6 & 0.90 & -6.7 & 0.87 & -20.0 & 0.86 & -6.1 & 0.85 & -15.7 \\ 
& Combined & 0.61 & -1.3 & 0.50 & -0.4 & 0.49 & -0.4 & 0.51 & -0.5 & 0.55 & -0.8 & 0.73 & -8.1 \\ 
\bottomrule
\end{tabular}

    }
\end{table}

\subsection{Imperceptibility}

\begin{figure*}[b!]
    \centering
    \scriptsize
    \newcommand{\imwidth}{0.14\textwidth}
    \setlength{\tabcolsep}{0pt}
    \begin{tabular}{c@{\hskip 2pt} c@{\hskip 1pt}c@{\hskip 1pt}c@{\hskip 1pt}c@{\hskip 1pt}c@{\hskip 1pt}c}
    \toprule
    Original & \shortstack{HiDDeN} & \shortstack{MBRS} & \shortstack{CIN} & \shortstack{TrustMark} & \shortstack{WAM} & \shortstack{\ours} \\
    \midrule
    \includegraphics[width=\imwidth]{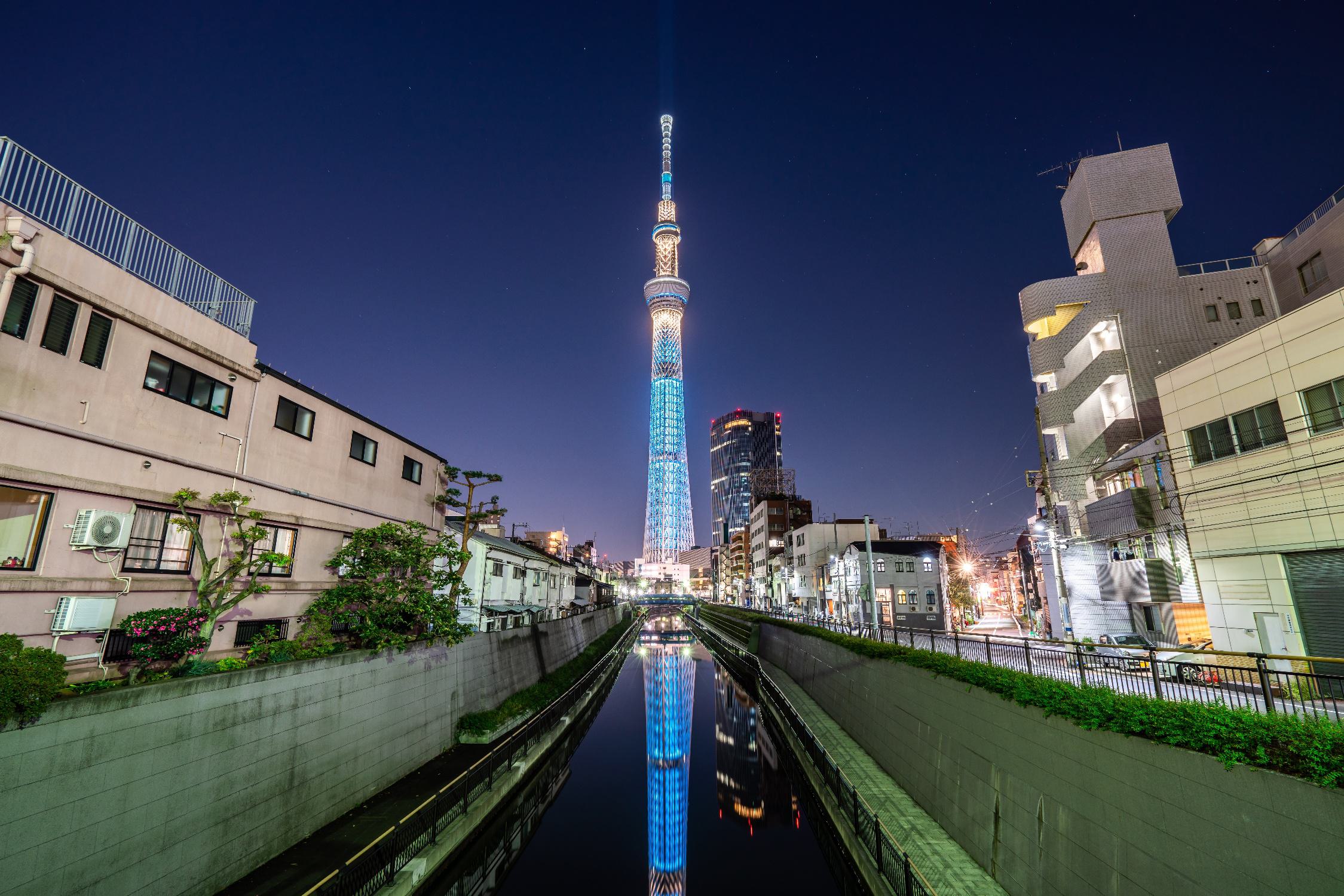} & 
    \includegraphics[width=\imwidth]{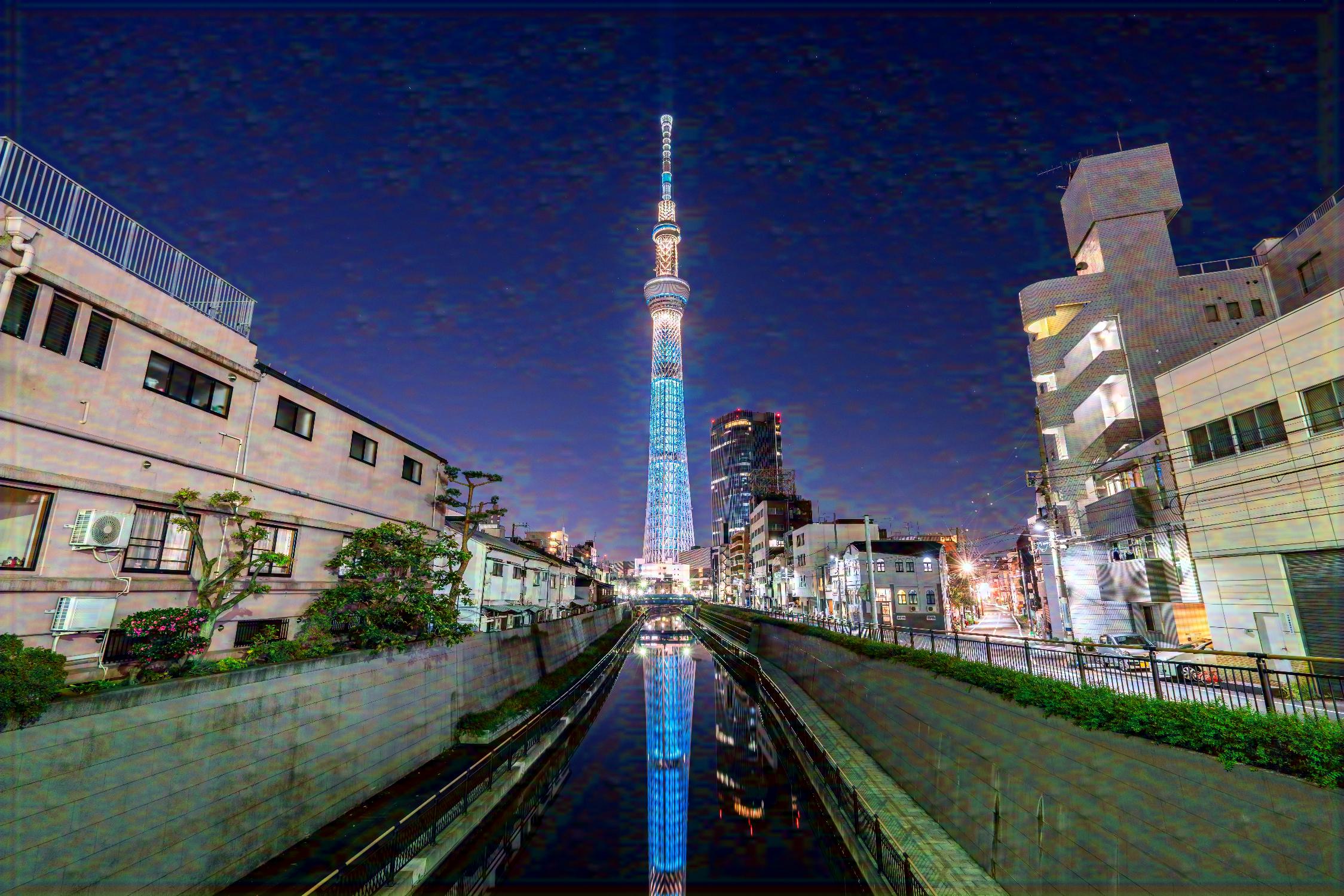} & 
    \includegraphics[width=\imwidth]{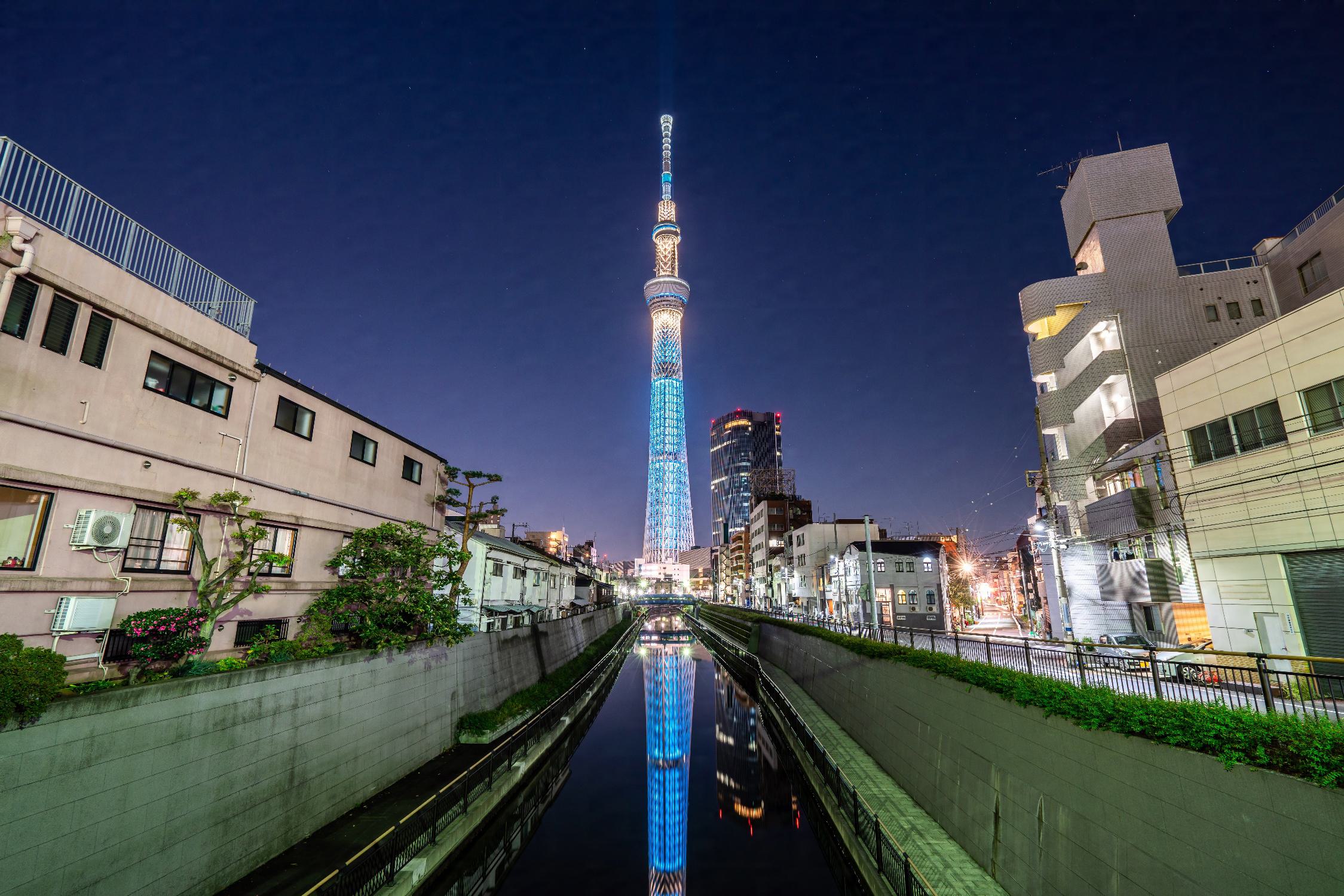} & 
    \includegraphics[width=\imwidth]{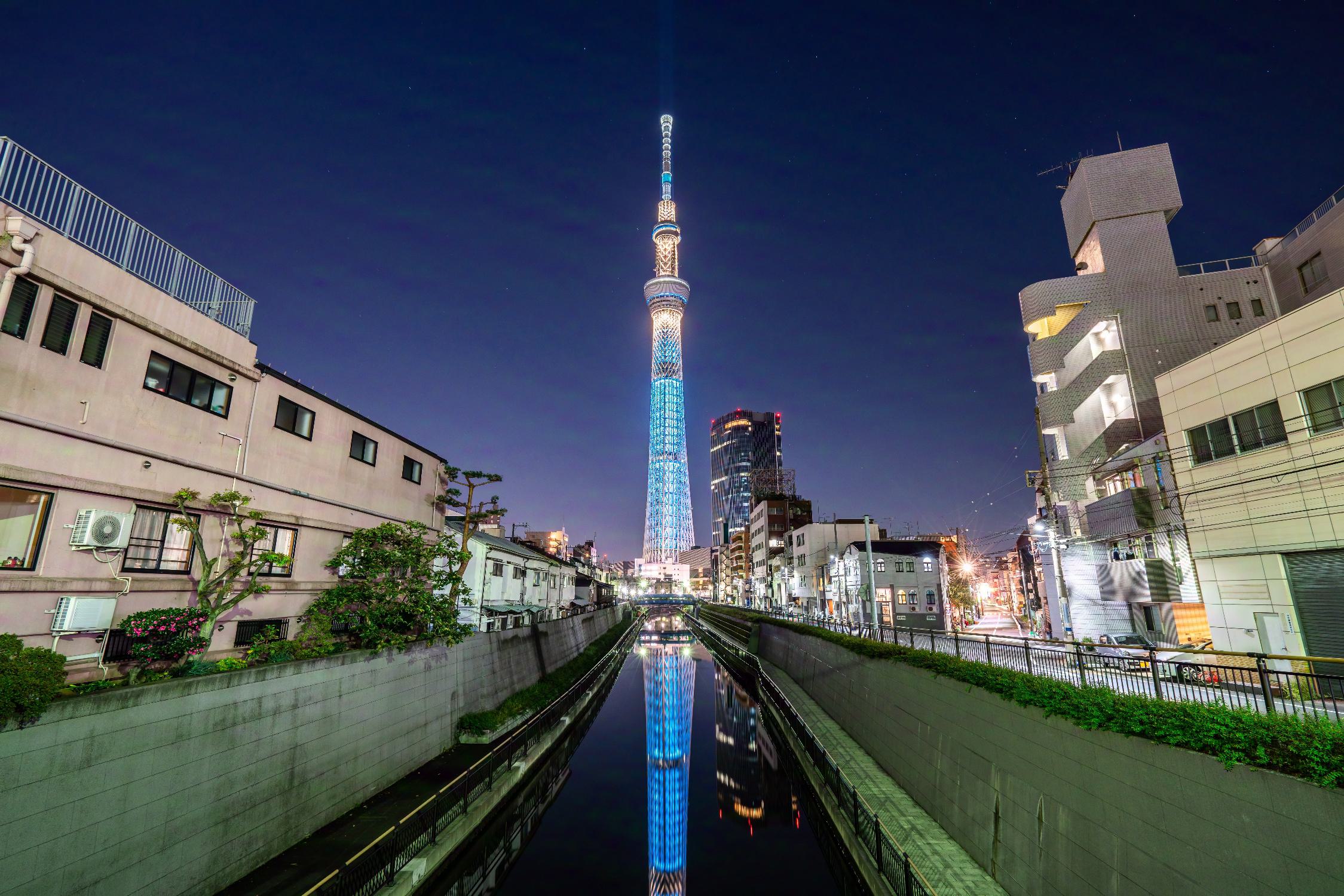} & 
    \includegraphics[width=\imwidth]{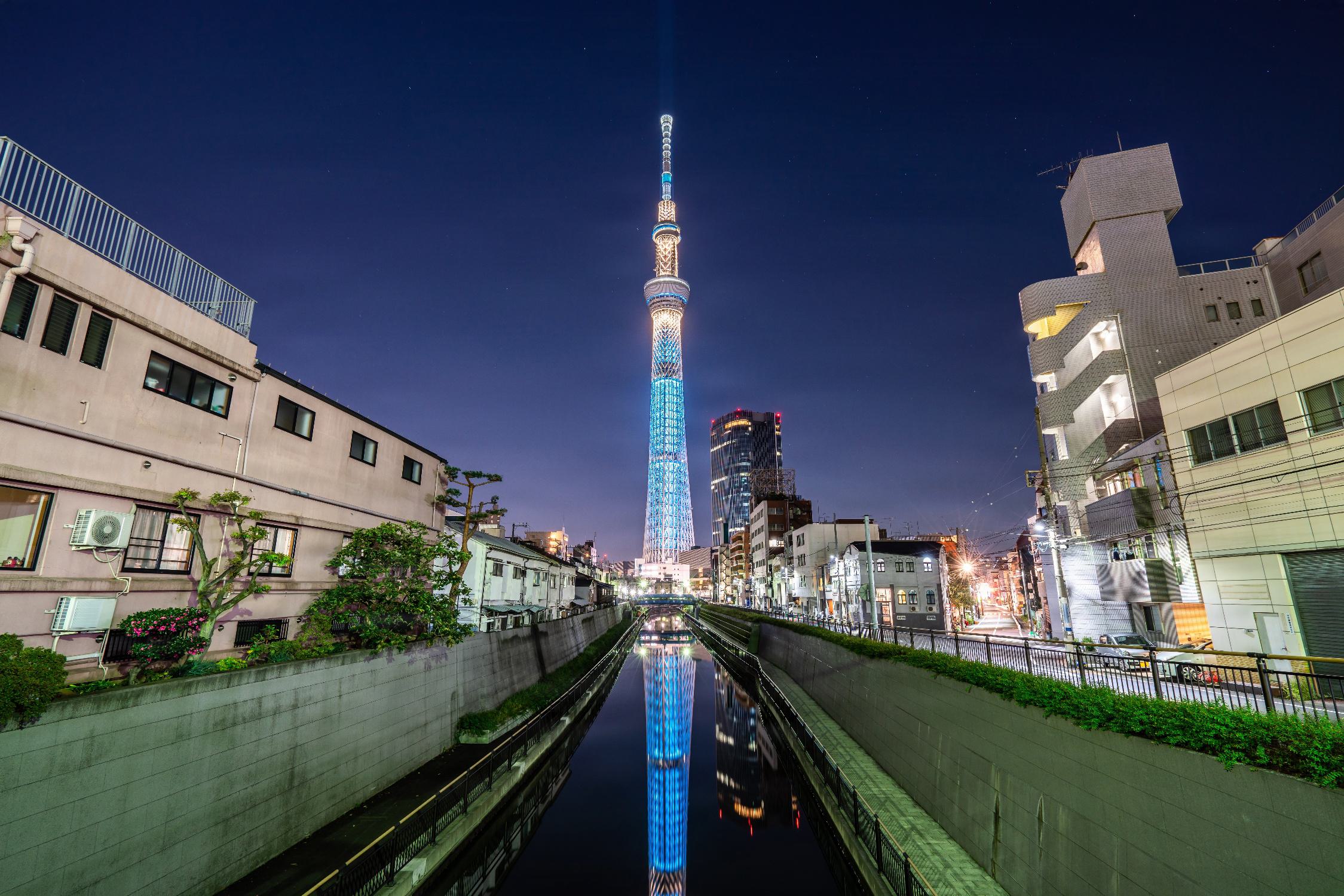} & 
    \includegraphics[width=\imwidth]{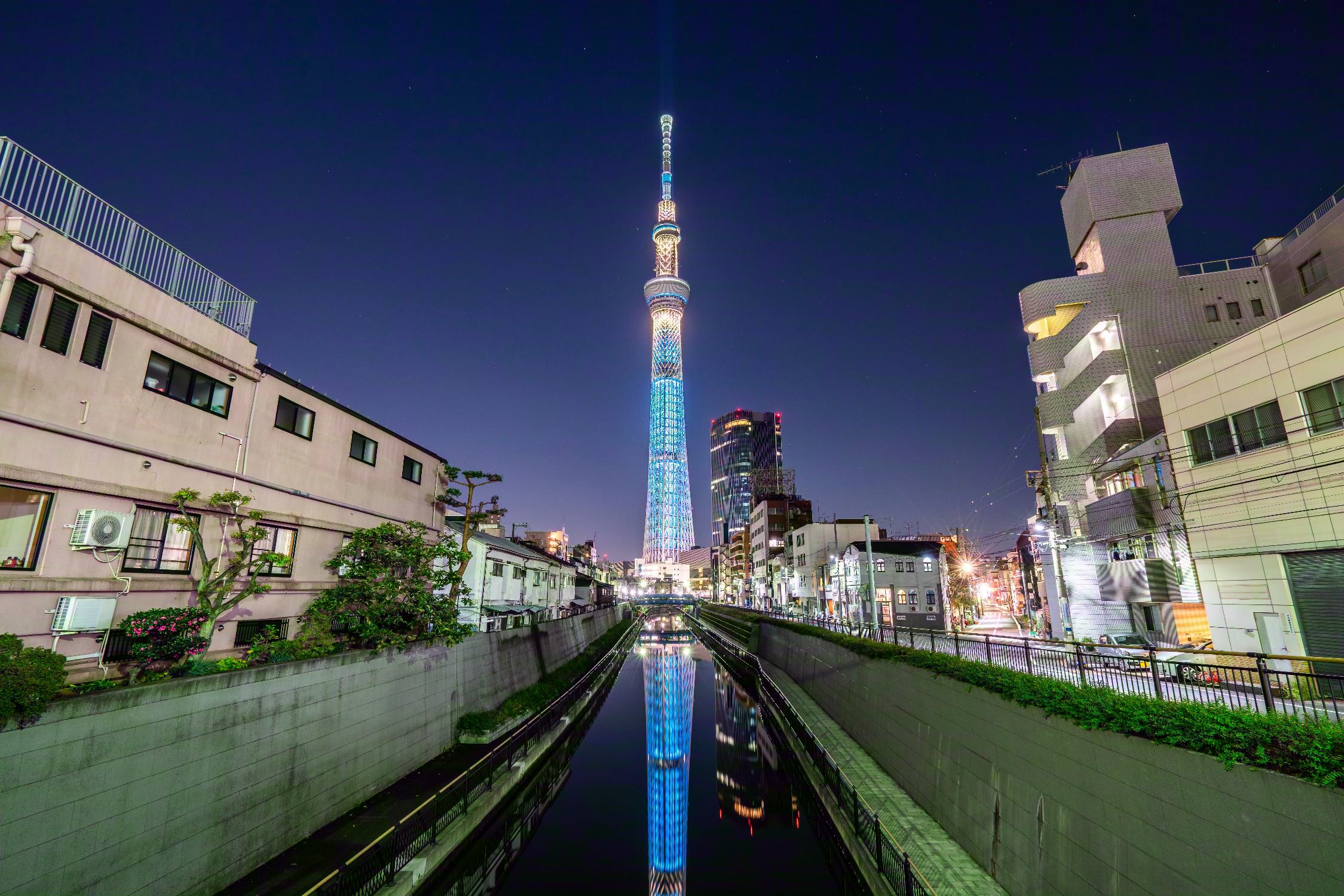} & 
    \includegraphics[width=\imwidth]{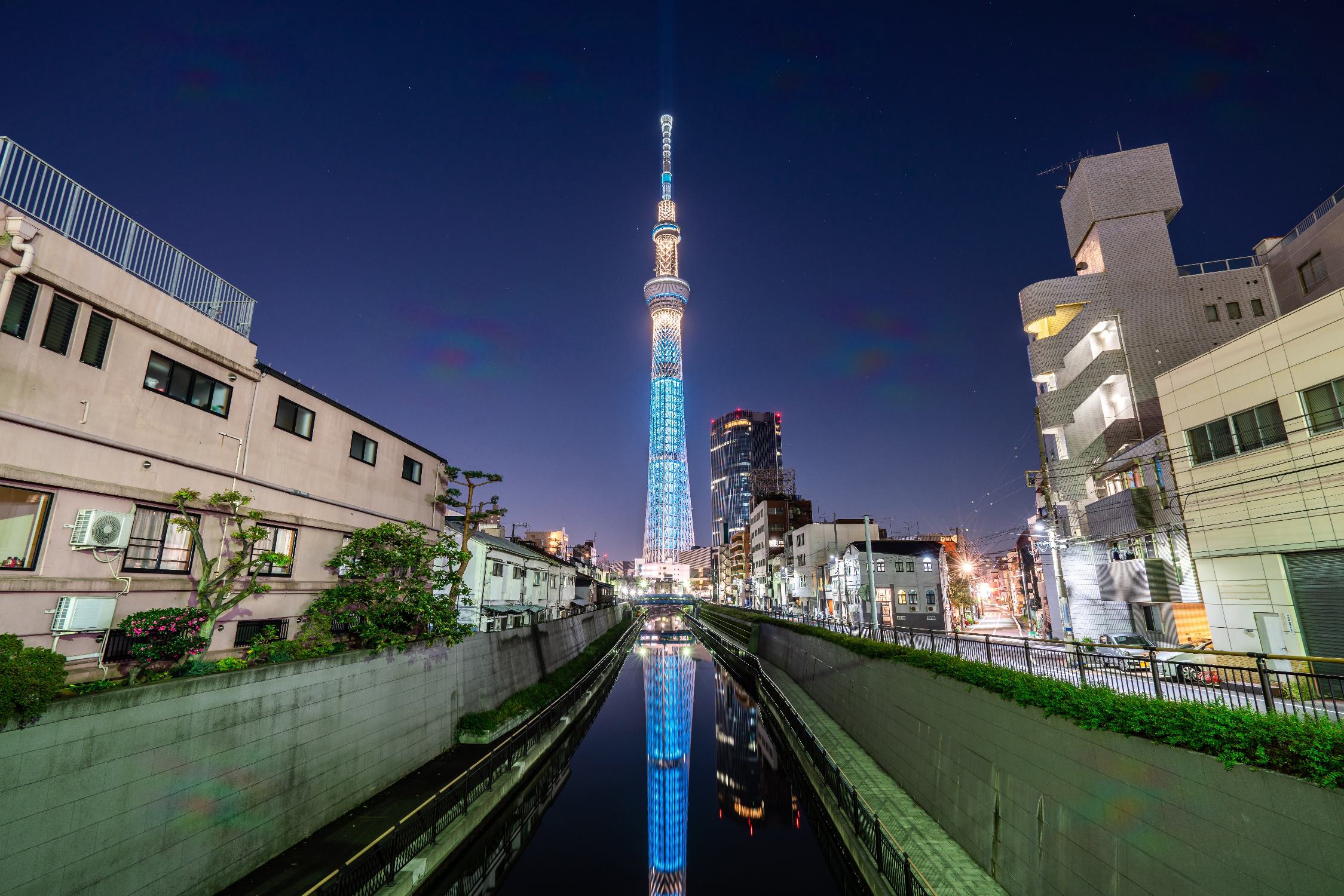} \\ & 
    \includegraphics[width=\imwidth]{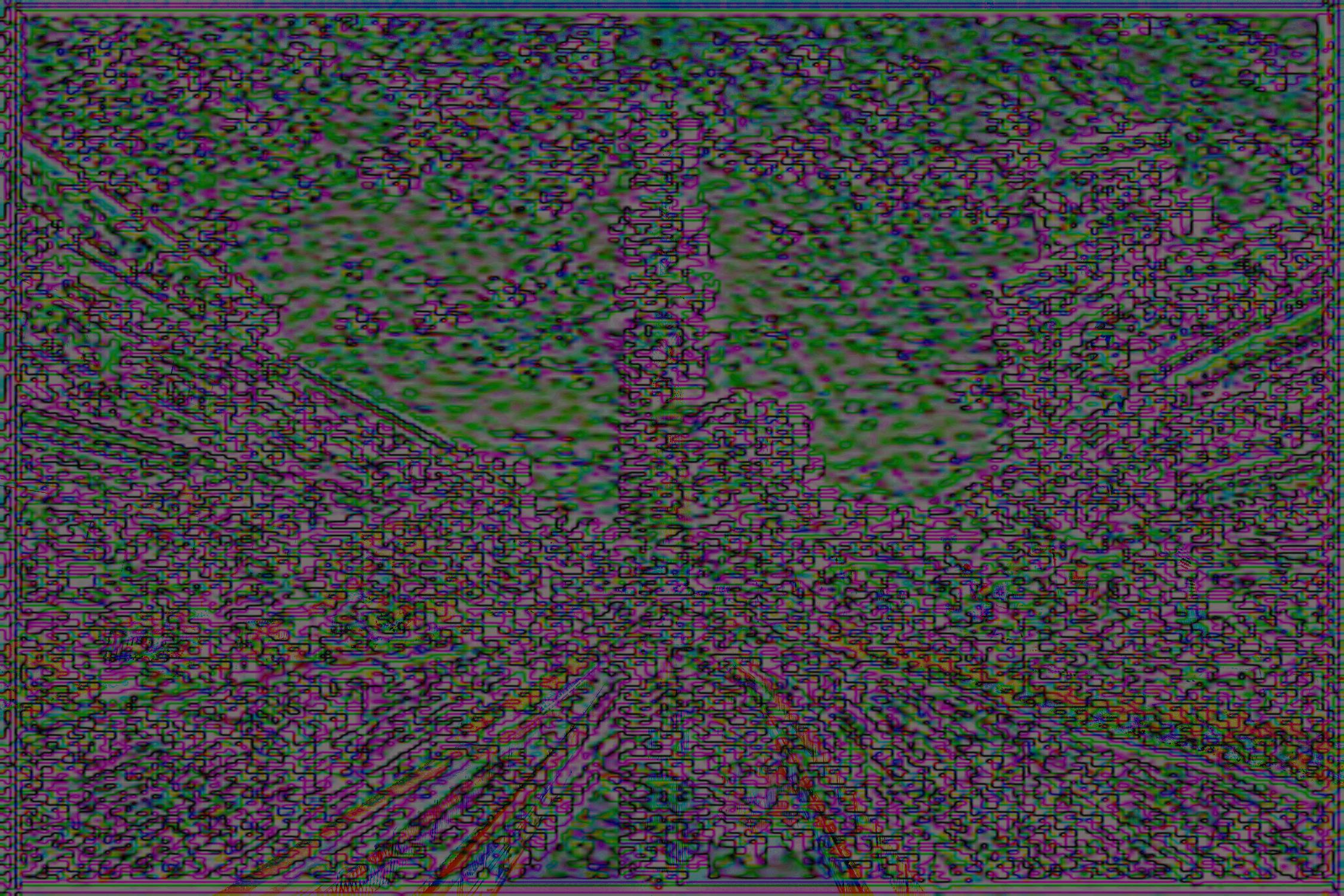} & 
    \includegraphics[width=\imwidth]{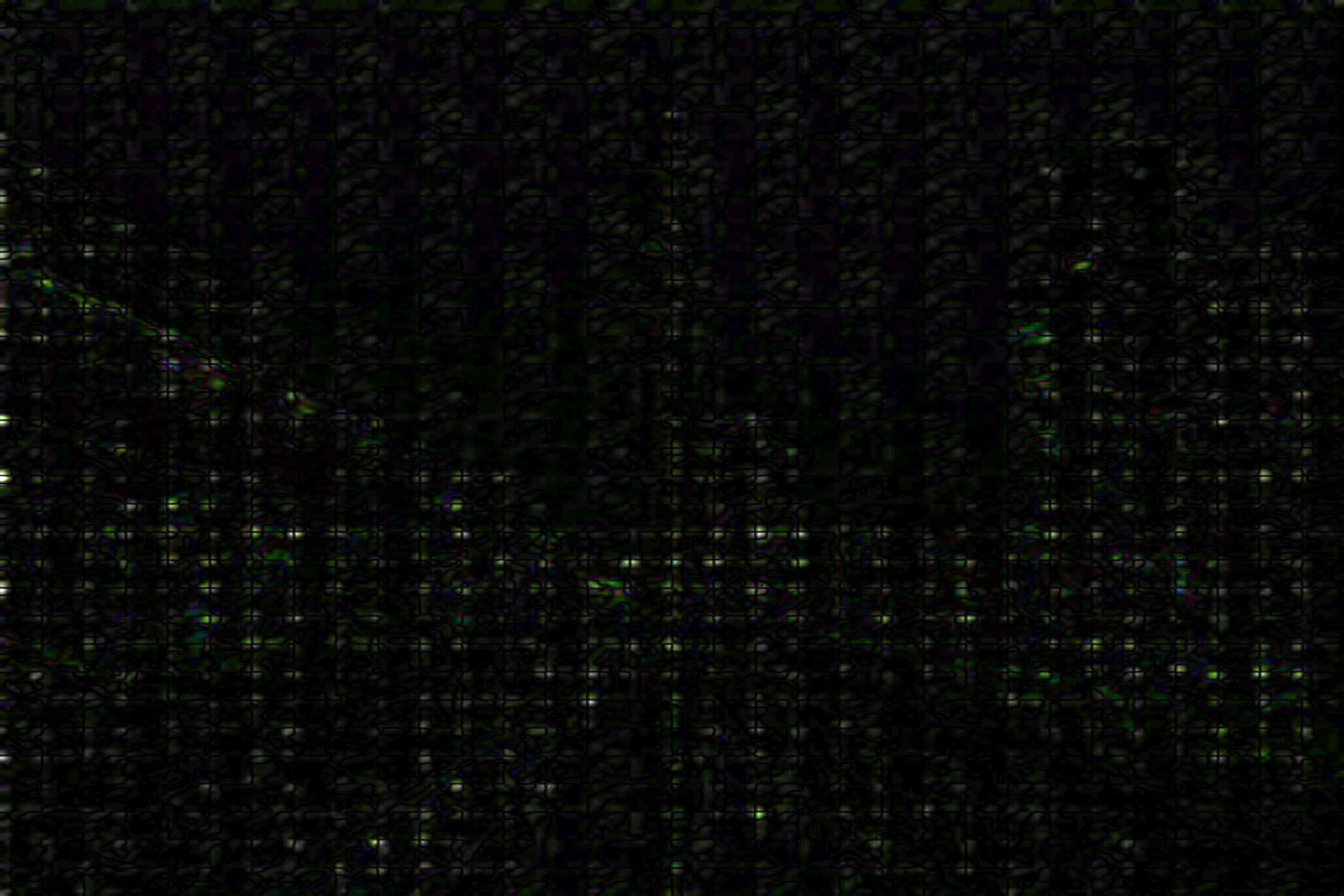} & 
    \includegraphics[width=\imwidth]{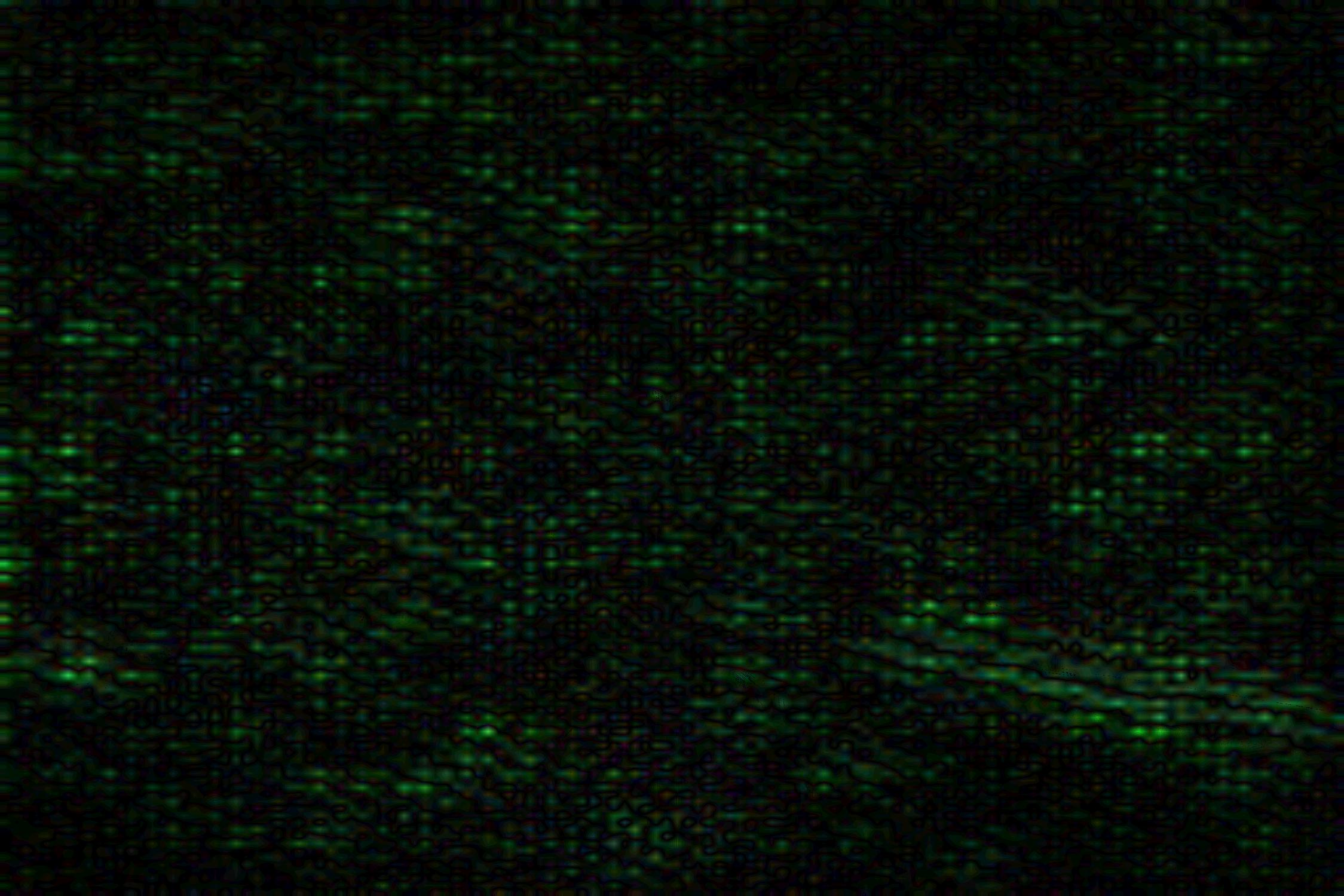} & 
    \includegraphics[width=\imwidth]{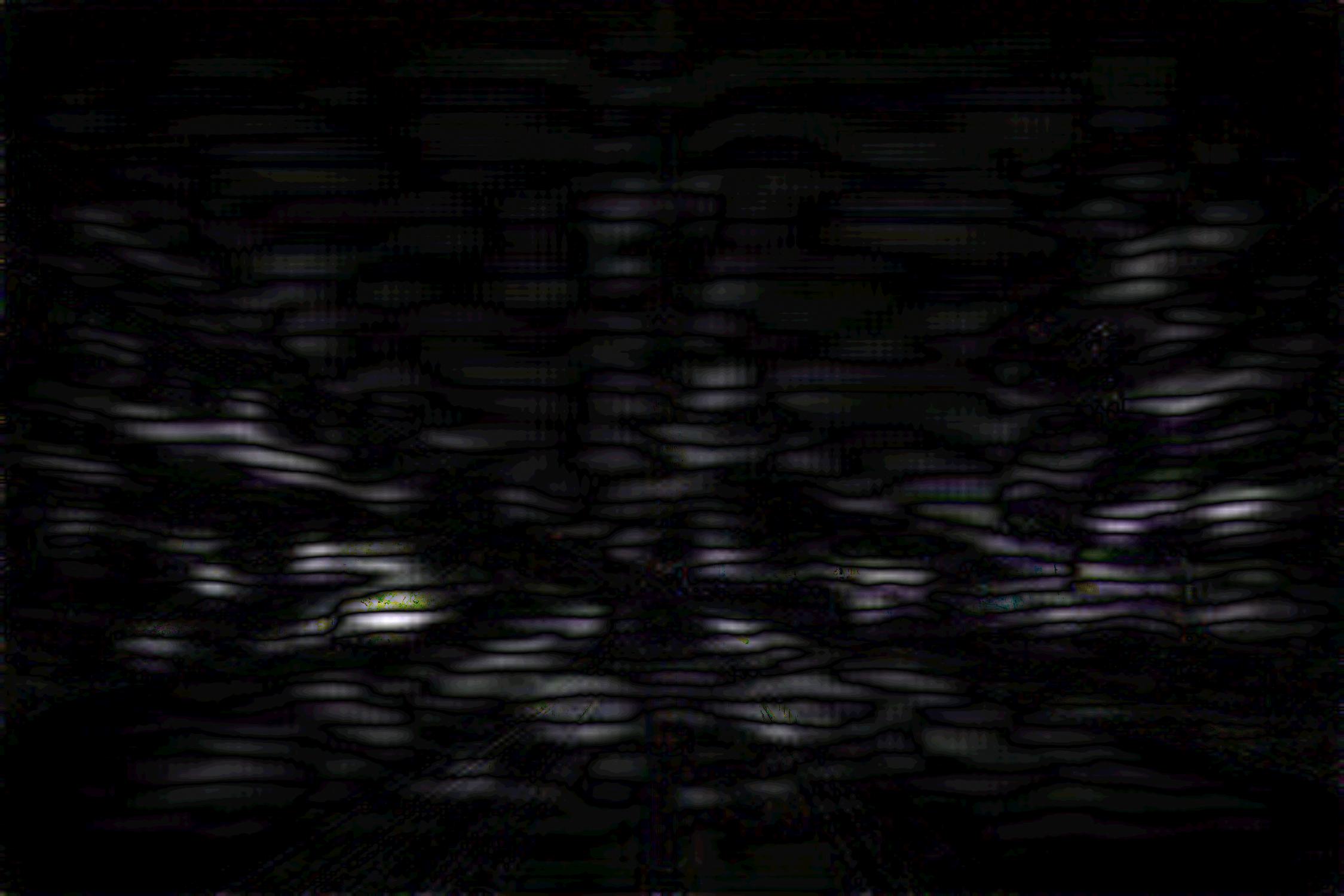} & 
    \includegraphics[width=\imwidth]{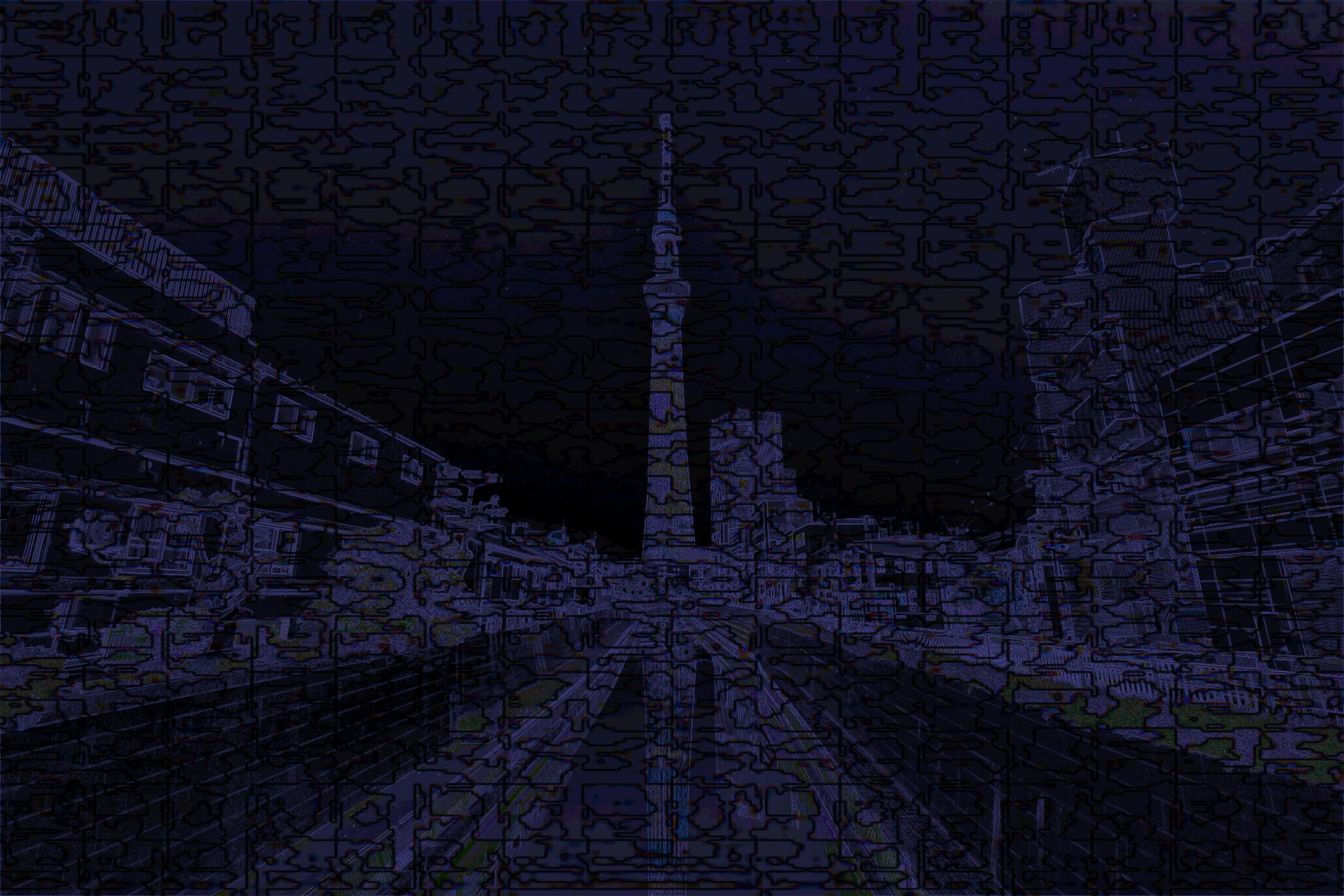} & 
    \includegraphics[width=\imwidth]{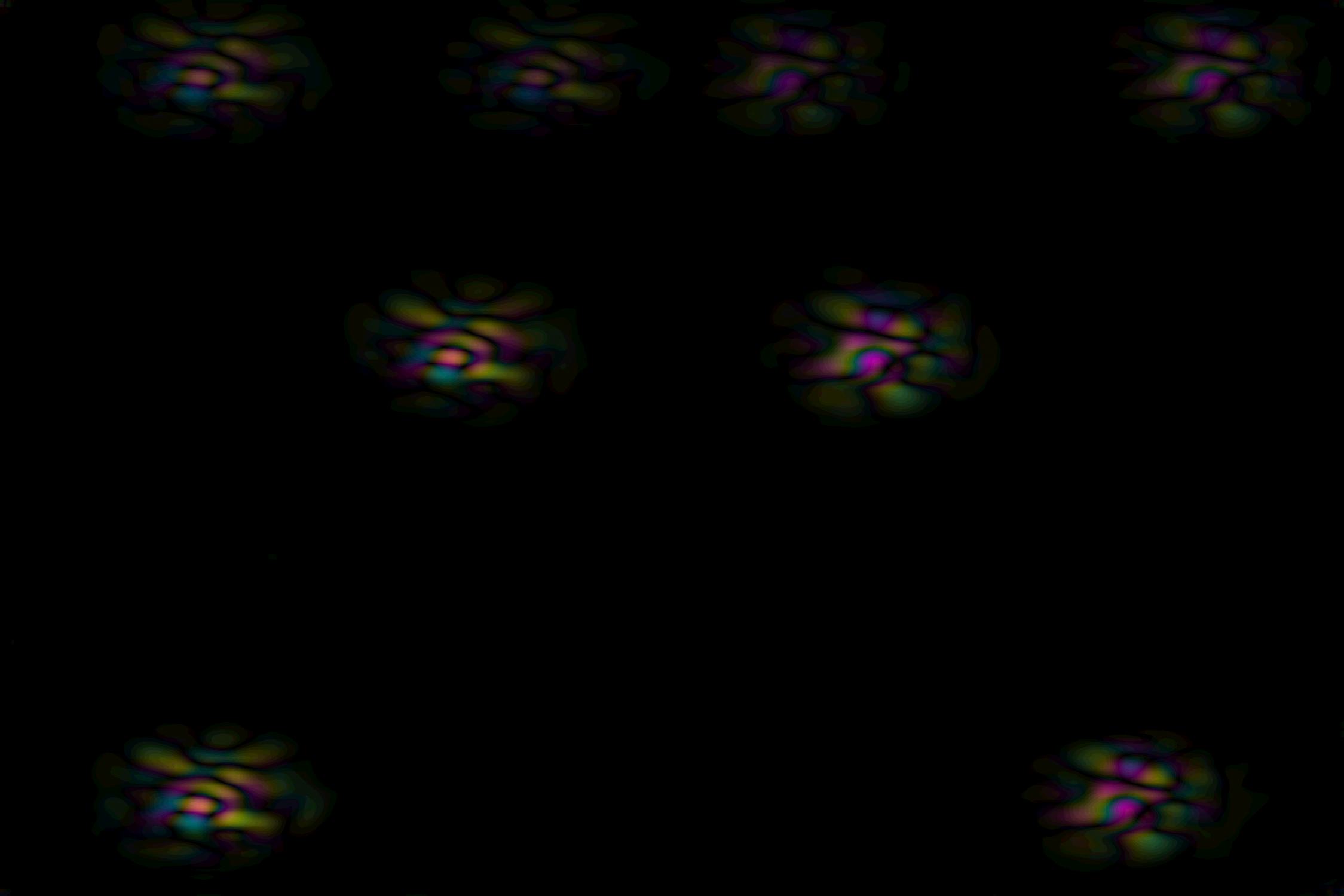} \\\rule{0pt}{6ex}\includegraphics[width=\imwidth]{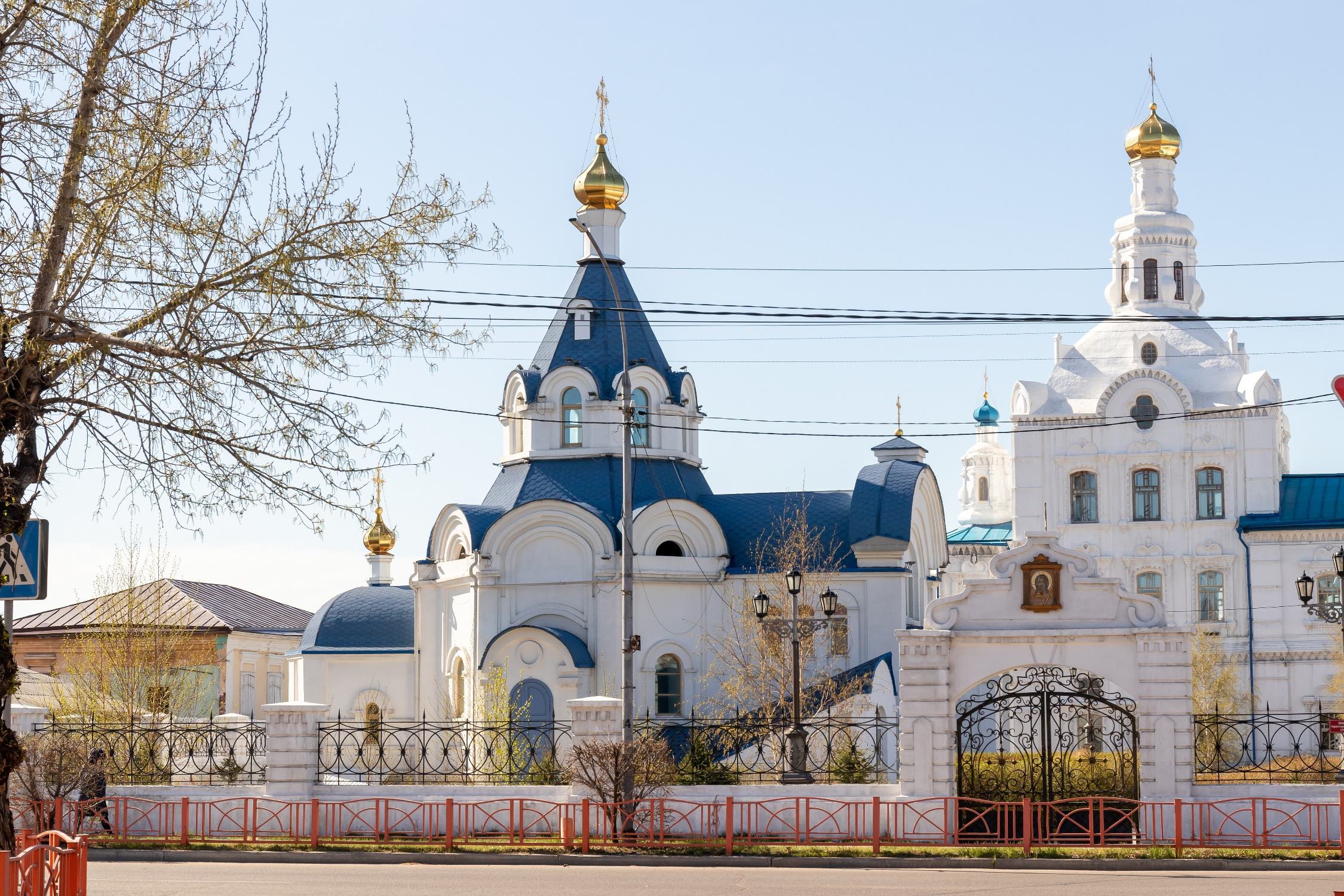} & 
    \includegraphics[width=\imwidth]{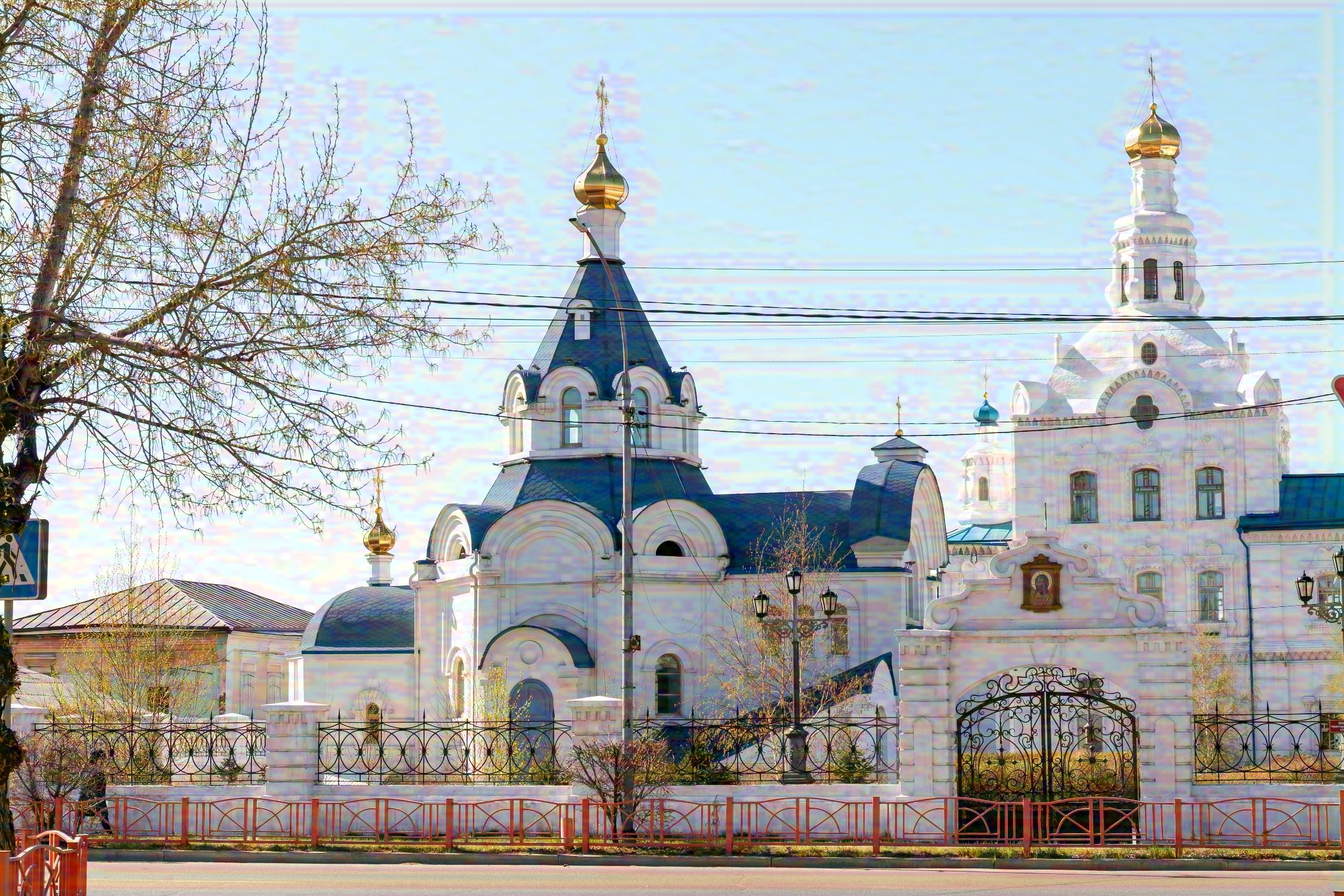} & 
    \includegraphics[width=\imwidth]{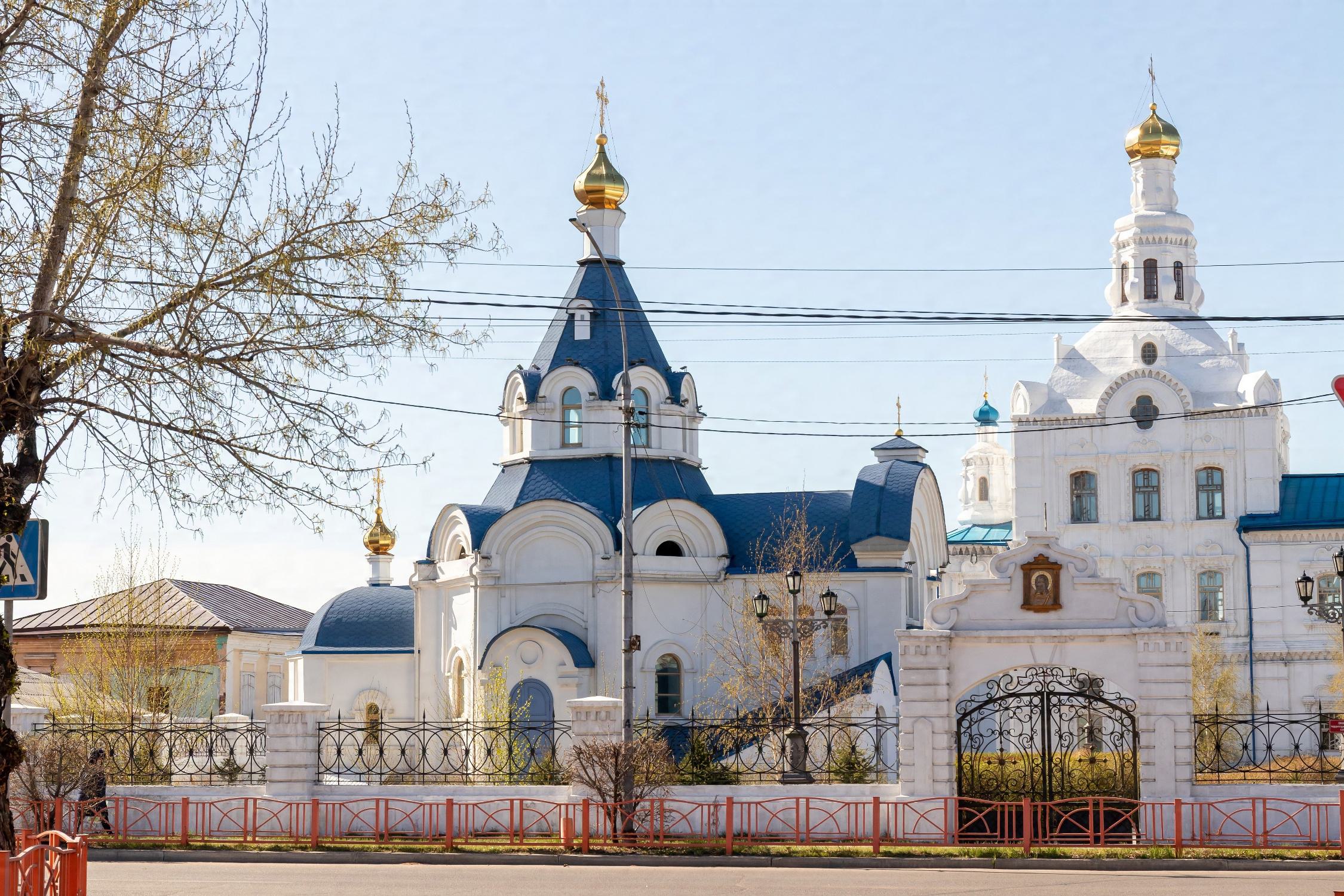} & 
    \includegraphics[width=\imwidth]{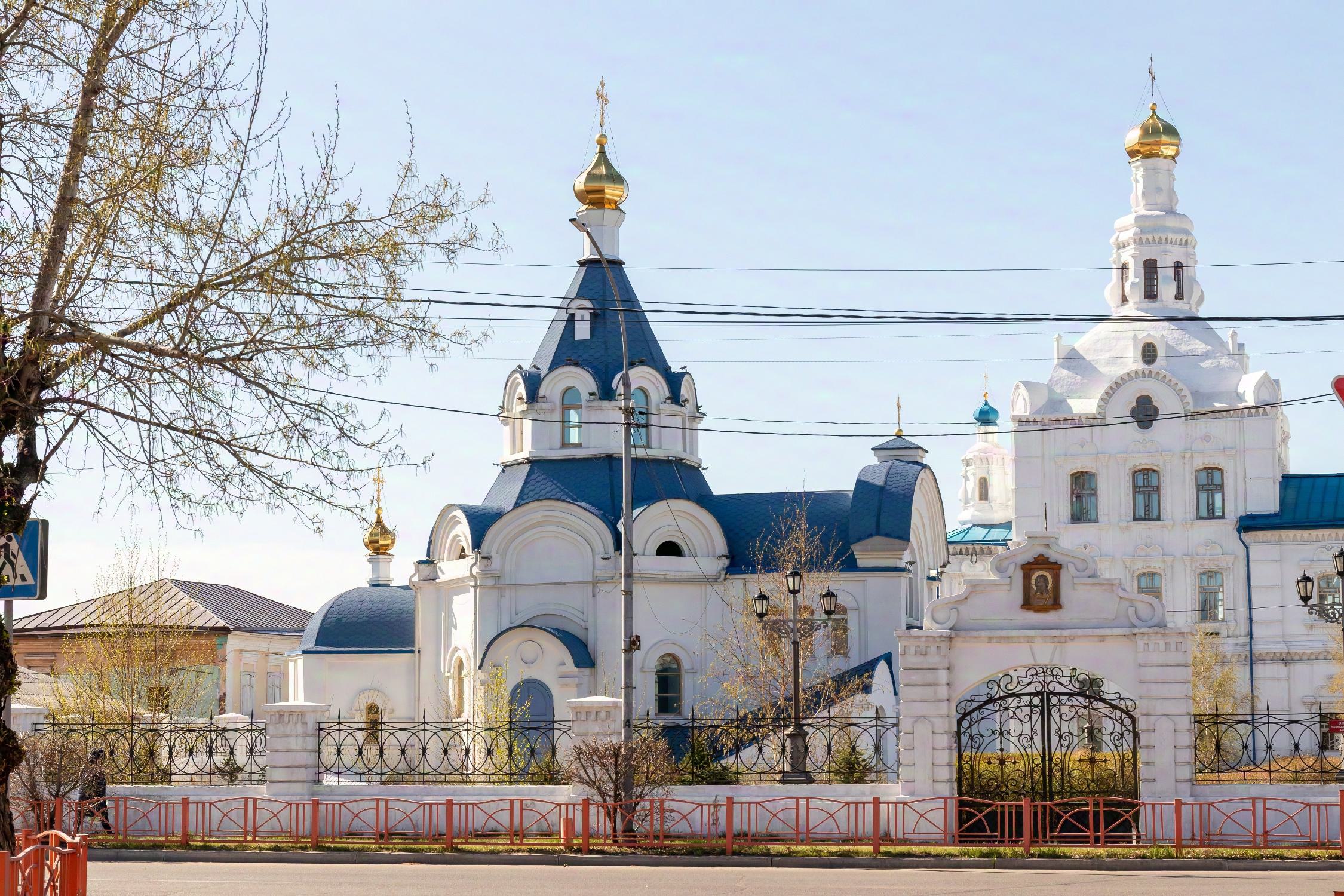} & 
    \includegraphics[width=\imwidth]{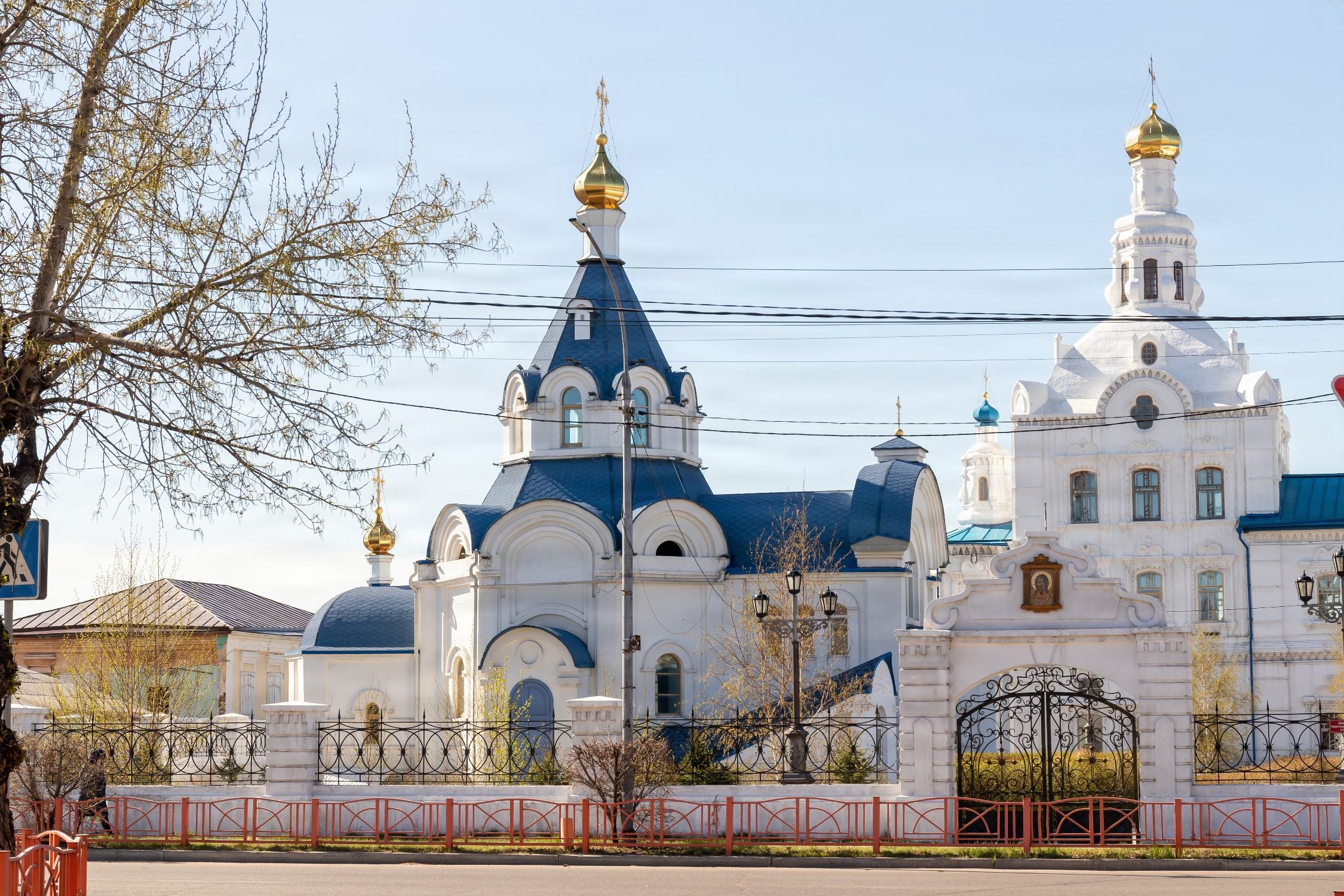} & 
    \includegraphics[width=\imwidth]{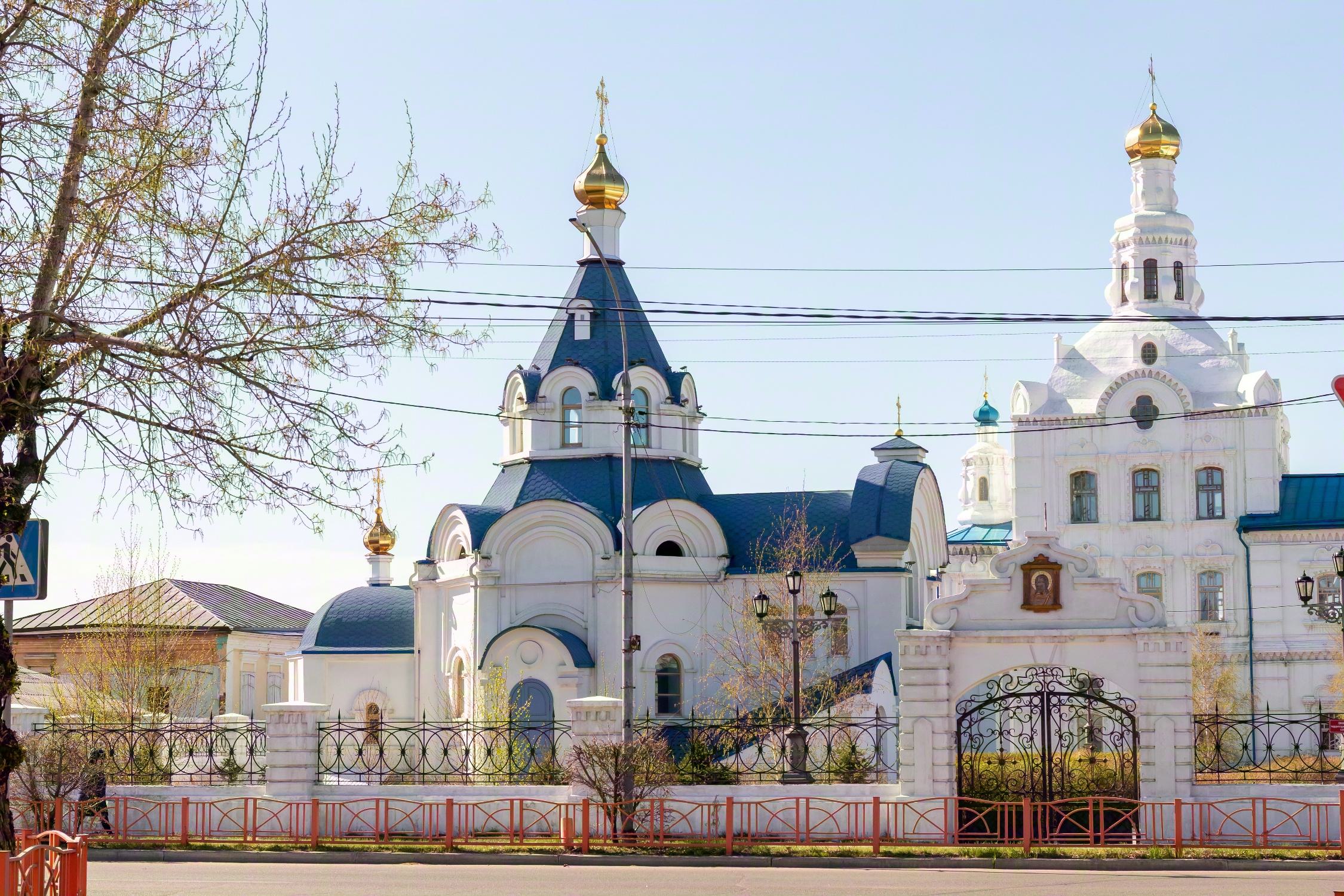} & 
    \includegraphics[width=\imwidth]{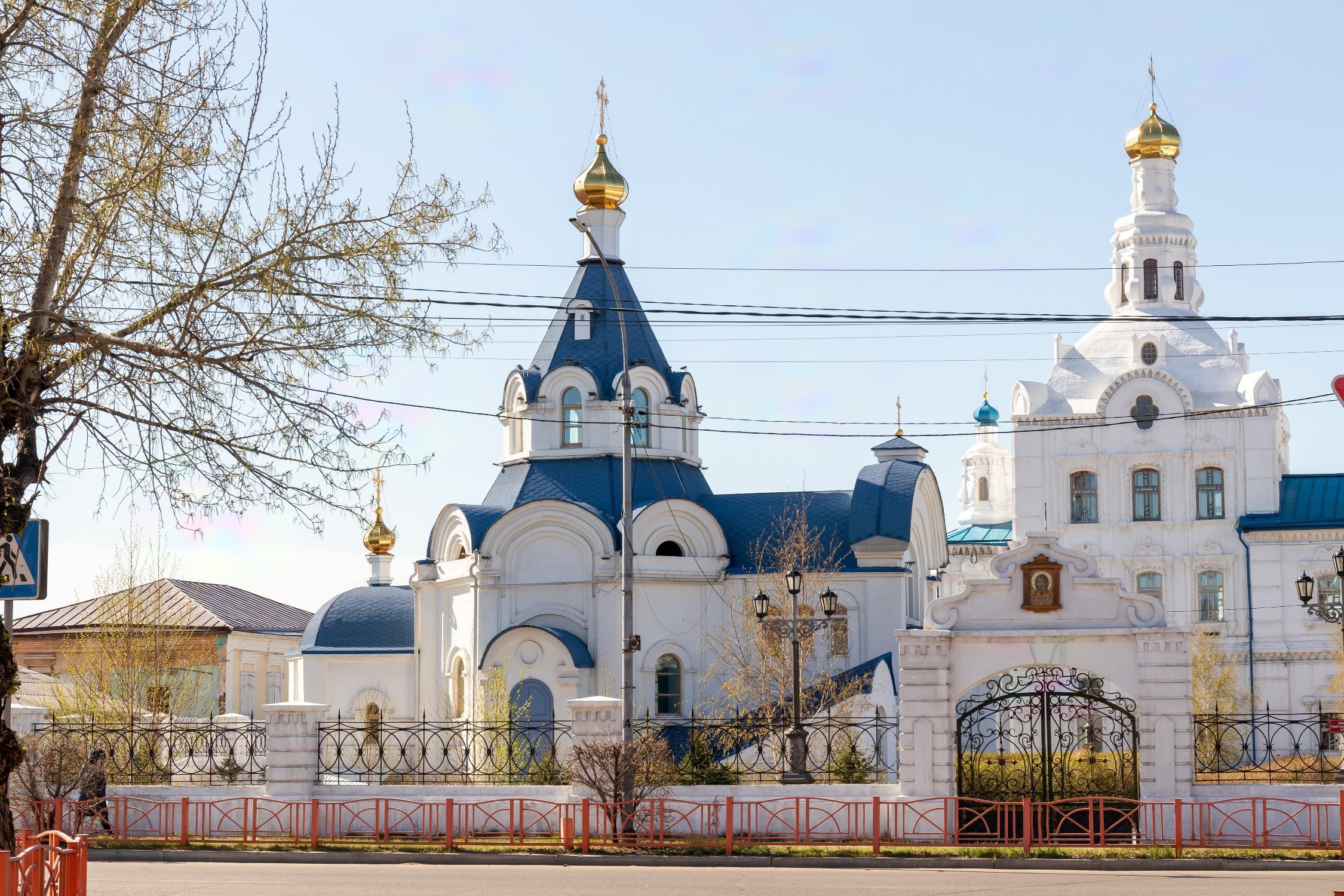} \\ & 
    \includegraphics[width=\imwidth]{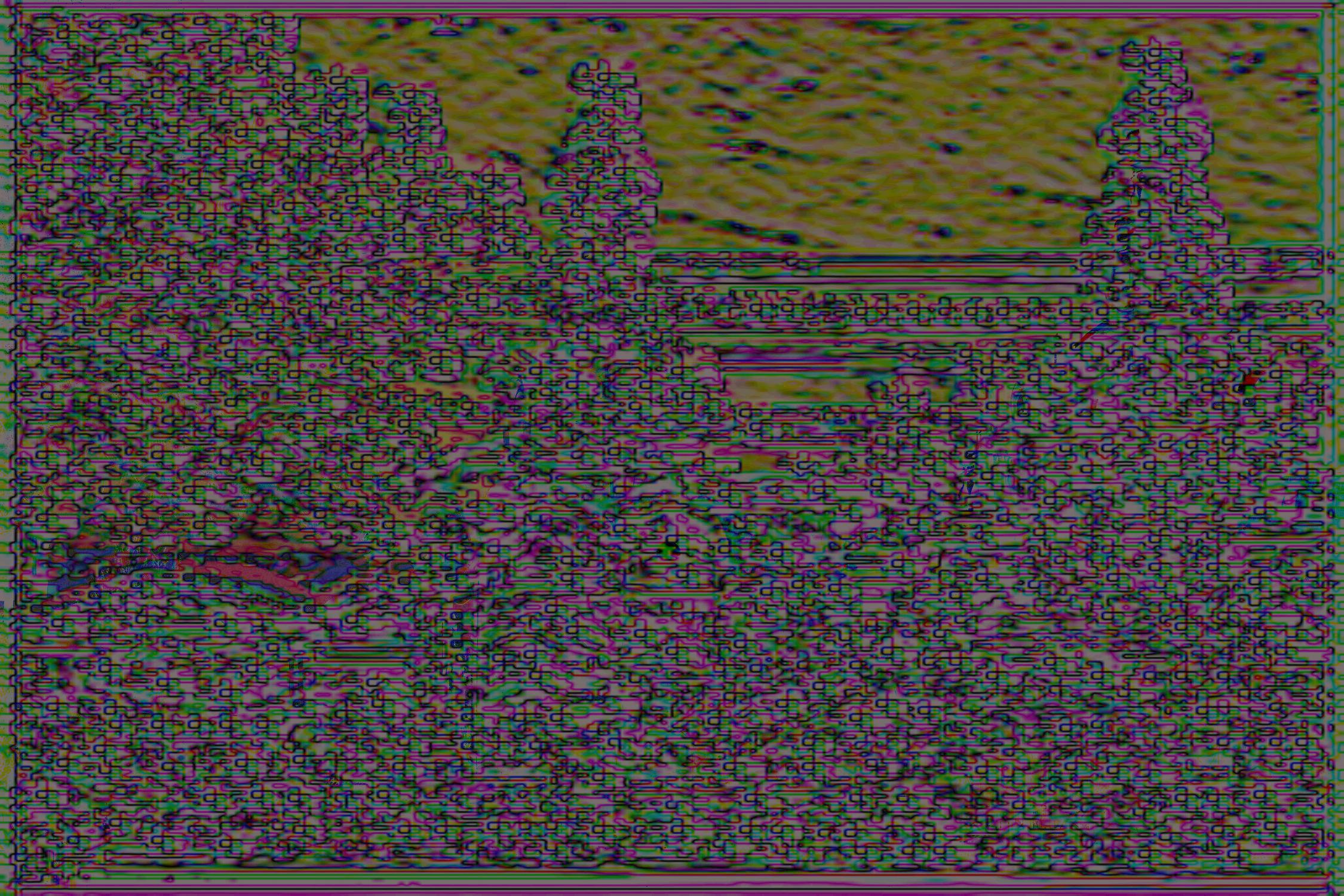} & 
    \includegraphics[width=\imwidth]{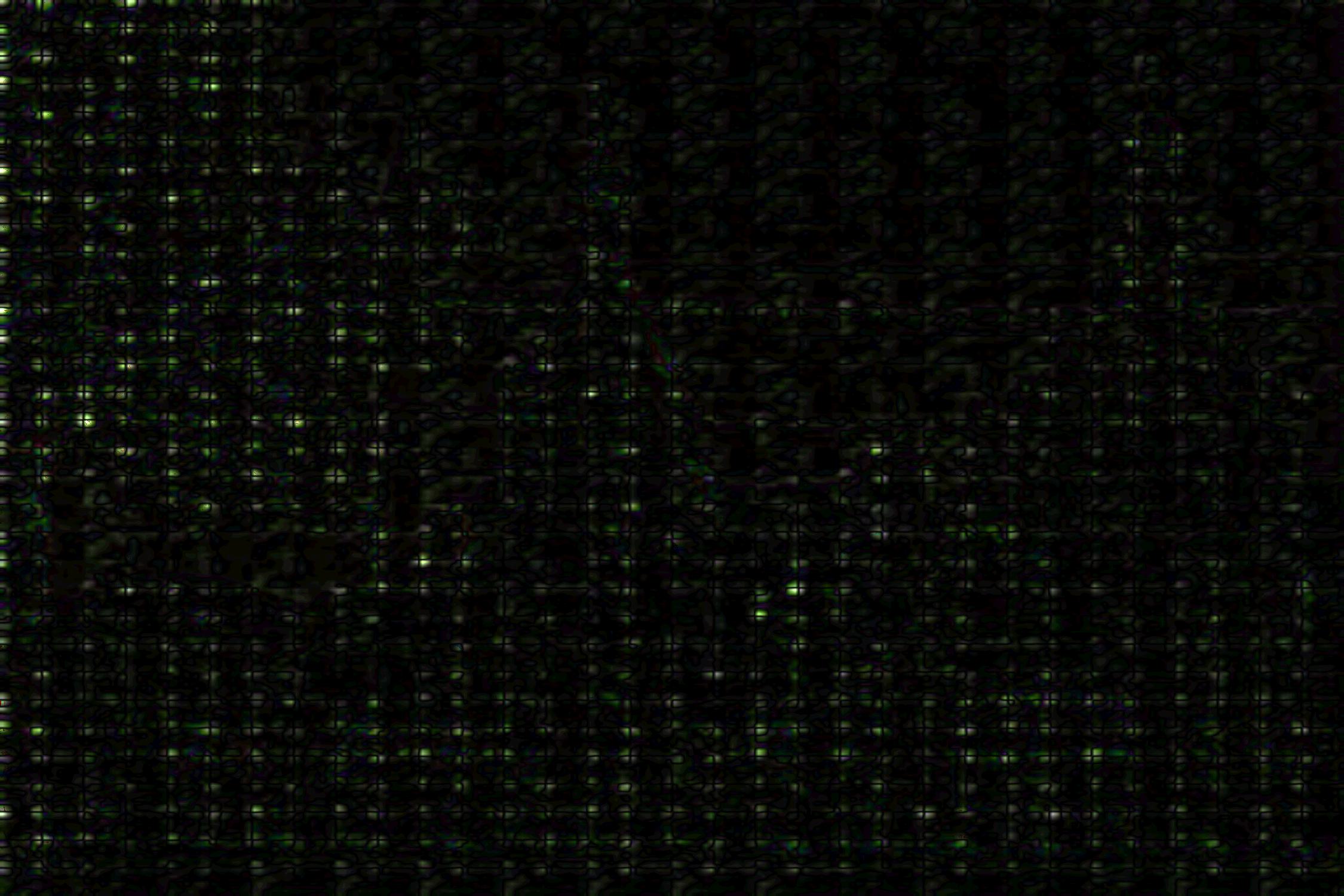} & 
    \includegraphics[width=\imwidth]{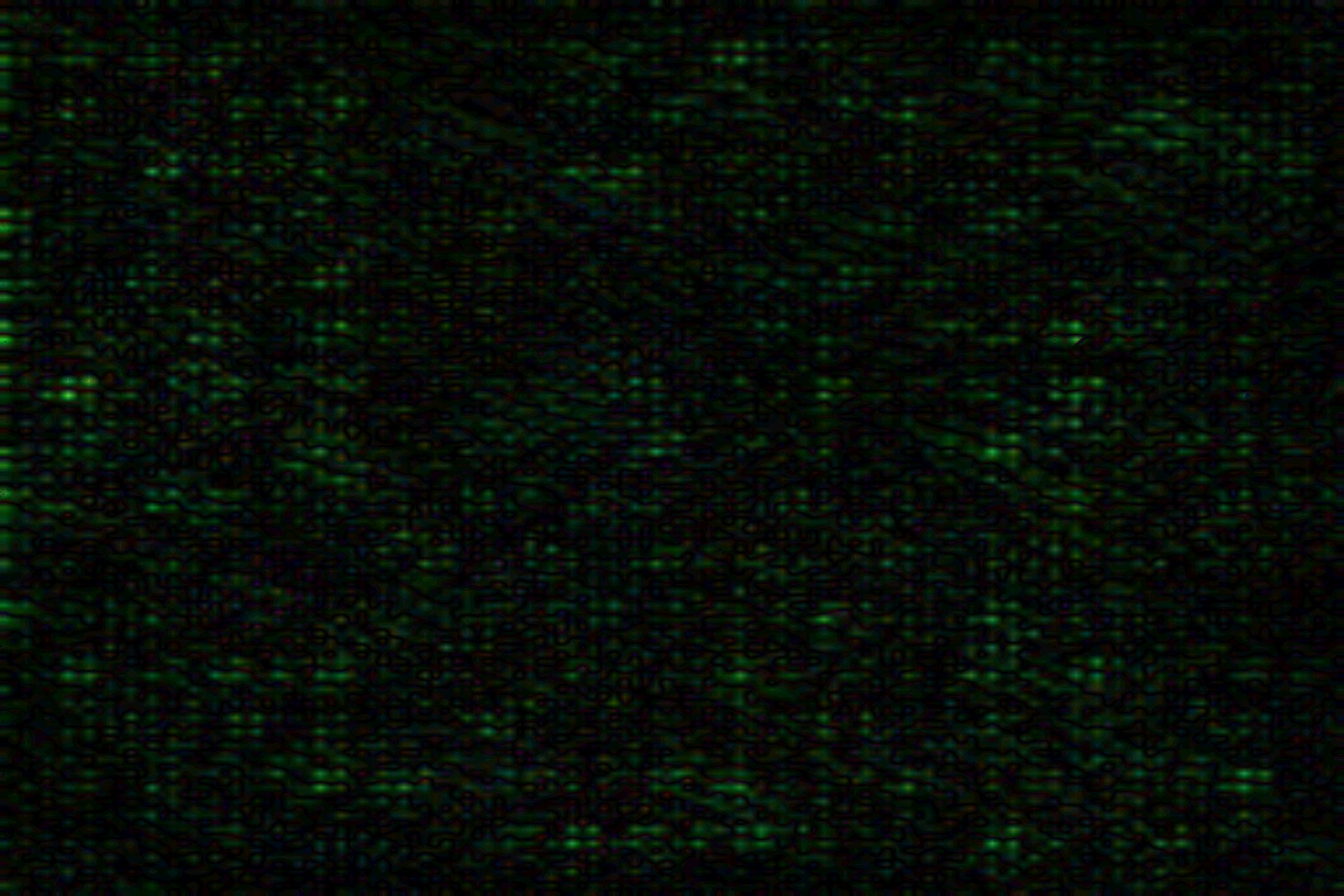} & 
    \includegraphics[width=\imwidth]{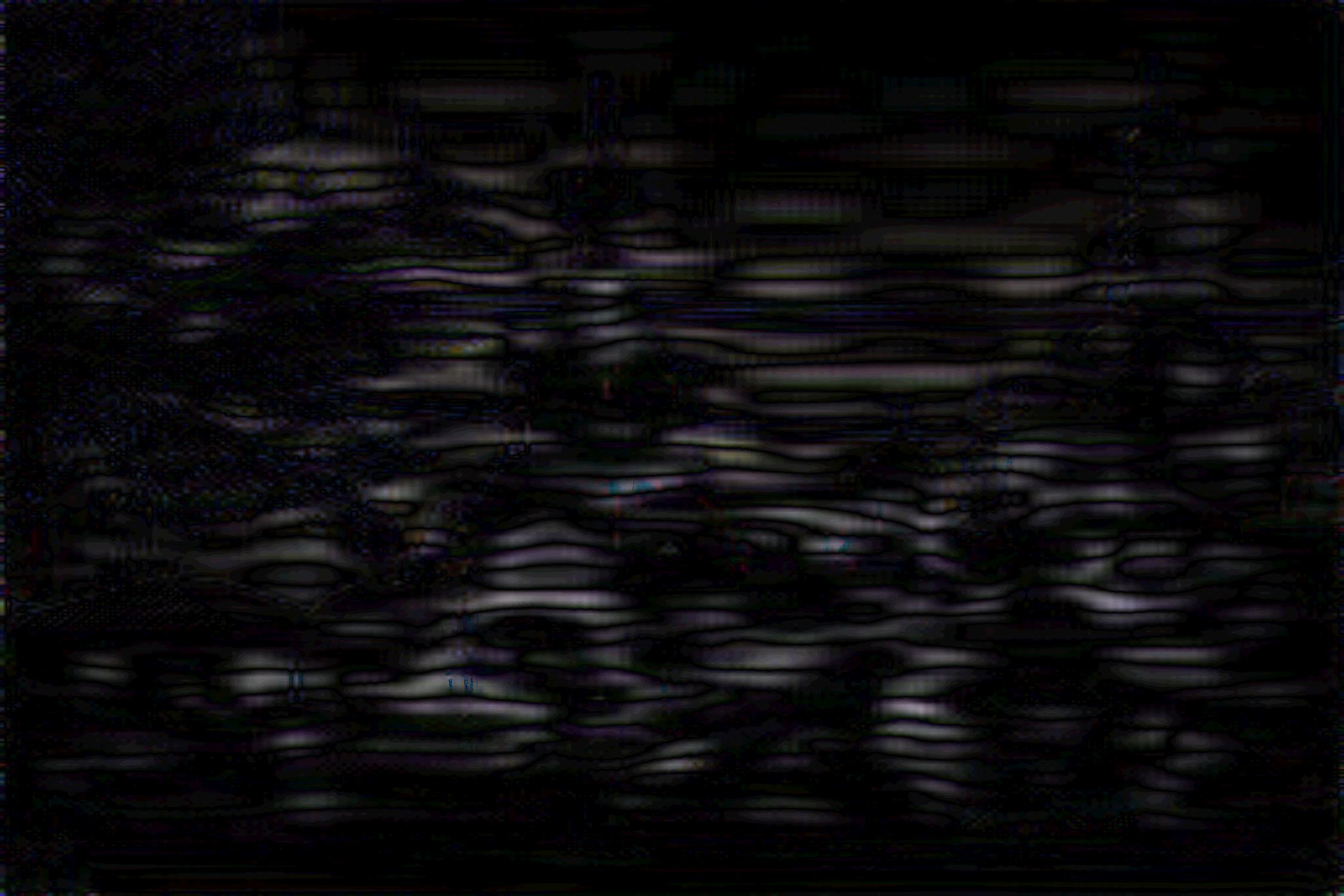} & 
    \includegraphics[width=\imwidth]{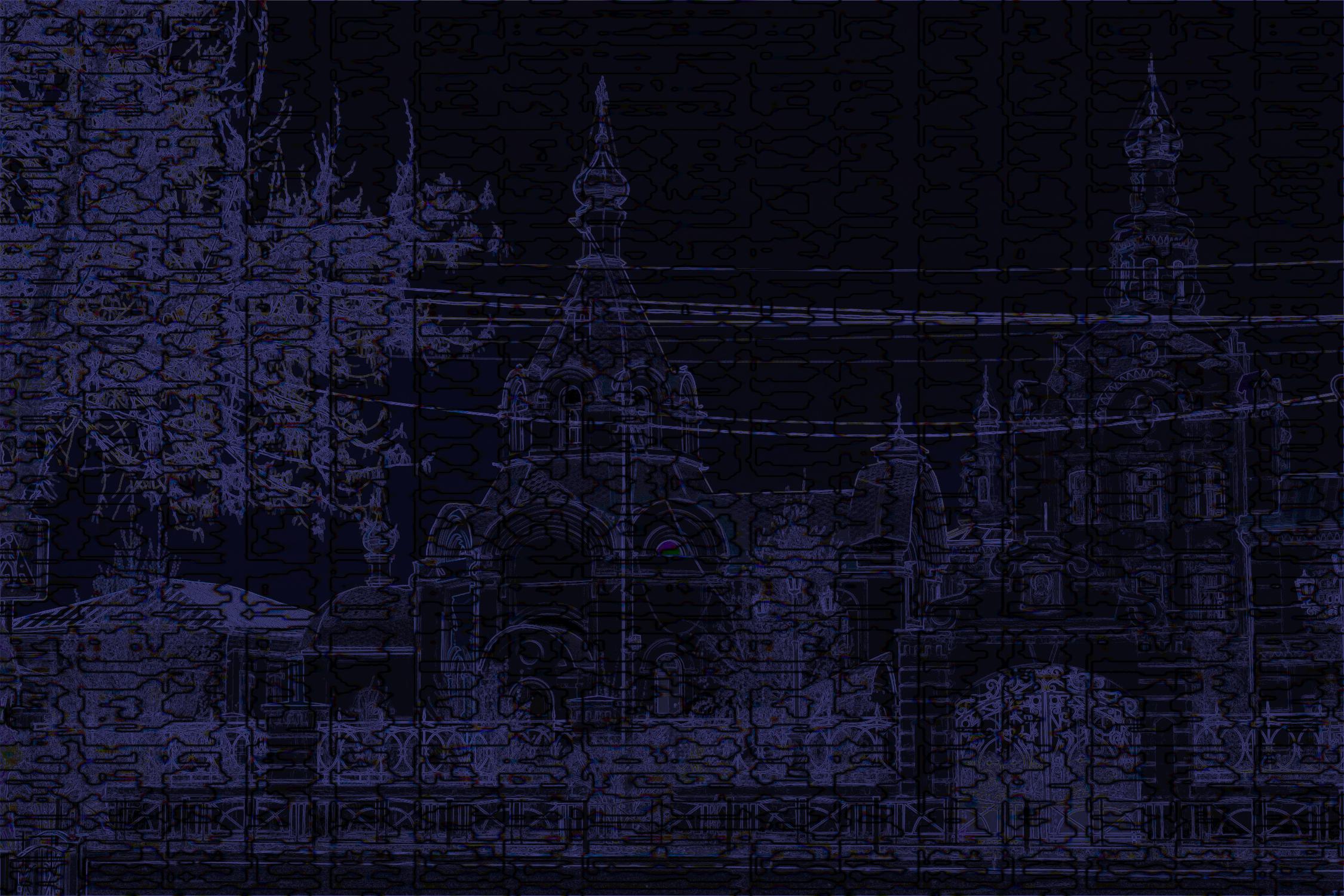} & 
    \includegraphics[width=\imwidth]{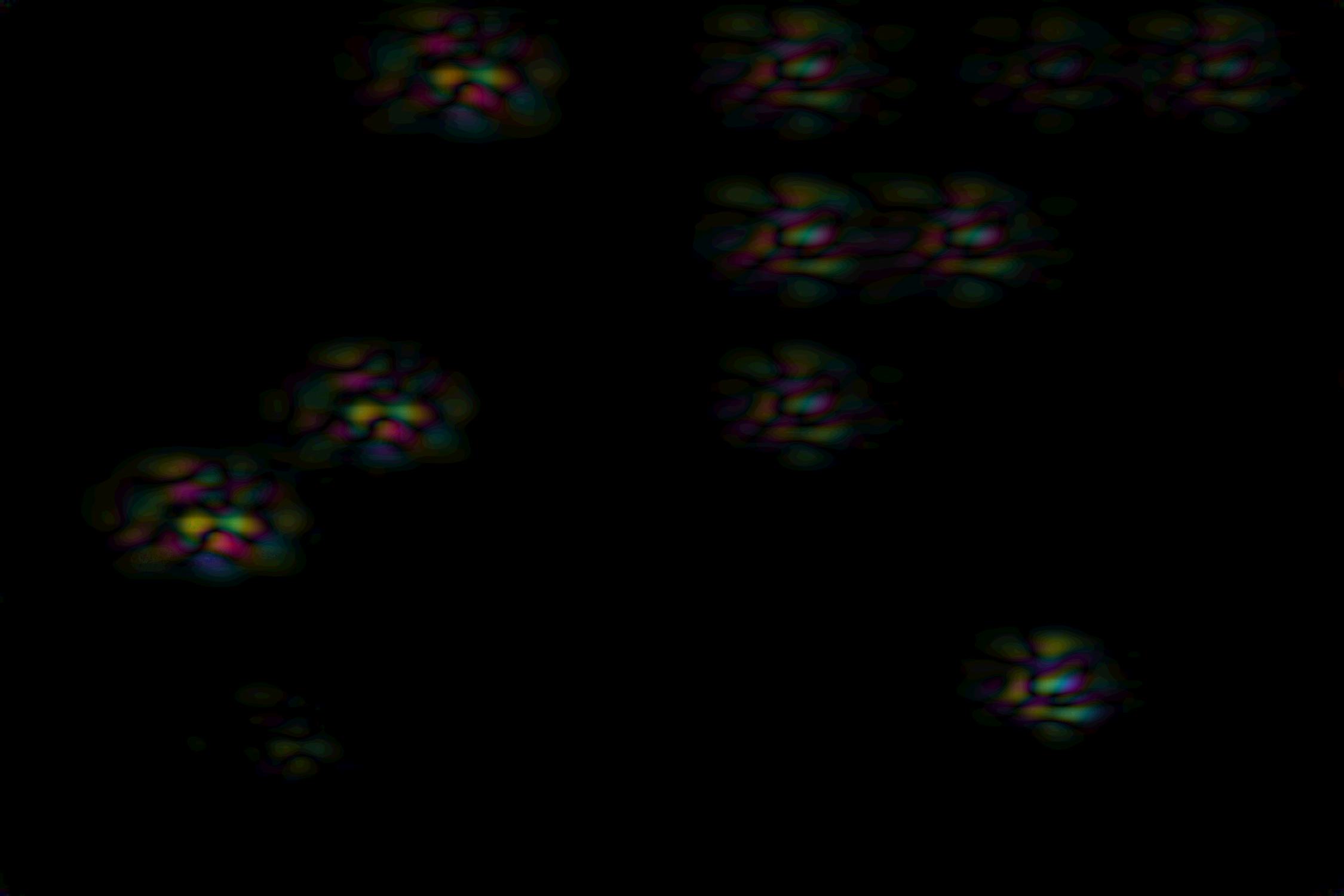} \\
    \end{tabular}
    \caption{
        Qualitative results for different watermarking methods.
        Images are from the SA-1b dataset at their original resolution ($\approx$2k $\times$ 1k), and we show more examples in App.~\ref{app:qualitative}.
        Although watermarks are imperceptible at first glance, most are visible under close inspection, especially in the flat areas, like the skies in both images.
        They are also of very different nature between the methods.
    }\label{fig:main-qualitative} 
\end{figure*}

We first show some examples of watermarked images in Fig.~\ref{fig:main-qualitative}, and of video frames in App.~\ref{app:qualitative}.
We observe that the watermarks are imperceptible at first glance, but most are visible under close inspection, especially in flat areas, like the skies in both images.
Different methods, which employ various perceptual losses and architectures, result in watermarks of distinct characteristics.
For instance, MBRS and CIN tend to create grid-like patterns, while TrustMark and \ours\ tend to create wavier patterns.

We also quantitatively evaluate the imperceptibility of the watermarking models on the image datasets COCO and SA-1b and the video dataset SA-V, and report results in Tab.~\ref{tab:eval_quality}.
For every baseline, we use their nominal strength (most of the time $\alpha_w=1$ in Eq.~\ref{eq:highres}), although they could be adapted to control the imperceptibility/robustness trade-off as done in Sec.~\ref{sec:adaptable-inference}.
We report the PSNR, SSIM, and LPIPS between the watermarked and original images of the SA-1b dataset, as well as the same metrics for videos of the SA-V dataset (cut to 5s), with the addition of VMAF for videos (note that the PSNR is computed on the whole video, and not as an average of the frames as for SSIM and LPIPS).
We observe that \ours\ achieves the highest PSNR and SSIM scores, while MBRS achieves better VMAF and TrusMark achieves better LPIPS, closely followed by \ours.

\begin{table}[t!]
    \caption{
      Evaluation of the watermark imperceptibility. 
      We report the average PSNR, SSIM, and LPIPS between watermarked and original images of the SA-1b and COCO datasets, as well as the same metrics for videos of the SA-V dataset (cut to 5s), with the addition of VMAF~\citep{vmaf} for videos.
    }
    \label{tab:eval_quality}
    \centering
    {\footnotesize
    
\begin{tabular}{rr *{6}{c}}
    \toprule
    & & \rotatebox[origin=l]{0}{\shortstack[c]{HiDDeN}} & \rotatebox[origin=l]{0}{\shortstack[c]{MBRS}} & \rotatebox[origin=l]{0}{\shortstack[c]{CIN}} & \rotatebox[origin=l]{0}{\shortstack[c]{TrustMark}} & \rotatebox[origin=l]{0}{\shortstack[c]{WAM}} & \rotatebox[origin=l]{0}{\shortstack[c]{Video Seal (ours)}} \\
    \midrule
    \multirow{3}{*}{\rotatebox[origin=c]{90}{Image}} 
    & SSIM ($\uparrow$) & 0.927 & 0.997 & 0.997 & 0.995 & 0.989 & 0.999 \\
    & LPIPS ($\downarrow$) & 0.229 & 0.003 & 0.019 & 0.002 & 0.031 & 0.009 \\
    & PSNR ($\uparrow$) & 30.36 & 45.60 & 44.90 & 42.09 & 39.86 & 47.39 \\
    \midrule
    \multirow{4}{*}{\rotatebox[origin=c]{90}{Video}}     
    & SSIM ($\uparrow$) & 0.857 & 0.995 & 0.994 & 0.995 & 0.981 & 0.998 \\
    & LPIPS ($\downarrow$) & 0.362 & 0.008 & 0.032 & 0.003 & 0.047 & 0.013 \\
    & PSNR ($\uparrow$) & 30.19 & 46.55 & 45.80 & 43.07 & 40.72 & 48.02 \\
    & VMAF ($\uparrow$) & 74.61 & 94.10 & 92.93 & 89.36 & 89.78 & 93.77 \\
    \bottomrule
\end{tabular}

    }
\end{table}

It is important to note that video imperceptibility is not fully captured in these examples and in these metrics.
In practice, a watermark that is imperceptible in an image may not necessarily be imperceptible in a video, particularly when the watermark lacks consistency across frames. 
For instance, we found that TrustMark can produce shadowy artifacts as the watermark tracks the motion of the video, making it more visible.
This is less pronounced for \ours, which tends to produce blobs that do not follow objects.
However, clear metrics to evaluate this are still lacking, and would require a more comprehensive study on the perception of watermarks in videos.
Notably, we observe that even at very high PSNR, SSIM or VMAF, the artifacts produced by \ours\ may be annoying to the human eye and highly depend on the cover videos.

\subsection{Latency}

\begin{table}[t!]
    \caption{
        Efficiency of watermark embedding and extraction.
        We report the number of GFlops for embedding and extraction for models at their nominal resolution ($256 \times 256$ for all methods but CIN which is $128 \times 128$).
        Additionally, we report the processing time per second of video for embedding and extraction on CPU and GPU, averaged over 20 videos from the SA-V dataset.
        We use the video inference framework of Sec.~\ref{sec:video-inference} to fairly compare all models.
    }\label{tab:latency}
    \centering
    \resizebox{1.0\linewidth}{!}{
    \begin{tabular}{rr *{6}{c}}
    \toprule
    & & \rotatebox[origin=l]{0}{\shortstack[c]{HiDDeN\\ {\footnotesize \citep{zhu2018hidden}}}} & \rotatebox[origin=l]{0}{\shortstack[c]{MBRS\\ {\footnotesize \citep{jia2021mbrs}}}} & \rotatebox[origin=l]{0}{\shortstack[c]{CIN\\ {\footnotesize \citep{ma2022towards}}}} & \rotatebox[origin=l]{0}{\shortstack[c]{TrustMark\\ {\footnotesize \citep{bui2023trustmark}}}} & \rotatebox[origin=l]{0}{\shortstack[c]{WAM\\ {\footnotesize \citep{sander2024watermark}}}} & \rotatebox[origin=l]{0}{\shortstack[c]{\ours\\ {\footnotesize (ours)}}} \\
    \midrule
    \multirow{3}{*}{\rotatebox[origin=c]{90}{Embed}} 
    & GFlops & 22.4 & 32.2 & 16.6 & 10.3 & 42.6 & 42.0 \\
    & CPU - Time (s) & 0.67 & 1.99 & 1.04 & 0.64 & 3.64 & 1.14 \\
    & GPU - Time (s) & 0.42 & 0.47 & 0.47 & 0.42 & 3.19 & 0.42 \\
    \midrule
    \multirow{3}{*}{\rotatebox[origin=c]{90}{Extract}} 
    & GFlops         & 39.0 & 27.0 & 17.9 & ~4.1 & 68.7 & ~3.1 \\
    & CPU - Time (s) & 1.64 & 2.31 & 3.49 & 0.41 & 2.52 & 0.69 \\
    & GPU - Time (s) & 0.19 & 0.29 & 0.77 & 0.11 & 0.47 & 0.11 \\
    \bottomrule
    \end{tabular}
    }
\end{table}

We evaluate the latency of \ours\ compared to the image watermarking models repurposed for video watermarking. 
We use the video inference framework introduced in Sec.~\ref{sec:video-inference}, with the downscale/upscale of the watermark signal and temporal watermark propagation with $k=4$ -- to ensure a fair comparison across all models and see if the inference efficiency generalizes the same way across all models.
Each model was compiled using TorchScript to optimize performance.
Experiments are conducted on video clips from the SA-V dataset (full length, with a duration ranging from 10 to 24 seconds), with 2 Intel(R) Xeon(R) 6230 @ 2.10GHz and 480GB of RAM as CPU, and (optionally) a Tesla V100-SXM2-16GB as GPU. 
We evaluate the time needed for embedding and extraction in two scenarios: using only the CPU and using both the CPU and GPU (we do not consider video loading and saving times in the following).

We report the GFlops and time in seconds for both CPU and GPU configurations in Tab.~\ref{tab:latency}.
The GFlops required for embedding are consistent across models within a range of 10 to 43, while the GFlops required for extraction vary more widely from 3 to 69.
In terms of GPU time, WAM is the slowest at embedding because it uses a heatmap to attenuate the watermark, which is computationally expensive at high resolution (high resolution images are never sent to the GPU to reduce memory constraints, so the compute of the heatmap is done on the CPU).
The other models are much faster (around 0.5-2 seconds on CPU), but quite similar to each other.
On GPU in particular, the transfer time from CPU to GPU and the CPU operations on high-resolution videos seem to be the bottleneck.
For extraction, all the models are in the same ballpark.

\subsection{Imperceptibility/Robustness trade-off}\label{sec:adaptable-inference}

\begin{figure}[b!]
    \centering
    \includegraphics[width=\linewidth]{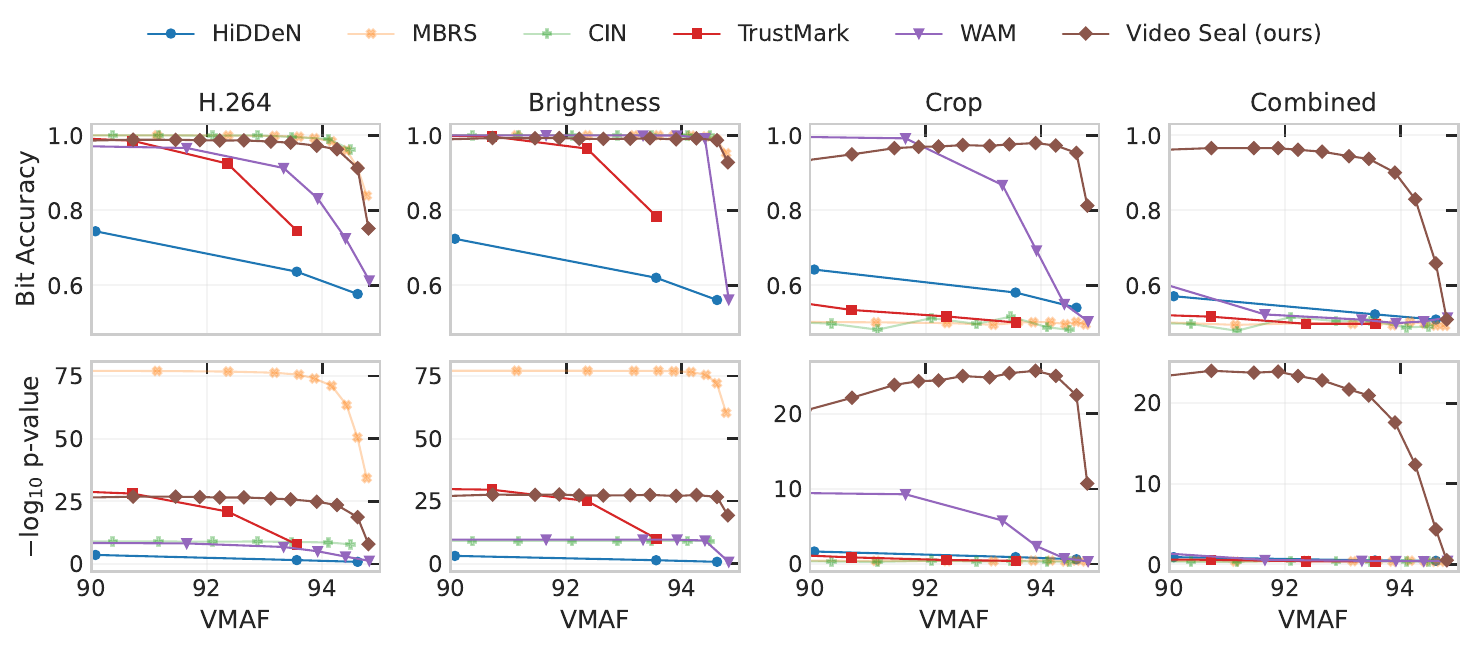}
    \caption{
    Robustness/quality trade-off across transformations for various models on 5s videos from SA-V.
    We compare the performance of six watermarking methods under H.264 compression (CRF=30), brightness adjustments (strength 0.5), cropping (50\% area-wise), and the combination of the 3 transformations.
    (MBRS and CIN are palished because of their lack of robustness to geometric operations).
    We report for each transformation type \emph{(top)} the bit accuracy and \emph{(bottom)} the $-\logpval$, which accounts for the total number of bits, against the VMAF between the watermarked and the original videos. 
    \ours\ achieves higher robustness compared to baselines especially under challenging transformations combining geometric transformations and video compression.
    }\label{fig:trade-off}
\end{figure}

We previously reported the robustness and imperceptibility of the watermarking models at their nominal strength.
In practice, one may want to adapt the strength $\alpha_w$ to control the imperceptibility/robustness trade-off.
We investigate this trade-off by varying the strength of the watermark for each model.
We report in Fig.~\ref{fig:trade-off} the bit accuracy and $\logpval$ for various models, under different transformations, against the VMAF between the watermarked and the original videos.
This is done on 3-seconds clips from SA-V.
We observe that MBRS and TrustMark obtain higher values for $-\logpval$ for a good range of VMAF since they hide more bits (256 and 100 respectively).
However, these methods fall short on more challenging transformations, especially when combining geometric transformations and video compression where \ours\ achieves higher robustness, in particular at very high PSNR ($> 50$~dB) or VMAF ($> 94$).

\section{Ablation Studies}

\subsection{Video training}\label{ablation:video_training}

In this section, we investigate whether training a video watermarking model with frame propagation and differentiable video compression is beneficial, or if applying an image watermarking model to videos during inference is sufficient.
We also investigate if it is beneficial to pre-train on images and then to continue training on a mix of images and videos. 
This could potentially leverage the faster training times of image-based models while adapting to video-specific transformations.

To test this we design three main scenarios:
\begin{enumerate}
    \item Image-only training, where the model is trained solely on images; 
    \item Video-only training, where the model is trained exclusively on videos;
    \item Mixed training, where the model is first pre-trained on images and then further trained on a mix of images and videos using a scheduled approach.
\end{enumerate}

When video training is activated, we further explore two sub-cases: 
\begin{enumerate}[label=\Alph*.]
    \item With all augmentations, including video compression,
    \item Without video compression augmentations.    
\end{enumerate}

This allows us to isolate the impact of video compression on the training process, as opposed to relying solely on differentiable frame propagation of the watermark. 
We report the mean bit accuracy over different compressions and the PSNR during training, across multiple seeds for each experiment. 
In this experiment, $\nbits=16$ to facilitate training and focus on the impact of video training.
During the video training phase, we employ a balanced schedule, alternating between images and videos with a 1:1 ratio (\ie, one epoch for images followed by one epoch for videos) from our experiments we found that this helps stabilizing the training.

\begin{figure}[b!]
    \centering
    \includegraphics[width=\linewidth]{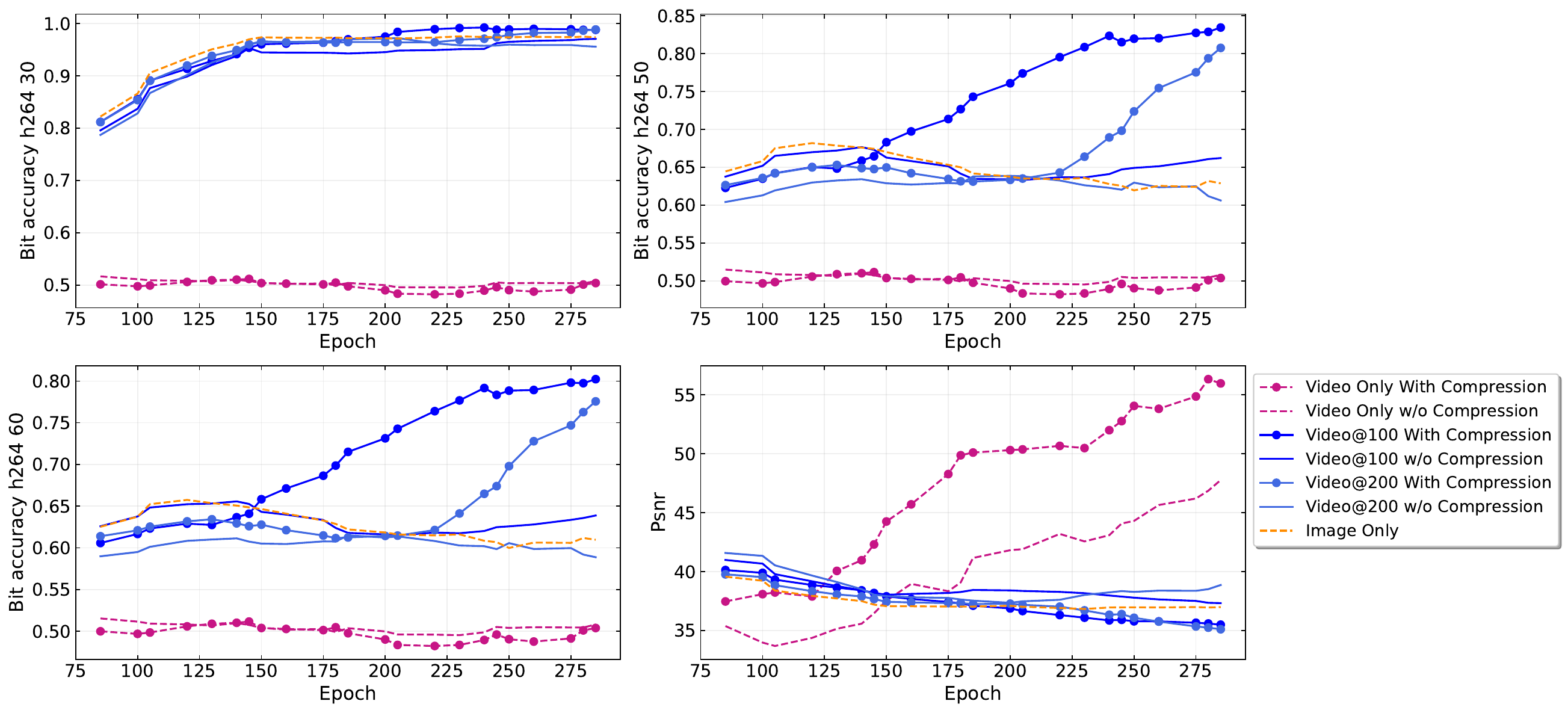}
    \caption{
    Video training with compression augmentation after image pre-training (100-200 epochs) yields the most successful training regimen, rapidly increasing bit accuracy, especially with stronger compressions (CRF 50-60), without sacrificing PSNR. 
    This approach outperforms video training alone that seems to be insufficient for a stable training, demonstrating the effectiveness of our mixed approach with image pre-training for the model optimization. 
    }\label{fig:videotraining}
\end{figure}

The results, as shown in Fig.~\ref{fig:videotraining}, reveal that the most effective combination involves pre-training on images, followed by video training with compression augmentation. This approach yields significant improvements in bit accuracy, particularly at higher compression rates. 
Notably, when video training commences (epoch 100 or 200) after image pre-training, the bit accuracy increases rapidly, especially for stronger compressions (CRF 40 and 50). This suggests that the incorporation of differentiable compression augmentation provides a robust optimization signal to the model. 
Furthermore, this improvement in robustness does not come at the cost of lower PSNR values compared to other ablations, underscoring the effectiveness of the proposed approach.

In contrast, video training alone without image pre-training proves ineffective, resulting in a very low bit accuracy. This highlights the importance of the mixed approach, which leverages image pre-training to initialize the network before training on videos. The scheduled training strategy employed in this study demonstrates the benefits of combining the efficiency of image-based models with the adaptability to video-specific transformations afforded by video training.

\subsection{Extractor fine-tuning}\label{ablation:extractor_finetuning}

\begin{figure}[b!]
    \centering
    \includegraphics[width=\linewidth]{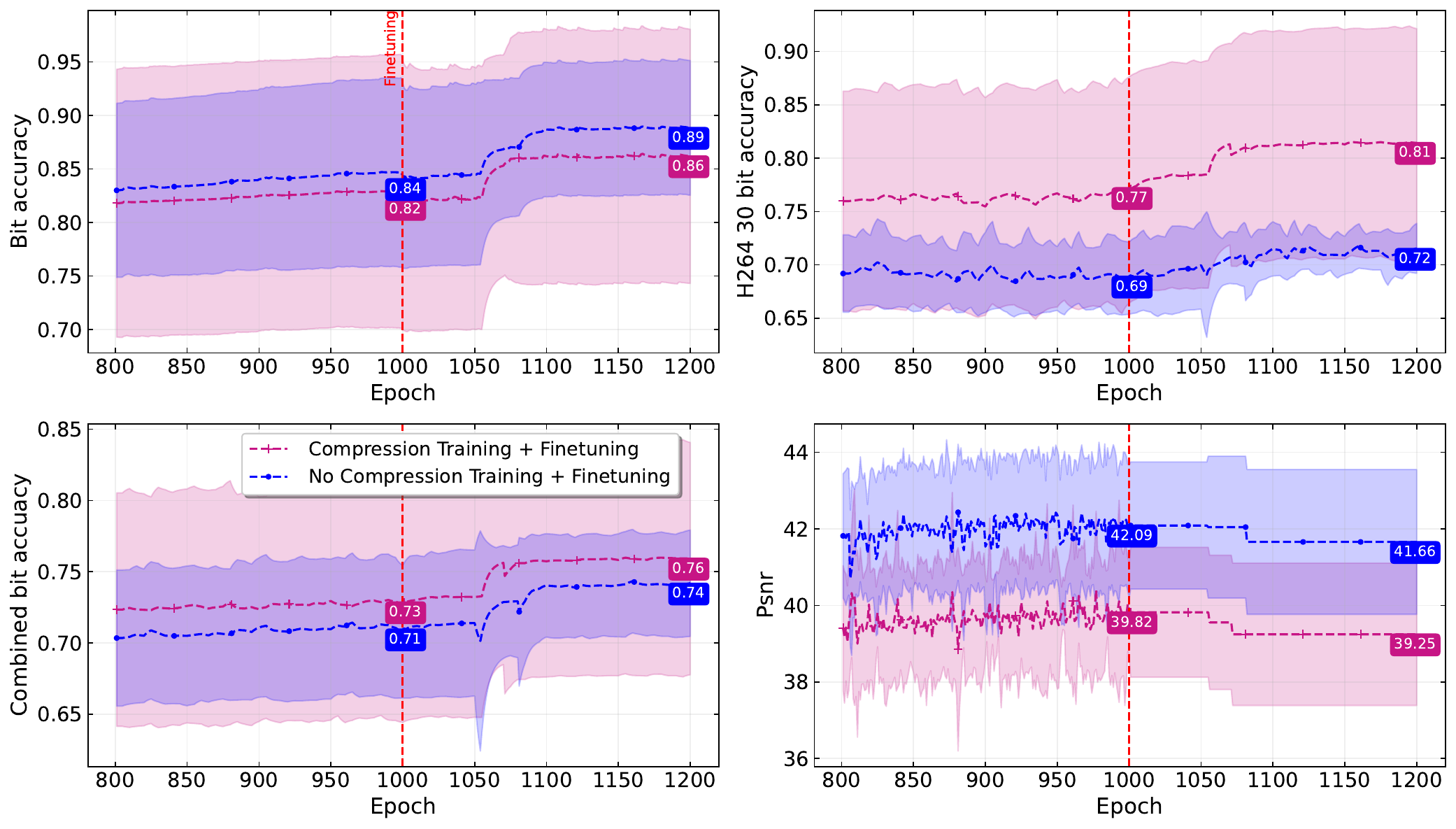}
    \caption{
    Extractor fine-tuning results. 
    Fine-tuning boosts the average training bit accuracy (top-left), bit accuracy on H.264 (CRF=30) (top-right), and on a combined augmentation with H.264, crop and brightness change (bottom-left), without influencing the PSNR (bottom-right), as the generated watermark remains unchanged. 
    All models are trained to convergence for 1000 epochs, followed by 200 epochs of fine-tuning (red dotted line).
    }\label{fig:extractor-finetuning}
\end{figure}

In this section, we investigate the impact of fine-tuning the extractor of the watermark while freezing the generator as a method to break free from the trade-off between imperceptibility and robustness. 
We expect fine-tuning to provide additional gains in bit accuracy for some models, particularly towards augmentations that have not been seen enough during training or models that haven't achieved full convergence. 
To investigate this, we train multiple models with varying parameters including numbers of bits (64 and 128) and video training start epoch (200, 500, and 1000). We train all models to convergence for 1000 epochs, then freeze the generator and fine-tune the extractor for an additional 200 epochs. We then compare two scenarios:
\begin{enumerate}
\item Training and fine-tuning with compression augmentations, where the models are trained on lightweight augmentations and leaving robustness to compressions to the end.
\item Training on all augmentations, with compression augmentations left to fine-tuning time.
\end{enumerate}

The rationale behind scenario 2. is that compression augmentations introduce instabilities in training due to the slow compression times and the small batch size needed to fit in memory.
Therefore, we investigate the benefits of leaving the compression augmentations only when the embedder is frozen.   

Our results, shown in Fig.~\ref{ablation:extractor_finetuning}, indicate that fine-tuning allows for extra gains in the average bit accuracy overall, without compromising the PSNR. 
Fine-tuning can therefore be a viable solution to enhance the robustness of the extractor without suffering from the imperceptibility/robustness trade-off.
Interestingly, our results also show that there is no significant difference in the effect of pre-training with or without compression augmentations. 
In fact, the results suggest that it is better to start with all augmentations, including compression, from the beginning.

\subsection{Video inference parameters}\label{sec:ablation-video-inference}

\paragraph{Step-size at embedding time.}
To efficiently embed the watermark in videos, we use temporal propagation presented in Sec.~\ref{subsec:video-inference}.
It involves embedding the watermark every $k$ frames, where $k$ is the step-size, and copying the watermark distortion onto the next frames.
We investigate the impact of the step-size on the watermark robustness and the speed of the embedding.
We report the bit accuracy on the same combined augmentation as in Fig.~\ref{fig:trade-off}, \ie, for an H.264 compression with CRF=30, a crop removing half of the video, and a brightness change, as well as the time taken to embed the watermark on both CPU and GPU.
We observe that the step-size $k$ does not significantly impact the watermark robustness, while greatly increasing the speed of the embedding.
However, it empirically introduces shadowy or blinkering artifacts in the video.
Therefore, the step-size should still be kept small to ensure the watermark is imperceptible when the video is moving fast (\eg, $k=4$ in our experiments).
We leave the exploration of more advanced temporal propagation techniques for future work.

\begin{figure}[b!]
    \centering
    \begin{subfigure}[b]{0.46\textwidth}
        \centering
        \includegraphics[width=\linewidth, clip, trim={0 0.4in 0 0}]{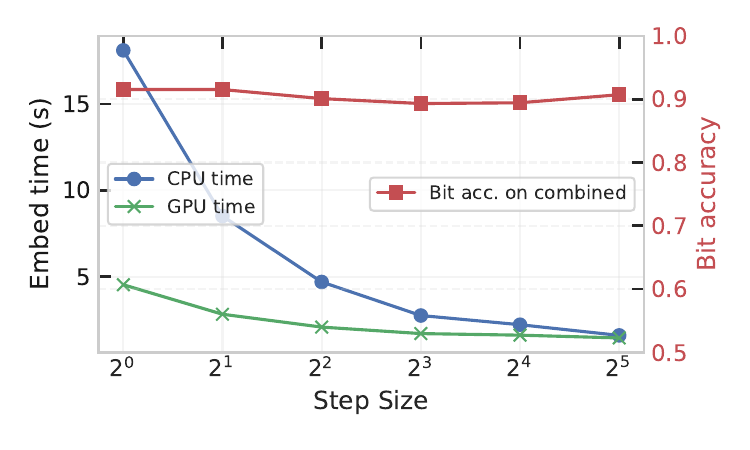}
        \caption{Step-size.}
        \label{fig:extract_time_bit_acc_vs_step_size}
    \end{subfigure}
    \hfill
    \begin{subfigure}[b]{0.46\textwidth}
        \centering
        \includegraphics[width=\linewidth, clip, trim={0 0.4in 0 0}]{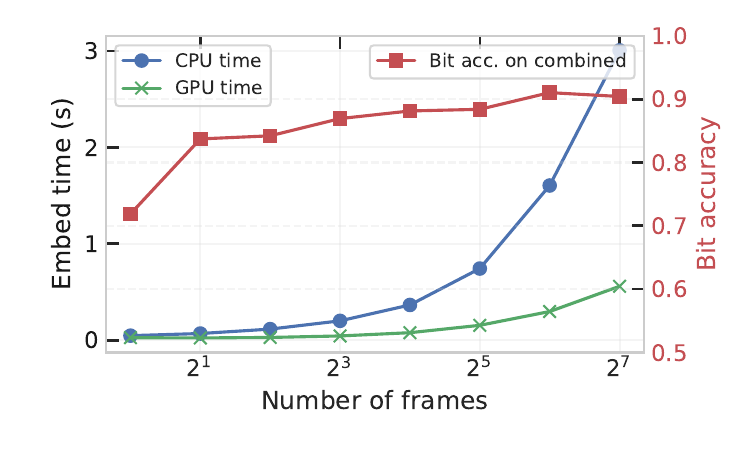}
        \caption{Number of frames.}
        \label{fig:extract_time_bit_acc_vs_num_frames}
    \end{subfigure}
    \caption{
        Ablation study on the step-size at embedding time and the number of frames at extraction time.
        Embedding and extraction are done on 5s clips. 
        The reported bit accuracy is on the same combined augmentation as in Fig.~\ref{fig:trade-off}, \ie, H.264, crop and brightness change.
        We observe that the step-size $k$ in the temporal propagation does not significantly impact the watermark robustness, while greatly increasing the speed of the embedding -- although it sometimes introduces shadow of glitter artifacts in the video.
        The number of frames $T$ at extraction time has a more significant impact on both the watermark extraction performance and the speed of the extraction.
    }\label{fig:ablation-extract-time}
\end{figure}

\paragraph{Number of frames at extraction time.}
At extraction time, we predict a soft message for each frame $i \in [1, T]$ and aggregate them into a single message.
We investigate the impact of the number of frames $T$ on the watermark extraction performance and the speed of the extraction, with the same setup as in the previous ablation.
As shown in Fig.~\ref{fig:ablation-extract-time}, the number of frames $T$ at extraction time has a more significant impact on both the watermark extraction performance and the speed of the extraction.
Notably, the bit accuracy increases with the number of frames, as the model has more information to predict the binary message.

\paragraph{Aggregation at extraction time.}
As previously stated, the extractor predicts one soft message $\tilde{\mathbf{m}_i}$ per frame $i$, which is aggregated into a single message for the entire video.
By default, the aggregation averages all the messages bit-wise, as explained in Eq.~\ref{eq:video-extraction}.
We experimentally observed that when the extraction predicts a logit $\tilde{\mathbf{m}}_{i,k}$ for a given frame $i$ and bit $k$, the logit is likely to be higher for the correct bit than for the incorrect ones.
We therefore investigate the impact of different aggregation methods on the watermark extraction performance.
We define the following ones:
\begin{itemize}[leftmargin=*, itemsep=0pt]
    \item Average, the default method, which averages the messages bit-wise: $\tilde{m}_k = \frac{1}{T} \sum_{i=1}^T \tilde{\mathbf{m}_i}_{,k}$.
    \item Squared average, which rescales each bit by its absolute value before averaging:
    $\tilde{m}_k = \frac{1}{T} \sum_{i=1}^T \abs{\tilde{\mathbf{m}_i}_{,k}}\tilde{\mathbf{m}_i}_{,k}$.
    \item L1 average, which computes the L1 norm of the frame-wise logits before averaging:
    $\tilde{m}_k = \frac{1}{T} \sum_{i=1}^T \| \tilde{\mathbf{m}_i} \|_1  \tilde{\mathbf{m}_i}_{,k}$.
    \item L2 average, which computes the L2 norm of the frame-wise logits before averaging:
    $\tilde{m}_k = \frac{1}{T} \sum_{i=1}^T \| \tilde{\mathbf{m}_i} \|_2  \tilde{\mathbf{m}_i}_{,k}$.
\end{itemize}
The final bit at position $k$ is then thresholded to obtain the hard message: $\hat{m}_k = \ind{\tilde{m}_k > 0}$.

We report in Tab.~\ref{tab:aggregation} the bit accuracy and $\logpval$ for the different aggregation methods.
The experimental setup is the same as in Sec.~\ref{sec:adaptable-inference}, \ie, we watermark 3s videos from the SA-V dataset, and run the extraction on the entire clip.
The bit accuracy and $\logpval$ are similar across the different methods, with a small improvement for the ``L1 average'', but not significant enough to justify the increased complexity.

\begin{table}[t!]
    \caption{
        Ablation study on the aggregation method for watermarking extraction on video.
        We use the same setup as in Sec.~\ref{sec:adaptable-inference}, \ie, 100 3s videos from the SA-V dataset.
        Identity, Valuemetric, Geometric, Compression, and Combined refer to the different types of transformations applied before extraction, on which the bit accuracy and $\logpval$ are averaged.
        We observe that the aggregation method does not significantly impact the watermark extraction performance.
    }
    \label{tab:aggregation}
    \centering
    {\footnotesize
    \begin{tabular}{r *{10}{c}}
        \toprule
         & \multicolumn{2}{c}{Identity} & \multicolumn{2}{c}{Valuemetric} & \multicolumn{2}{c}{Geometric} & \multicolumn{2}{c}{Compression} & \multicolumn{2}{c}{Combined} \\
         Aggregation & \multicolumn{2}{c}{Bit~acc.~/~$\logpval$} & \multicolumn{2}{c}{Bit~acc.~/~$\logpval$} & \multicolumn{2}{c}{Bit~acc.~/~$\logpval$} & \multicolumn{2}{c}{Bit~acc.~/~$\logpval$} & \multicolumn{2}{c}{Bit~acc.~/~$\logpval$} \\
        \cmidrule(lr){2-3} \cmidrule(lr){4-5} \cmidrule(lr){6-7} \cmidrule(lr){8-9} \cmidrule(lr){10-11} 
        Avg & 0.992 & -27.6 & 0.904 & -20.6 & 0.863 & -18.3 & 0.837 & -15.3 & 0.730 & -8.6 \\
        L1 avg & 0.994 & -27.9 & 0.908 & -21.1 & 0.867 & -18.6 & 0.844 & -16.0 & 0.742 & -9.3 \\
        L2 avg & 0.993 & -27.7 & 0.907 & -20.8 & 0.861 & -17.8 & 0.842 & -15.7 & 0.742 & -9.2 \\
        Squared avg & 0.989 & -27.2 & 0.906 & -20.5 & 0.857 & -17.5 & 0.843 & -15.6 & 0.742 & -9.2 \\

        \bottomrule
    \end{tabular}
    }
\end{table}

\section{Related Work}\label{sec:related}

\paragraph{Traditional video watermarking} operates within the framework of video codecs like H.264/AVC and HEVC which utilize entropy coding and motion estimation as part of their compression techniques. They can be broadly categorized into two main approaches. The first approach involves exploiting the Reversible Variable Length Codes (RVLC), which are a type of entropy coding used in video compression to represent frequently occurring symbols with shorter codes. 
In RVLC-based watermarking~\citep{biswas2005-RVLC, noorkami2007-RVLC, mobasseri2004-RVLC}, the watermark is embedded by modifying the variable length codes in a way that is reversible, meaning the original video content can be restored after extraction of the watermark. The second approach~\citep{mohaghegh2008h-motionvec} focuses on manipulating motion vectors, which are used to describe the movement of objects or blocks between frames in a video sequence. In motion vector-based watermarking, the watermark is embedded by slightly altering the motion vectors, typically those with larger magnitudes, in a way that is imperceptible to the human eye.

\paragraph{Deep-learning-based video watermarking.}

Early work on deep learning-based video watermarking models, such as VStegNet~\citep{mishra2019vstegnet} and RivaGan~\citep{zhang2019robust}, have been proposed to address the limitations of traditional methods. VStegNet introduced a deep learning architecture that achieves high embedding capacity and visual quality but lacks robustness to video distortions or compression. In contrast, RivaGan employed a GAN training architecture with an attention-based mechanism and adversarial networks to optimize for robustness. However, its use of 4D video tensors raises concerns about efficiency and usability. To simulate non-differentiable compression algorithms, RivaGan incorporated a noise layer mimicking H.264 compression using Discrete Cosine Transform (DCT). While RivaGan's open-sourced training code is available, the trained models are not, making comparisons challenging. 
\cite{weng2019high} is mostly concerned with video steganography. It focuses on hiding data in the less complex inter-frame residuals rather than directly within the more dense video frames. This work also does not consider robustness to distortions. 

DVMark~\citep{luo2023dvmark} enhances robustness in video watermarking through a multiscale design in both the encoder and decoder. This approach embeds messages across multiple spatio-temporal scales, resulting in improved robustness compared to single-scale networks. The model operates on 4D video tensors and can support variable resolutions, similar to \cite{zhu2018hidden}, without requiring downsampling or upsampling. However, this raises concerns about its efficiency and usability in practice, particularly for long videos.
To address the challenge of compression, DVMark and VHNet~\citep{shen2023vhnet} introduce a trainable CompressionNet that simulates video compression. 
This allows their networks to be optimized for robustness to compression in a differentiable way.
Other approaches include REVMark~\citep{zhang2023novel} which also uses a differentiable approximation of H.264 to simulate video compression and achieves robust watermarking for 128$\times$128 videos with a 96-bit payload, the works of \citep{zhang2024hide} and \citep{chang2024dnn}, which apply deep watermarking in the frequency domain using either DCT and Dual-Tree Complex Wavelet Transform (DT-CWT), respectively, and RIVIE~\citep{jia2022rivie}, which simulates real-world camera imaging distortions and adds temporal loss functions and a distortion network. 
Lastly, V$^2$A-Mark~\citep{zhang2024v2a} embeds two watermarks, one for tamper localization and the other to hide a 32-bits payload, but it does not report any results on geometric transformations.

ItoV~\citep{ye2023itov} is the most similar to our work.
It adapts image watermarking architectures to process videos by merging the temporal dimension with the channel dimension, allowing 2D CNNs to treat videos as images. 
This approach aims to reduce computational resources and leverage faster convergence speeds compared to 3D CNNs. 
However, it still requires feeding the entire video at once, raising questions about its efficiency and ability to handle longer videos.
Notably, ItoV employs a skip gradient trick to enable direct training on video codec augmentations, achieving good robustness against H.264 compressions at CRF=22. However, the lack of reproducibility assets limits further assessment of its robustness.

\paragraph{Image watermarking} has also been a long-standing research topic, very much intertwined with video watermarking.
Early works date back to the spatial domain methods of \cite{van1994digital}, \cite{nikolaidis1998robust}, \cite{bas2002geometrically}, as well as to the ones applying the watermark in a frequency domain such as DFT~\citep{urvoy2014perceptual}, QFT~\citep{ouyang2015qdft}, DCT~\citep{bors1996image, piva1997dct, barni1998dct}, and DWT~\citep{xia1998wavelet, barni2001improved, furon2008broken}. 
The focus has since then shifted towards deep learning, pioneered by HiDDeN~\citep{zhu2018hidden}, which has been extended by the incorporation of adversarial training~\citep{luo2020distortion}, attention filters~\citep{zhang2020robust, yu2020attention}, robust optimization~\citep{wen2019romark} or invertible networks~\citep{ma2022towards, fang2023flow}.
More recent works include new features such as the option to embed the watermark at any resolution~\citep{bui2023trustmark}, robustness to diffusion purification~\citep{pan2024jigmark} or localized extraction of one or several messages from the same image~\citep{sander2024watermark}.
A parallel line of research has recently emerged, focusing on watermarking specific to AI-generated content~\citep{yu2020attention, yu2021responsible}, with notable works including Stable Signature~\citep{fernandez2023stable}, Tree-Ring~\citep{wen2023tree}, and their follow-ups~\citep{kim2023wouaf, hong2024exact, ci2024ringid}. 
These methods aim to embed watermarks during the generation process, often providing a more robust and/or secure way to track AI-generated content.
On the other hand, \ours\ is post-hoc, meaning that, to apply it to AI-generated content, we would need to watermark after the generation, making it more flexible, but also less secure, \eg, in the case of open-sourcing the generative model.

\section{Conclusion}

In this paper, we introduce \ours, a comprehensive and efficient framework for video watermarking. 
Our work addresses the need for robust, efficient and flexible watermarking solutions coming with the increasing ease of access of video generative models and sophisticated video editing tools. 
It provides a strong open foundation for researchers and practitioners to test and iterate on.
It also highlights some open challenges of video watermarking. For instance, the need for better metrics~\citep{mantiuk2024colorvideovdp} to evaluate imperceptibility and better training objectives for it.
Future work could focus on ensuring visual consistency across watermarked frames, 
embedding in a domain better suited for video compression (\eg, YUV or YCbCr),
increasing the payload and the robustness of the watermarks, as well as exploring the security of the framework.

\clearpage

\bibliographystyle{assets/plainnat}
\bibliography{references.bib}

\begin{thebibliography}{79}
\providecommand{\natexlab}[1]{#1}
\providecommand{\url}[1]{\texttt{#1}}
\expandafter\ifx\csname urlstyle\endcsname\relax
  \providecommand{\doi}[1]{doi: #1}\else
  \providecommand{\doi}{doi: \begingroup \urlstyle{rm}\Url}\fi

\bibitem[Chi(2023)]{ChineseAIGovernance}
Chinese ai governance rules, 2023.
\newblock \url{http://www.cac.gov.cn/2023-07/13/c_1690898327029107.htm}.
\newblock Accessed on August 29, 2023.

\bibitem[Eur(2023)]{EuropeanAIAct}
European ai act, 2023.
\newblock \url{https://artificialintelligenceact.eu/}.
\newblock Accessed on August 29, 2023.

\bibitem[{Alliance for Open Media}(2018)]{AV12018}
{Alliance for Open Media}.
\newblock Av1 bitstream \& decoding process specification, 2018.
\newblock \url{https://aomediacodec.github.io/av1-spec/av1-spec.pdf}.

\bibitem[Barni et~al.(1998)Barni, Bartolini, Cappellini, and Piva]{barni1998dct}
Mauro Barni, Franco Bartolini, Vito Cappellini, and Alessandro Piva.
\newblock A dct-domain system for robust image watermarking.
\newblock \emph{Signal processing}, 66\penalty0 (3):\penalty0 357--372, 1998.

\bibitem[Barni et~al.(2001)Barni, Bartolini, and Piva]{barni2001improved}
Mauro Barni, Franco Bartolini, and Alessandro Piva.
\newblock Improved wavelet-based watermarking through pixel-wise masking.
\newblock \emph{IEEE transactions on image processing}, 10\penalty0 (5):\penalty0 783--791, 2001.

\bibitem[Bas et~al.(2002)Bas, Chassery, and Macq]{bas2002geometrically}
Patrick Bas, J-M Chassery, and Benoit Macq.
\newblock Geometrically invariant watermarking using feature points.
\newblock \emph{IEEE transactions on image Processing}, 11\penalty0 (9):\penalty0 1014--1028, 2002.

\bibitem[Bengio et~al.(2013)Bengio, L{\'e}onard, and Courville]{bengio2013estimating}
Yoshua Bengio, Nicholas L{\'e}onard, and Aaron Courville.
\newblock Estimating or propagating gradients through stochastic neurons for conditional computation.
\newblock \emph{arXiv preprint arXiv:1308.3432}, 2013.

\bibitem[Biswas et~al.(2005)Biswas, Das, and Petriu]{biswas2005-RVLC}
Satyendra Biswas, Sunil~R Das, and Emil~M Petriu.
\newblock An adaptive compressed mpeg-2 video watermarking scheme.
\newblock \emph{IEEE transactions on Instrumentation and Measurement}, 54\penalty0 (5):\penalty0 1853--1861, 2005.

\bibitem[Bors and Pitas(1996)]{bors1996image}
Adrian~G Bors and Ioannis Pitas.
\newblock Image watermarking using dct domain constraints.
\newblock In \emph{ICIP}, 1996.

\bibitem[Brooks et~al.(2024)Brooks, Peebles, Holmes, DePue, Guo, Jing, Schnurr, Taylor, Luhman, Luhman, Ng, Wang, and Ramesh]{brooks2024video}
Tim Brooks, Bill Peebles, Connor Holmes, Will DePue, Yufei Guo, Li~Jing, David Schnurr, Joe Taylor, Troy Luhman, Eric Luhman, Clarence Ng, Ricky Wang, and Aditya Ramesh.
\newblock Video generation models as world simulators, 2024.
\newblock \url{https://openai.com/research/video-generation-models-as-world-simulators}.

\bibitem[Bui et~al.(2023)Bui, Agarwal, and Collomosse]{bui2023trustmark}
Tu~Bui, Shruti Agarwal, and John Collomosse.
\newblock Trustmark: Universal watermarking for arbitrary resolution images.
\newblock \emph{arXiv preprint arXiv:2311.18297}, 2023.

\bibitem[{California State Leg.}(2024)]{ca_ab3211_2024}
{California State Leg.}
\newblock Amendment to california assembly bill 3211.
\newblock California State Legislature, April 2024.
\newblock \url{https://legiscan.com/CA/text/AB3211/id/2984195}.
\newblock Amended in Assembly.

\bibitem[Chang et~al.(2024)Chang, Chen, Ding, and Liao]{chang2024dnn}
Xuanming Chang, Beijing Chen, Weiping Ding, and Xin Liao.
\newblock A dnn robust video watermarking method in dual-tree complex wavelet transform domain.
\newblock \emph{Journal of Information Security and Applications}, 85:\penalty0 103868, 2024.

\bibitem[Ci et~al.(2024)Ci, Yang, Song, and Shou]{ci2024ringid}
Hai Ci, Pei Yang, Yiren Song, and Mike~Zheng Shou.
\newblock Ringid: Rethinking tree-ring watermarking for enhanced multi-key identification.
\newblock \emph{arXiv preprint arXiv:2404.14055}, 2024.

\bibitem[Cover(1999)]{cover1999elements}
Thomas~M Cover.
\newblock \emph{Elements of information theory}.
\newblock John Wiley \& Sons, 1999.

\bibitem[Czolbe et~al.(2020)Czolbe, Krause, Cox, and Igel]{czolbe2020loss}
Steffen Czolbe, Oswin Krause, Ingemar Cox, and Christian Igel.
\newblock A loss function for generative neural networks based on watson’s perceptual model.
\newblock \emph{Advances in Neural Information Processing Systems}, 33:\penalty0 2051--2061, 2020.

\bibitem[D{\'e}fossez et~al.(2022)D{\'e}fossez, Copet, Synnaeve, and Adi]{defossez2022high}
Alexandre D{\'e}fossez, Jade Copet, Gabriel Synnaeve, and Yossi Adi.
\newblock High fidelity neural audio compression.
\newblock \emph{arXiv preprint arXiv:2210.13438}, 2022.

\bibitem[Dosovitskiy(2020)]{dosovitskiy2020image}
Alexey Dosovitskiy.
\newblock An image is worth 16x16 words: Transformers for image recognition at scale.
\newblock \emph{arXiv preprint arXiv:2010.11929}, 2020.

\bibitem[Elfwing et~al.(2018)Elfwing, Uchibe, and Doya]{elfwing2018sigmoid}
Stefan Elfwing, Eiji Uchibe, and Kenji Doya.
\newblock Sigmoid-weighted linear units for neural network function approximation in reinforcement learning.
\newblock \emph{Neural networks}, 107:\penalty0 3--11, 2018.

\bibitem[Fang et~al.(2023)Fang, Qiu, Chen, Zhang, Zhang, and Chang]{fang2023flow}
Han Fang, Yupeng Qiu, Kejiang Chen, Jiyi Zhang, Weiming Zhang, and Ee-Chien Chang.
\newblock Flow-based robust watermarking with invertible noise layer for black-box distortions.
\newblock In \emph{Proceedings of the AAAI conference on artificial intelligence}, volume~37, pages 5054--5061, 2023.

\bibitem[Fernandez et~al.(2023)Fernandez, Couairon, J{\'e}gou, Douze, and Furon]{fernandez2023stable}
Pierre Fernandez, Guillaume Couairon, Herv{\'e} J{\'e}gou, Matthijs Douze, and Teddy Furon.
\newblock The stable signature: Rooting watermarks in latent diffusion models.
\newblock In \emph{International Conference on Computer Vision}, pages 22466--22477, 2023.

\bibitem[Furon and Bas(2008)]{furon2008broken}
Teddy Furon and Patrick Bas.
\newblock Broken arrows.
\newblock \emph{EURASIP Journal on Information Security}, 2008:\penalty0 1--13, 2008.

\bibitem[Hong et~al.(2024)Hong, Lee, Jeon, Bae, and Chun]{hong2024exact}
Seongmin Hong, Kyeonghyun Lee, Suh~Yoon Jeon, Hyewon Bae, and Se~Young Chun.
\newblock On exact inversion of dpm-solvers.
\newblock In \emph{Proceedings of the IEEE/CVF Conference on Computer Vision and Pattern Recognition}, pages 7069--7078, 2024.

\bibitem[Isola et~al.(2017)Isola, Zhu, Zhou, and Efros]{isola2017image}
Phillip Isola, Jun-Yan Zhu, Tinghui Zhou, and Alexei~A Efros.
\newblock Image-to-image translation with conditional adversarial networks.
\newblock In \emph{Proceedings of the IEEE conference on computer vision and pattern recognition}, pages 1125--1134, 2017.

\bibitem[Jia et~al.(2022)Jia, Gao, Zhu, Min, Hu, and Zhai]{jia2022rivie}
Jun Jia, Zhongpai Gao, Dandan Zhu, Xiongkuo Min, Menghan Hu, and Guangtao Zhai.
\newblock Rivie: Robust inherent video information embedding.
\newblock \emph{IEEE Transactions on Multimedia}, 25:\penalty0 7364--7377, 2022.

\bibitem[Jia et~al.(2021)Jia, Fang, and Zhang]{jia2021mbrs}
Zhaoyang Jia, Han Fang, and Weiming Zhang.
\newblock Mbrs: Enhancing robustness of dnn-based watermarking by mini-batch of real and simulated jpeg compression.
\newblock In \emph{Proceedings of the 29th ACM international conference on multimedia}, pages 41--49, 2021.

\bibitem[Kim et~al.(2023)Kim, Min, Patel, Cheng, and Yang]{kim2023wouaf}
Changhoon Kim, Kyle Min, Maitreya Patel, Sheng Cheng, and Yezhou Yang.
\newblock Wouaf: Weight modulation for user attribution and fingerprinting in text-to-image diffusion models.
\newblock \emph{arXiv preprint arXiv:2306.04744}, 2023.

\bibitem[Kirillov et~al.(2023)Kirillov, Mintun, Ravi, Mao, Rolland, Gustafson, Xiao, Whitehead, Berg, Lo, et~al.]{kirillov2023segment}
Alexander Kirillov, Eric Mintun, Nikhila Ravi, Hanzi Mao, Chloe Rolland, Laura Gustafson, Tete Xiao, Spencer Whitehead, Alexander~C Berg, Wan-Yen Lo, et~al.
\newblock Segment anything.
\newblock In \emph{Proceedings of the IEEE/CVF International Conference on Computer Vision}, pages 4015--4026, 2023.

\bibitem[Lim and Ye(2017)]{lim2017geometric}
Jae~Hyun Lim and Jong~Chul Ye.
\newblock Geometric gan.
\newblock \emph{arXiv preprint arXiv:1705.02894}, 2017.

\bibitem[Lin et~al.(2014)Lin, Maire, Belongie, Hays, Perona, Ramanan, Doll{\'a}r, and Zitnick]{lin2014microsoft}
Tsung-Yi Lin, Michael Maire, Serge Belongie, James Hays, Pietro Perona, Deva Ramanan, Piotr Doll{\'a}r, and C~Lawrence Zitnick.
\newblock Microsoft coco: Common objects in context.
\newblock In \emph{Computer Vision--ECCV 2014: 13th European Conference, Zurich, Switzerland, September 6-12, 2014, Proceedings, Part V 13}, pages 740--755. Springer, 2014.

\bibitem[Loshchilov and Hutter(2016)]{loshchilov2016sgdr}
Ilya Loshchilov and Frank Hutter.
\newblock Sgdr: Stochastic gradient descent with warm restarts.
\newblock \emph{arXiv preprint arXiv:1608.03983}, 2016.

\bibitem[Loshchilov and Hutter(2018)]{loshchilov2017decoupled}
Ilya Loshchilov and Frank Hutter.
\newblock Decoupled weight decay regularization.
\newblock In \emph{ICLR}, 2018.

\bibitem[Luo et~al.(2020)Luo, Zhan, Chang, Yang, and Milanfar]{luo2020distortion}
Xiyang Luo, Ruohan Zhan, Huiwen Chang, Feng Yang, and Peyman Milanfar.
\newblock Distortion agnostic deep watermarking.
\newblock In \emph{CVPR}, 2020.

\bibitem[Luo et~al.(2023)Luo, Li, Chang, Liu, Milanfar, and Yang]{luo2023dvmark}
Xiyang Luo, Yinxiao Li, Huiwen Chang, Ce~Liu, Peyman Milanfar, and Feng Yang.
\newblock Dvmark: a deep multiscale framework for video watermarking.
\newblock \emph{IEEE Transactions on Image Processing}, 2023.

\bibitem[Ma et~al.(2022)Ma, Guo, Hou, Yang, Li, Jia, and Xie]{ma2022towards}
Rui Ma, Mengxi Guo, Yi~Hou, Fan Yang, Yuan Li, Huizhu Jia, and Xiaodong Xie.
\newblock Towards blind watermarking: Combining invertible and non-invertible mechanisms.
\newblock In \emph{Proceedings of the 30th ACM International Conference on Multimedia}, pages 1532--1542, 2022.

\bibitem[Mantiuk et~al.(2024)Mantiuk, Hanji, Ashraf, Asano, and Chapiro]{mantiuk2024colorvideovdp}
Rafal~K Mantiuk, Param Hanji, Maliha Ashraf, Yuta Asano, and Alexandre Chapiro.
\newblock Colorvideovdp: A visual difference predictor for image, video and display distortions.
\newblock \emph{arXiv preprint arXiv:2401.11485}, 2024.

\bibitem[Marcel and Rodriguez(2010)]{marcel2010torchvision}
S{\'e}bastien Marcel and Yann Rodriguez.
\newblock Torchvision the machine-vision package of torch.
\newblock In \emph{International Conference on Multimedia}. ACM, 2010.

\bibitem[Mishra et~al.(2019)Mishra, Kumar, Nigam, and Islam]{mishra2019vstegnet}
Aayush Mishra, Suraj Kumar, Aditya Nigam, and Saiful Islam.
\newblock Vstegnet: Video steganography network using spatio-temporal features and micro-bottleneck.
\newblock In \emph{The British Machine Vision Conference}, page 274, 2019.

\bibitem[Mobasseri and Cinalli(2004)]{mobasseri2004-RVLC}
Bijan~G Mobasseri and Domenick Cinalli.
\newblock Reversible watermarking using two-way decodable codes.
\newblock In \emph{Security, Steganography, and Watermarking of Multimedia Contents VI}, volume 5306, pages 397--404. SPIE, 2004.

\bibitem[Mohaghegh and Fatemi(2008)]{mohaghegh2008h-motionvec}
Najla Mohaghegh and Omid Fatemi.
\newblock H. 264 copyright protection with motion vector watermarking.
\newblock In \emph{2008 International Conference on Audio, Language and Image Processing}, pages 1384--1389. IEEE, 2008.

\bibitem[Netflix(2016)]{vmaf}
Netflix.
\newblock Vmaf - video multi-method assessment fusion.
\newblock \url{https://github.com/Netflix/vmaf}, 2016.
\newblock Accessed: 2024-11-18.

\bibitem[Nikolaidis and Pitas(1998)]{nikolaidis1998robust}
Nikos Nikolaidis and Ioannis Pitas.
\newblock Robust image watermarking in the spatial domain.
\newblock \emph{Signal processing}, 1998.

\bibitem[Noorkami and Mersereau(2007)]{noorkami2007-RVLC}
Maneli Noorkami and Russell~M Mersereau.
\newblock A framework for robust watermarking of h. 264-encoded video with controllable detection performance.
\newblock \emph{IEEE Transactions on information forensics and security}, 2\penalty0 (1):\penalty0 14--23, 2007.

\bibitem[Odena et~al.(2016)Odena, Dumoulin, and Olah]{odena2016deconvolution}
Augustus Odena, Vincent Dumoulin, and Chris Olah.
\newblock Deconvolution and checkerboard artifacts.
\newblock \emph{Distill}, 1\penalty0 (10):\penalty0 e3, 2016.

\bibitem[Ouyang et~al.(2015)Ouyang, Coatrieux, Chen, and Shu]{ouyang2015qdft}
Junlin Ouyang, Gouenou Coatrieux, Beijing Chen, and Huazhong Shu.
\newblock Color image watermarking based on quaternion fourier transform and improved uniform log-polar mapping.
\newblock \emph{Computers \& Electrical Engineering}, 2015.

\bibitem[Pan et~al.(2024)Pan, Zeng, Lin, Yu, Hsieh, Henderson, and Jia]{pan2024jigmark}
Minzhou Pan, Yi~Zeng, Xue Lin, Ning Yu, Cho-Jui Hsieh, Peter Henderson, and Ruoxi Jia.
\newblock Jigmark: A black-box approach for enhancing image watermarks against diffusion model edits.
\newblock \emph{arXiv preprint arXiv:2406.03720}, 2024.

\bibitem[Piva et~al.(1997)Piva, Barni, Bartolini, and Cappellini]{piva1997dct}
Alessandro Piva, Mauro Barni, Franco Bartolini, and Vito Cappellini.
\newblock Dct-based watermark recovering without resorting to the uncorrupted original image.
\newblock In \emph{Proceedings of international conference on image processing}, volume~1, pages 520--523. IEEE, 1997.

\bibitem[Polyak et~al.(2024)Polyak, Zohar, Brown, Tjandra, Sinha, Lee, Vyas, Shi, Ma, Chuang, et~al.]{polyak2024movie}
Adam Polyak, Amit Zohar, Andrew Brown, Andros Tjandra, Animesh Sinha, Ann Lee, Apoorv Vyas, Bowen Shi, Chih-Yao Ma, Ching-Yao Chuang, et~al.
\newblock Movie gen: A cast of media foundation models.
\newblock \emph{arXiv preprint arXiv:2410.13720}, 2024.

\bibitem[Ramesh et~al.(2022)Ramesh, Dhariwal, Nichol, Chu, and Chen]{ramesh2022dalle2}
Aditya Ramesh, Prafulla Dhariwal, Alex Nichol, Casey Chu, and Mark Chen.
\newblock Hierarchical text-conditional image generation with clip latents.
\newblock \emph{arXiv preprint arXiv:2204.06125}, 2022.

\bibitem[Ravi et~al.(2024)Ravi, Gabeur, Hu, Hu, Ryali, Ma, Khedr, R{\"a}dle, Rolland, Gustafson, Mintun, Pan, Alwala, Carion, Wu, Girshick, Doll{\'a}r, and Feichtenhofer]{ravi2024sam2}
Nikhila Ravi, Valentin Gabeur, Yuan-Ting Hu, Ronghang Hu, Chaitanya Ryali, Tengyu Ma, Haitham Khedr, Roman R{\"a}dle, Chloe Rolland, Laura Gustafson, Eric Mintun, Junting Pan, Kalyan~Vasudev Alwala, Nicolas Carion, Chao-Yuan Wu, Ross Girshick, Piotr Doll{\'a}r, and Christoph Feichtenhofer.
\newblock Sam 2: Segment anything in images and videos.
\newblock \emph{arXiv preprint arXiv:2408.00714}, 2024.
\newblock \url{https://arxiv.org/abs/2408.00714}.

\bibitem[Richardson(2010)]{richardson2010}
Iain~E. Richardson.
\newblock \emph{The H.264 Advanced Video Compression Standard}.
\newblock John Wiley \& Sons, 2nd edition, 2010.

\bibitem[Rombach et~al.(2022)Rombach, Blattmann, Lorenz, Esser, and Ommer]{rombach2022high}
Robin Rombach, Andreas Blattmann, Dominik Lorenz, Patrick Esser, and Bj{\"o}rn Ommer.
\newblock High-resolution image synthesis with latent diffusion models.
\newblock In \emph{Proceedings of the IEEE/CVF Conference on Computer Vision and Pattern Recognition}, pages 10684--10695, 2022.

\bibitem[Ronneberger et~al.(2015)Ronneberger, Fischer, and Brox]{ronneberger2015u}
Olaf Ronneberger, Philipp Fischer, and Thomas Brox.
\newblock U-net: Convolutional networks for biomedical image segmentation.
\newblock In \emph{Medical image computing and computer-assisted intervention--MICCAI 2015: 18th international conference, Munich, Germany, October 5-9, 2015, proceedings, part III 18}, pages 234--241. Springer, 2015.

\bibitem[Saharia et~al.(2022)Saharia, Chan, Saxena, Li, Whang, Denton, Ghasemipour, Gontijo~Lopes, Karagol~Ayan, Salimans, et~al.]{saharia2022photorealistic}
Chitwan Saharia, William Chan, Saurabh Saxena, Lala Li, Jay Whang, Emily~L Denton, Kamyar Ghasemipour, Raphael Gontijo~Lopes, Burcu Karagol~Ayan, Tim Salimans, et~al.
\newblock Photorealistic text-to-image diffusion models with deep language understanding.
\newblock \emph{Advances in neural information processing systems}, 35:\penalty0 36479--36494, 2022.

\bibitem[San~Roman et~al.(2024)San~Roman, Fernandez, Elsahar, D{\'e}fossez, Furon, and Tran]{san2024proactive}
Robin San~Roman, Pierre Fernandez, Hady Elsahar, Alexandre D{\'e}fossez, Teddy Furon, and Tuan Tran.
\newblock Proactive detection of voice cloning with localized watermarking.
\newblock In \emph{International Conference on Machine Learning}, volume 235, 2024.

\bibitem[Sander et~al.(2024)Sander, Fernandez, Durmus, Furon, and Douze]{sander2024watermark}
Tom Sander, Pierre Fernandez, Alain Durmus, Teddy Furon, and Matthijs Douze.
\newblock Watermark anything with localized messages.
\newblock \emph{arXiv preprint arXiv:2411.07231}, 2024.

\bibitem[Shen et~al.(2023)Shen, Yao, Tan, and Qin]{shen2023vhnet}
Xiaofeng Shen, Heng Yao, Shunquan Tan, and Chuan Qin.
\newblock Vhnet: A video hiding network with robustness to video coding.
\newblock \emph{Journal of Information Security and Applications}, 75:\penalty0 103515, 2023.

\bibitem[Touvron et~al.(2021)Touvron, Cord, Douze, Massa, Sablayrolles, and J{\'e}gou]{touvron2021training}
Hugo Touvron, Matthieu Cord, Matthijs Douze, Francisco Massa, Alexandre Sablayrolles, and Herv{\'e} J{\'e}gou.
\newblock Training data-efficient image transformers \& distillation through attention.
\newblock In \emph{International conference on machine learning}, pages 10347--10357. PMLR, 2021.

\bibitem[Urvoy et~al.(2014)Urvoy, Goudia, and Autrusseau]{urvoy2014perceptual}
Matthieu Urvoy, Dalila Goudia, and Florent Autrusseau.
\newblock Perceptual dft watermarking with improved detection and robustness to geometrical distortions.
\newblock \emph{IEEE Transactions on Information Forensics and Security}, 2014.

\bibitem[{USA}(2023)]{USAIAnnouncement}
{USA}.
\newblock Ensuring safe, secure, and trustworthy ai.
\newblock \url{https://www.whitehouse.gov/wp-content/uploads/2023/07/Ensuring-Safe-Secure-and-Trustworthy-AI.pdf}, July 2023.
\newblock Accessed: [july 2023].

\bibitem[Van~Schyndel et~al.(1994)Van~Schyndel, Tirkel, and Osborne]{van1994digital}
Ron~G Van~Schyndel, Andrew~Z Tirkel, and Charles~F Osborne.
\newblock A digital watermark.
\newblock In \emph{Proceedings of 1st international conference on image processing}, volume~2, pages 86--90. IEEE, 1994.

\bibitem[Wang et~al.(2004)Wang, Bovik, Sheikh, and Simoncelli]{wang2004image}
Zhou Wang, Alan~C Bovik, Hamid~R Sheikh, and Eero~P Simoncelli.
\newblock Image quality assessment: from error visibility to structural similarity.
\newblock \emph{IEEE transactions on image processing}, 13\penalty0 (4):\penalty0 600--612, 2004.

\bibitem[Wen and Aydore(2019)]{wen2019romark}
Bingyang Wen and Sergul Aydore.
\newblock Romark: A robust watermarking system using adversarial training.
\newblock \emph{arXiv preprint arXiv:1910.01221}, 2019.

\bibitem[Wen et~al.(2023)Wen, Kirchenbauer, Geiping, and Goldstein]{wen2023tree}
Yuxin Wen, John Kirchenbauer, Jonas Geiping, and Tom Goldstein.
\newblock Tree-ring watermarks: Fingerprints for diffusion images that are invisible and robust.
\newblock \emph{arXiv preprint arXiv:2305.20030}, 2023.

\bibitem[Weng et~al.(2019)Weng, Li, Chi, and Mu]{weng2019high}
Xinyu Weng, Yongzhi Li, Lu~Chi, and Yadong Mu.
\newblock High-capacity convolutional video steganography with temporal residual modeling.
\newblock In \emph{Proceedings of the 2019 on international conference on multimedia retrieval}, pages 87--95, 2019.

\bibitem[Xia et~al.(1998)Xia, Boncelet, and Arce]{xia1998wavelet}
Xiang-Gen Xia, Charles~G Boncelet, and Gonzalo~R Arce.
\newblock Wavelet transform based watermark for digital images.
\newblock \emph{Optics Express}, 1998.

\bibitem[Xian et~al.(2024)Xian, Wang, Bi, Srinivasa, Kundu, Hong, and Ding]{xian2024raw}
Xun Xian, Ganghua Wang, Xuan Bi, Jayanth Srinivasa, Ashish Kundu, Mingyi Hong, and Jie Ding.
\newblock Raw: A robust and agile plug-and-play watermark framework for ai-generated images with provable guarantees.
\newblock \emph{arXiv preprint arXiv:2403.18774}, 2024.

\bibitem[Ye et~al.(2023)Ye, Gao, Wang, Song, and Wei]{ye2023itov}
Guanhui Ye, Jiashi Gao, Yuchen Wang, Liyan Song, and Xuetao Wei.
\newblock Itov: efficiently adapting deep learning-based image watermarking to video watermarking.
\newblock In \emph{2023 International Conference on Culture-Oriented Science and Technology (CoST)}, pages 192--197. IEEE, 2023.

\bibitem[Yu(2020)]{yu2020attention}
Chong Yu.
\newblock Attention based data hiding with generative adversarial networks.
\newblock In \emph{AAAI}, 2020.

\bibitem[Yu et~al.(2021)Yu, Skripniuk, Chen, Davis, and Fritz]{yu2021responsible}
Ning Yu, Vladislav Skripniuk, Dingfan Chen, Larry~S Davis, and Mario Fritz.
\newblock Responsible disclosure of generative models using scalable fingerprinting.
\newblock In \emph{International Conference on Learning Representations}, 2021.

\bibitem[Zhang and Sennrich(2019)]{zhang2019root}
Biao Zhang and Rico Sennrich.
\newblock Root mean square layer normalization.
\newblock \emph{Advances in Neural Information Processing Systems}, 32, 2019.

\bibitem[Zhang et~al.(2021)Zhang, Karjauv, Benz, and Kweon]{zhang2021asl}
Chaoning Zhang, Adil Karjauv, Philipp Benz, and In~So Kweon.
\newblock Towards robust deep hiding under non-differentiable distortions for practical blind watermarking.
\newblock In \emph{Proceedings of the 29th ACM International Conference on Multimedia}, pages 5158--5166, 2021.

\bibitem[Zhang et~al.(2020)Zhang, Wang, Cao, Shen, and Li]{zhang2020robust}
Honglei Zhang, Hu~Wang, Yuanzhouhan Cao, Chunhua Shen, and Yidong Li.
\newblock Robust watermarking using inverse gradient attention.
\newblock \emph{arXiv preprint arXiv:2011.10850}, 2020.

\bibitem[Zhang et~al.(2019)Zhang, Xu, Cuesta-Infante, and Veeramachaneni]{zhang2019robust}
Kevin~Alex Zhang, Lei Xu, Alfredo Cuesta-Infante, and Kalyan Veeramachaneni.
\newblock Robust invisible video watermarking with attention.
\newblock \emph{arXiv preprint arXiv:1909.01285}, 2019.

\bibitem[Zhang et~al.(2018)Zhang, Isola, Efros, Shechtman, and Wang]{zhang2018perceptual}
Richard Zhang, Phillip Isola, Alexei~A Efros, Eli Shechtman, and Oliver Wang.
\newblock The unreasonable effectiveness of deep features as a perceptual metric.
\newblock In \emph{CVPR}, 2018.

\bibitem[Zhang et~al.(2024{\natexlab{a}})Zhang, Xu, Li, Yu, Li, Xu, and Zhang]{zhang2024v2a}
Xuanyu Zhang, Youmin Xu, Runyi Li, Jiwen Yu, Weiqi Li, Zhipei Xu, and Jian Zhang.
\newblock V2a-mark: Versatile deep visual-audio watermarking for manipulation localization and copyright protection.
\newblock \emph{arXiv preprint arXiv:2404.16824}, 2024{\natexlab{a}}.

\bibitem[Zhang et~al.(2023)Zhang, Ni, Su, and Liao]{zhang2023novel}
Yulin Zhang, Jiangqun Ni, Wenkang Su, and Xin Liao.
\newblock A novel deep video watermarking framework with enhanced robustness to h. 264/avc compression.
\newblock In \emph{Proceedings of the 31st ACM International Conference on Multimedia}, pages 8095--8104, 2023.

\bibitem[Zhang et~al.(2024{\natexlab{b}})Zhang, Wang, Wang, and Wu]{zhang2024hide}
Zhiwei Zhang, Han Wang, Guisong Wang, and Xinxiao Wu.
\newblock Hide and track: Towards blind video watermarking network in frequency domain.
\newblock \emph{Neurocomputing}, 579:\penalty0 127435, 2024{\natexlab{b}}.

\bibitem[Zhu et~al.(2018)Zhu, Kaplan, Johnson, and Fei-Fei]{zhu2018hidden}
Jiren Zhu, Russell Kaplan, Justin Johnson, and Li~Fei-Fei.
\newblock Hidden: Hiding data with deep networks.
\newblock In \emph{Proceedings of the European conference on computer vision (ECCV)}, pages 657--672, 2018.

\end{thebibliography}

\clearpage
\beginappendix
\section{Theoretical Analyses}

\subsection{Comparing at different payloads}\label{app:payload}

We consider a binary message $m \in \{0, 1\}^{\nbits}$
and its estimate $\hat{m}$ after the process of watermark embedding, edition and watermark extraction.
This transmission is measured with a certain accuracy $\bitacc(m, \hat{m})$, which does not take into account the payload $\nbits$.
We thus introduce two metrics to be able to compare the performance of models operating at different payloads $\nbits$. %

We consider that each bit is a binary symmetric channel (BSC) with a probability of error $p$.
Its entropy is given by $h(p) = -p \log_2 p - (1-p) \log_2 (1-p)$, and its capacity is $c(p) = 1 - h(p)$.
If $\nbits$ such channels exist, the total capacity is $c(p) \times \nbits$.
In our case, we assume that, given an observed bit accuracy $\bitacc(m, \hat{m})$, each bit is a BSC with a probability of error defined $p = 1 - \bitacc(m, \hat{m})$.
We define the expected capacity as:
\begin{equation}
    C(p) = \nbits \times  \left( 1 - \left( -p \log_2 p - (1-p) \log_2 p \right) \right),
\end{equation}
where $p = \bitacc(m, \hat{m})$.
It represents the number of bits that would be theoretically transmittable from a Shannon perspective, if we assumed that the observed bit accuracy is the true probability of error.

Another way to approach the problem is to consider it as a statistical detection test.
We consider the null hypothesis $H_0$ that each bit of the output binary message $\hat{m}$ is independent and distributed as a Bernoulli variable with probability of success $0.5$, and the alternative hypothesis $H_1$ which is that $\hat{m} = m$.
Given an observed bit accuracy $\bitacc(m, \hat{m})$, 
the $p$-value is the probability of observing a bit accuracy at least as extreme as the one obtained under the null hypothesis.
It is given by the cumulative distribution function of the binomial distribution:
\begin{equation}
    \pval(m, \hat{m}) = \sum_{k \geq \nbits p }^{\nbits} \binom{\nbits}{k} 1/2^{\nbits} = I_{1/2} ( \nbits\,p, \nbits\,(1 - p) + 1), 
\end{equation}
where $p = \bitacc(m, \hat{m})$, and where the c.d.f. of the binomial is expressed by $I_x(a, b)$, the regularized incomplete Beta function.

\begin{figure}[h!]
    \centering
    \includegraphics[width=0.9\linewidth]{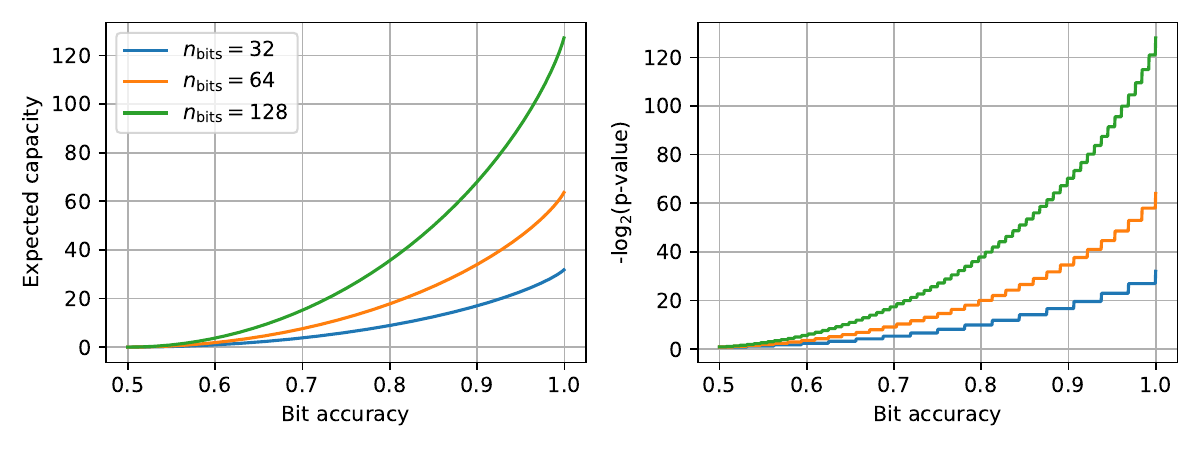}
    \caption{
      Expected capacity and $p$-value as a function of the number of bits.
    }\label{fig:payload}
\end{figure}

In Fig. \ref{fig:payload}, we show the expected capacity and the $\log_2$ of the $p$-value, as a function of the number of bits and the bit accuracy.
Interestingly, we observe that both metrics follow the exact same trend, with discontinuities for the $p$-value due to the discrete nature of the binomial distribution.
In these plots, we can for instance see that a bit accuracy of $0.9$ for a payload of $64$ bits would be approximately equivalent to a bit accuracy of $0.8$ for a payload of $128$ bits, in terms of expected capacity or $p$-value.

Note that the $p$-value and capacity discussed in this context are part of a theoretical analysis aimed at evaluating methods in binary message transmission. 
Unlike the traditional $p$-value used in statistical hypothesis testing, which assesses the likelihood of observing a bit accuracy as extreme as the observed one under $H_0$, this $p$-value is not directly related to the actual outcomes of a statistical test. 
It is purely a conceptual tool to analyze and compare different scenarios of bit accuracy and payload sizes.

\section{Additional Details and Results}

\subsection{More qualitative results}\label{app:qualitative}

\begin{figure*}[b!]
    \centering
    \scriptsize
    \newcommand{\imwidth}{0.14\textwidth}
    \setlength{\tabcolsep}{0pt}
    \begin{tabular}{c@{\hskip 2pt} c@{\hskip 1pt}c@{\hskip 1pt}c@{\hskip 1pt}c@{\hskip 1pt}c@{\hskip 1pt}c}
    \toprule
    Original & \shortstack{HiDDeN} & \shortstack{MBRS} & \shortstack{CIN} & \shortstack{TrustMark} & \shortstack{WAM} & \shortstack{\ours} \\
    \midrule
    \includegraphics[width=\imwidth]{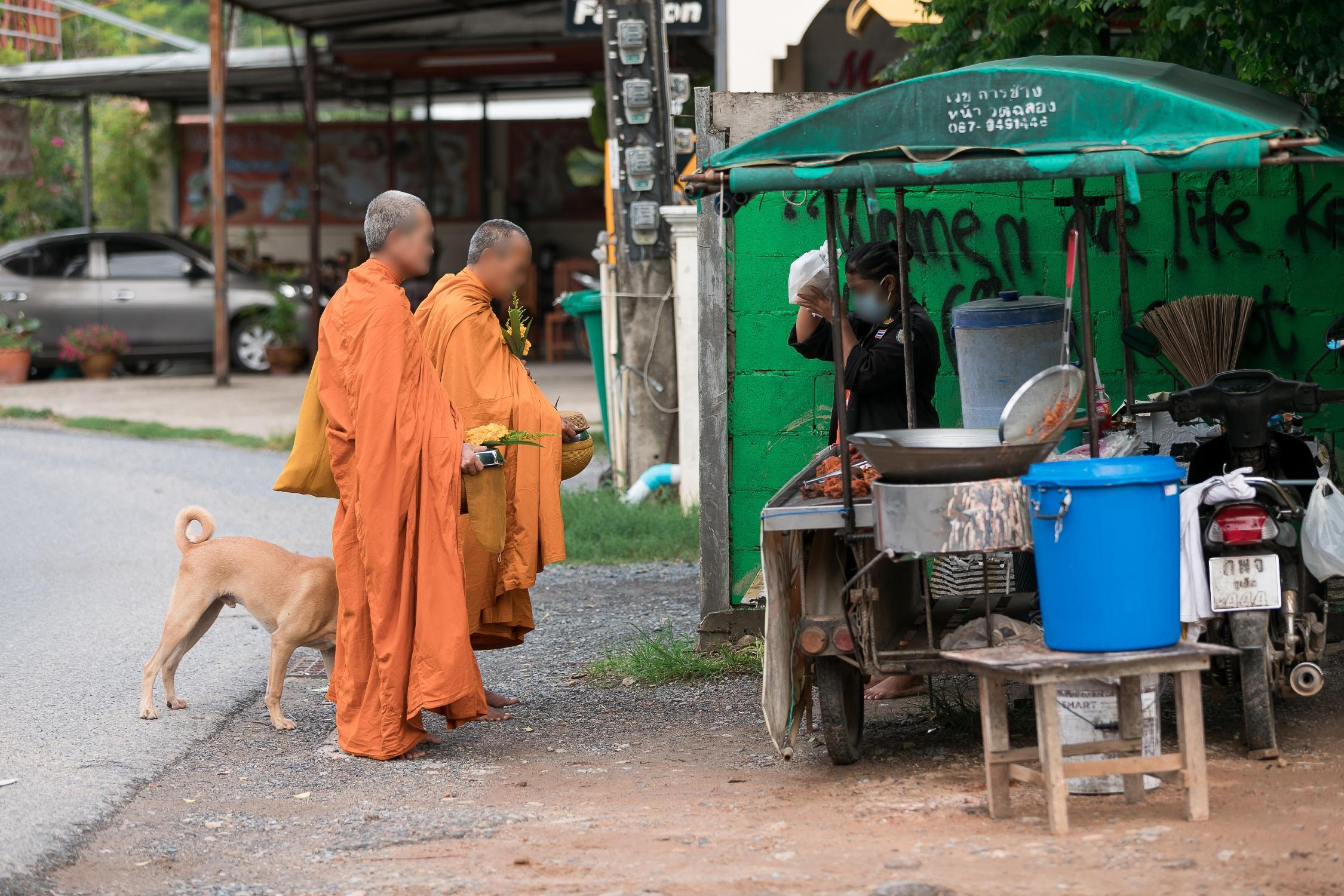} & 
    \includegraphics[width=\imwidth]{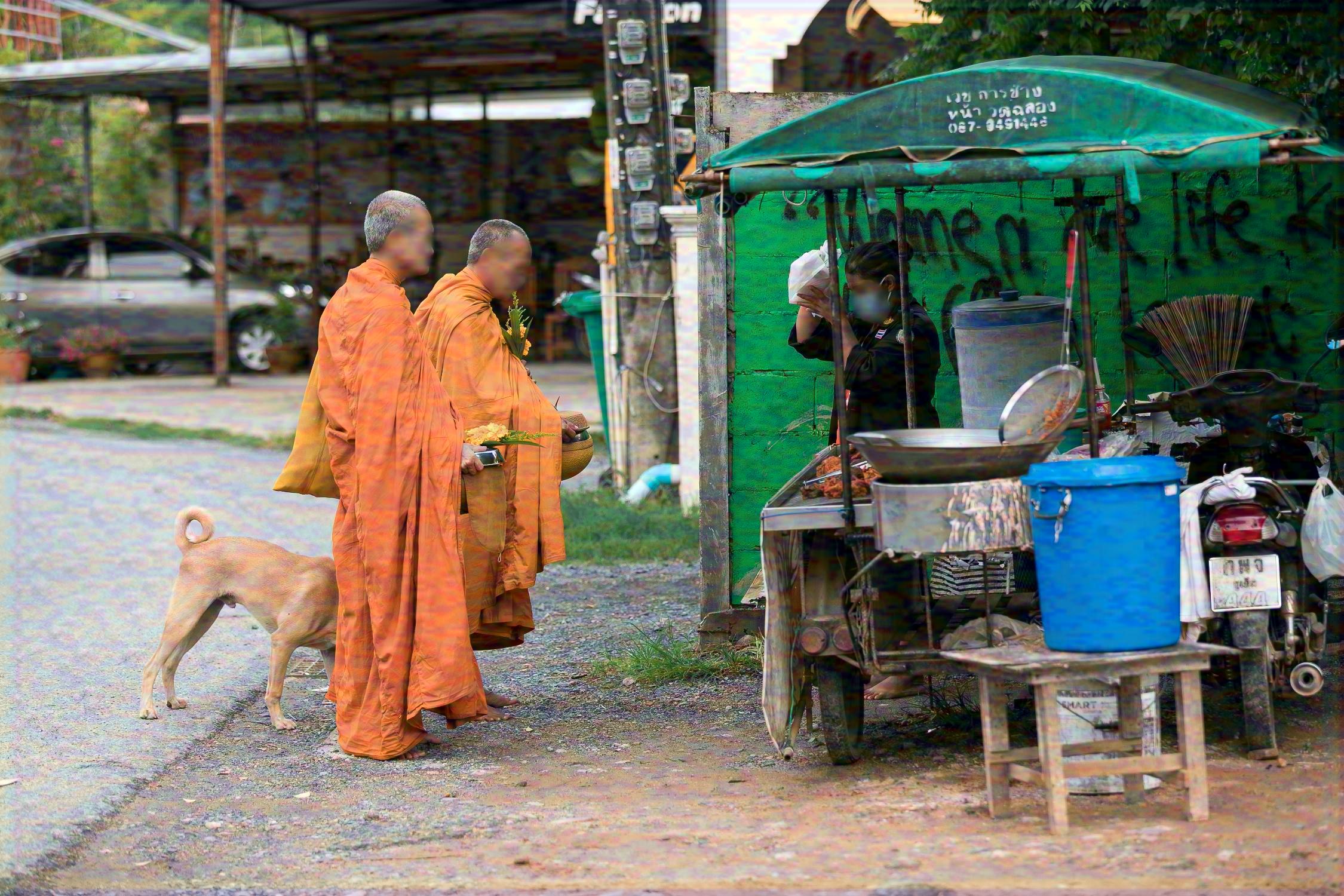} & 
    \includegraphics[width=\imwidth]{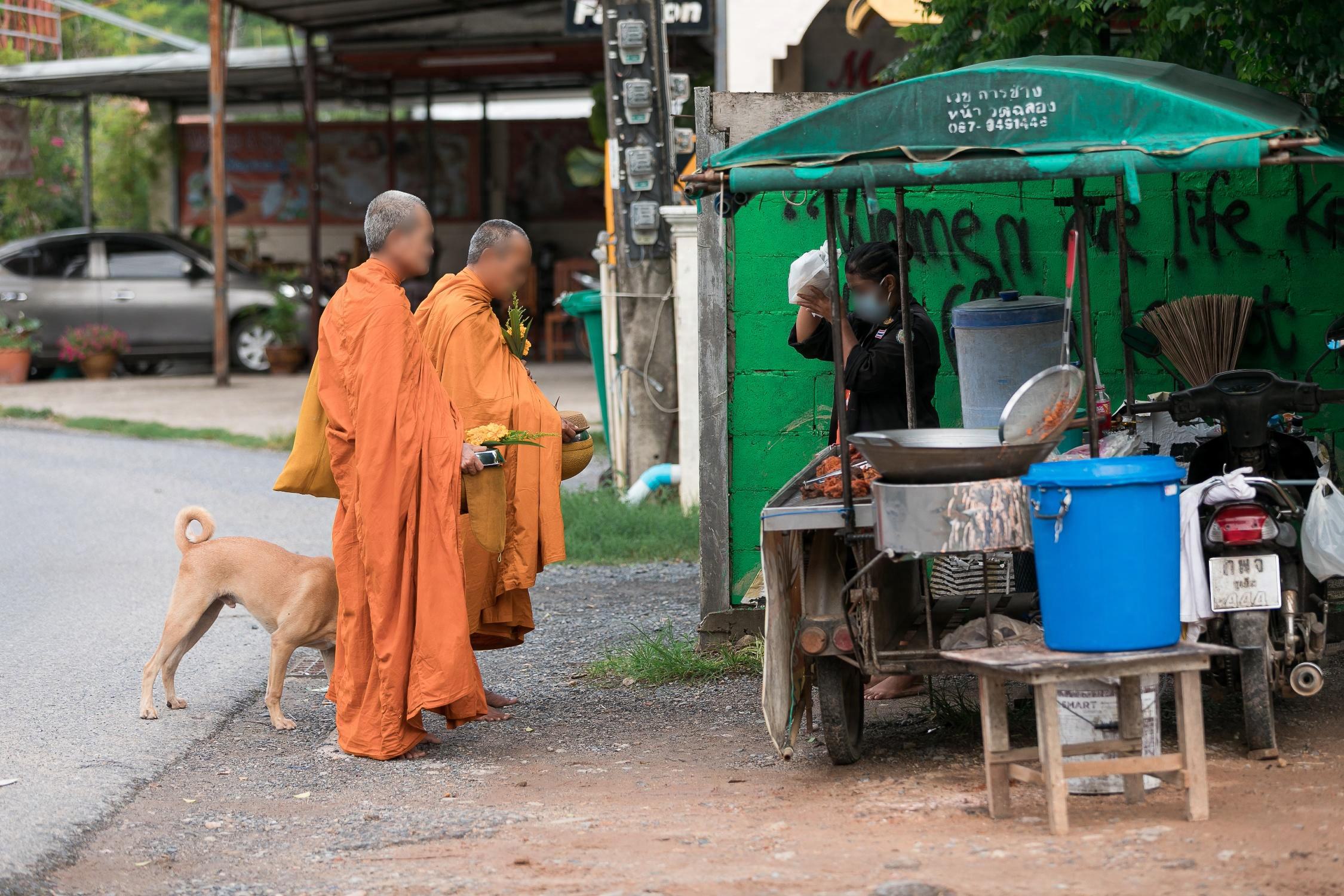} & 
    \includegraphics[width=\imwidth]{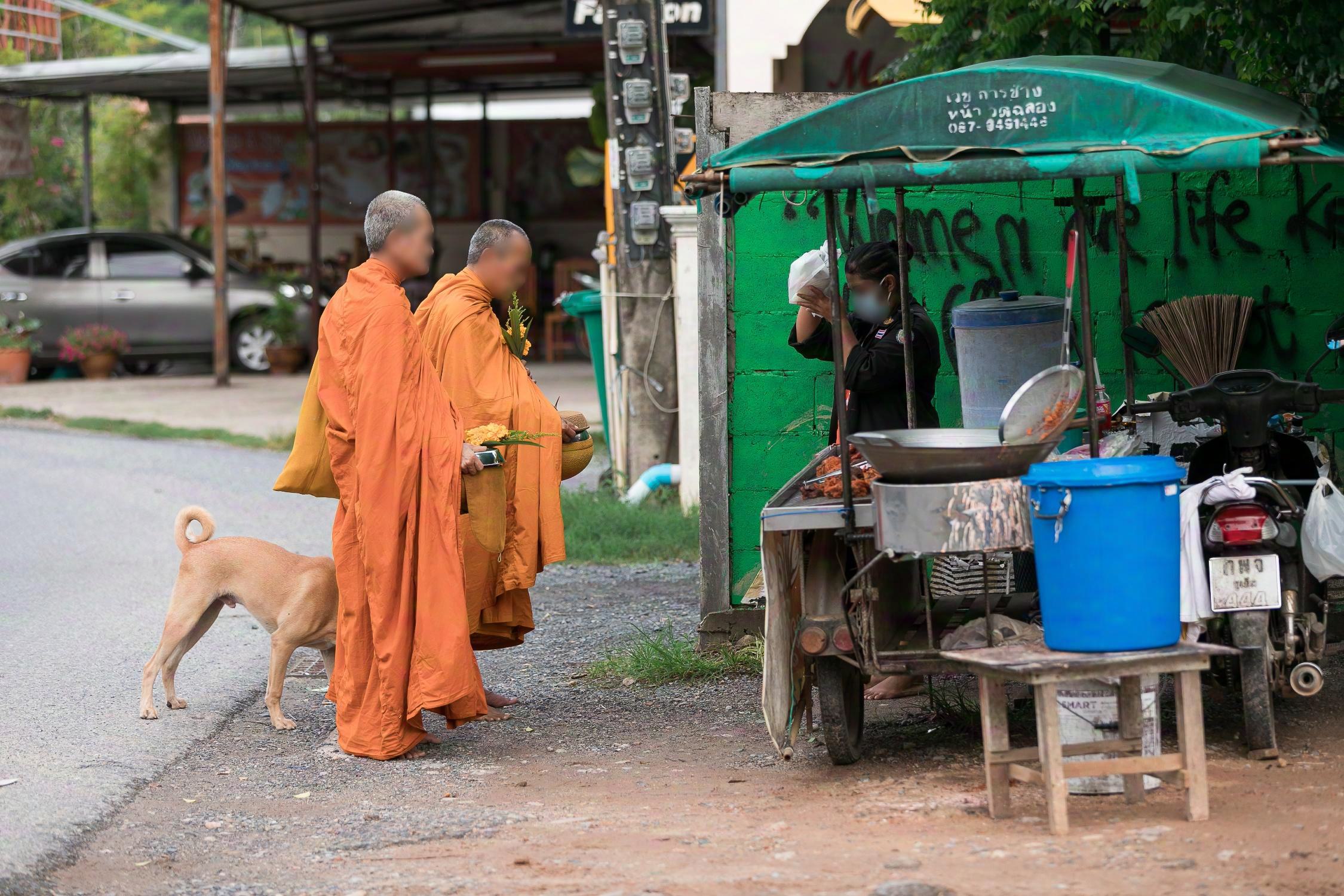} & 
    \includegraphics[width=\imwidth]{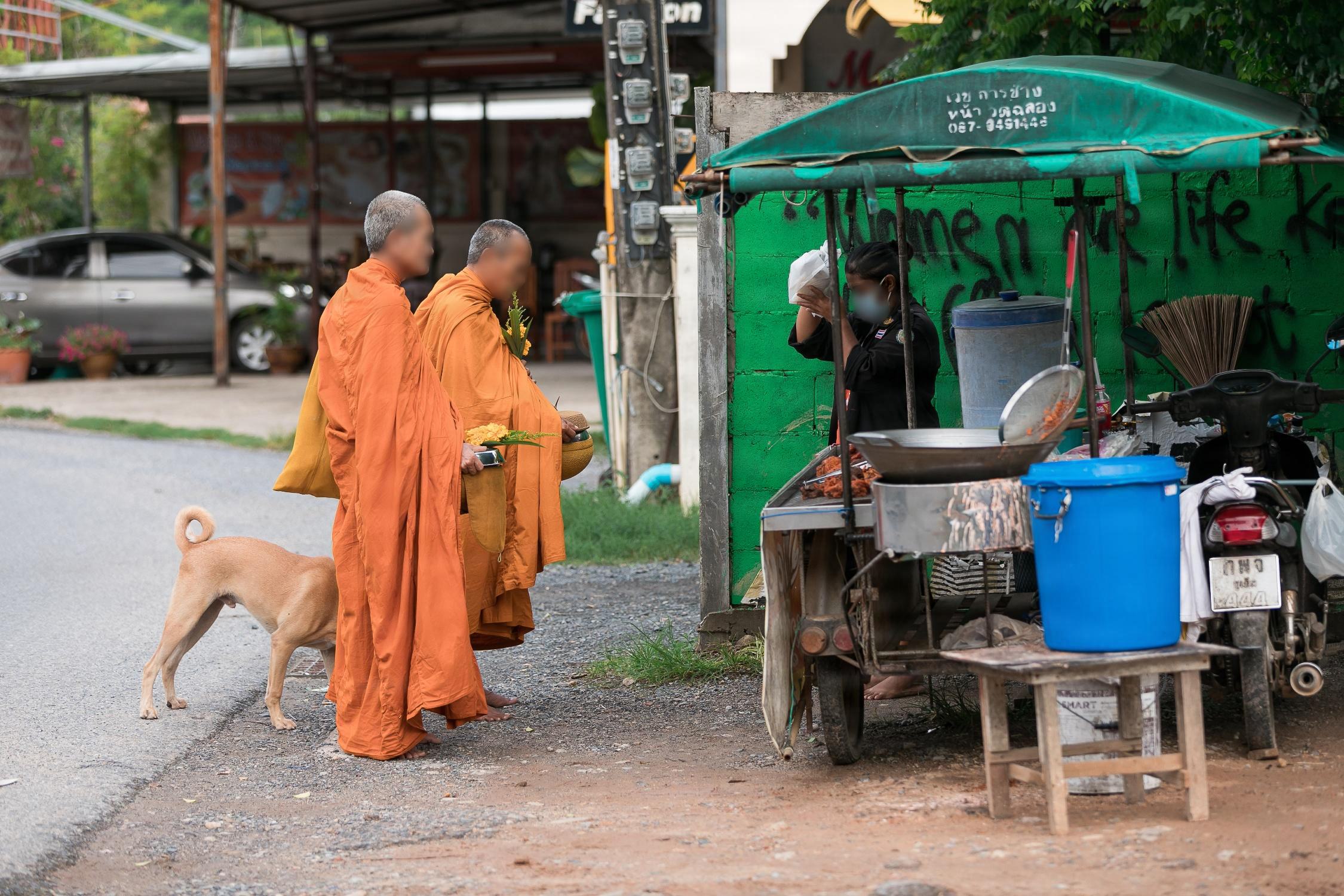} & 
    \includegraphics[width=\imwidth]{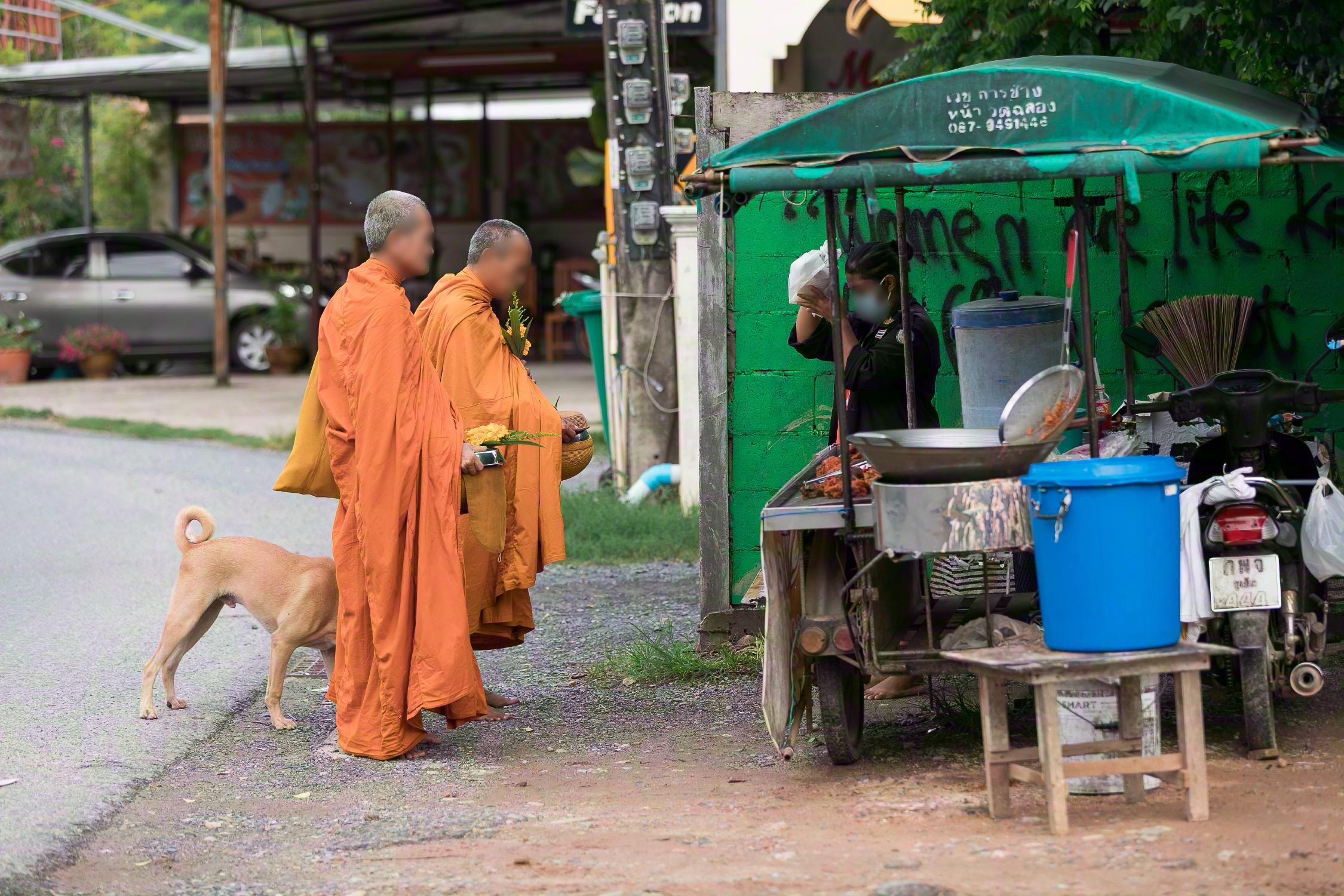} & 
    \includegraphics[width=\imwidth]{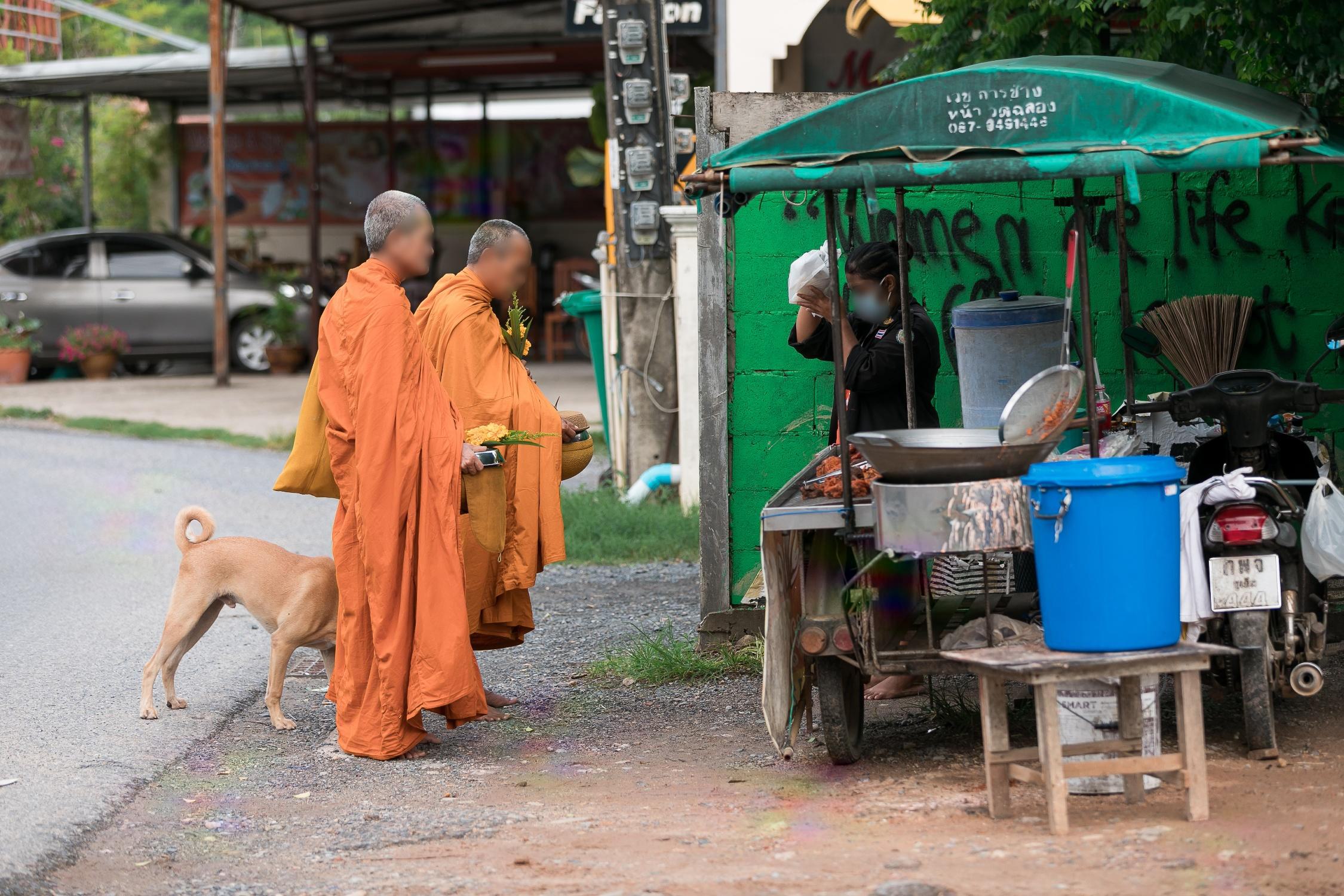} \\ & 
    \includegraphics[width=\imwidth]{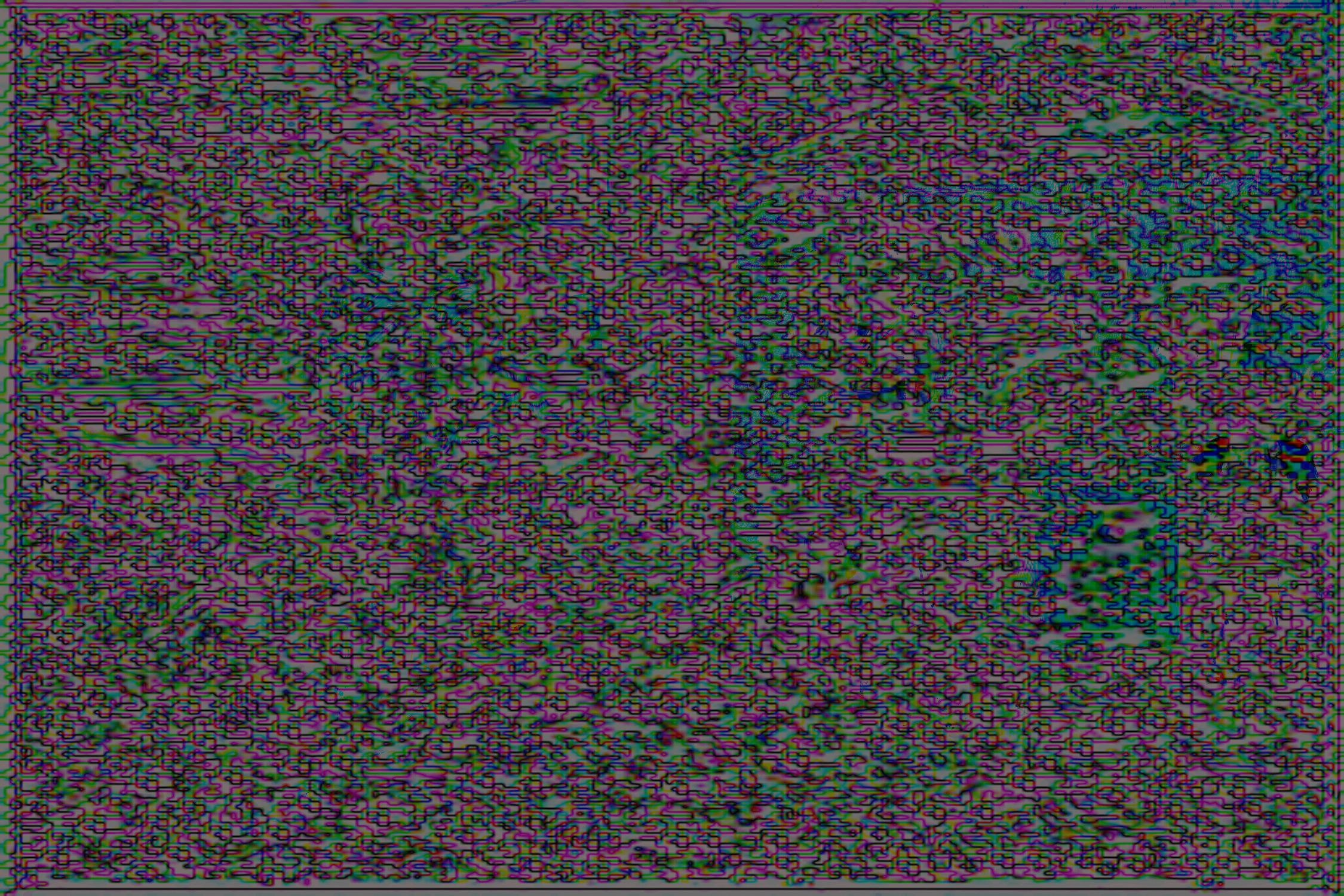} & 
    \includegraphics[width=\imwidth]{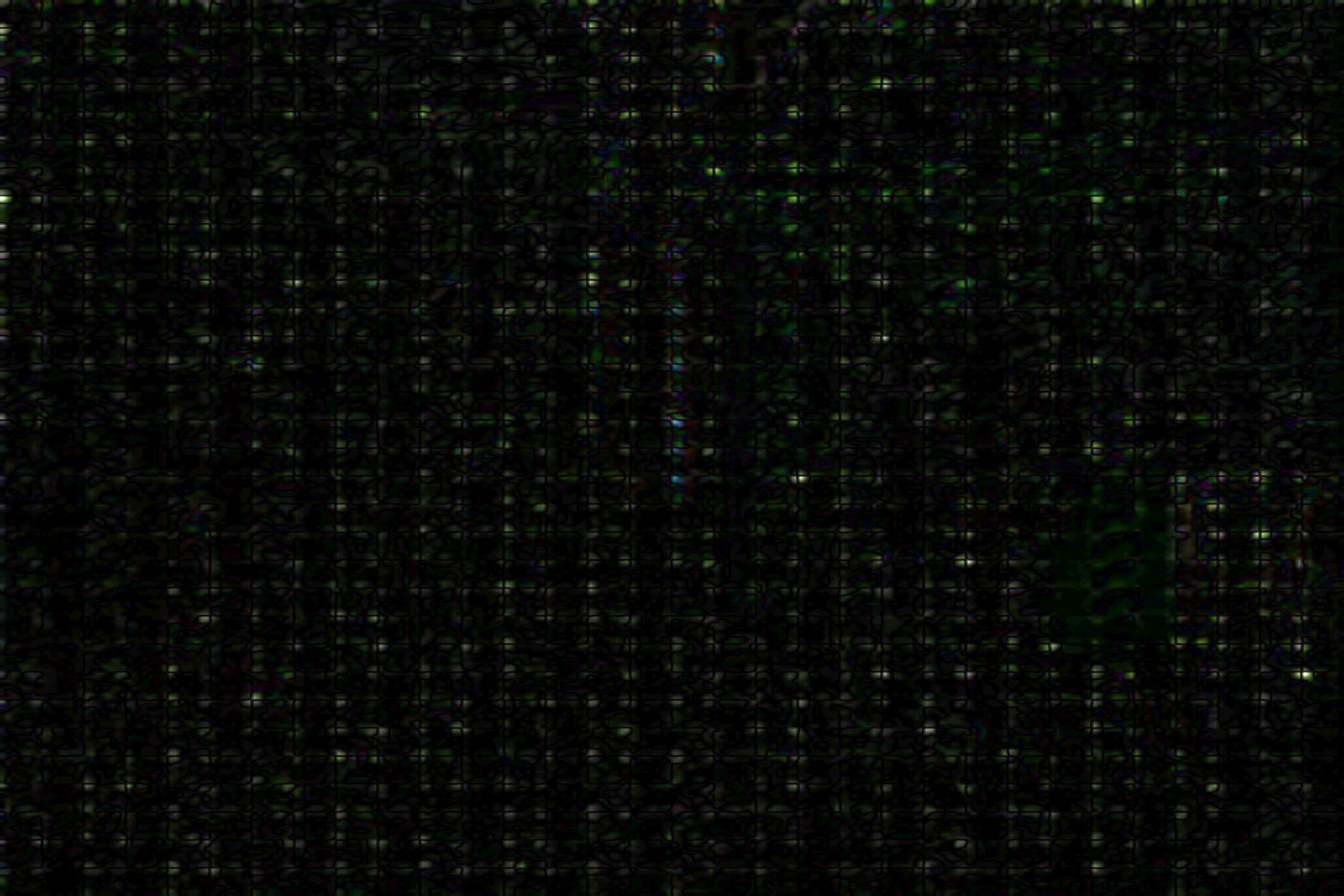} & 
    \includegraphics[width=\imwidth]{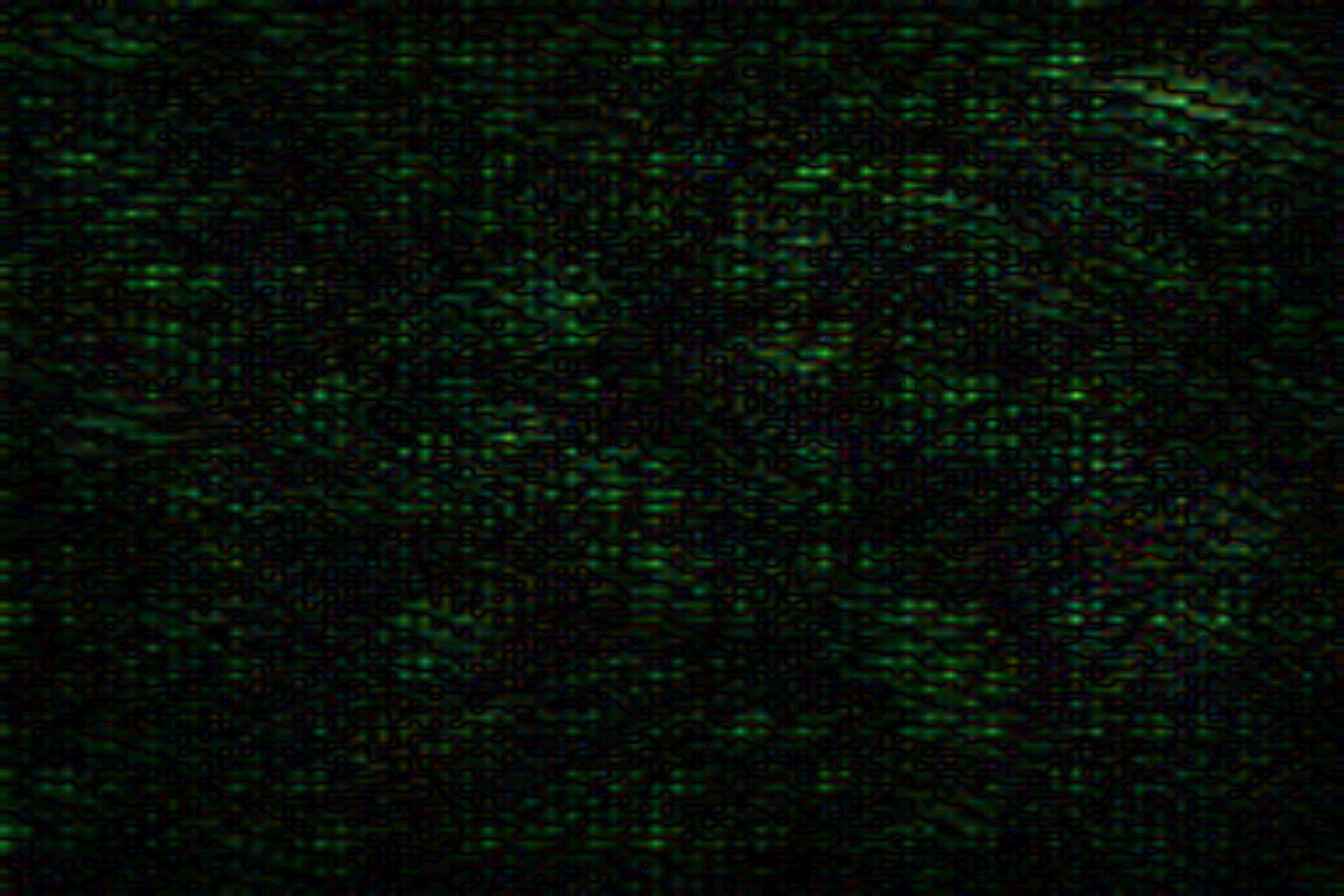} & 
    \includegraphics[width=\imwidth]{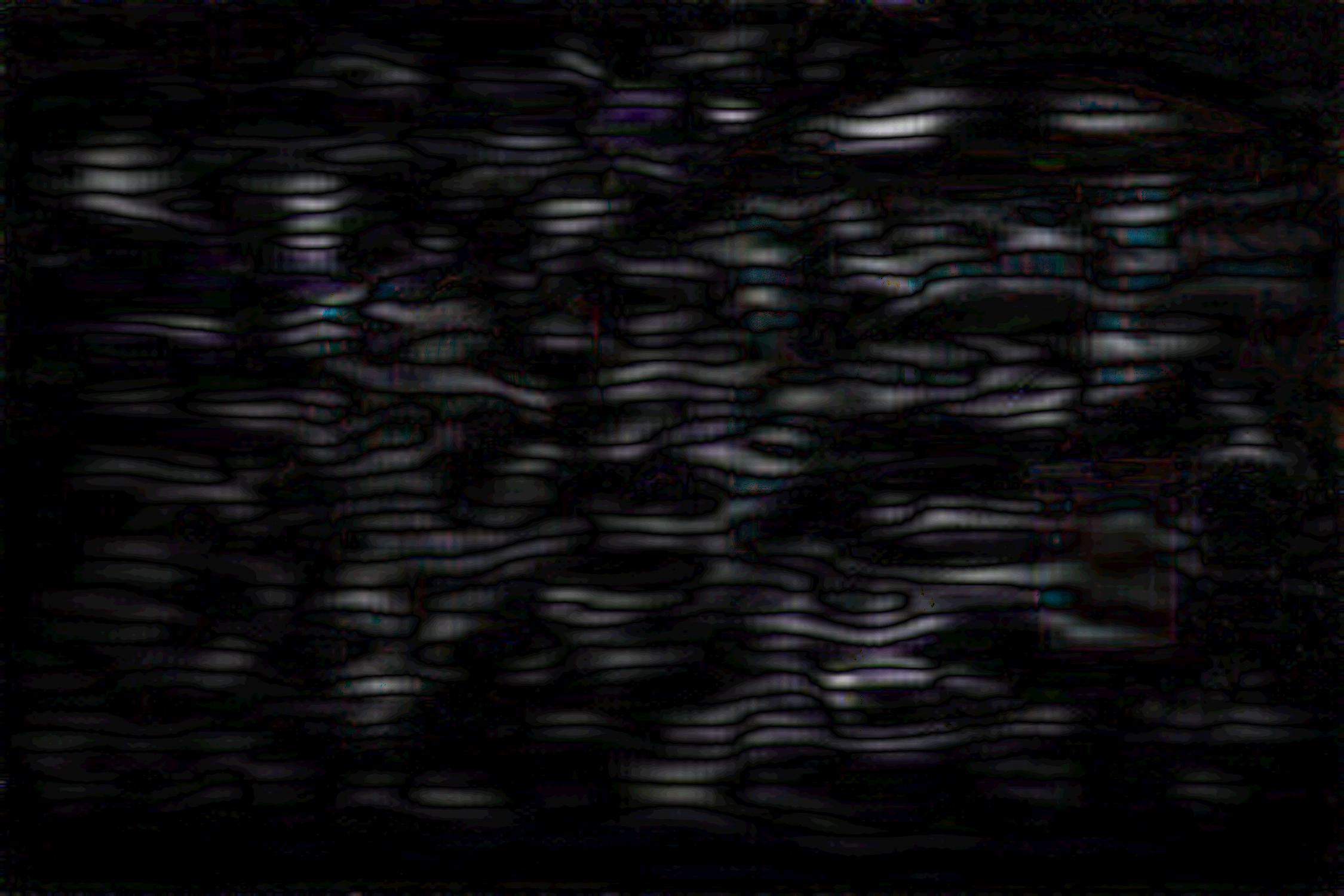} & 
    \includegraphics[width=\imwidth]{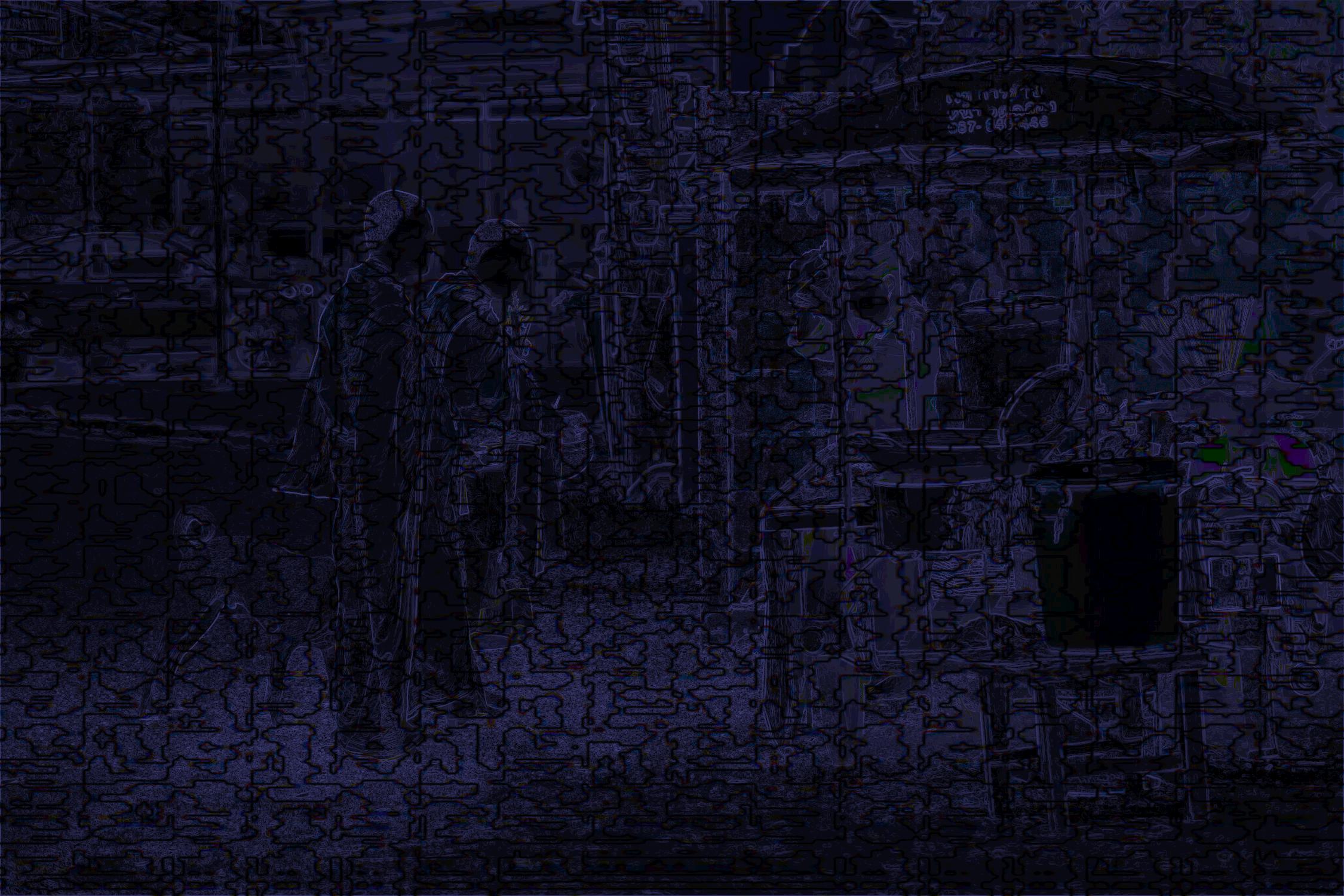} & 
    \includegraphics[width=\imwidth]{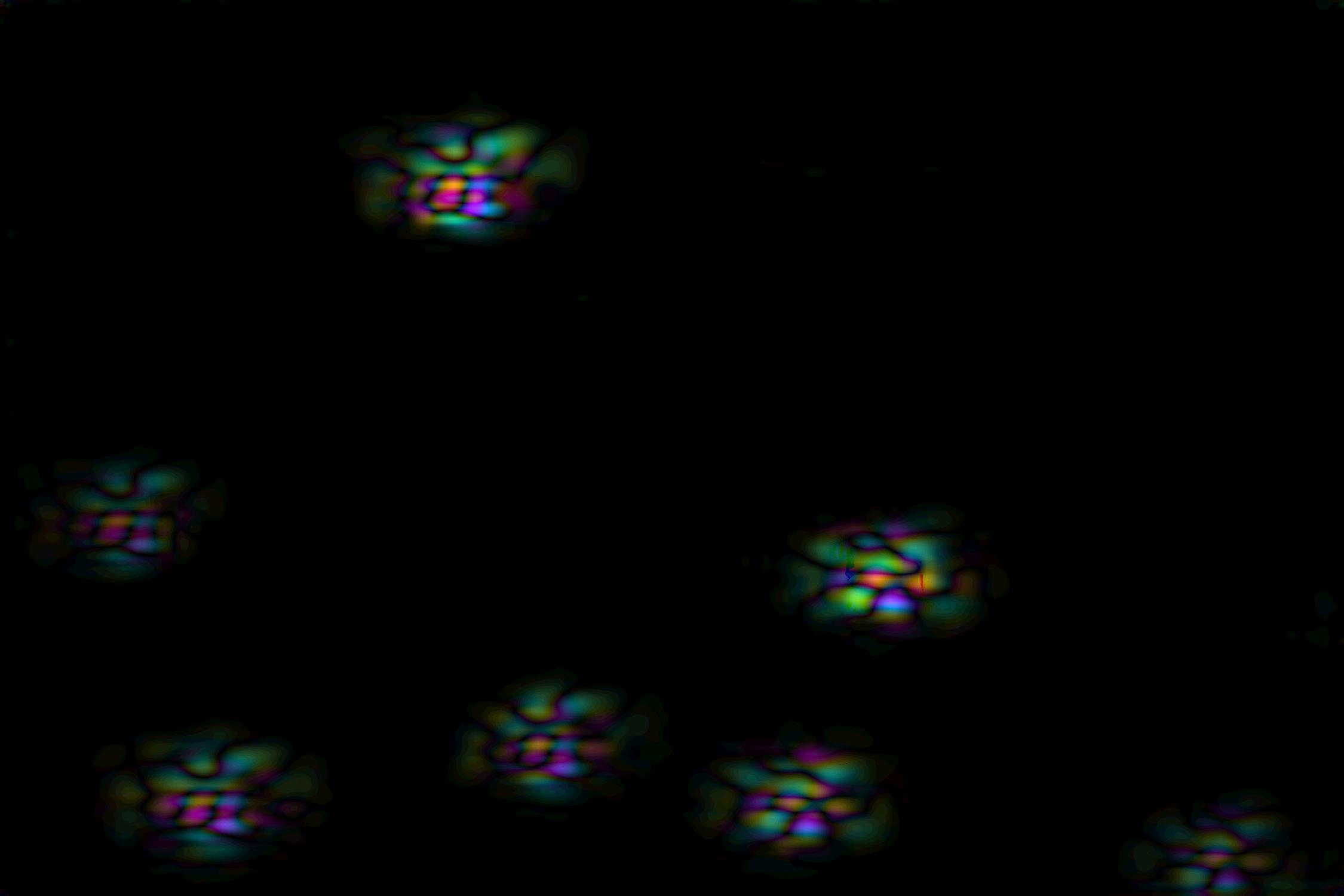} \\ \rule{0pt}{6ex}\includegraphics[width=\imwidth]{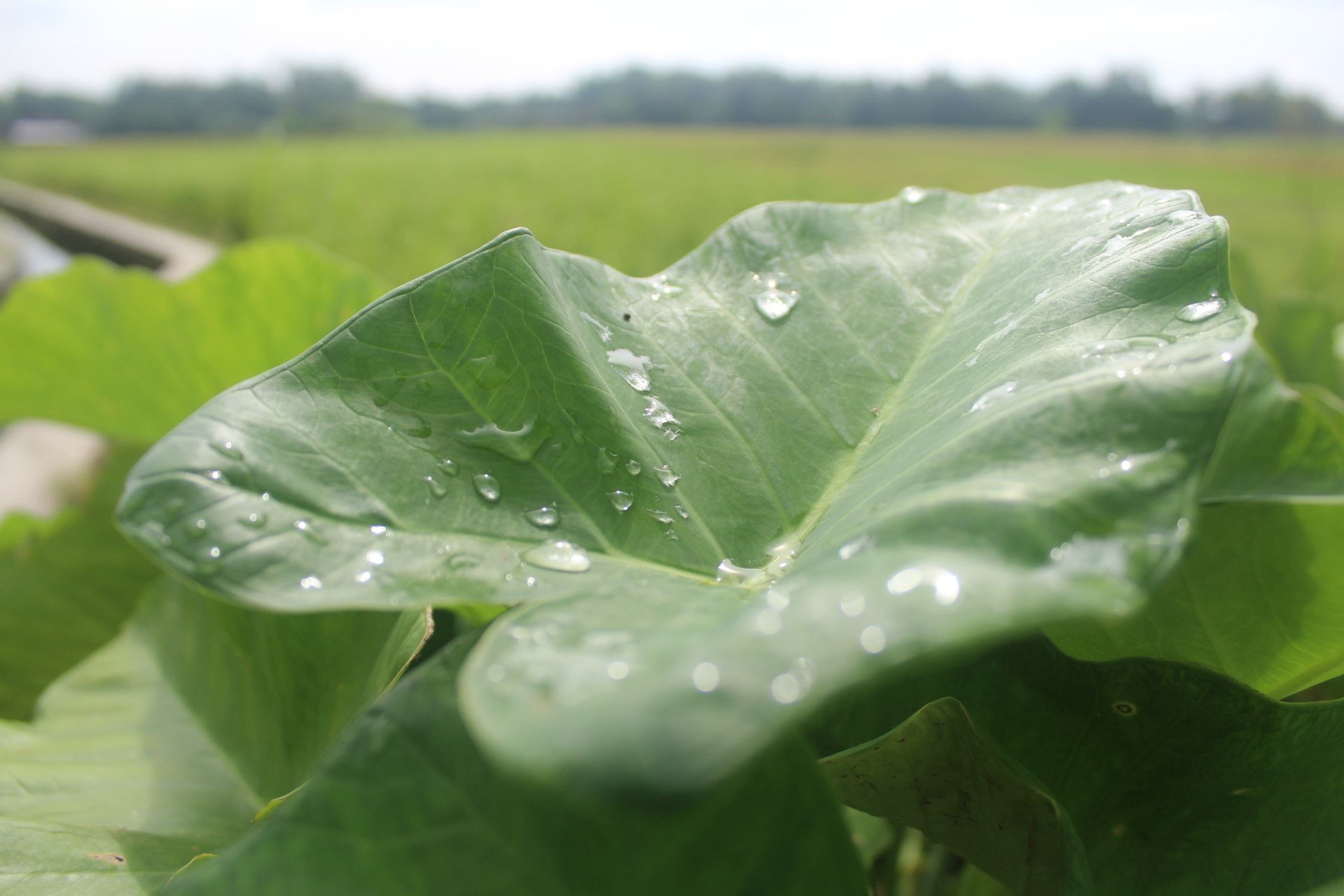} & 
    \includegraphics[width=\imwidth]{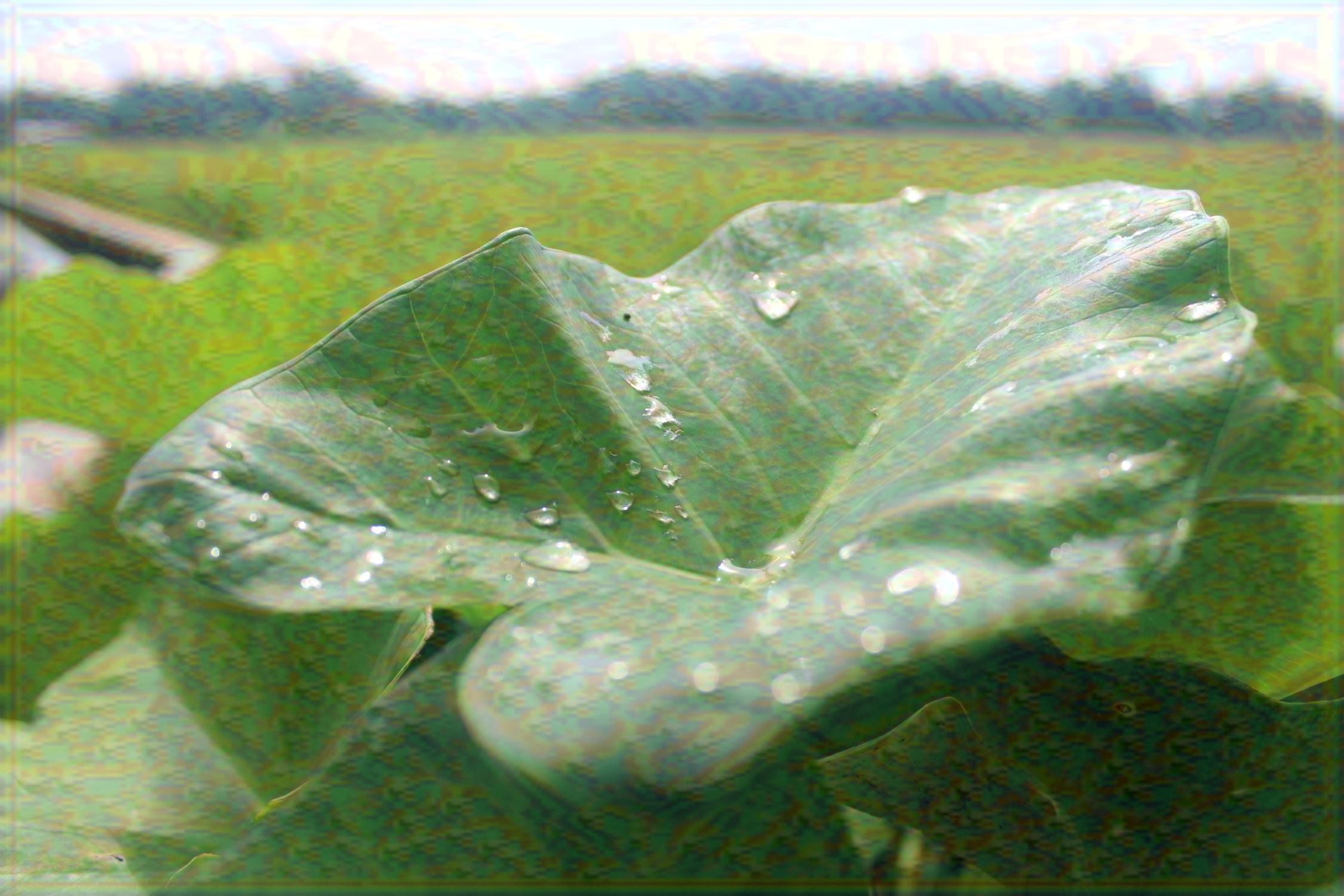} & 
    \includegraphics[width=\imwidth]{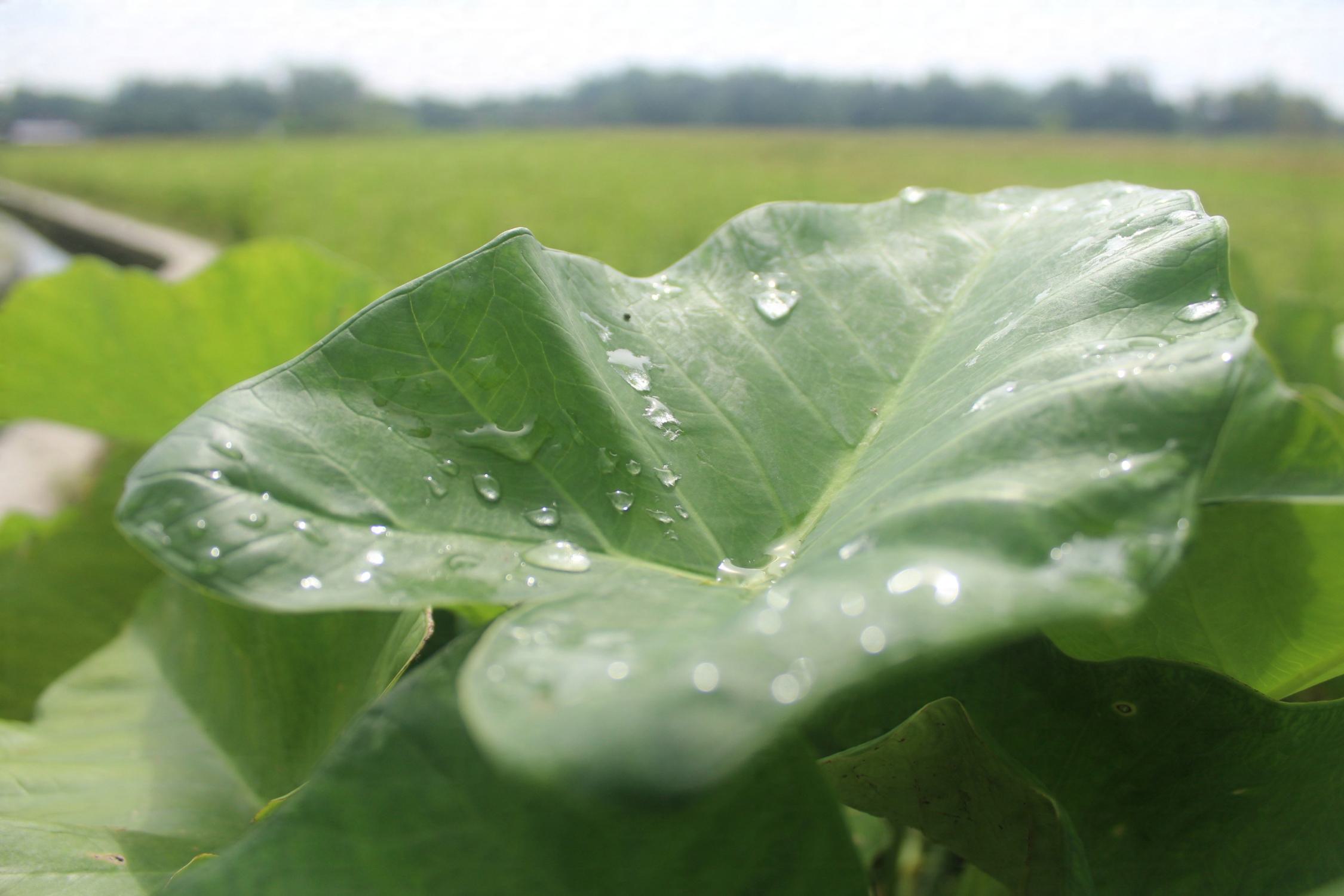} & 
    \includegraphics[width=\imwidth]{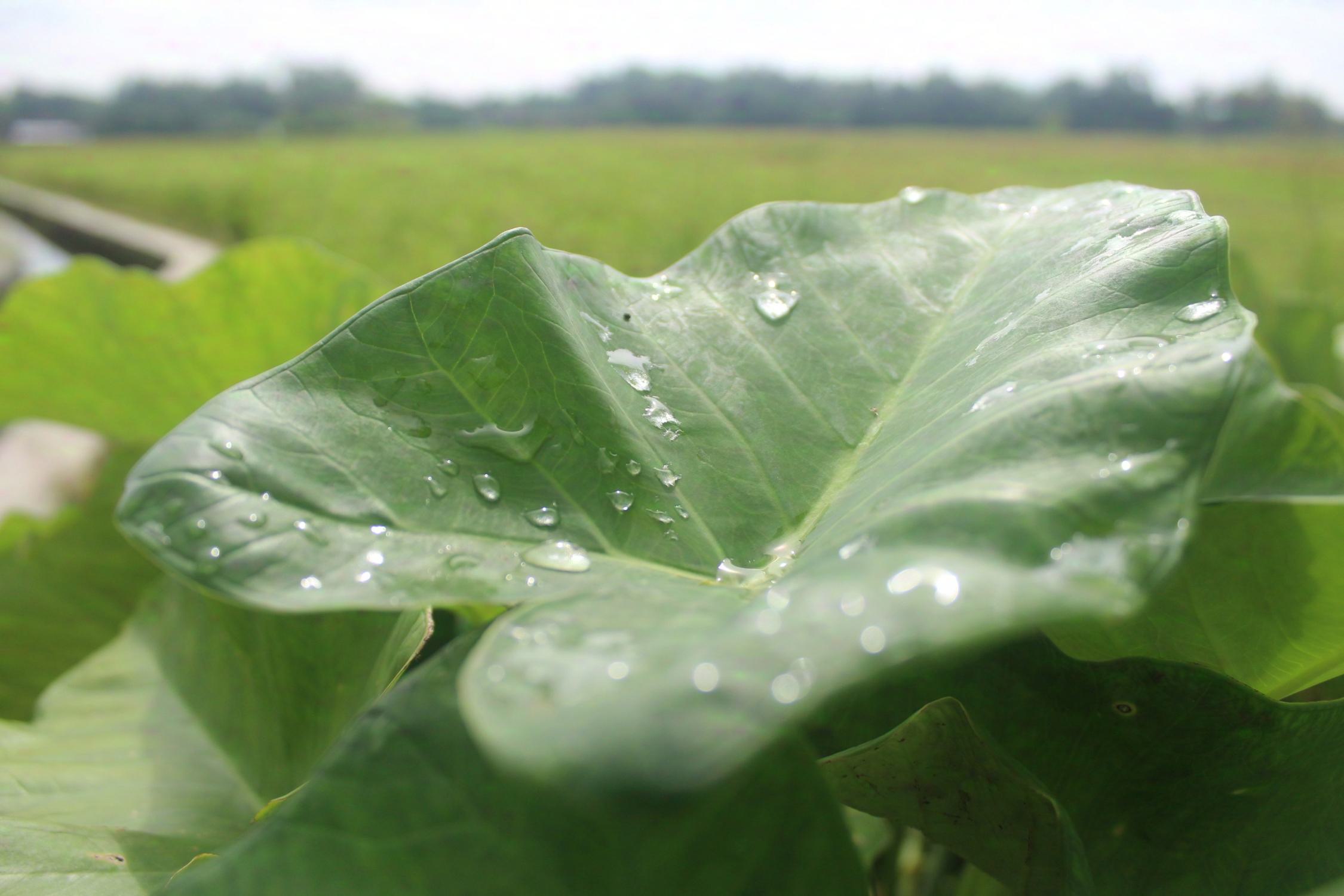} & 
    \includegraphics[width=\imwidth]{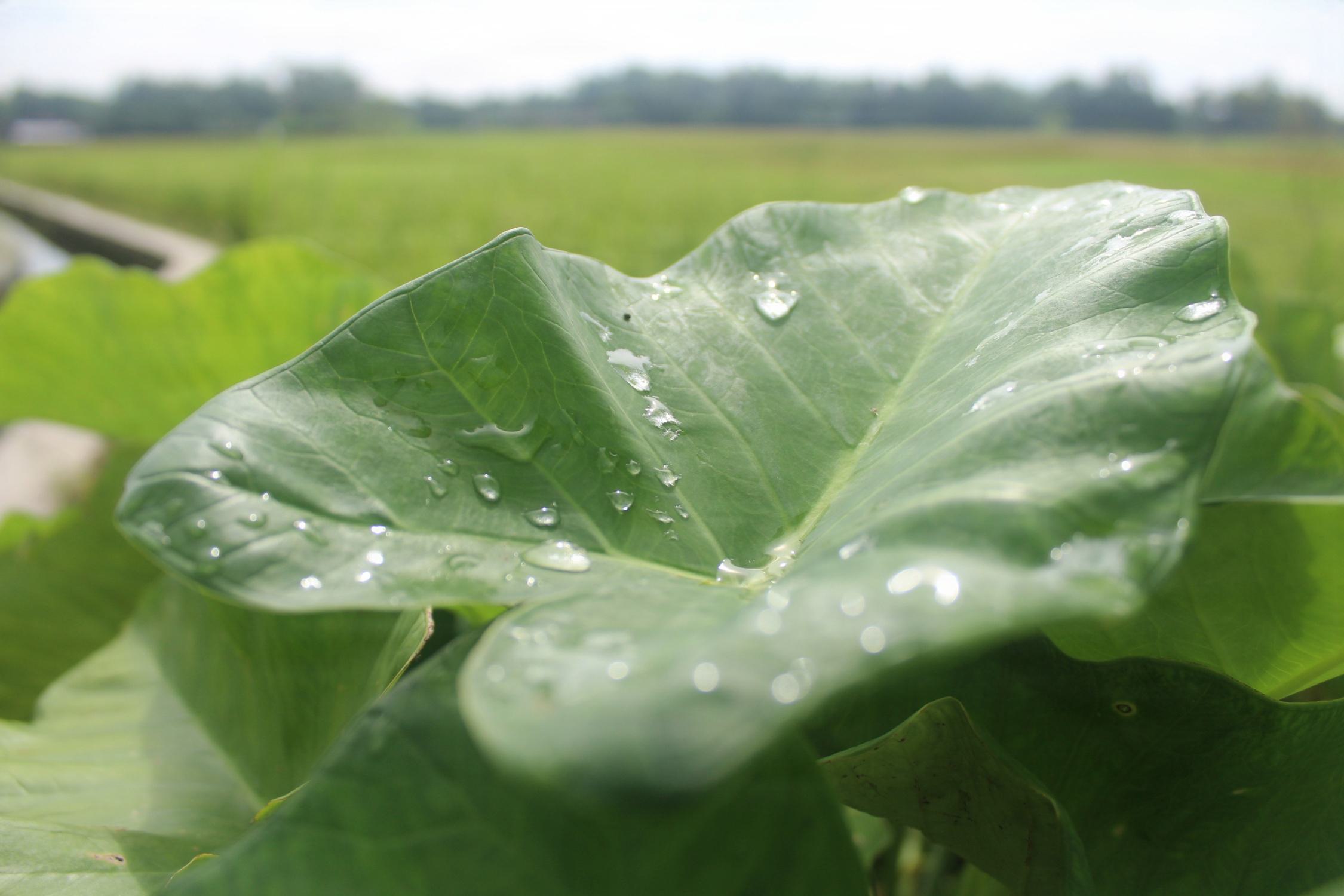} & 
    \includegraphics[width=\imwidth]{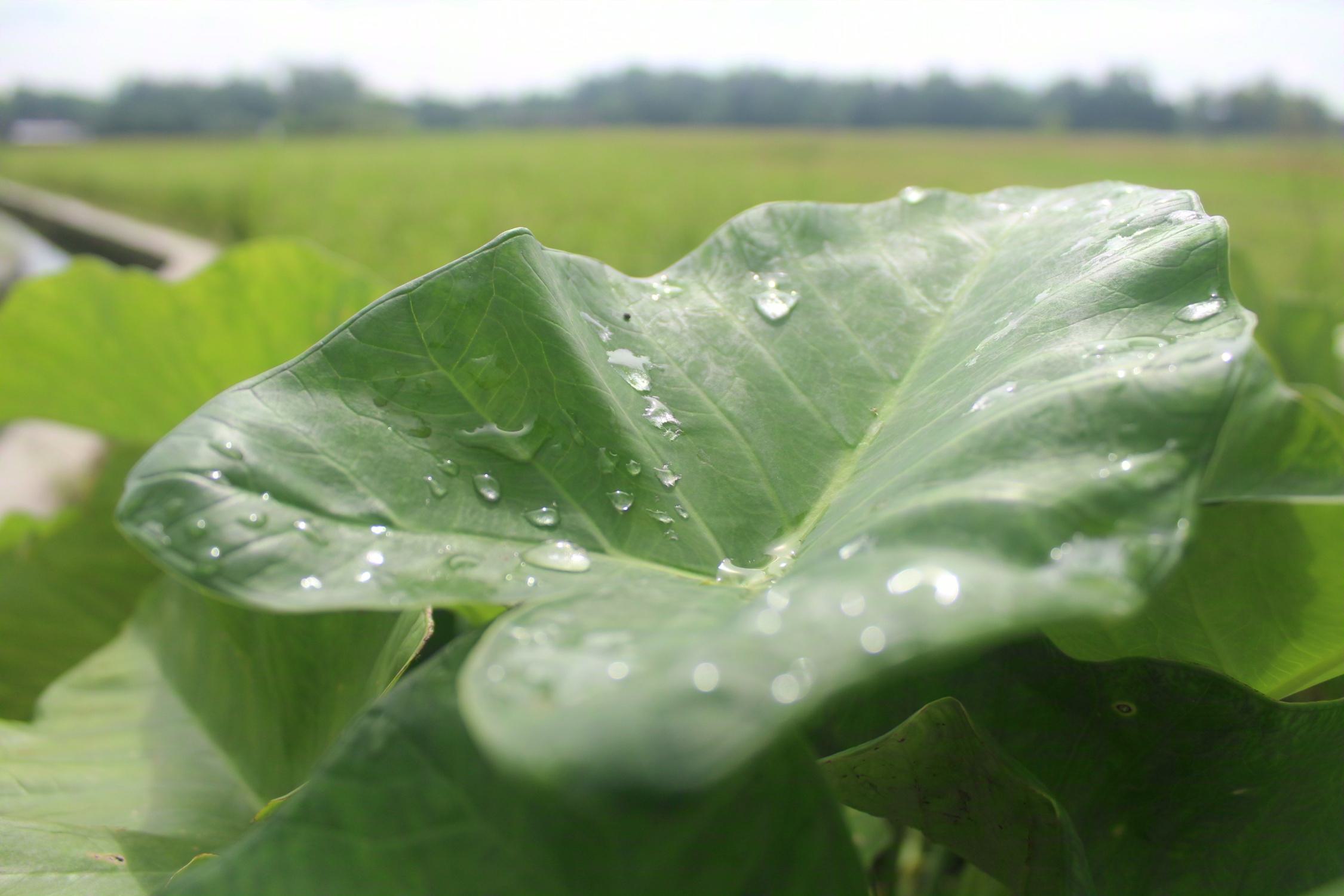} & 
    \includegraphics[width=\imwidth]{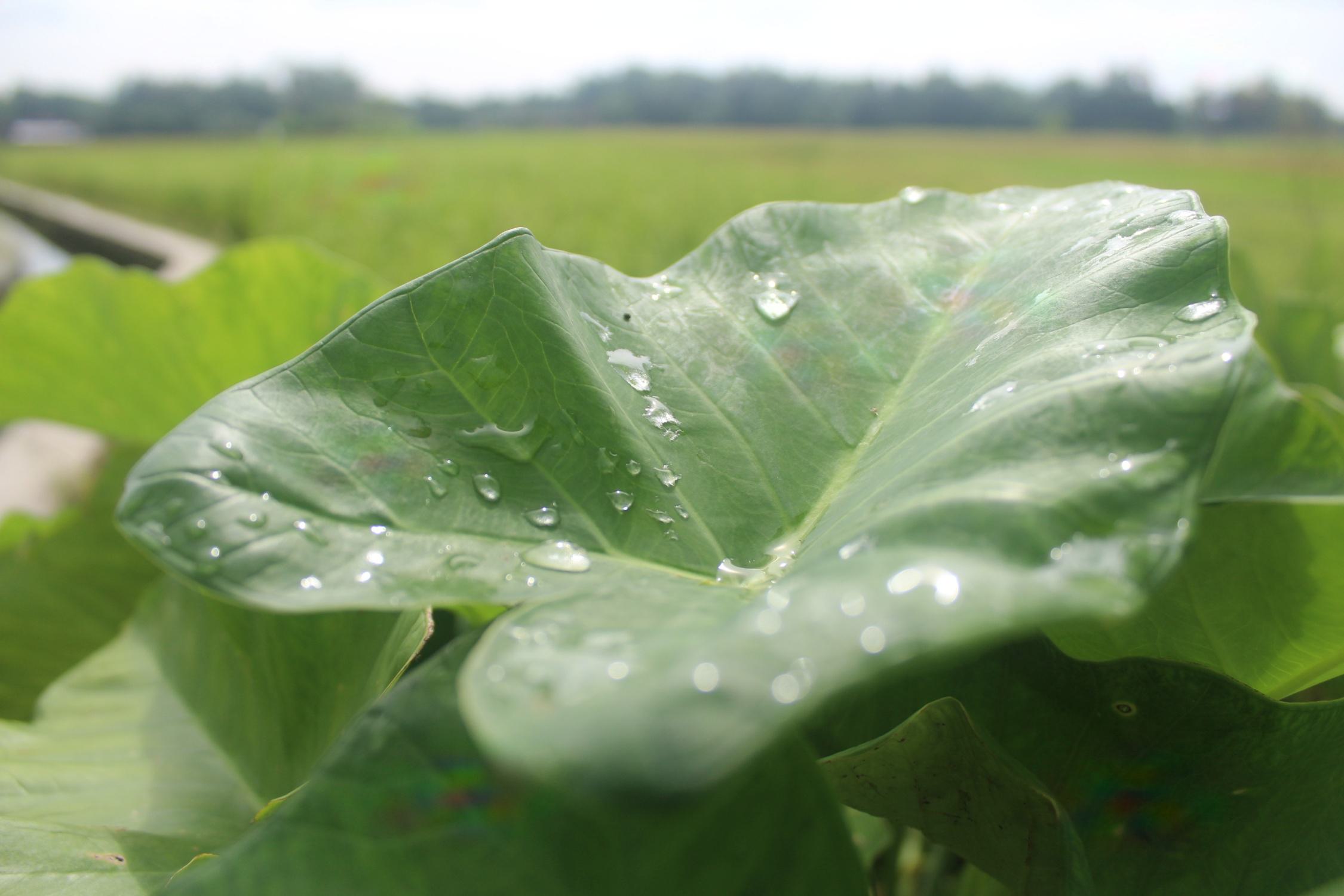} \\ & 
    \includegraphics[width=\imwidth]{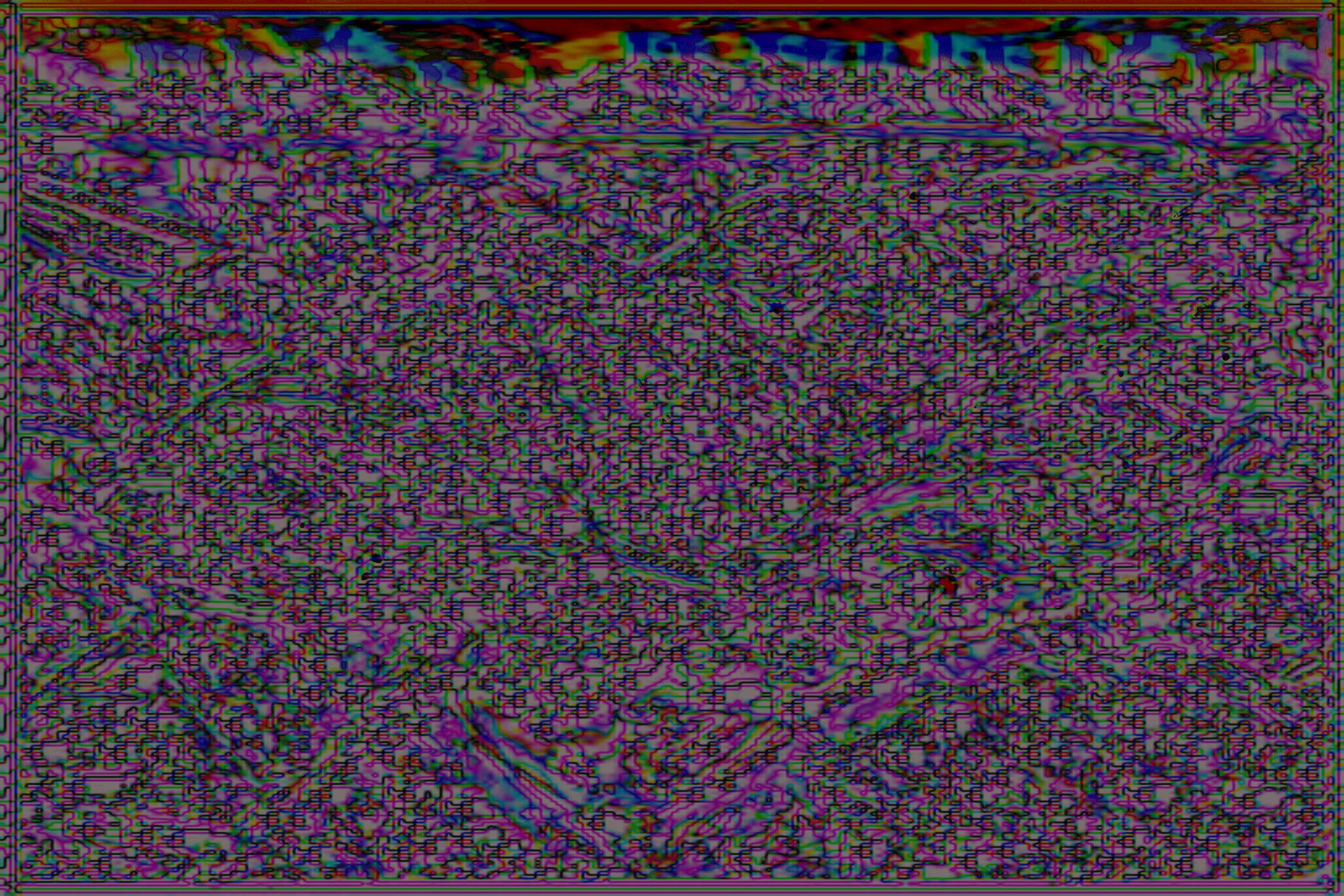} & 
    \includegraphics[width=\imwidth]{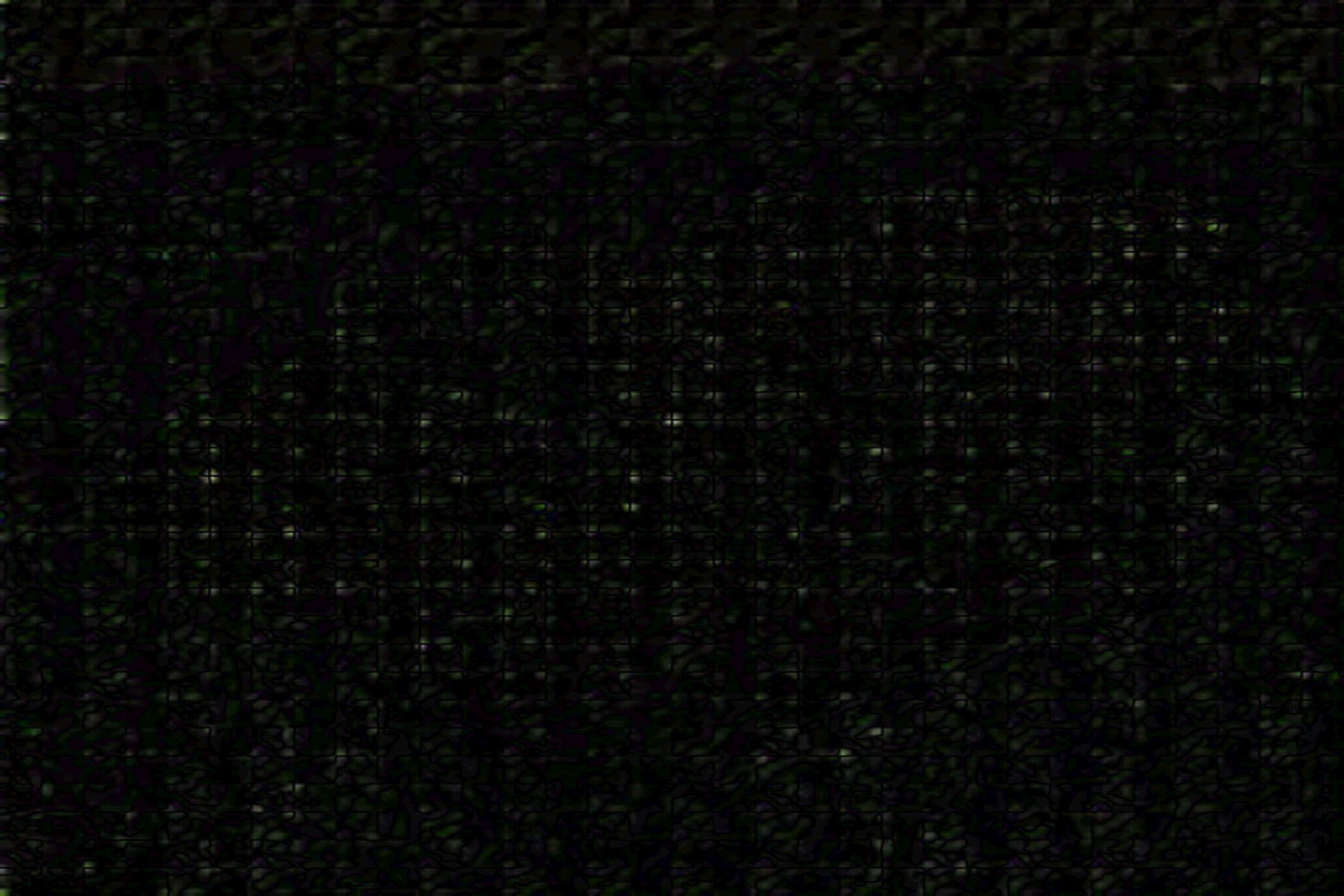} & 
    \includegraphics[width=\imwidth]{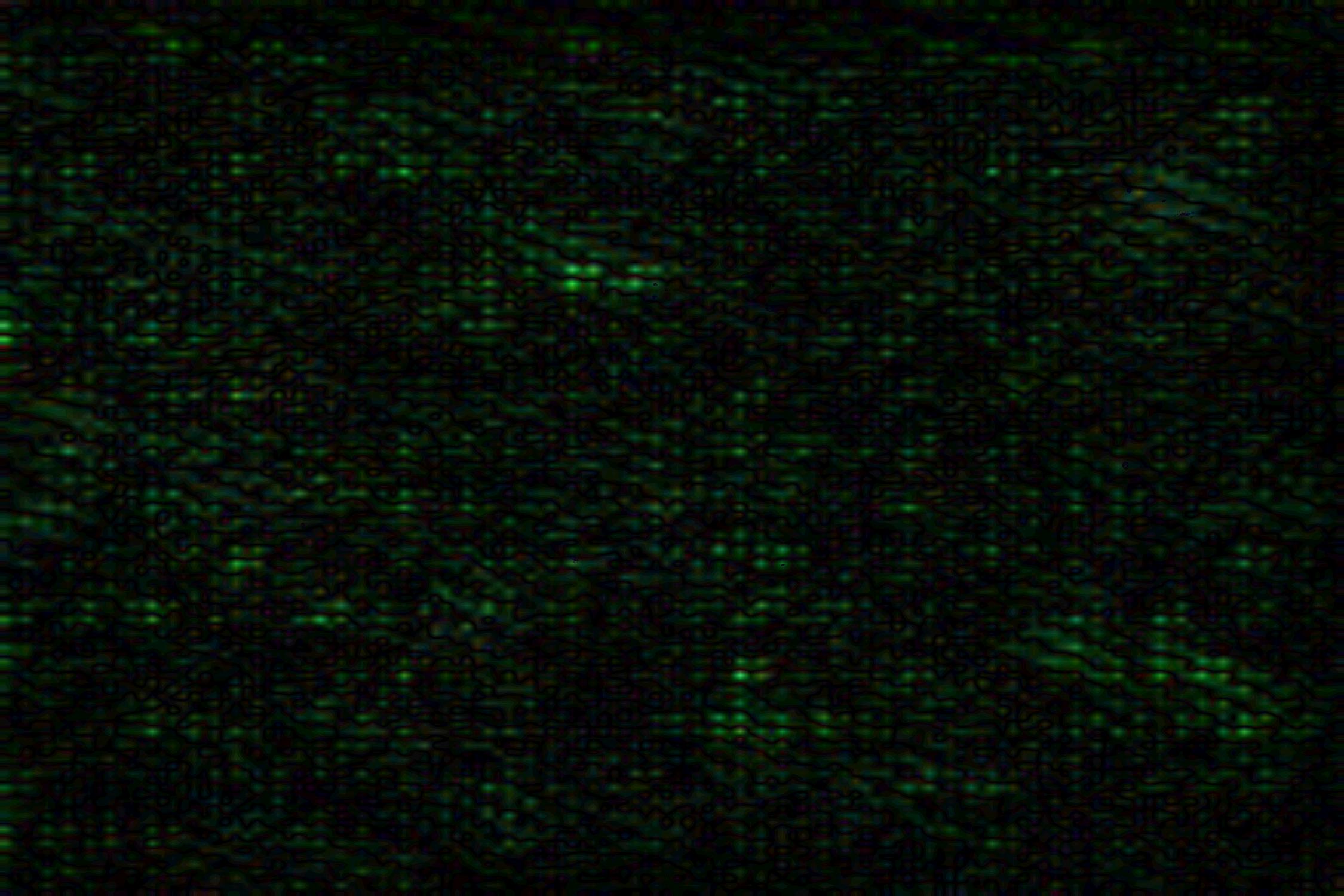} & 
    \includegraphics[width=\imwidth]{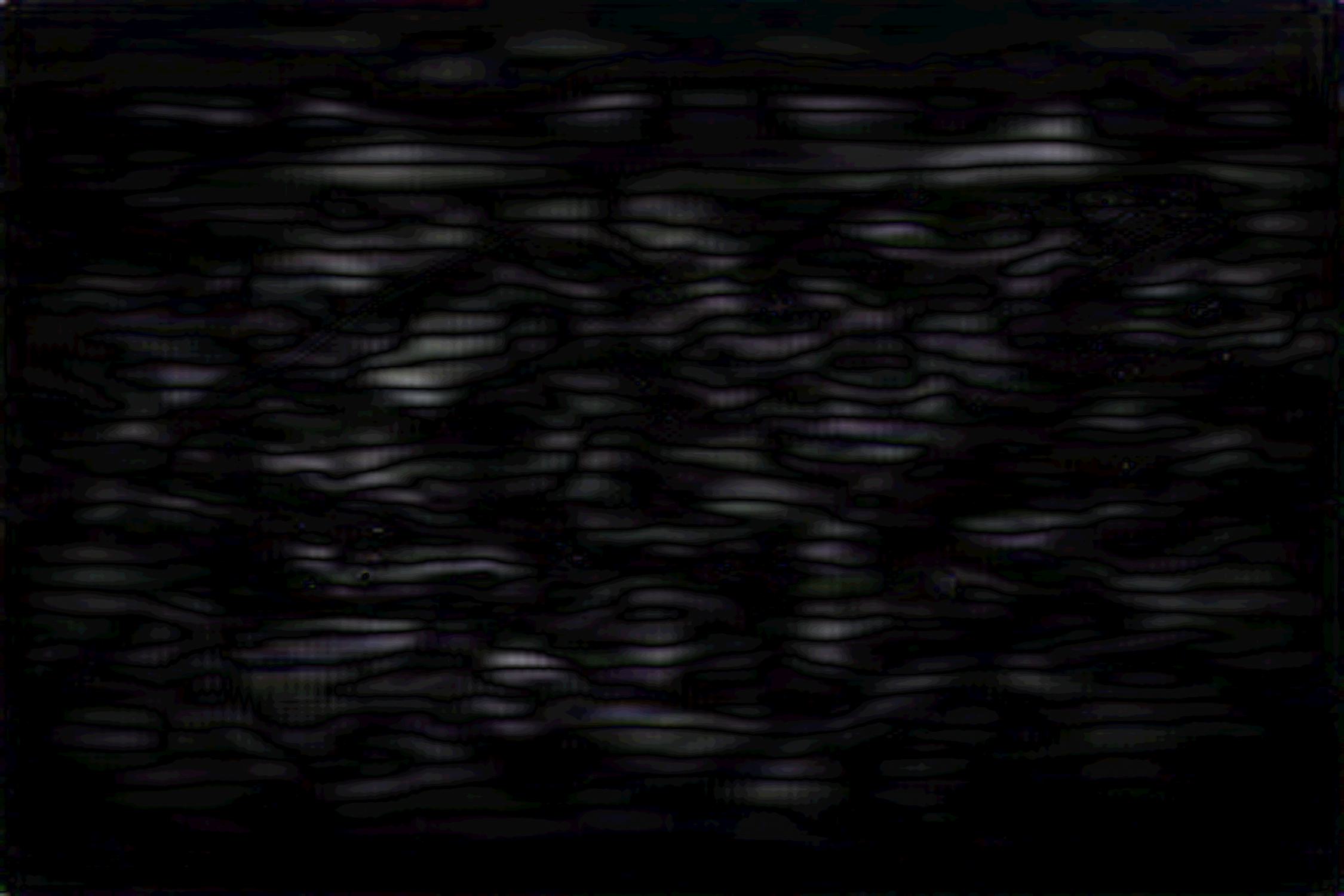} & 
    \includegraphics[width=\imwidth]{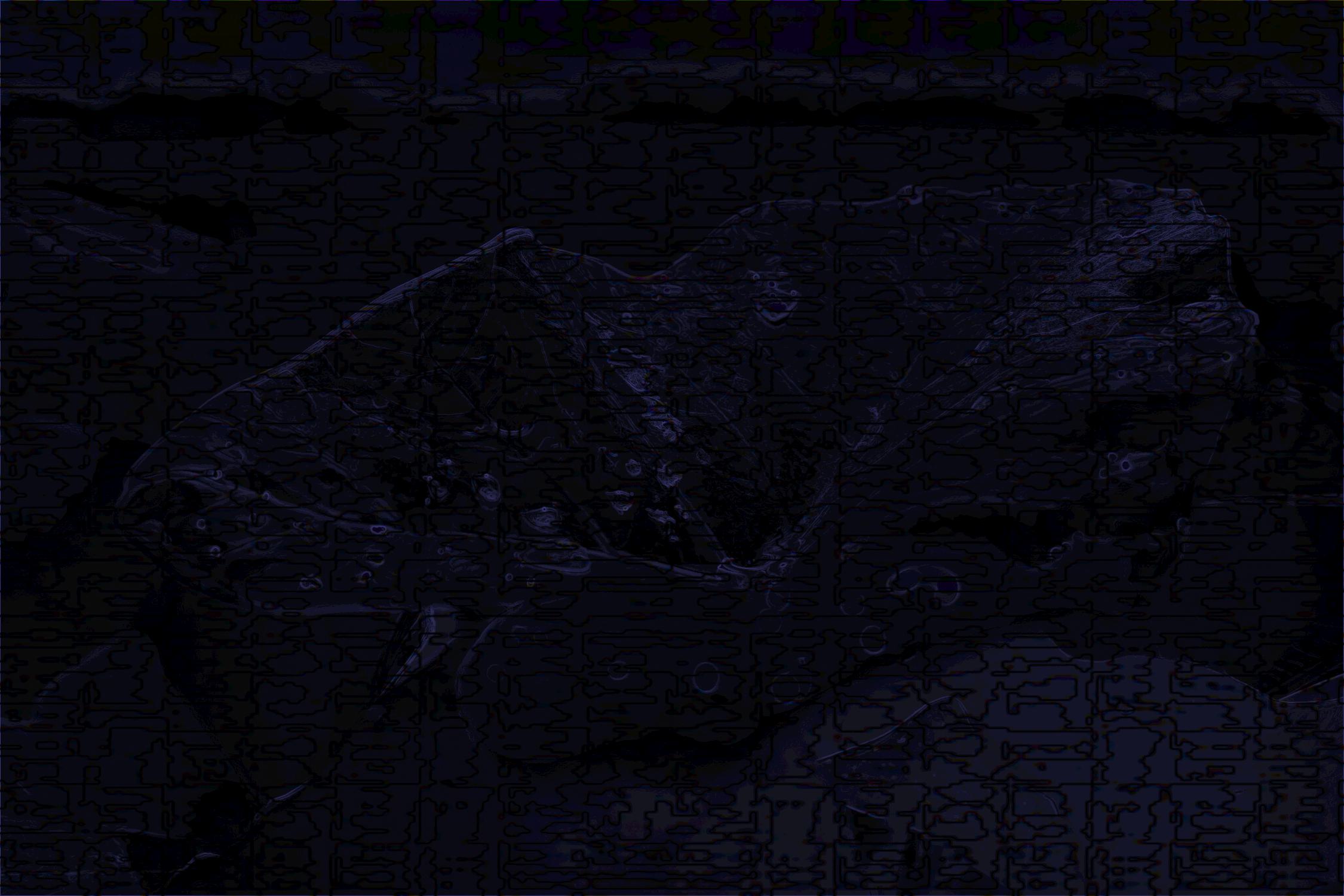} & 
    \includegraphics[width=\imwidth]{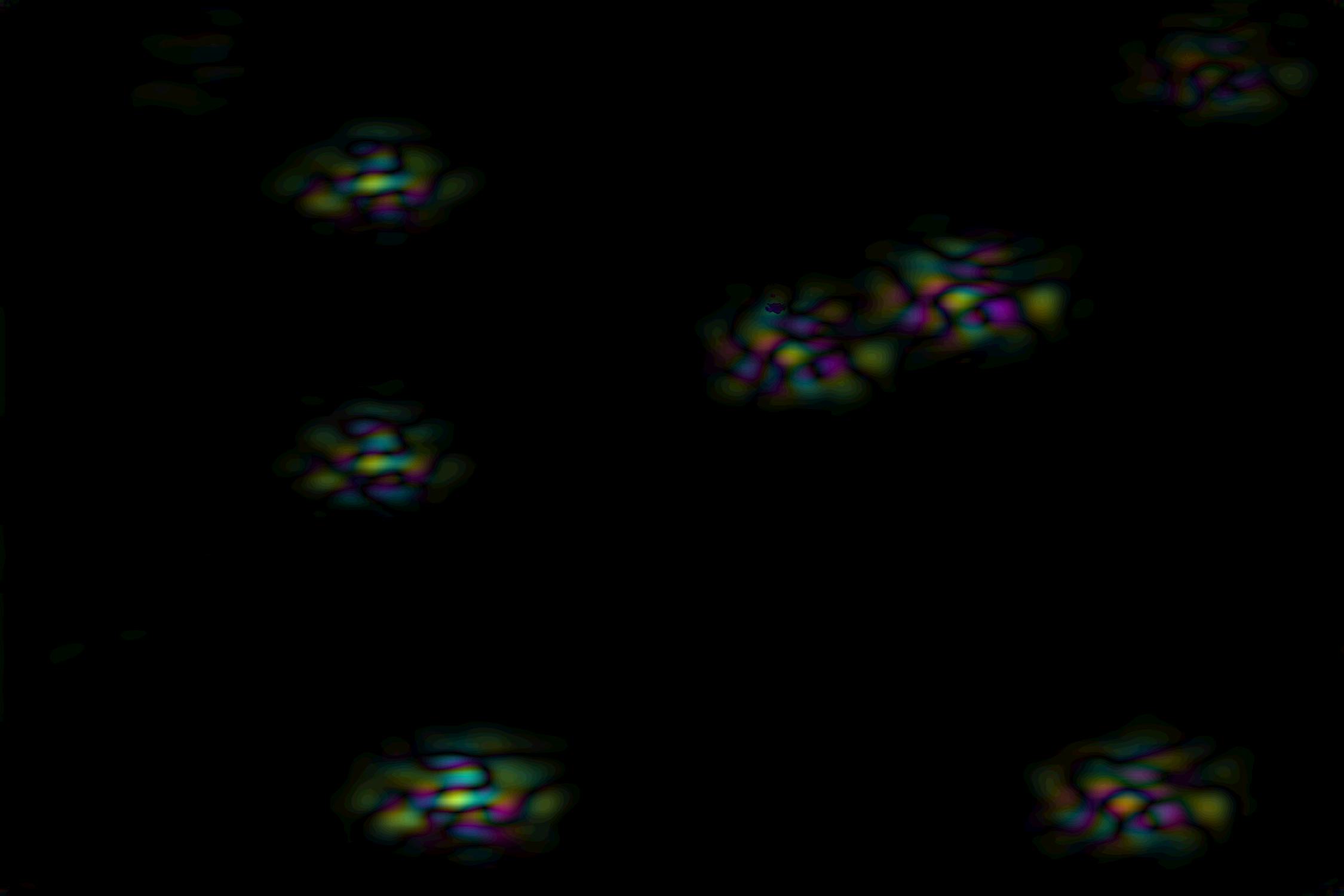} \\ \rule{0pt}{6ex}
    \includegraphics[width=\imwidth]{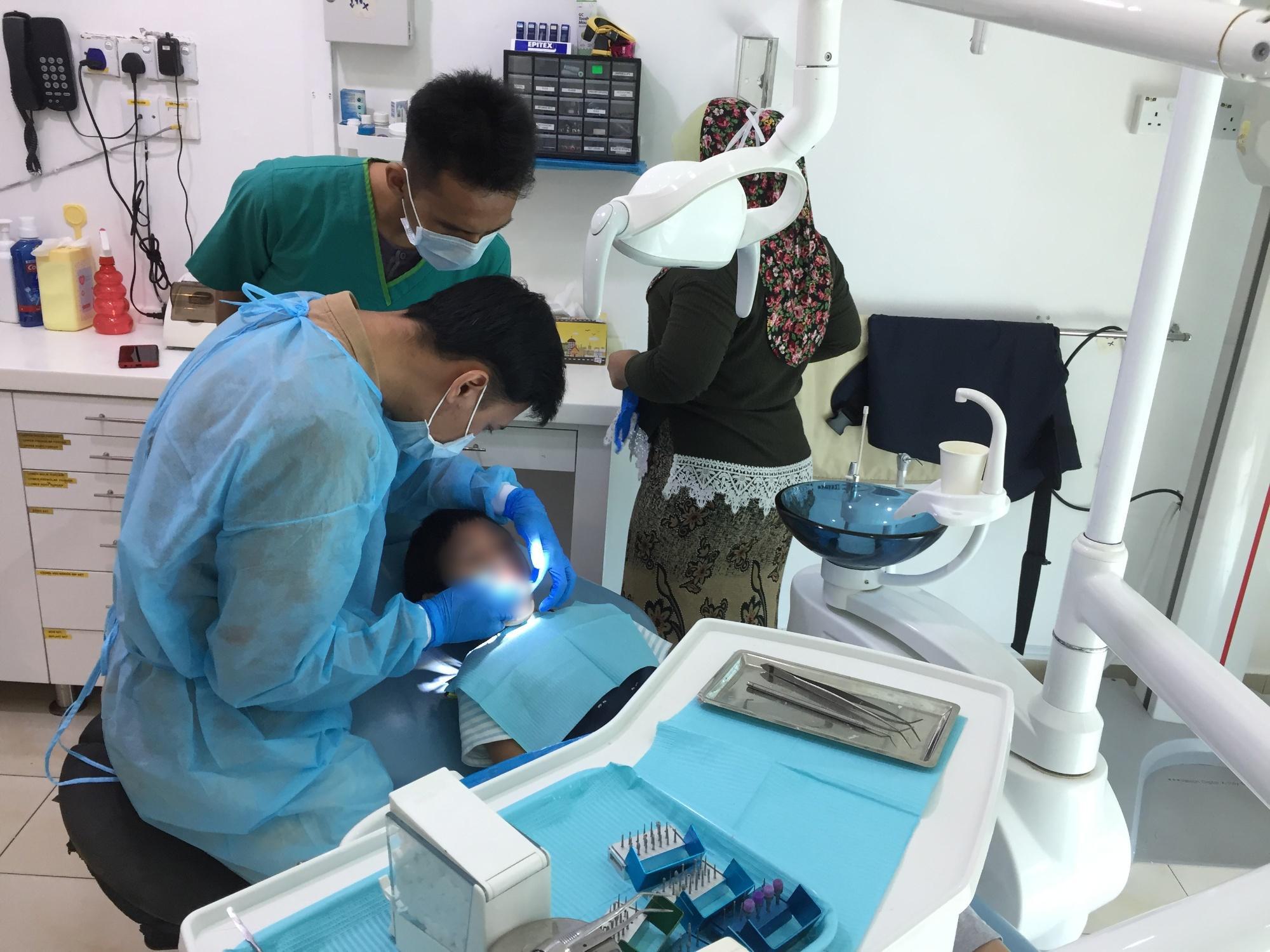} & 
    \includegraphics[width=\imwidth]{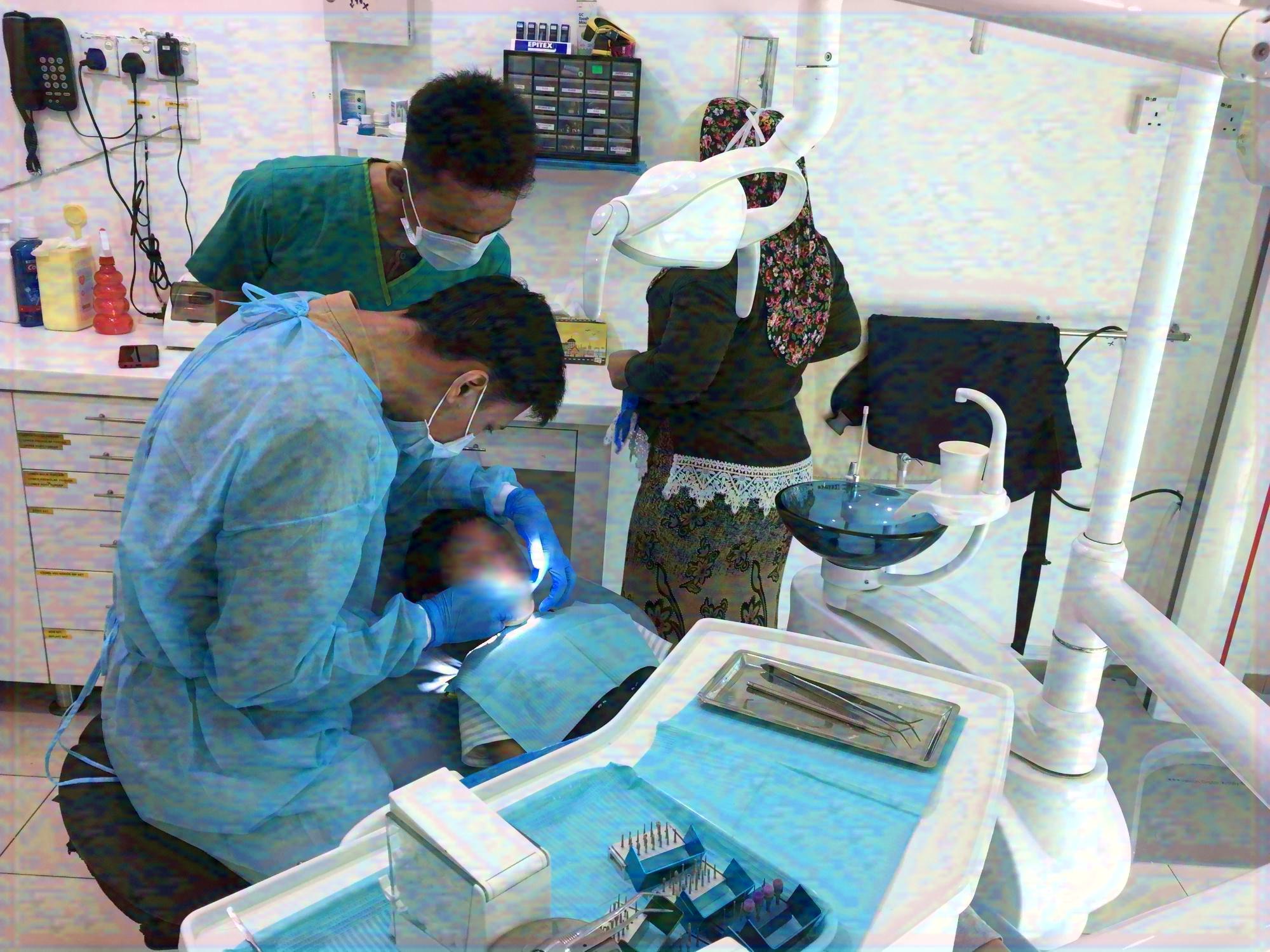} & 
    \includegraphics[width=\imwidth]{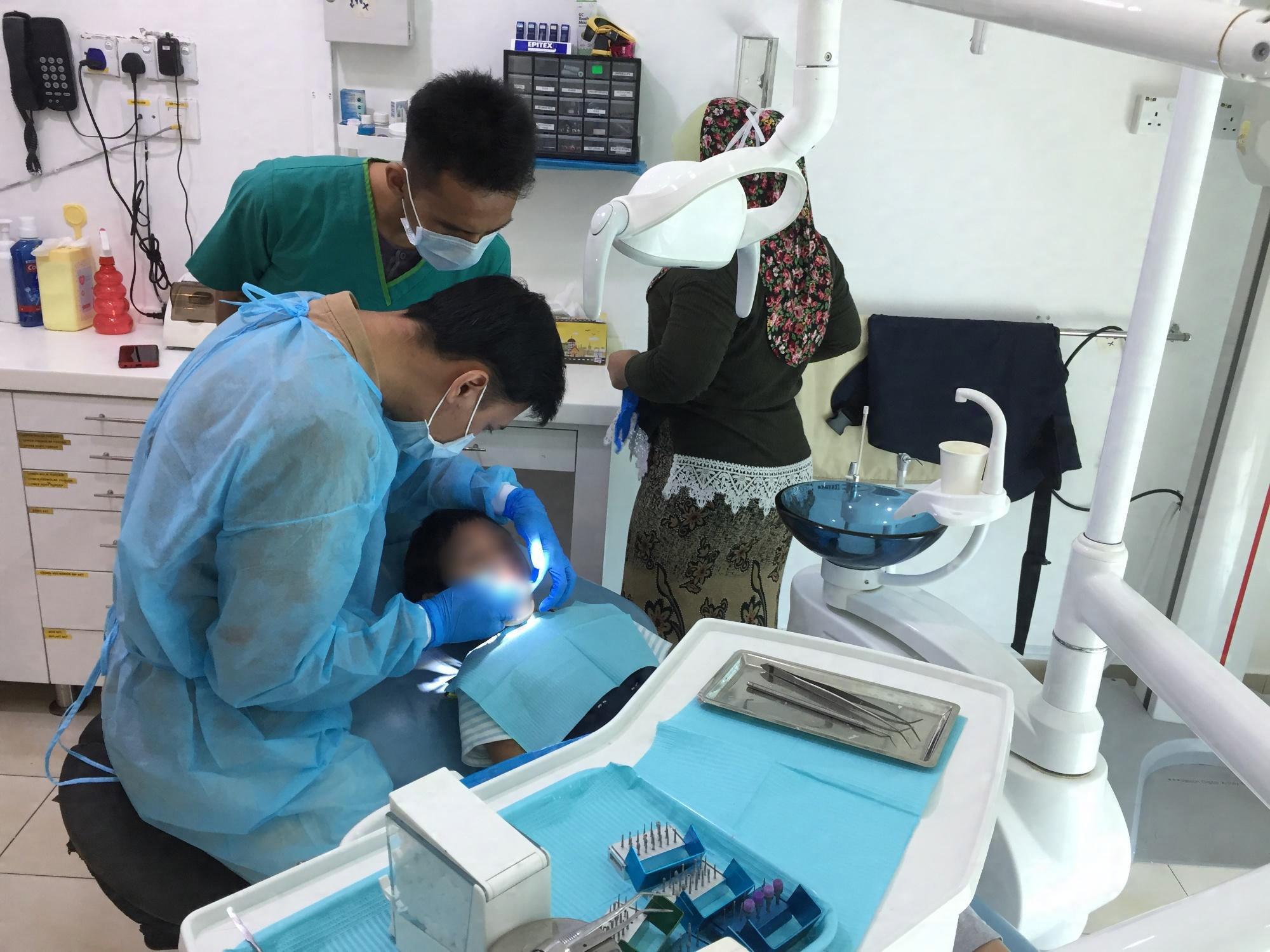} & 
    \includegraphics[width=\imwidth]{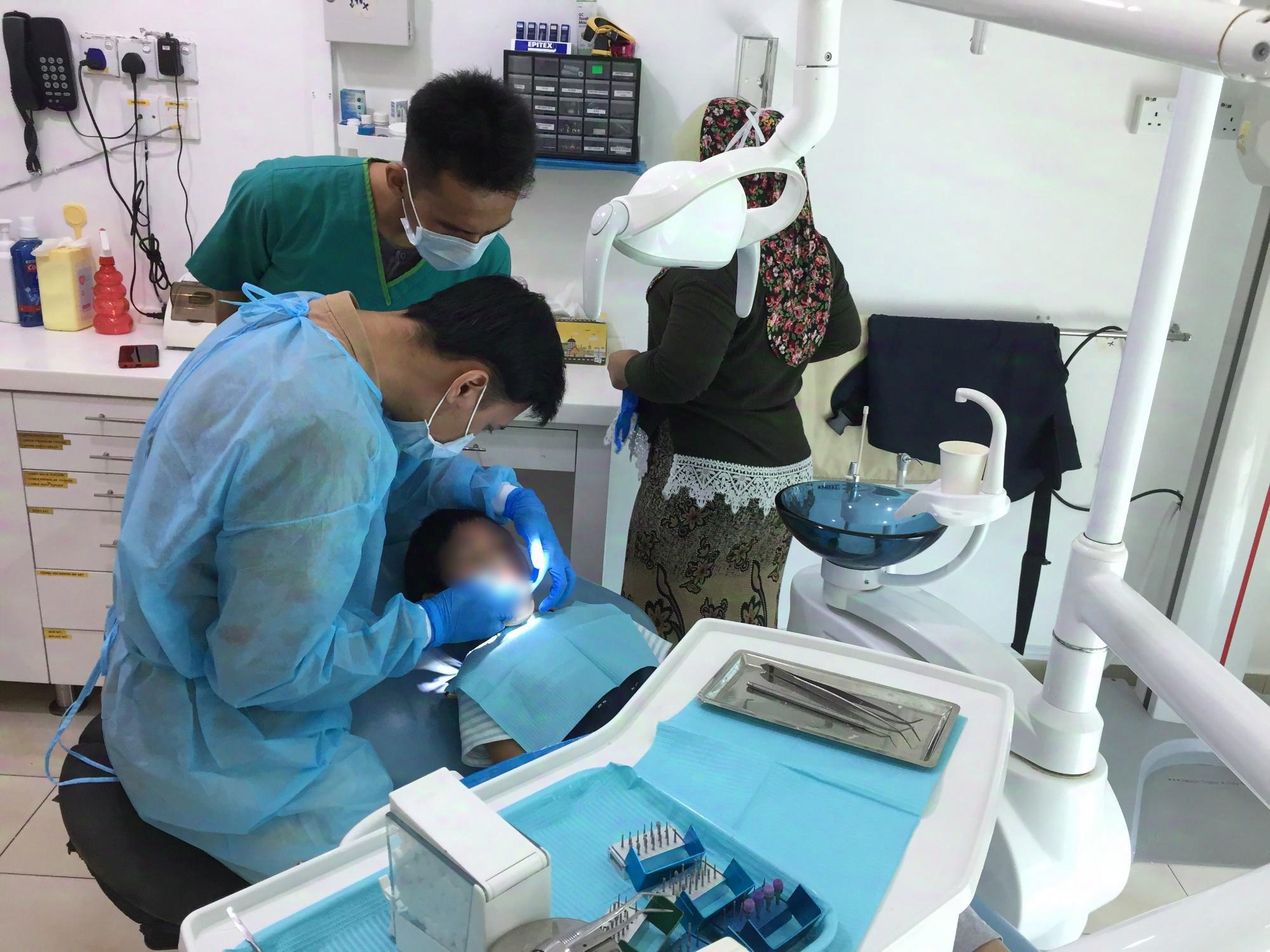} & 
    \includegraphics[width=\imwidth]{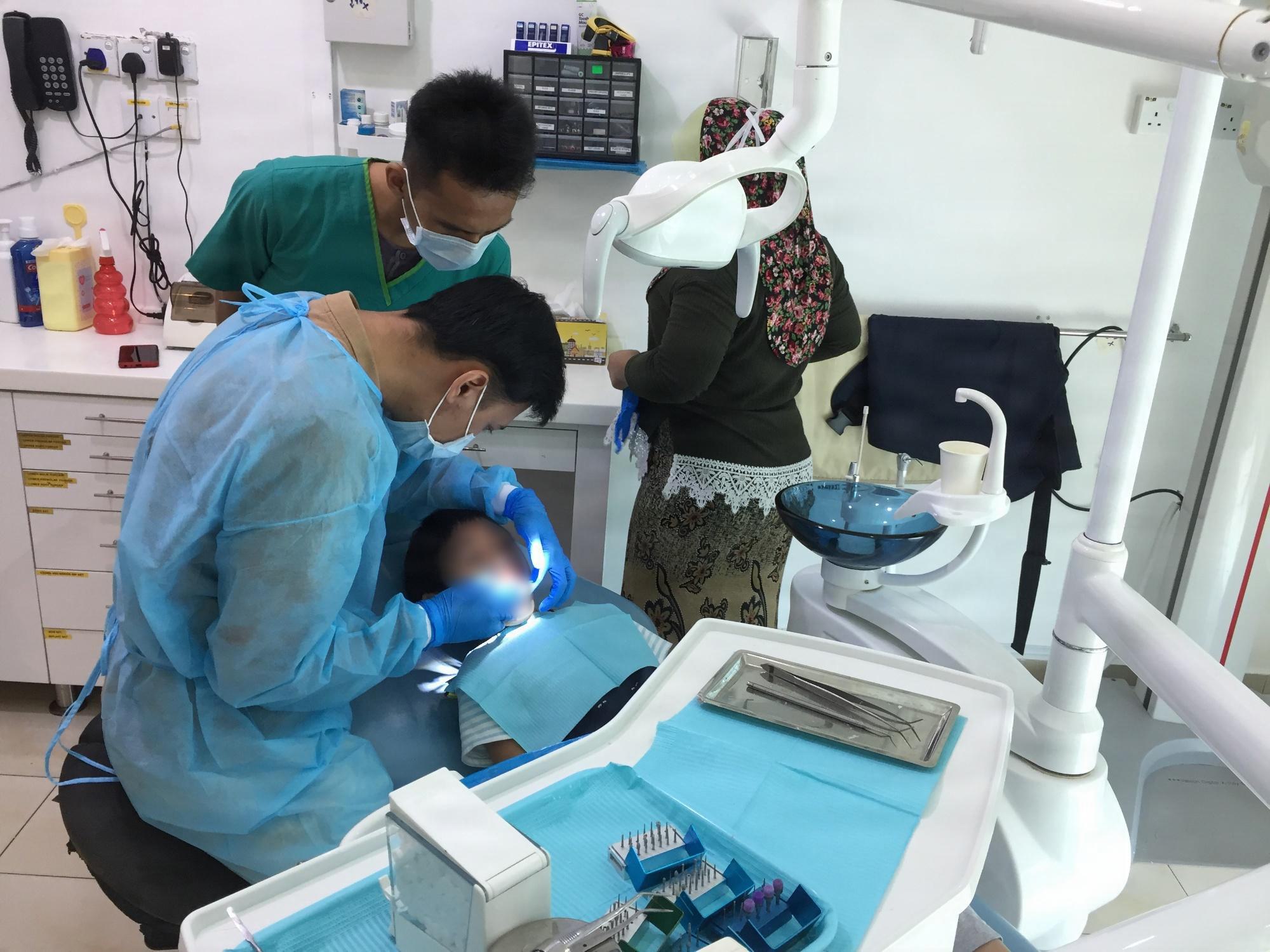} & 
    \includegraphics[width=\imwidth]{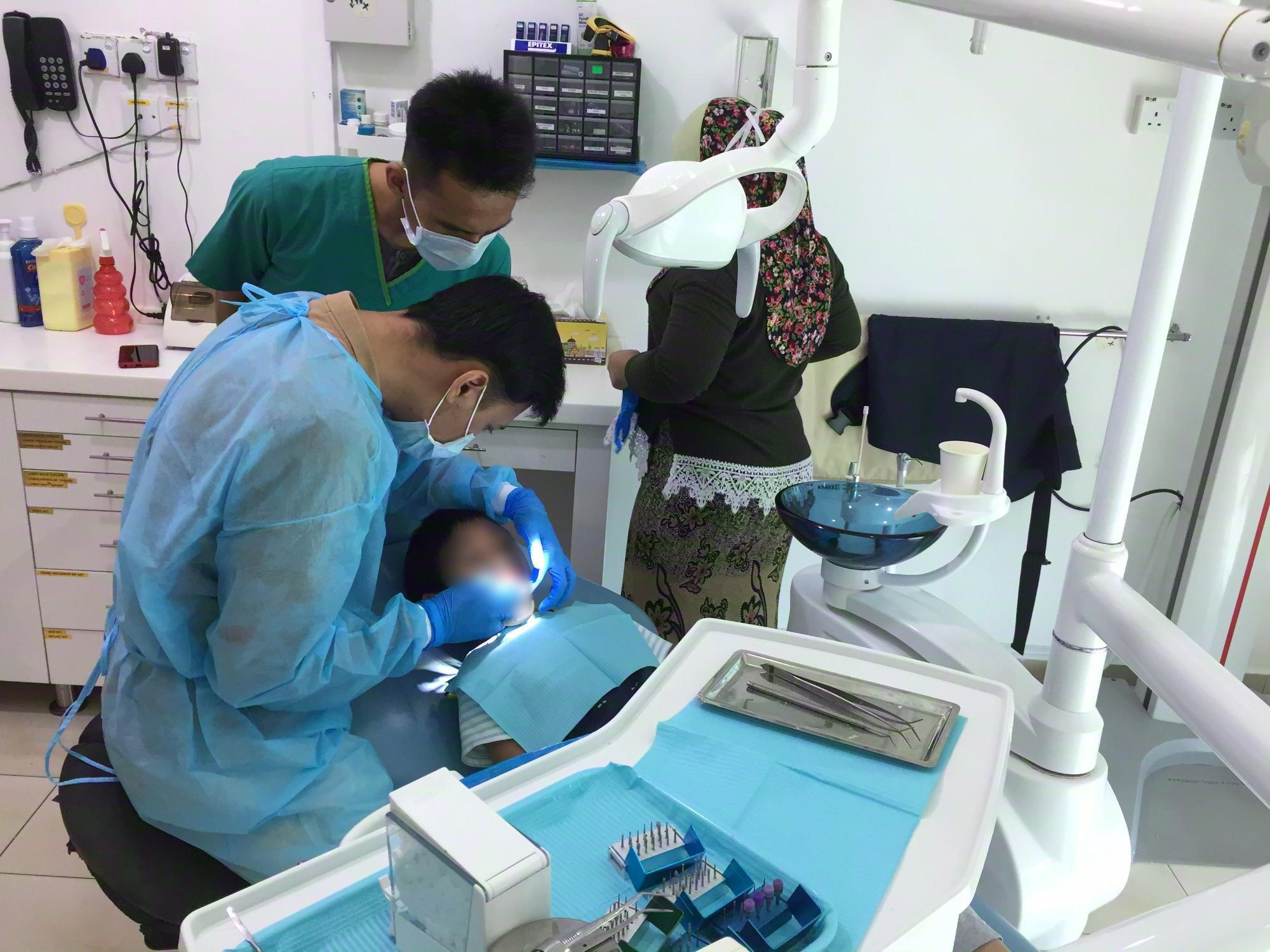} & 
    \includegraphics[width=\imwidth]{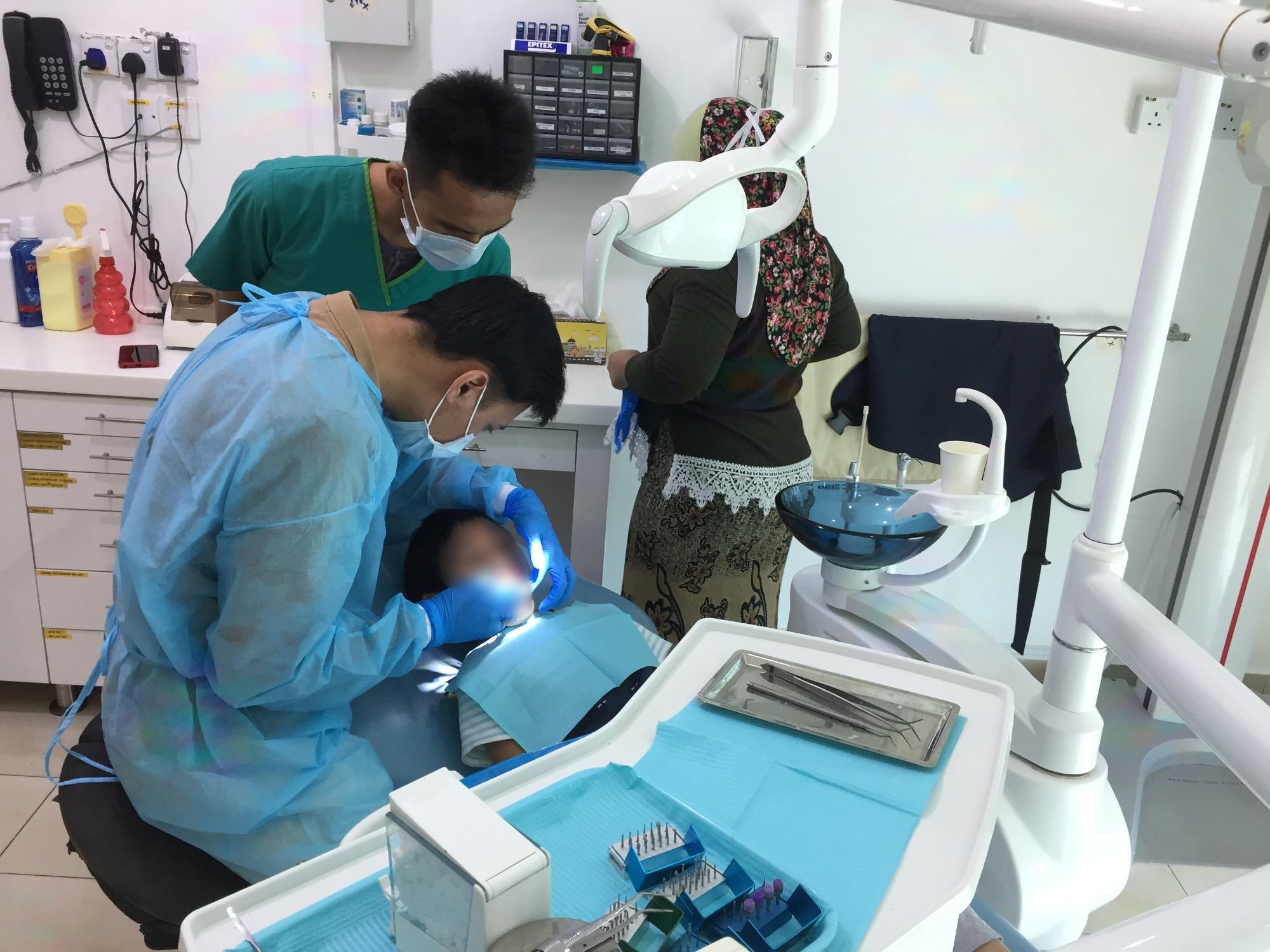} \\ & 
    \includegraphics[width=\imwidth]{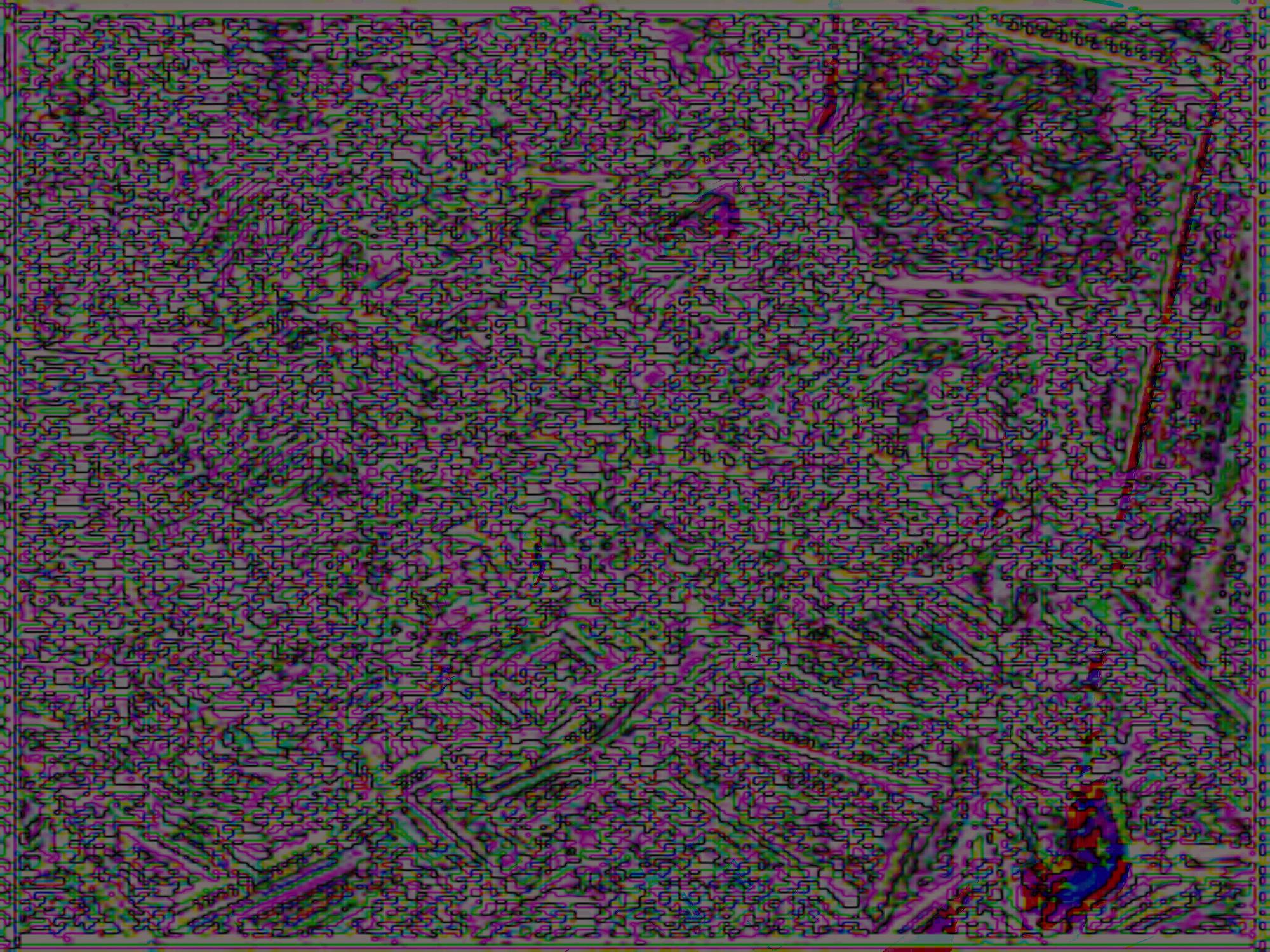} & 
    \includegraphics[width=\imwidth]{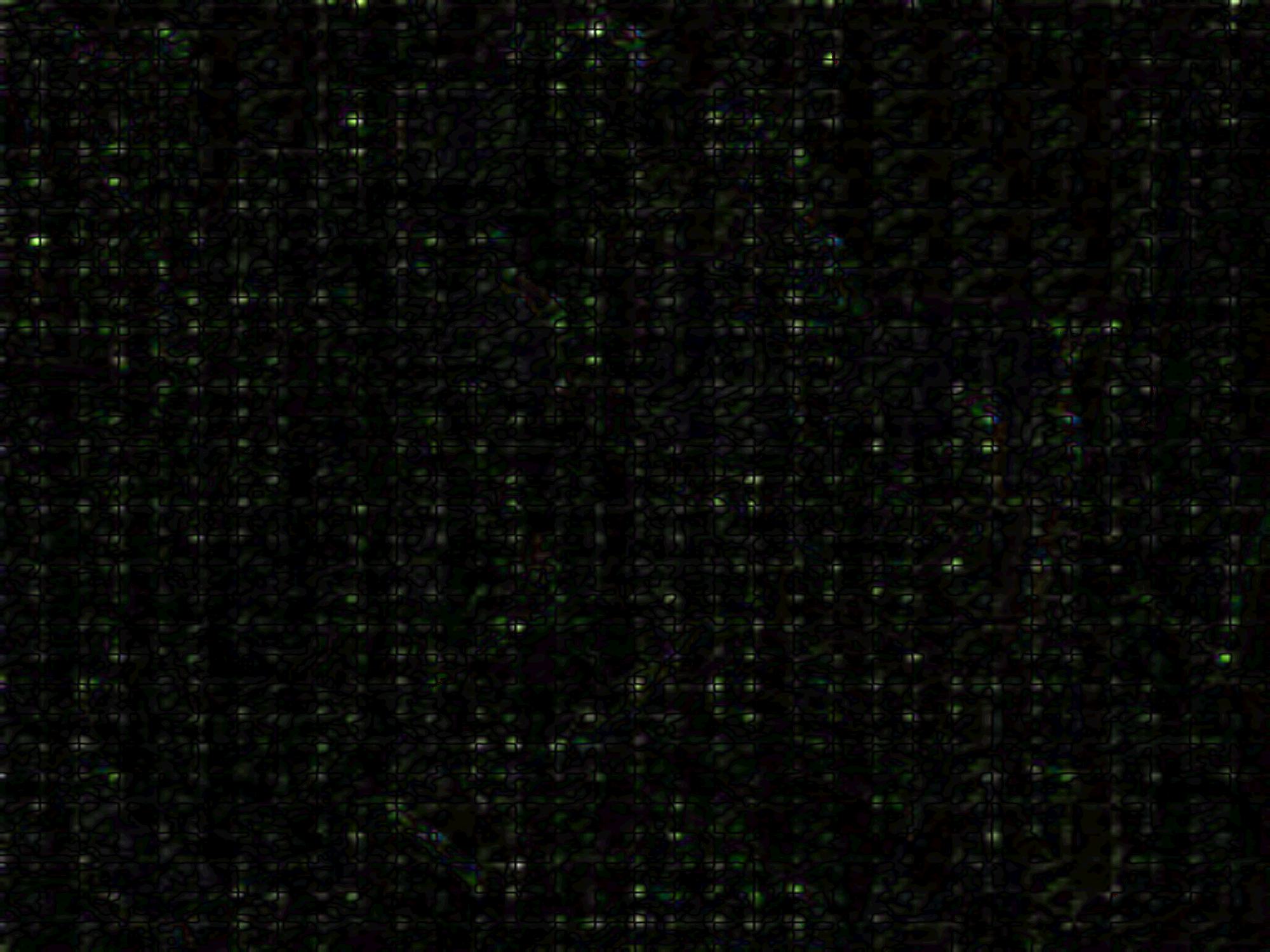} & 
    \includegraphics[width=\imwidth]{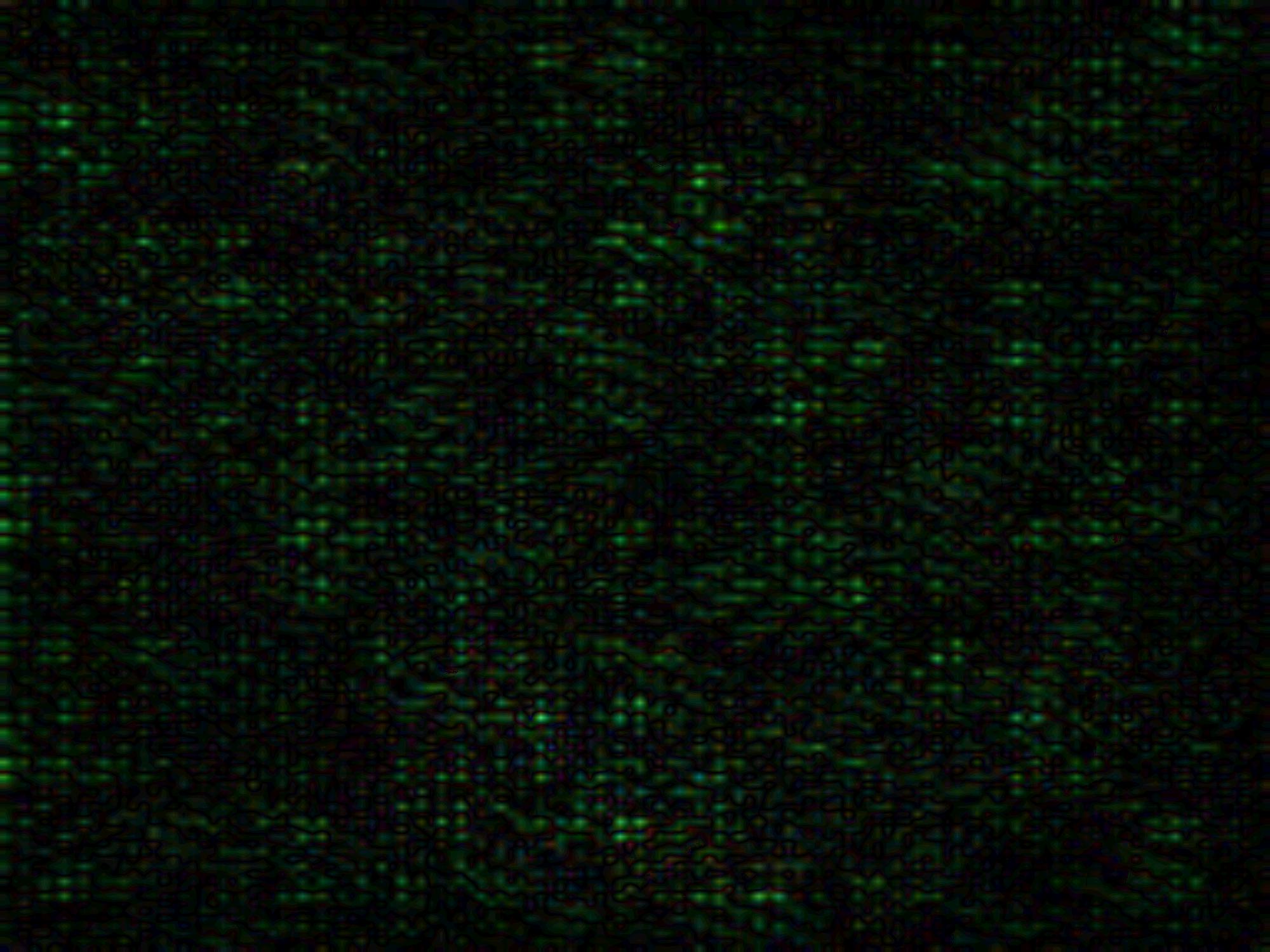} & 
    \includegraphics[width=\imwidth]{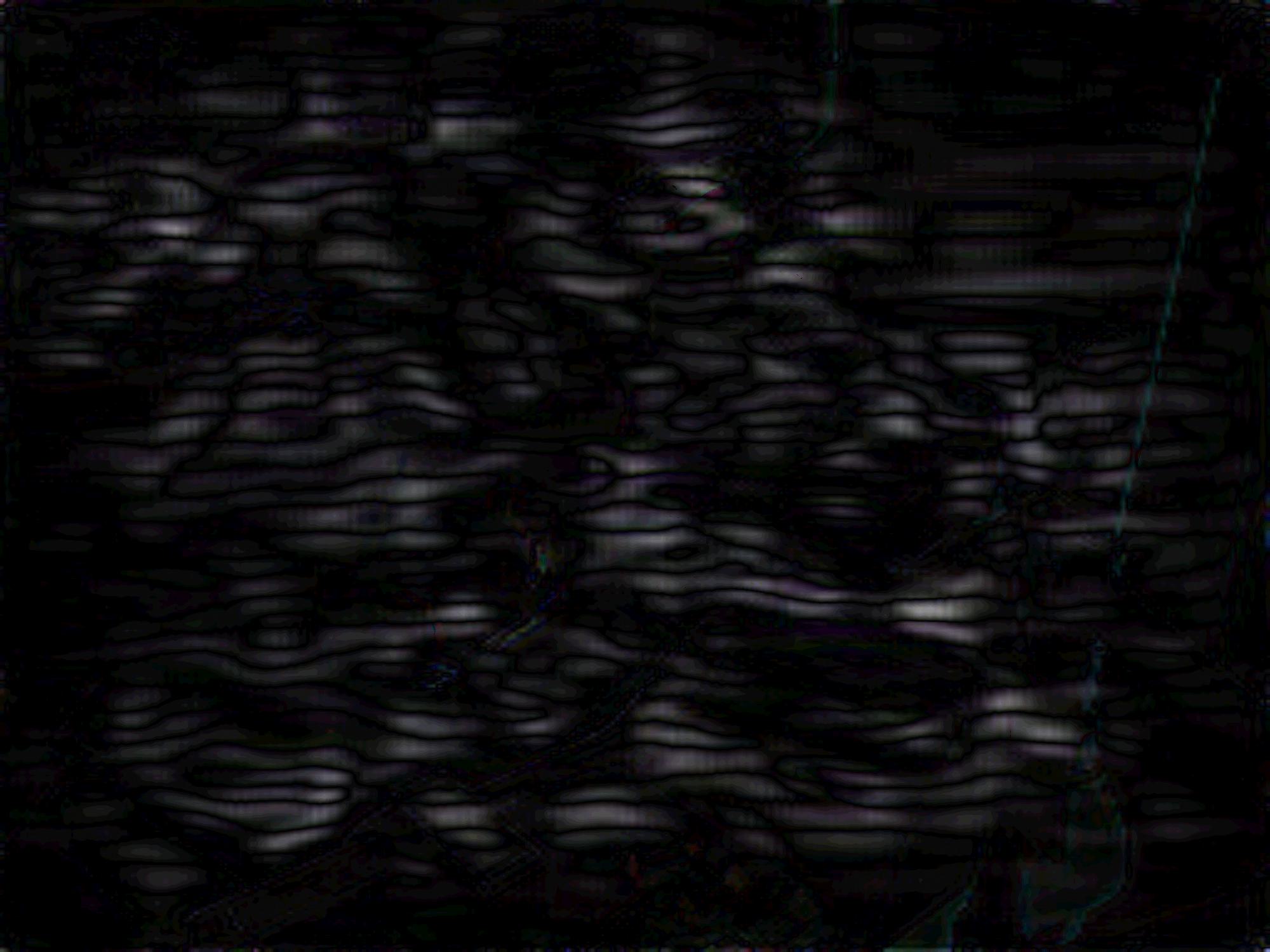} & 
    \includegraphics[width=\imwidth]{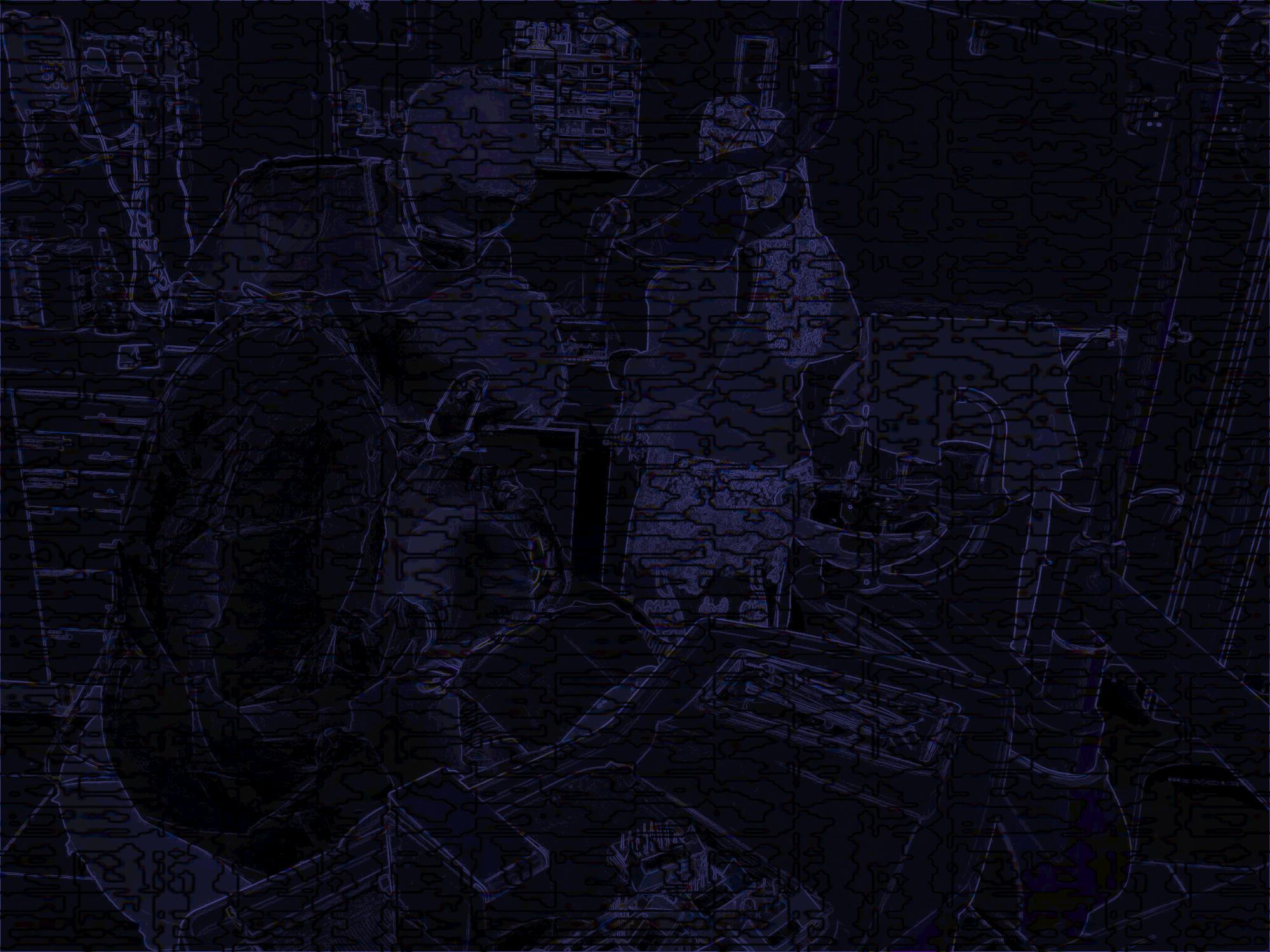} & 
    \includegraphics[width=\imwidth]{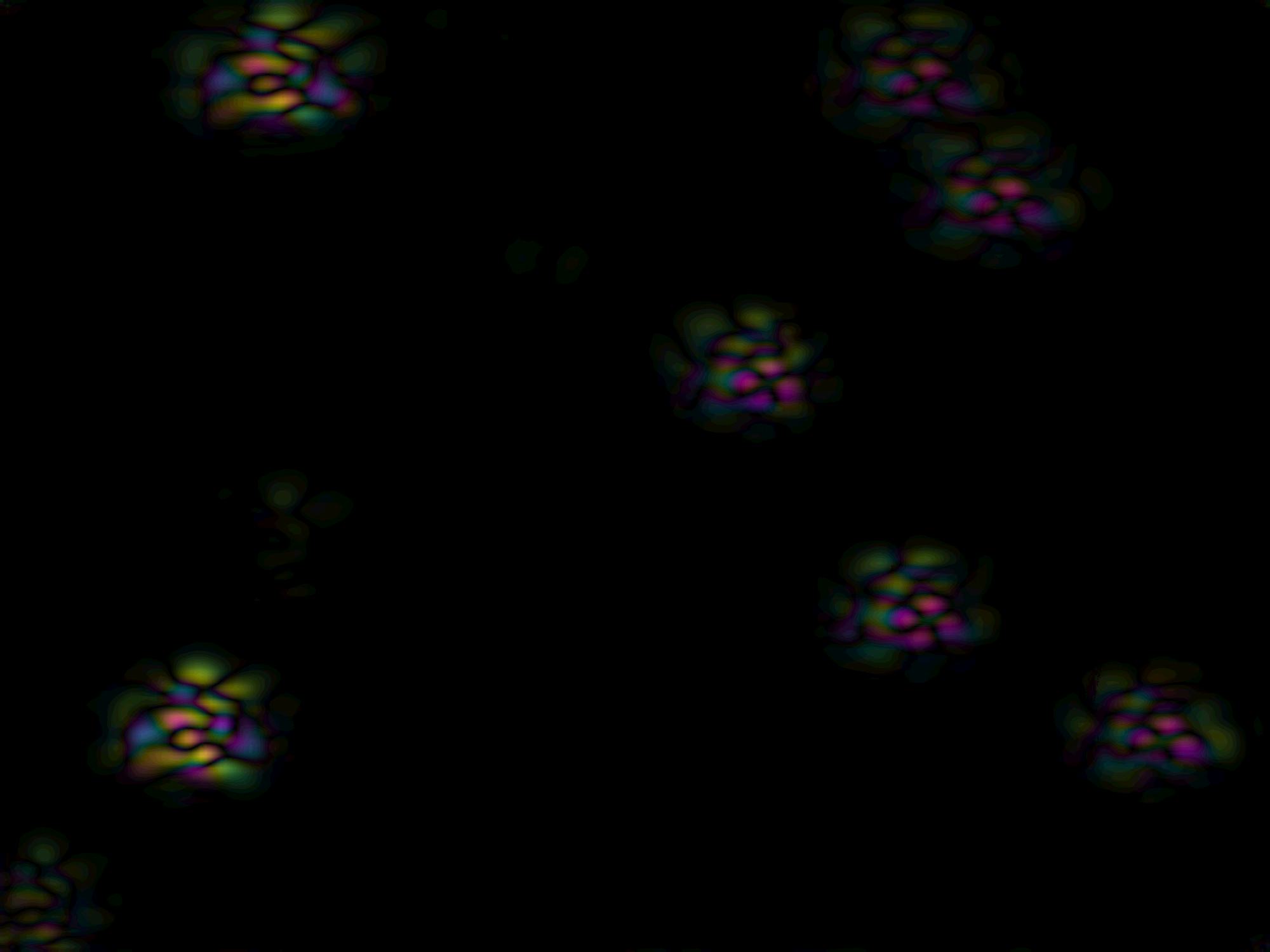} \\ \rule{0pt}{6ex}
    \end{tabular}
    \caption{
        Qualitative results for different watermarking methods.
        Images are from the SA-1b dataset at their original resolution ($\approx$2k $\times$ 1k).
    }\label{fig:app-qualitative-imgs} 
\end{figure*}

\begin{figure*}[h]
    \centering
    \scriptsize
    \setlength{\tabcolsep}{0pt}
    \begin{tabular}{c@{\hskip 2pt}c}
    \toprule
    \rotatebox[origin=l]{90}{Original}  
    &  \includegraphics[width=0.95\textwidth]{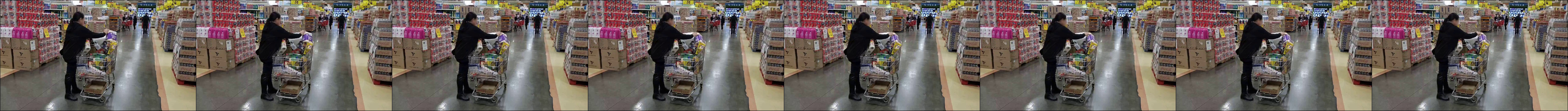} \\
    \midrule \multirow{2}{*}{\rotatebox[origin=c]{90}{HiDDeN}}  
    &  \includegraphics[width=0.95\textwidth]{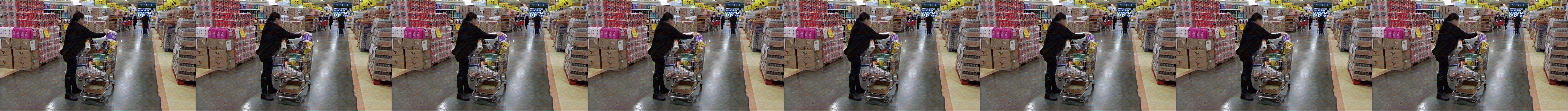} \\
    &  \includegraphics[width=0.95\textwidth]{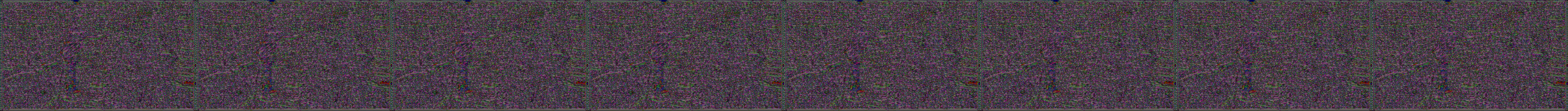} \\
    \midrule \multirow{2}{*}{\rotatebox[origin=c]{90}{MBRS}}  
    &  \includegraphics[width=0.95\textwidth]{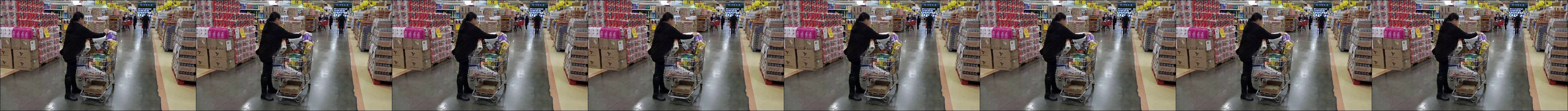} \\
    &  \includegraphics[width=0.95\textwidth]{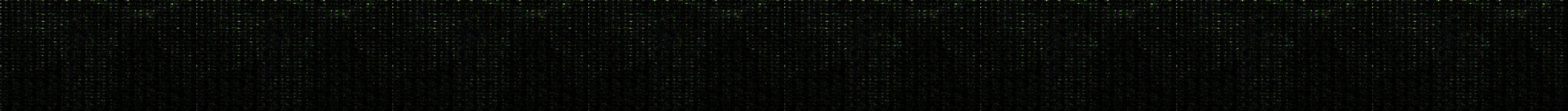} \\
    \midrule \multirow{2}{*}{\rotatebox[origin=c]{90}{CIN}}  
    &  \includegraphics[width=0.95\textwidth]{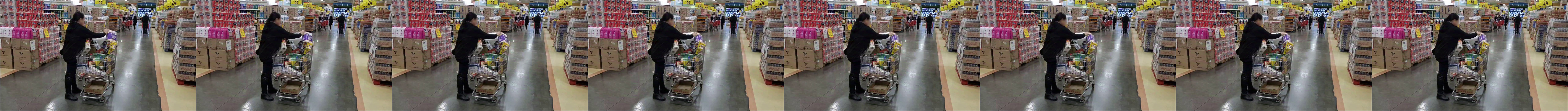} \\
    &  \includegraphics[width=0.95\textwidth]{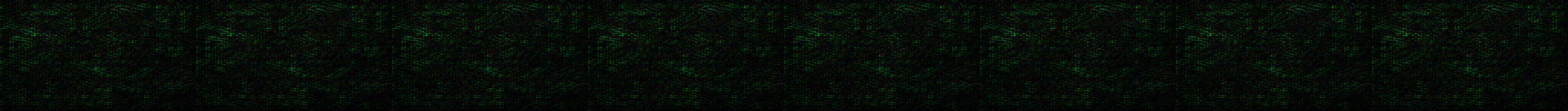} \\
    \midrule \multirow{2}{*}{\rotatebox[origin=c]{90}{TrustMark}}  
    &  \includegraphics[width=0.95\textwidth]{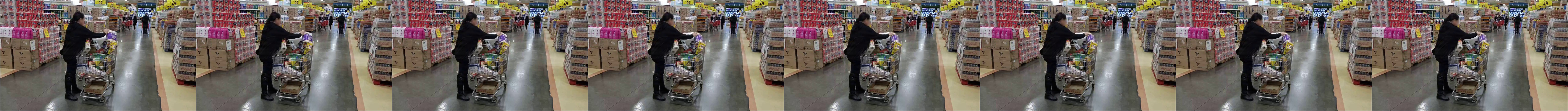} \\
    &  \includegraphics[width=0.95\textwidth]{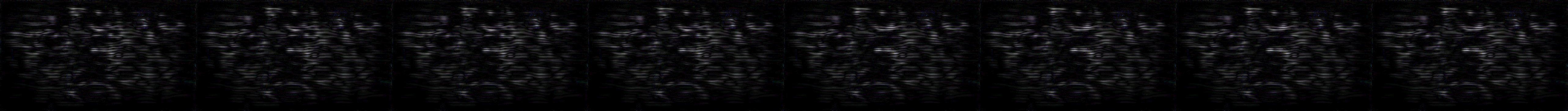} \\
    \midrule \multirow{2}{*}{\rotatebox[origin=c]{90}{WAM}}  
    &  \includegraphics[width=0.95\textwidth]{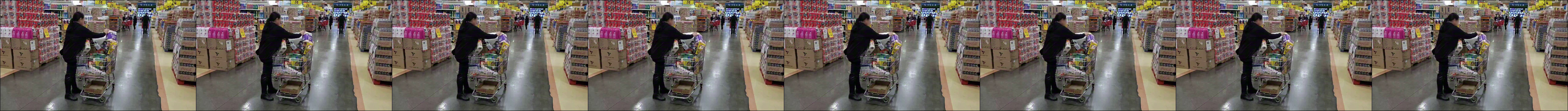} \\
    &  \includegraphics[width=0.95\textwidth]{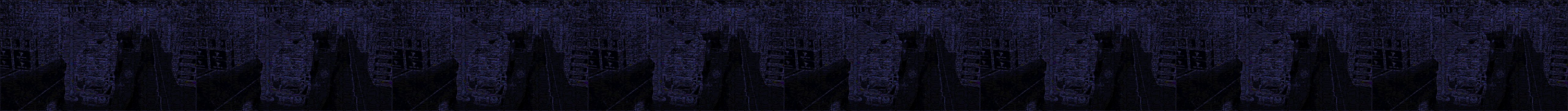} \\
    \midrule \multirow{2}{*}{\rotatebox[origin=c]{90}{\ours}}  
    &  \includegraphics[width=0.95\textwidth]{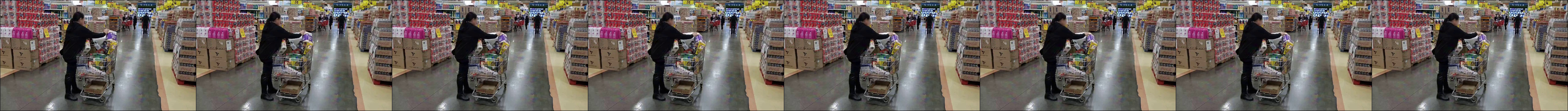} \\
    &  \includegraphics[width=0.95\textwidth]{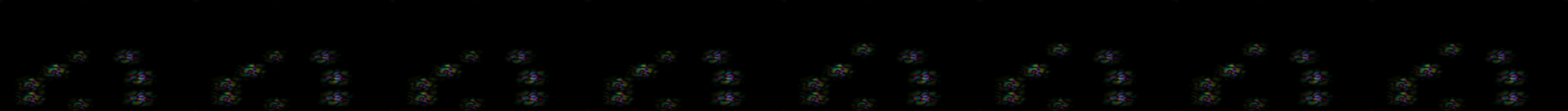} \\
    \bottomrule
    \end{tabular}
    \caption{
        Qualitative results for different watermarking methods.
        Frames are from the SA-V dataset at their original resolution ($\approx$2k $\times$ 1k).
    }\label{fig:app-qualitative-vids}
\end{figure*}

We show in Fig.~\ref{fig:app-qualitative-imgs} additional examples of watermarked images from SA-1b, and in Fig.~\ref{fig:app-qualitative-vids} watermarked frames from videos from SA-V.
They extend results of Fig.~\ref{fig:main-qualitative}.

\subsection{Full robustness results}\label{app:robustness}\label{app:transformations}

We report the robustness of watermark extraction across many transformations, and for various models, on the SA-1b, COCO, and SA-V datasets.
We report for each transformation type the bit accuracy and the $\logpval$, which accounts for the total number of bits, against the PSNR between the watermarked and the original videos.
When averaging categories of transformations, as done in Tab.~\ref{tab:robustness-sa-1b-sa-v}, we consider:
\begin{itemize}
    \item Identity: only the identity;
    \item Valuemetric: brightness, contrast, hue, saturation, Gaussian blur, median filter;
    \item Compression: JPEG (for images), H.264, H.265 (for videos)
    \item Geometric: horizontal flip, crop, resize, rotation, perspective;
    \item Combined: Compression (different CRFs) followed by a crop and a brightness change.
\end{itemize}

\clearpage
\newpage

\newpage

\clearpage
\newpage
\begin{table}[H]
    \centering
    \caption{
        Full results for the robustness of watermark extraction on the SA-1b dataset.
    }\label{tab:robustness-all-sa-1b}
    \resizebox{\linewidth}{0.45\textheight}{
        
    \begin{tabular}{c *{2}{>{\centering\arraybackslash}p{1.3cm}} *{2}{>{\centering\arraybackslash}p{1.3cm}} *{2}{>{\centering\arraybackslash}p{1.3cm}} *{2}{>{\centering\arraybackslash}p{1.3cm}} *{2}{>{\centering\arraybackslash}p{1.3cm}} *{2}{>{\centering\arraybackslash}p{1.3cm}}}
    \toprule
    & \multicolumn{2}{c}{\shortstack{HiDDeN}} & \multicolumn{2}{c}{\shortstack{MBRS}} & \multicolumn{2}{c}{\shortstack{CIN}} & \multicolumn{2}{c}{\shortstack{TrustMark}} & \multicolumn{2}{c}{\shortstack{WAM}} & \multicolumn{2}{c}{\shortstack{Video Seal (ours)}} \\
    \cmidrule(lr){2-3} \cmidrule(lr){4-5} \cmidrule(lr){6-7} \cmidrule(lr){8-9} \cmidrule(lr){10-11} \cmidrule(lr){12-13}
    & \multicolumn{2}{p{1cm}}{Bit~acc.~\footnotesize{($\uparrow$)}/~$\logpval$~\footnotesize{($\downarrow$)}} & \multicolumn{2}{p{1cm}}{Bit~acc.~\footnotesize{($\uparrow$)}/~$\logpval$~\footnotesize{($\downarrow$)}} & \multicolumn{2}{p{1cm}}{Bit~acc.~\footnotesize{($\uparrow$)}/~$\logpval$~\footnotesize{($\downarrow$)}} & \multicolumn{2}{p{1cm}}{Bit~acc.~\footnotesize{($\uparrow$)}/~$\logpval$~\footnotesize{($\downarrow$)}} & \multicolumn{2}{p{1cm}}{Bit~acc.~\footnotesize{($\uparrow$)}/~$\logpval$~\footnotesize{($\downarrow$)}} & \multicolumn{2}{p{1cm}}{Bit~acc.~\footnotesize{($\uparrow$)}/~$\logpval$~\footnotesize{($\downarrow$)}} \\
    \midrule
    Identity & 1.00 & -14.2 & 0.99 & -70.6 & 1.00 & -9.0 & 1.00 & -29.9 & 1.00 & -9.6 & 0.99 & -27.3 \\
HorizontalFlip & 0.69 & -2.4 & 0.50 & -0.5 & 0.50 & -0.4 & 1.00 & -29.9 & 1.00 & -9.6 & 0.99 & -27.2 \\
Rotate 5 & 0.93 & -10.1 & 0.50 & -0.4 & 0.50 & -0.4 & 0.61 & -3.4 & 0.98 & -8.7 & 0.98 & -26.0 \\
Rotate 10 & 0.83 & -5.9 & 0.50 & -0.4 & 0.50 & -0.3 & 0.51 & -0.5 & 0.72 & -2.3 & 0.96 & -23.3 \\
Rotate 30 & 0.55 & -0.7 & 0.50 & -0.4 & 0.50 & -0.4 & 0.50 & -0.4 & 0.50 & -0.4 & 0.58 & -1.4 \\
Rotate 45 & 0.50 & -0.4 & 0.50 & -0.4 & 0.50 & -0.3 & 0.50 & -0.4 & 0.50 & -0.4 & 0.51 & -0.5 \\
Rotate 90 & 0.49 & -0.4 & 0.50 & -0.4 & 0.50 & -0.4 & 0.50 & -0.4 & 0.50 & -0.4 & 0.50 & -0.4 \\
Resize 0.32 & 0.99 & -14.0 & 0.98 & -69.1 & 1.00 & -9.0 & 1.00 & -29.9 & 1.00 & -9.6 & 0.99 & -27.3 \\
Resize 0.45 & 1.00 & -14.1 & 0.99 & -70.1 & 1.00 & -9.0 & 1.00 & -29.9 & 1.00 & -9.6 & 0.99 & -27.3 \\
Resize 0.55 & 1.00 & -14.2 & 0.99 & -70.2 & 1.00 & -9.0 & 1.00 & -29.9 & 1.00 & -9.6 & 0.99 & -27.3 \\
Resize 0.63 & 1.00 & -14.2 & 0.99 & -70.4 & 1.00 & -9.0 & 1.00 & -29.9 & 1.00 & -9.6 & 0.99 & -27.3 \\
Resize 0.71 & 1.00 & -14.2 & 0.99 & -70.4 & 1.00 & -9.0 & 1.00 & -29.9 & 1.00 & -9.6 & 0.99 & -27.3 \\
Resize 0.77 & 1.00 & -14.2 & 0.99 & -70.4 & 1.00 & -9.0 & 1.00 & -29.9 & 1.00 & -9.6 & 0.99 & -27.3 \\
Resize 0.84 & 1.00 & -14.2 & 0.99 & -70.5 & 1.00 & -9.0 & 1.00 & -29.9 & 1.00 & -9.6 & 0.99 & -27.3 \\
Resize 0.89 & 1.00 & -14.2 & 0.99 & -70.5 & 1.00 & -9.0 & 1.00 & -29.9 & 1.00 & -9.6 & 0.99 & -27.3 \\
Resize 0.95 & 1.00 & -14.2 & 0.99 & -70.5 & 1.00 & -9.0 & 1.00 & -29.9 & 1.00 & -9.6 & 0.99 & -27.3 \\
Resize 1.0 & 1.00 & -14.2 & 0.99 & -70.6 & 1.00 & -9.0 & 1.00 & -29.9 & 1.00 & -9.6 & 0.99 & -27.3 \\
Crop 0.32 & 0.48 & -0.3 & 0.50 & -0.4 & 0.50 & -0.4 & 0.50 & -0.4 & 0.79 & -3.8 & 0.50 & -0.4 \\
Crop 0.45 & 0.50 & -0.4 & 0.50 & -0.4 & 0.50 & -0.4 & 0.50 & -0.4 & 0.94 & -7.5 & 0.52 & -0.6 \\
Crop 0.55 & 0.58 & -0.9 & 0.50 & -0.4 & 0.50 & -0.4 & 0.50 & -0.3 & 0.97 & -8.4 & 0.70 & -5.8 \\
Crop 0.63 & 0.66 & -2.0 & 0.50 & -0.4 & 0.50 & -0.4 & 0.50 & -0.4 & 0.98 & -8.8 & 0.84 & -13.4 \\
Crop 0.71 & 0.77 & -4.1 & 0.50 & -0.4 & 0.50 & -0.4 & 0.55 & -1.2 & 0.99 & -9.0 & 0.92 & -20.2 \\
Crop 0.77 & 0.86 & -6.8 & 0.50 & -0.4 & 0.50 & -0.4 & 0.92 & -21.0 & 0.99 & -9.2 & 0.96 & -23.2 \\
Crop 0.84 & 0.93 & -9.9 & 0.50 & -0.4 & 0.49 & -0.3 & 1.00 & -30.0 & 0.99 & -9.3 & 0.98 & -25.2 \\
Crop 0.89 & 0.95 & -11.0 & 0.50 & -0.4 & 0.49 & -0.3 & 1.00 & -30.0 & 0.99 & -9.4 & 0.98 & -26.2 \\
Crop 0.95 & 0.96 & -11.6 & 0.50 & -0.5 & 0.51 & -0.4 & 1.00 & -30.1 & 0.99 & -9.4 & 0.99 & -26.4 \\
Crop 1.0 & 1.00 & -14.2 & 0.99 & -70.6 & 1.00 & -9.0 & 1.00 & -29.9 & 1.00 & -9.6 & 0.99 & -27.3 \\
Perspective 0.1 & 0.94 & -10.8 & 0.51 & -0.5 & 0.51 & -0.6 & 0.92 & -20.6 & 0.99 & -9.2 & 0.98 & -26.3 \\
Perspective 0.2 & 0.93 & -9.8 & 0.50 & -0.4 & 0.50 & -0.4 & 0.61 & -3.2 & 0.92 & -7.0 & 0.98 & -25.8 \\
Perspective 0.3 & 0.89 & -8.3 & 0.50 & -0.4 & 0.50 & -0.4 & 0.52 & -0.7 & 0.79 & -3.8 & 0.97 & -24.7 \\
Perspective 0.4 & 0.85 & -6.7 & 0.50 & -0.4 & 0.51 & -0.4 & 0.51 & -0.5 & 0.68 & -2.1 & 0.95 & -22.2 \\
Perspective 0.5 & 0.80 & -5.3 & 0.50 & -0.4 & 0.50 & -0.4 & 0.50 & -0.4 & 0.61 & -1.2 & 0.91 & -18.4 \\
Perspective 0.6 & 0.76 & -4.1 & 0.50 & -0.4 & 0.50 & -0.4 & 0.50 & -0.4 & 0.57 & -0.9 & 0.85 & -13.4 \\
Perspective 0.7 & 0.71 & -3.1 & 0.50 & -0.4 & 0.50 & -0.4 & 0.50 & -0.4 & 0.54 & -0.6 & 0.77 & -8.9 \\
Perspective 0.8 & 0.66 & -2.2 & 0.50 & -0.4 & 0.50 & -0.4 & 0.50 & -0.4 & 0.52 & -0.5 & 0.69 & -5.5 \\
Brightness 0.1 & 0.59 & -1.2 & 0.62 & -4.5 & 1.00 & -9.0 & 0.81 & -11.3 & 0.95 & -7.7 & 0.96 & -22.8 \\
Brightness 0.25 & 0.82 & -5.7 & 0.84 & -30.0 & 1.00 & -9.0 & 0.97 & -25.4 & 1.00 & -9.6 & 0.99 & -27.0 \\
Brightness 0.5 & 0.95 & -11.2 & 0.95 & -57.4 & 1.00 & -9.0 & 1.00 & -29.3 & 1.00 & -9.6 & 0.99 & -27.3 \\
Brightness 0.75 & 0.99 & -13.6 & 0.98 & -67.4 & 1.00 & -9.0 & 1.00 & -29.8 & 1.00 & -9.6 & 0.99 & -27.3 \\
Brightness 1.0 & 1.00 & -14.2 & 0.99 & -70.6 & 1.00 & -9.0 & 1.00 & -29.9 & 1.00 & -9.6 & 0.99 & -27.3 \\
Brightness 1.25 & 1.00 & -14.2 & 0.95 & -57.7 & 1.00 & -9.0 & 0.98 & -26.9 & 1.00 & -9.6 & 0.97 & -25.5 \\
Brightness 1.5 & 0.99 & -13.9 & 0.90 & -47.2 & 1.00 & -9.0 & 0.94 & -23.3 & 1.00 & -9.6 & 0.94 & -23.0 \\
Brightness 1.75 & 0.99 & -13.6 & 0.88 & -40.5 & 1.00 & -8.9 & 0.91 & -20.3 & 1.00 & -9.6 & 0.91 & -20.6 \\
Brightness 2.0 & 0.98 & -13.2 & 0.85 & -36.1 & 0.99 & -8.8 & 0.88 & -17.8 & 1.00 & -9.5 & 0.89 & -18.6 \\
Contrast 0.1 & 0.55 & -0.7 & 0.75 & -16.5 & 1.00 & -9.0 & 0.75 & -8.1 & 1.00 & -9.5 & 0.97 & -23.7 \\
Contrast 0.25 & 0.79 & -4.8 & 0.89 & -41.7 & 1.00 & -9.0 & 0.95 & -23.6 & 1.00 & -9.6 & 0.99 & -27.1 \\
Contrast 0.5 & 0.95 & -11.1 & 0.96 & -61.4 & 1.00 & -9.0 & 0.99 & -29.0 & 1.00 & -9.6 & 0.99 & -27.3 \\
Contrast 0.75 & 0.99 & -13.6 & 0.98 & -68.3 & 1.00 & -9.0 & 1.00 & -29.8 & 1.00 & -9.6 & 0.99 & -27.3 \\
Contrast 1.0 & 1.00 & -14.2 & 0.99 & -70.6 & 1.00 & -9.0 & 1.00 & -29.9 & 1.00 & -9.6 & 0.99 & -27.3 \\
Contrast 1.25 & 1.00 & -14.3 & 0.96 & -61.1 & 1.00 & -9.0 & 0.98 & -27.8 & 1.00 & -9.6 & 0.98 & -26.2 \\
Contrast 1.5 & 1.00 & -14.2 & 0.93 & -53.6 & 1.00 & -9.0 & 0.96 & -24.6 & 1.00 & -9.6 & 0.96 & -24.5 \\
Contrast 1.75 & 1.00 & -14.1 & 0.91 & -47.9 & 1.00 & -9.0 & 0.93 & -21.7 & 1.00 & -9.6 & 0.95 & -23.0 \\
Contrast 2.0 & 0.99 & -14.1 & 0.89 & -43.5 & 1.00 & -9.0 & 0.90 & -19.2 & 1.00 & -9.6 & 0.93 & -21.5 \\
Hue -0.4 & 0.25 & -0.0 & 0.92 & -49.0 & 0.03 & -0.0 & 0.97 & -26.5 & 0.49 & -0.3 & 0.50 & -0.4 \\
Hue -0.3 & 0.45 & -0.2 & 0.94 & -56.0 & 0.26 & -0.2 & 0.98 & -27.6 & 0.48 & -0.3 & 0.49 & -0.4 \\
Hue -0.2 & 0.61 & -1.3 & 0.95 & -58.5 & 0.89 & -6.3 & 0.99 & -27.9 & 0.98 & -8.7 & 0.88 & -14.7 \\
Hue -0.1 & 0.78 & -4.6 & 0.97 & -64.6 & 1.00 & -9.0 & 0.99 & -29.1 & 1.00 & -9.6 & 0.98 & -25.9 \\
Hue 0.0 & 1.00 & -14.2 & 0.99 & -70.6 & 1.00 & -9.0 & 1.00 & -29.9 & 1.00 & -9.6 & 0.99 & -27.3 \\
Hue 0.1 & 0.85 & -6.8 & 0.97 & -64.8 & 1.00 & -9.0 & 0.99 & -29.1 & 1.00 & -9.6 & 0.98 & -26.0 \\
Hue 0.2 & 0.62 & -1.4 & 0.96 & -59.9 & 0.95 & -7.6 & 0.99 & -28.6 & 0.98 & -9.1 & 0.85 & -13.2 \\
Hue 0.3 & 0.44 & -0.2 & 0.95 & -58.7 & 0.13 & -0.0 & 0.99 & -28.3 & 0.48 & -0.3 & 0.50 & -0.4 \\
Hue 0.4 & 0.28 & -0.0 & 0.93 & -52.3 & 0.02 & -0.0 & 0.97 & -26.6 & 0.49 & -0.3 & 0.50 & -0.4 \\
Hue 0.5 & 0.03 & -0.0 & 0.89 & -43.0 & 0.00 & -0.0 & 0.95 & -23.5 & 0.49 & -0.3 & 0.51 & -0.5 \\
JPEG 40 & 1.00 & -14.1 & 0.98 & -69.1 & 1.00 & -9.0 & 1.00 & -29.4 & 1.00 & -9.6 & 0.99 & -26.9 \\
JPEG 50 & 1.00 & -14.1 & 0.99 & -70.6 & 1.00 & -9.0 & 1.00 & -29.8 & 1.00 & -9.6 & 0.99 & -27.0 \\
JPEG 60 & 1.00 & -14.2 & 0.98 & -69.7 & 1.00 & -9.0 & 1.00 & -29.7 & 1.00 & -9.6 & 0.99 & -27.1 \\
JPEG 70 & 1.00 & -14.2 & 0.98 & -69.6 & 1.00 & -9.0 & 1.00 & -29.7 & 1.00 & -9.6 & 0.99 & -27.2 \\
JPEG 80 & 1.00 & -14.2 & 0.99 & -70.1 & 1.00 & -9.0 & 1.00 & -29.8 & 1.00 & -9.6 & 0.99 & -27.2 \\
JPEG 90 & 1.00 & -14.2 & 0.99 & -70.5 & 1.00 & -9.0 & 1.00 & -29.8 & 1.00 & -9.6 & 0.99 & -27.3 \\
GaussianBlur 3 & 1.00 & -14.2 & 0.99 & -70.3 & 1.00 & -9.0 & 1.00 & -29.9 & 1.00 & -9.6 & 0.99 & -27.3 \\
GaussianBlur 5 & 1.00 & -14.1 & 0.99 & -69.9 & 1.00 & -9.0 & 1.00 & -29.9 & 1.00 & -9.6 & 0.99 & -27.3 \\
GaussianBlur 9 & 0.99 & -13.8 & 0.98 & -68.7 & 1.00 & -9.0 & 1.00 & -29.9 & 1.00 & -9.6 & 0.99 & -27.3 \\
GaussianBlur 13 & 0.98 & -13.1 & 0.98 & -66.8 & 1.00 & -9.0 & 1.00 & -29.9 & 1.00 & -9.6 & 0.99 & -27.3 \\
GaussianBlur 17 & 0.97 & -12.2 & 0.97 & -63.9 & 1.00 & -9.0 & 1.00 & -29.9 & 1.00 & -9.6 & 0.99 & -27.3 \\
MedianFilter 3 & 1.00 & -14.2 & 0.99 & -70.2 & 1.00 & -9.0 & 1.00 & -29.9 & 1.00 & -9.6 & 0.99 & -27.3 \\
MedianFilter 5 & 0.99 & -14.0 & 0.98 & -69.3 & 1.00 & -9.0 & 1.00 & -29.8 & 1.00 & -9.6 & 0.99 & -27.3 \\
MedianFilter 9 & 0.98 & -13.1 & 0.97 & -65.6 & 1.00 & -9.0 & 1.00 & -29.8 & 1.00 & -9.6 & 0.99 & -27.3 \\
MedianFilter 13 & 0.95 & -11.1 & 0.96 & -59.6 & 1.00 & -9.0 & 1.00 & -29.6 & 1.00 & -9.6 & 0.99 & -27.3 \\
MedianFilter 17 & 0.90 & -8.5 & 0.93 & -50.9 & 1.00 & -9.0 & 1.00 & -29.5 & 1.00 & -9.6 & 0.99 & -27.3 \\
(JPEG Crop Brightness) (40 0.71 0.5) & 0.70 & -2.6 & 0.50 & -0.4 & 0.51 & -0.4 & 0.53 & -0.8 & 0.83 & -5.2 & 0.89 & -17.2 \\
(JPEG Crop Brightness) (60 0.71 0.5) & 0.70 & -2.6 & 0.50 & -0.4 & 0.49 & -0.3 & 0.53 & -0.8 & 0.87 & -5.9 & 0.91 & -18.6 \\
(JPEG Crop Brightness) (80 0.71 0.5) & 0.70 & -2.6 & 0.50 & -0.4 & 0.50 & -0.4 & 0.53 & -0.8 & 0.89 & -6.6 & 0.92 & -19.5 \\
        \bottomrule
        \end{tabular}

    }
\end{table}

\newpage
\begin{table}[H]
    \centering
    \caption{
        Full results for the robustness of watermark extraction on the COCO dataset.
    }\label{tab:robustness-all-coco}
    \resizebox{\linewidth}{0.45\textheight}{
        
    \begin{tabular}{c *{2}{>{\centering\arraybackslash}p{1.3cm}} *{2}{>{\centering\arraybackslash}p{1.3cm}} *{2}{>{\centering\arraybackslash}p{1.3cm}} *{2}{>{\centering\arraybackslash}p{1.3cm}} *{2}{>{\centering\arraybackslash}p{1.3cm}} *{2}{>{\centering\arraybackslash}p{1.3cm}}}
    \toprule
    & \multicolumn{2}{c}{\shortstack{HiDDeN}} & \multicolumn{2}{c}{\shortstack{MBRS}} & \multicolumn{2}{c}{\shortstack{CIN}} & \multicolumn{2}{c}{\shortstack{TrustMark}} & \multicolumn{2}{c}{\shortstack{WAM}} & \multicolumn{2}{c}{\shortstack{Video Seal (ours)}} \\
    \cmidrule(lr){2-3} \cmidrule(lr){4-5} \cmidrule(lr){6-7} \cmidrule(lr){8-9} \cmidrule(lr){10-11} \cmidrule(lr){12-13}
    & \multicolumn{2}{p{1cm}}{Bit~acc.~\footnotesize{($\uparrow$)}/~$\logpval$~\footnotesize{($\downarrow$)}} & \multicolumn{2}{p{1cm}}{Bit~acc.~\footnotesize{($\uparrow$)}/~$\logpval$~\footnotesize{($\downarrow$)}} & \multicolumn{2}{p{1cm}}{Bit~acc.~\footnotesize{($\uparrow$)}/~$\logpval$~\footnotesize{($\downarrow$)}} & \multicolumn{2}{p{1cm}}{Bit~acc.~\footnotesize{($\uparrow$)}/~$\logpval$~\footnotesize{($\downarrow$)}} & \multicolumn{2}{p{1cm}}{Bit~acc.~\footnotesize{($\uparrow$)}/~$\logpval$~\footnotesize{($\downarrow$)}} & \multicolumn{2}{p{1cm}}{Bit~acc.~\footnotesize{($\uparrow$)}/~$\logpval$~\footnotesize{($\downarrow$)}} \\
    \midrule
    Identity & 1.00 & -14.2 & 0.99 & -71.6 & 1.00 & -9.0 & 1.00 & -30.0 & 1.00 & -9.6 & 0.99 & -27.3 \\
HorizontalFlip & 0.69 & -2.5 & 0.50 & -0.5 & 0.50 & -0.4 & 1.00 & -29.9 & 1.00 & -9.6 & 0.99 & -27.1 \\
Rotate 5 & 0.93 & -9.9 & 0.50 & -0.5 & 0.50 & -0.4 & 0.64 & -4.6 & 0.99 & -9.0 & 0.98 & -26.0 \\
Rotate 10 & 0.84 & -6.1 & 0.50 & -0.4 & 0.50 & -0.4 & 0.51 & -0.5 & 0.77 & -3.1 & 0.97 & -23.9 \\
Rotate 30 & 0.55 & -0.7 & 0.50 & -0.4 & 0.50 & -0.3 & 0.50 & -0.4 & 0.50 & -0.4 & 0.59 & -1.5 \\
Rotate 45 & 0.49 & -0.3 & 0.50 & -0.4 & 0.50 & -0.4 & 0.50 & -0.4 & 0.50 & -0.4 & 0.51 & -0.5 \\
Rotate 90 & 0.50 & -0.4 & 0.50 & -0.4 & 0.50 & -0.4 & 0.50 & -0.4 & 0.49 & -0.3 & 0.50 & -0.4 \\
Resize 0.32 & 0.83 & -6.1 & 0.90 & -44.5 & 1.00 & -9.0 & 1.00 & -29.9 & 1.00 & -9.6 & 0.99 & -27.1 \\
Resize 0.45 & 0.93 & -10.3 & 0.96 & -62.1 & 1.00 & -9.0 & 1.00 & -29.9 & 1.00 & -9.6 & 0.99 & -27.3 \\
Resize 0.55 & 0.97 & -12.2 & 0.97 & -65.4 & 1.00 & -9.0 & 1.00 & -29.9 & 1.00 & -9.6 & 0.99 & -27.3 \\
Resize 0.63 & 0.98 & -12.9 & 0.98 & -67.7 & 1.00 & -9.0 & 1.00 & -29.9 & 1.00 & -9.6 & 0.99 & -27.3 \\
Resize 0.71 & 0.98 & -13.3 & 0.98 & -68.9 & 1.00 & -9.0 & 1.00 & -29.9 & 1.00 & -9.6 & 0.99 & -27.3 \\
Resize 0.77 & 0.99 & -13.5 & 0.98 & -69.5 & 1.00 & -9.0 & 1.00 & -29.9 & 1.00 & -9.6 & 0.99 & -27.3 \\
Resize 0.84 & 0.99 & -13.7 & 0.99 & -69.9 & 1.00 & -9.0 & 1.00 & -30.0 & 1.00 & -9.6 & 0.99 & -27.3 \\
Resize 0.89 & 0.99 & -13.8 & 0.99 & -70.1 & 1.00 & -9.0 & 1.00 & -30.0 & 1.00 & -9.6 & 0.99 & -27.3 \\
Resize 0.95 & 0.99 & -13.8 & 0.99 & -70.4 & 1.00 & -9.0 & 1.00 & -30.0 & 1.00 & -9.6 & 0.99 & -27.3 \\
Resize 1.0 & 1.00 & -14.2 & 0.99 & -71.6 & 1.00 & -9.0 & 1.00 & -30.0 & 1.00 & -9.6 & 0.99 & -27.3 \\
Crop 0.32 & 0.48 & -0.3 & 0.50 & -0.4 & 0.50 & -0.4 & 0.50 & -0.4 & 0.81 & -4.2 & 0.50 & -0.4 \\
Crop 0.45 & 0.50 & -0.4 & 0.50 & -0.4 & 0.50 & -0.4 & 0.50 & -0.4 & 0.94 & -7.7 & 0.52 & -0.6 \\
Crop 0.55 & 0.57 & -0.9 & 0.50 & -0.4 & 0.50 & -0.4 & 0.50 & -0.4 & 0.98 & -8.8 & 0.70 & -5.6 \\
Crop 0.63 & 0.66 & -2.0 & 0.50 & -0.4 & 0.50 & -0.3 & 0.50 & -0.4 & 0.99 & -9.1 & 0.85 & -14.3 \\
Crop 0.71 & 0.77 & -4.1 & 0.50 & -0.4 & 0.50 & -0.4 & 0.55 & -1.1 & 0.99 & -9.3 & 0.93 & -21.1 \\
Crop 0.77 & 0.86 & -6.8 & 0.50 & -0.4 & 0.49 & -0.3 & 0.92 & -21.1 & 0.99 & -9.4 & 0.96 & -23.8 \\
Crop 0.84 & 0.93 & -9.8 & 0.50 & -0.4 & 0.50 & -0.4 & 1.00 & -30.0 & 1.00 & -9.4 & 0.98 & -25.6 \\
Crop 0.89 & 0.95 & -11.0 & 0.50 & -0.4 & 0.50 & -0.4 & 1.00 & -30.1 & 1.00 & -9.5 & 0.98 & -26.1 \\
Crop 0.95 & 0.96 & -11.7 & 0.51 & -0.5 & 0.51 & -0.4 & 1.00 & -30.0 & 1.00 & -9.5 & 0.99 & -26.4 \\
Crop 1.0 & 1.00 & -14.2 & 0.99 & -71.6 & 1.00 & -9.0 & 1.00 & -30.0 & 1.00 & -9.6 & 0.99 & -27.3 \\
Perspective 0.1 & 0.94 & -10.3 & 0.50 & -0.5 & 0.51 & -0.6 & 0.92 & -21.1 & 0.99 & -9.4 & 0.98 & -26.3 \\
Perspective 0.2 & 0.92 & -9.3 & 0.50 & -0.4 & 0.50 & -0.4 & 0.61 & -3.3 & 0.94 & -7.5 & 0.98 & -25.9 \\
Perspective 0.3 & 0.89 & -8.0 & 0.50 & -0.4 & 0.50 & -0.4 & 0.52 & -0.7 & 0.81 & -4.1 & 0.97 & -24.7 \\
Perspective 0.4 & 0.84 & -6.5 & 0.50 & -0.4 & 0.50 & -0.4 & 0.51 & -0.5 & 0.70 & -2.3 & 0.95 & -22.4 \\
Perspective 0.5 & 0.80 & -5.1 & 0.50 & -0.4 & 0.50 & -0.4 & 0.50 & -0.4 & 0.63 & -1.4 & 0.91 & -18.4 \\
Perspective 0.6 & 0.75 & -3.9 & 0.50 & -0.4 & 0.50 & -0.4 & 0.50 & -0.4 & 0.58 & -1.0 & 0.85 & -13.9 \\
Perspective 0.7 & 0.70 & -2.9 & 0.50 & -0.4 & 0.50 & -0.4 & 0.50 & -0.4 & 0.55 & -0.7 & 0.77 & -8.9 \\
Perspective 0.8 & 0.65 & -2.1 & 0.50 & -0.4 & 0.50 & -0.4 & 0.50 & -0.4 & 0.53 & -0.6 & 0.70 & -5.9 \\
Brightness 0.1 & 0.58 & -1.0 & 0.63 & -4.8 & 1.00 & -9.0 & 0.83 & -12.6 & 0.97 & -8.3 & 0.96 & -22.5 \\
Brightness 0.25 & 0.82 & -5.7 & 0.85 & -31.9 & 1.00 & -9.0 & 0.97 & -26.4 & 1.00 & -9.6 & 0.99 & -27.0 \\
Brightness 0.5 & 0.95 & -11.1 & 0.96 & -59.0 & 1.00 & -9.0 & 1.00 & -29.5 & 1.00 & -9.6 & 0.99 & -27.2 \\
Brightness 0.75 & 0.99 & -13.7 & 0.98 & -68.5 & 1.00 & -9.0 & 1.00 & -29.9 & 1.00 & -9.6 & 0.99 & -27.3 \\
Brightness 1.0 & 1.00 & -14.2 & 0.99 & -71.6 & 1.00 & -9.0 & 1.00 & -30.0 & 1.00 & -9.6 & 0.99 & -27.3 \\
Brightness 1.25 & 1.00 & -14.2 & 0.96 & -61.6 & 1.00 & -9.0 & 0.99 & -28.2 & 1.00 & -9.6 & 0.98 & -26.5 \\
Brightness 1.5 & 0.99 & -14.0 & 0.93 & -52.7 & 1.00 & -9.0 & 0.96 & -25.5 & 1.00 & -9.6 & 0.97 & -25.0 \\
Brightness 1.75 & 0.99 & -13.8 & 0.89 & -44.8 & 0.99 & -8.9 & 0.93 & -22.6 & 1.00 & -9.6 & 0.94 & -23.1 \\
Brightness 2.0 & 0.98 & -13.5 & 0.87 & -39.1 & 0.99 & -8.8 & 0.91 & -20.0 & 1.00 & -9.6 & 0.92 & -21.3 \\
Contrast 0.1 & 0.54 & -0.6 & 0.77 & -19.3 & 1.00 & -9.0 & 0.77 & -9.1 & 1.00 & -9.6 & 0.96 & -23.2 \\
Contrast 0.25 & 0.77 & -4.4 & 0.91 & -45.0 & 1.00 & -9.0 & 0.96 & -24.8 & 1.00 & -9.6 & 0.99 & -27.0 \\
Contrast 0.5 & 0.94 & -10.8 & 0.97 & -63.1 & 1.00 & -9.0 & 1.00 & -29.2 & 1.00 & -9.6 & 0.99 & -27.3 \\
Contrast 0.75 & 0.99 & -13.7 & 0.98 & -69.3 & 1.00 & -9.0 & 1.00 & -29.9 & 1.00 & -9.6 & 0.99 & -27.3 \\
Contrast 1.0 & 1.00 & -14.2 & 0.99 & -71.6 & 1.00 & -9.0 & 1.00 & -30.0 & 1.00 & -9.6 & 0.99 & -27.3 \\
Contrast 1.25 & 1.00 & -14.2 & 0.96 & -60.7 & 1.00 & -9.0 & 0.99 & -28.1 & 1.00 & -9.6 & 0.98 & -26.4 \\
Contrast 1.5 & 1.00 & -14.2 & 0.93 & -53.2 & 1.00 & -9.0 & 0.96 & -25.3 & 1.00 & -9.6 & 0.97 & -25.2 \\
Contrast 1.75 & 1.00 & -14.1 & 0.91 & -47.7 & 1.00 & -9.0 & 0.94 & -22.5 & 1.00 & -9.6 & 0.96 & -23.9 \\
Contrast 2.0 & 0.99 & -14.0 & 0.89 & -43.6 & 1.00 & -9.0 & 0.91 & -20.1 & 1.00 & -9.6 & 0.95 & -22.5 \\
Hue -0.4 & 0.25 & -0.0 & 0.93 & -52.4 & 0.03 & -0.0 & 0.97 & -26.5 & 0.49 & -0.3 & 0.50 & -0.4 \\
Hue -0.3 & 0.45 & -0.2 & 0.95 & -58.4 & 0.25 & -0.2 & 0.98 & -27.6 & 0.48 & -0.3 & 0.50 & -0.4 \\
Hue -0.2 & 0.61 & -1.3 & 0.96 & -60.6 & 0.92 & -7.1 & 0.99 & -28.2 & 0.98 & -9.0 & 0.88 & -15.1 \\
Hue -0.1 & 0.79 & -4.8 & 0.98 & -66.8 & 1.00 & -9.0 & 1.00 & -29.3 & 1.00 & -9.6 & 0.98 & -26.0 \\
Hue 0.0 & 1.00 & -14.2 & 0.99 & -71.6 & 1.00 & -9.0 & 1.00 & -30.0 & 1.00 & -9.6 & 0.99 & -27.3 \\
Hue 0.1 & 0.86 & -7.2 & 0.98 & -67.1 & 1.00 & -9.0 & 1.00 & -29.4 & 1.00 & -9.6 & 0.98 & -25.9 \\
Hue 0.2 & 0.64 & -1.7 & 0.97 & -63.2 & 0.96 & -7.9 & 0.99 & -29.0 & 0.99 & -9.2 & 0.86 & -13.3 \\
Hue 0.3 & 0.45 & -0.2 & 0.96 & -61.9 & 0.14 & -0.1 & 0.99 & -28.7 & 0.48 & -0.3 & 0.50 & -0.4 \\
Hue 0.4 & 0.28 & -0.0 & 0.94 & -55.2 & 0.01 & -0.0 & 0.98 & -26.8 & 0.49 & -0.3 & 0.50 & -0.4 \\
Hue 0.5 & 0.03 & -0.0 & 0.91 & -47.0 & 0.00 & -0.0 & 0.95 & -24.4 & 0.49 & -0.3 & 0.51 & -0.5 \\
JPEG 40 & 0.99 & -13.4 & 0.96 & -60.2 & 0.97 & -7.8 & 0.98 & -27.6 & 0.97 & -8.6 & 0.91 & -17.8 \\
JPEG 50 & 0.99 & -13.5 & 0.97 & -63.5 & 0.98 & -8.3 & 0.99 & -28.4 & 0.99 & -9.1 & 0.94 & -20.9 \\
JPEG 60 & 0.99 & -13.6 & 0.98 & -66.5 & 0.99 & -8.7 & 0.99 & -28.8 & 0.99 & -9.4 & 0.96 & -23.2 \\
JPEG 70 & 0.99 & -13.6 & 0.98 & -66.9 & 0.99 & -8.8 & 0.99 & -29.2 & 1.00 & -9.5 & 0.97 & -24.8 \\
JPEG 80 & 0.99 & -13.7 & 0.98 & -69.0 & 1.00 & -9.0 & 1.00 & -29.5 & 1.00 & -9.6 & 0.98 & -26.4 \\
JPEG 90 & 0.99 & -13.9 & 0.98 & -69.3 & 1.00 & -9.0 & 1.00 & -29.7 & 1.00 & -9.6 & 0.99 & -27.0 \\
GaussianBlur 3 & 0.98 & -12.7 & 0.98 & -67.5 & 1.00 & -9.0 & 1.00 & -29.9 & 1.00 & -9.6 & 0.99 & -27.3 \\
GaussianBlur 5 & 0.93 & -10.3 & 0.96 & -61.4 & 1.00 & -9.0 & 1.00 & -29.9 & 1.00 & -9.6 & 0.99 & -27.3 \\
GaussianBlur 9 & 0.83 & -6.0 & 0.90 & -45.5 & 1.00 & -9.0 & 1.00 & -29.9 & 1.00 & -9.6 & 0.99 & -27.2 \\
GaussianBlur 13 & 0.75 & -3.6 & 0.84 & -31.3 & 1.00 & -9.0 & 1.00 & -29.7 & 1.00 & -9.6 & 0.99 & -27.0 \\
GaussianBlur 17 & 0.69 & -2.5 & 0.77 & -21.1 & 1.00 & -9.0 & 0.99 & -29.2 & 1.00 & -9.5 & 0.99 & -26.5 \\
MedianFilter 3 & 0.97 & -12.6 & 0.97 & -65.0 & 1.00 & -9.0 & 1.00 & -29.8 & 1.00 & -9.6 & 0.99 & -27.2 \\
MedianFilter 5 & 0.89 & -8.2 & 0.93 & -51.2 & 1.00 & -9.0 & 1.00 & -29.6 & 1.00 & -9.6 & 0.99 & -27.2 \\
MedianFilter 9 & 0.75 & -3.6 & 0.77 & -22.3 & 1.00 & -9.0 & 0.99 & -28.4 & 1.00 & -9.5 & 0.99 & -26.6 \\
MedianFilter 13 & 0.67 & -2.2 & 0.64 & -7.5 & 0.98 & -8.4 & 0.95 & -24.1 & 0.98 & -8.8 & 0.97 & -24.1 \\
MedianFilter 17 & 0.63 & -1.5 & 0.56 & -2.7 & 0.96 & -7.5 & 0.84 & -15.5 & 0.94 & -7.2 & 0.92 & -19.0 \\
(JPEG Crop Brightness) (40 0.71 0.5) & 0.65 & -1.7 & 0.50 & -0.4 & 0.50 & -0.4 & 0.52 & -0.6 & 0.58 & -0.9 & 0.74 & -6.7 \\
(JPEG Crop Brightness) (60 0.71 0.5) & 0.66 & -1.9 & 0.50 & -0.4 & 0.50 & -0.4 & 0.52 & -0.6 & 0.66 & -1.7 & 0.82 & -11.2 \\
(JPEG Crop Brightness) (80 0.71 0.5) & 0.67 & -2.1 & 0.50 & -0.4 & 0.51 & -0.4 & 0.53 & -0.7 & 0.78 & -3.9 & 0.89 & -16.9 \\
        \bottomrule
        \end{tabular}

    }
\end{table}

\newpage
\begin{table}[H]
    \centering
    \caption{
        Full results for the robustness of watermark extraction on the SA-V dataset.
    }\label{tab:robustness-all-sa-v}
    \resizebox{\linewidth}{!}{
        
    \begin{tabular}{c *{2}{>{\centering\arraybackslash}p{1.3cm}} *{2}{>{\centering\arraybackslash}p{1.3cm}} *{2}{>{\centering\arraybackslash}p{1.3cm}} *{2}{>{\centering\arraybackslash}p{1.3cm}} *{2}{>{\centering\arraybackslash}p{1.3cm}} *{2}{>{\centering\arraybackslash}p{1.3cm}}}
    \toprule
    & \multicolumn{2}{c}{\shortstack{HiDDeN}} & \multicolumn{2}{c}{\shortstack{MBRS}} & \multicolumn{2}{c}{\shortstack{CIN}} & \multicolumn{2}{c}{\shortstack{TrustMark}} & \multicolumn{2}{c}{\shortstack{WAM}} & \multicolumn{2}{c}{\shortstack{Video Seal (ours)}} \\
    \cmidrule(lr){2-3} \cmidrule(lr){4-5} \cmidrule(lr){6-7} \cmidrule(lr){8-9} \cmidrule(lr){10-11} \cmidrule(lr){12-13}
    & \multicolumn{2}{p{1cm}}{Bit~acc.~\footnotesize{($\uparrow$)}/~$\logpval$~\footnotesize{($\downarrow$)}} & \multicolumn{2}{p{1cm}}{Bit~acc.~\footnotesize{($\uparrow$)}/~$\logpval$~\footnotesize{($\downarrow$)}} & \multicolumn{2}{p{1cm}}{Bit~acc.~\footnotesize{($\uparrow$)}/~$\logpval$~\footnotesize{($\downarrow$)}} & \multicolumn{2}{p{1cm}}{Bit~acc.~\footnotesize{($\uparrow$)}/~$\logpval$~\footnotesize{($\downarrow$)}} & \multicolumn{2}{p{1cm}}{Bit~acc.~\footnotesize{($\uparrow$)}/~$\logpval$~\footnotesize{($\downarrow$)}} & \multicolumn{2}{p{1cm}}{Bit~acc.~\footnotesize{($\uparrow$)}/~$\logpval$~\footnotesize{($\downarrow$)}} \\
    \midrule
    Identity & 0.99 & -14.0 & 1.00 & -77.1 & 1.00 & -9.0 & 1.00 & -30.1 & 1.00 & -9.6 & 0.99 & -26.8 \\
HorizontalFlip & 0.70 & -2.7 & 0.50 & -0.5 & 0.49 & -0.3 & 1.00 & -30.1 & 1.00 & -9.6 & 0.99 & -26.5 \\
Rotate 10 & 0.82 & -5.7 & 0.51 & -0.6 & 0.49 & -0.3 & 0.53 & -0.8 & 0.73 & -2.3 & 0.94 & -20.9 \\
Rotate 90 & 0.49 & -0.3 & 0.50 & -0.3 & 0.52 & -0.5 & 0.50 & -0.4 & 0.49 & -0.3 & 0.49 & -0.3 \\
Resize 0.55 & 0.99 & -13.4 & 1.00 & -77.0 & 1.00 & -9.0 & 1.00 & -30.1 & 1.00 & -9.6 & 0.99 & -26.9 \\
Resize 0.71 & 0.99 & -13.6 & 1.00 & -77.1 & 1.00 & -9.0 & 1.00 & -30.1 & 1.00 & -9.6 & 0.99 & -26.8 \\
Crop 0.55 & 0.57 & -0.8 & 0.50 & -0.4 & 0.50 & -0.3 & 0.50 & -0.4 & 1.00 & -9.5 & 0.80 & -9.7 \\
Crop 0.71 & 0.74 & -3.4 & 0.50 & -0.4 & 0.50 & -0.4 & 0.55 & -1.1 & 1.00 & -9.6 & 0.97 & -24.0 \\
Perspective 0.5 & 0.76 & -4.2 & 0.50 & -0.4 & 0.51 & -0.4 & 0.50 & -0.4 & 0.63 & -1.4 & 0.94 & -20.5 \\
Brightness 0.5 & 0.91 & -9.2 & 1.00 & -76.7 & 1.00 & -9.0 & 1.00 & -30.1 & 1.00 & -9.6 & 0.99 & -26.8 \\
Brightness 1.5 & 1.00 & -14.1 & 0.98 & -70.9 & 1.00 & -9.0 & 0.99 & -29.0 & 1.00 & -9.6 & 0.99 & -26.6 \\
Contrast 0.5 & 0.90 & -8.6 & 1.00 & -76.8 & 1.00 & -9.0 & 1.00 & -30.1 & 1.00 & -9.6 & 0.99 & -26.8 \\
Contrast 1.5 & 1.00 & -14.2 & 0.98 & -71.1 & 1.00 & -9.0 & 0.99 & -28.4 & 1.00 & -9.6 & 0.98 & -26.0 \\
JPEG 40 & 0.98 & -13.0 & 0.99 & -75.0 & 1.00 & -9.0 & 1.00 & -29.6 & 1.00 & -9.5 & 0.98 & -25.6 \\
GaussianBlur 9 & 0.95 & -11.3 & 1.00 & -76.9 & 1.00 & -9.0 & 1.00 & -30.1 & 1.00 & -9.6 & 0.99 & -26.9 \\
MedianFilter 9 & 0.91 & -9.7 & 1.00 & -75.4 & 1.00 & -9.0 & 1.00 & -30.1 & 1.00 & -9.6 & 0.99 & -26.9 \\
Saturation 0.5 & 0.99 & -13.5 & 1.00 & -77.0 & 1.00 & -9.0 & 1.00 & -30.1 & 1.00 & -9.6 & 0.99 & -26.8 \\
Saturation 1.5 & 0.99 & -14.0 & 1.00 & -77.1 & 1.00 & -9.0 & 1.00 & -30.1 & 1.00 & -9.6 & 0.99 & -26.8 \\
Hue 0.25 & 0.54 & -0.7 & 1.00 & -76.9 & 0.63 & -2.3 & 1.00 & -29.6 & 0.89 & -6.9 & 0.61 & -2.5 \\
H264 30 & 0.95 & -11.2 & 0.98 & -69.4 & 1.00 & -9.0 & 1.00 & -29.5 & 0.97 & -8.5 & 0.97 & -24.3 \\
H264 40 & 0.85 & -7.0 & 0.80 & -28.9 & 0.95 & -7.4 & 0.94 & -22.9 & 0.86 & -5.4 & 0.89 & -16.7 \\
H264 50 & 0.69 & -2.7 & 0.59 & -3.6 & 0.84 & -4.9 & 0.59 & -2.0 & 0.71 & -2.6 & 0.72 & -6.6 \\
H264 60 & 0.67 & -2.3 & 0.58 & -3.2 & 0.84 & -4.9 & 0.56 & -1.4 & 0.69 & -2.4 & 0.70 & -5.5 \\
H264rgb 30 & 0.97 & -12.5 & 0.99 & -72.7 & 1.00 & -9.0 & 1.00 & -30.0 & 1.00 & -9.6 & 0.99 & -26.5 \\
H264rgb 40 & 0.90 & -8.9 & 0.88 & -43.9 & 1.00 & -8.9 & 0.99 & -29.0 & 0.99 & -9.3 & 0.97 & -24.5 \\
H264rgb 50 & 0.77 & -4.5 & 0.73 & -18.0 & 0.94 & -7.5 & 0.93 & -21.9 & 0.93 & -7.4 & 0.87 & -14.0 \\
H264rgb 60 & 0.76 & -4.2 & 0.73 & -17.2 & 0.93 & -7.3 & 0.92 & -21.0 & 0.93 & -7.2 & 0.84 & -12.6 \\
H265 30 & 0.95 & -11.3 & 0.97 & -67.7 & 0.98 & -8.5 & 0.99 & -29.1 & 0.96 & -8.1 & 0.96 & -22.8 \\
H265 40 & 0.86 & -7.2 & 0.71 & -13.9 & 0.75 & -3.0 & 0.92 & -21.2 & 0.73 & -2.9 & 0.76 & -8.6 \\
H265 50 & 0.64 & -1.8 & 0.56 & -2.0 & 0.60 & -1.0 & 0.60 & -2.4 & 0.53 & -0.6 & 0.54 & -1.0 \\
(H264 Crop Brightness) (30 0.71 0.5) & 0.64 & -1.6 & 0.50 & -0.4 & 0.49 & -0.4 & 0.52 & -0.6 & 0.61 & -1.3 & 0.89 & -16.6 \\
(H264 Crop Brightness) (40 0.71 0.5) & 0.62 & -1.4 & 0.50 & -0.4 & 0.49 & -0.3 & 0.51 & -0.5 & 0.53 & -0.5 & 0.72 & -6.1 \\
(H264 Crop Brightness) (50 0.71 0.5) & 0.57 & -0.9 & 0.50 & -0.3 & 0.50 & -0.4 & 0.50 & -0.4 & 0.51 & -0.4 & 0.58 & -1.6 \\
        \bottomrule
        \end{tabular}

    }
\end{table}

\end{document}